\documentclass[aps,prx,reprint,superscriptaddress,longbibliography,nofootinbib,floatfix]{revtex4-2}

\usepackage[utf8]{inputenc} 	
\usepackage[T1]{fontenc} 	
\usepackage{graphicx}
\usepackage{cancel} 
\usepackage{dcolumn}
\usepackage{bm}
\usepackage[english]{babel}
\usepackage{lipsum} 
\usepackage{amsmath}
\usepackage{hyperref} 
\hypersetup{colorlinks,%
citecolor=black,%
filecolor=black,%
linkcolor=black,%
urlcolor=black}
\bibpunct{(}{)}{;}{a}{,}{,}
\usepackage{times}
\usepackage{amsfonts,amssymb}
\usepackage{epsfig}
\usepackage{mathrsfs}
\usepackage{array}
\newcolumntype{C}[1]{>{\centering\arraybackslash}p{#1}}
\usepackage{color,soul}
\usepackage{bbold}
\usepackage{pstricks}
\usepackage{multirow}
\usepackage{textcomp}
\usepackage{dsfont}
\usepackage{braket}
\usepackage{empheq}
\hypersetup{colorlinks,%
citecolor=black,%
filecolor=black,%
linkcolor=black,%
urlcolor=black}
\usepackage{rotating}
\usepackage{amsthm}
\usepackage{eurosym}
\usepackage{siunitx}
\usepackage[babel]{csquotes}
\usepackage{mwe}
\usepackage[framemethod=default]{mdframed}
\usepackage{newfloat}

\DeclareFloatingEnvironment[name={S}]{suppfigure}
\DeclareFloatingEnvironment[name={S}]{supptable}
\usepackage{enumitem}
\usepackage{comment}
\usepackage{multirow} 

\usepackage[normalem]{ulem}
\input{Qcircuit}

\renewcommand{\arraystretch}{1.5}
\mathchardef\mhyphen="2D

\definecolor{jade}{rgb}{0.0, 0.66, 0.42}
\definecolor{gamboge}{rgb}{0.89, 0.61, 0.06}
\definecolor{bazaar}{rgb}{0.6, 0.47, 0.48}

\newcommand{\blockcomment}[1]{} 

\DeclareMathOperator{\tr}{Tr}

\newcommand{\id}{\mathds{1}}

\newcommand{\ketbra}[2]{\left|#1\middle\rangle\middle\langle#2\right|}
\newcommand{\proj}[1]{\left|#1\middle\rangle\middle\langle#1\right|}

\newcommand{\bbra}[1]{{ \langle \! \langle{#1}\vert }}
\newcommand{\kket}[1]{{ \vert {#1}  \rangle \!  \rangle}}
\newcommand{\kketbbra}[2]{{\kket{#1}\!\bbra{#2} }}

\newcommand{\BBra}[1]{{\left\langle \! \left\langle {#1} \right|\right.}}
\newcommand{\KKet}[1]{{\left. \left| {#1}  \right\rangle  \! \right\rangle}}

\newtheorem*{theorem*}{Theorem}
\newtheorem{theorem}{Theorem}

\newtheorem{definition}[theorem]{Definition}

\renewcommand{\H}{\mathcal{H}}
\renewcommand{\L}{\mathcal{L}}
\newcommand{\M}{\mathcal{M}}
\newcommand{\N}{\mathcal{N}}
\newcommand{\X}{\mathcal{X}}
\newcommand{\K}{\mathcal{K}}

\makeatletter

\def\section{%
  \@startsection{section}{1}{\z@}%
  {0.8cm plus 0.2cm minus 0.1cm}%
  {0.4cm plus 0.1cm minus 0.1cm}%
  {\normalfont\small\bfseries\raggedright}%
}

\def\subsection{%
  \@startsection{subsection}{2}{\z@}%
  {0.6cm plus 0.2cm minus 0.1cm}%
  {0.25cm plus 0.1cm minus 0.1cm}%
  {\normalfont\small\bfseries\raggedright}%
}

\def\subsubsection{%
  \@startsection{subsubsection}{3}{\z@}%
  {0.5cm plus 0.2cm minus 0.1cm}%
  {0.2cm plus 0.1cm minus 0.1cm}%
  {\normalfont\small\itshape\raggedright}%
}

\makeatletter
\@namedef{b@apsrev42Control}{{}{}{{}}{{}}}
\makeatother

\begin{document}

\title{Indefinite Quantum Causality} 





\begin{abstract}
In recent years, operational approaches to quantum foundations have been developed as a means of understanding the core principles and distinctive features of quantum theory. Such approaches typically view physical processes as sequences of operations, with earlier operations serving as causes of later effects. However, a growing literature is emerging on the possibility of relaxing this assumption and allowing for quantum indefiniteness in the causal order.
This development stems from a variety of motivations, both fundamental and applied, including exploring the role of causality in quantum theory, the interplay between quantum theory and general relativity, and higher-order quantum computing.
A prominent offshoot of this development is the emergence of indefinite causal order as a feasible resource for quantum information processing. This review provides an overview of the current state of the art in the field, covering the methodology underlying indefinite quantum causality within the so-called ``process matrix formalism'', outlining key results and experimental implementations, and discussing recent advances.
\end{abstract}

\author{Fabio Costa\textsuperscript{\dag}}
\affiliation{Nordita, Stockholm University and KTH Royal Institute of Technology, Hannes Alfv\'ens v\"ag 12 Stockholm, 106 91, Sweden}
\affiliation{School of Mathematics and Physics, The University of Queensland, St Lucia, QLD 4072, Australia}

\author{Giulia Rubino\textsuperscript{\dag}}
\altaffiliation{\href{mailto:giulia.rubino@bristol.ac.uk}{giulia.rubino@bristol.ac.uk}.}
\affiliation{Quantum Engineering Technology Labs,  H.~H.~Wills Physics Laboratory and School of Electrical,
Electronic and Mechanical Engineering, University of Bristol, BS8 1FD, United Kingdom}
\affiliation{H.~H.~Wills Physics Laboratory, University of Bristol, Tyndall Avenue, Bristol, BS8 1TL, United Kingdom}

\author{Cyril Branciard}
\affiliation{Univ. Grenoble Alpes, CNRS, Grenoble INP, Institut N\'eel, 38000 Grenoble, France}

\author{\v{C}aslav Brukner}
\affiliation{University of Vienna, Faculty of Physics, Vienna Doctoral School in Physics, and Vienna Center for Quantum Science and Technology (VCQ), Boltzmanngasse 5, A-1090 Vienna, Austria}
\affiliation{Institute for Quantum Optics and Quantum Information (IQOQI), Austrian Academy of Sciences, Boltzmanngasse 3, A-1090 Vienna, Austria}

\author{Marco T\'ulio Quintino}
\affiliation{Sorbonne Universit\'e, CNRS, LIP6, F-75005 Paris, France}




\maketitle 

\begingroup
\renewcommand\thefootnote{\dag}
\footnotetext{These authors contributed equally to this review.}
\endgroup


\tableofcontents

\section{Introduction}

\subsection{Motivation}

The causal structure of spacetime is often taken as the necessary starting point for describing physical processes in terms of states evolving in time, with causal influence going from past to future events within the light cone. From an information-processing perspective, this paradigm is well captured at an abstract level by the circuit model, where wires and boxes represent the propagation and transformation of physical systems, respectively, and causal influence is possible only between boxes connected by wires. The circuit model---and in particular, its quantum version---is widely used both in practical contexts, such as building quantum technologies, and in foundational studies of quantum theory, where circuits serve as a compact way to express the temporal evolution of a system within a well-defined causal structure.
However, several independent motivations have led to considering modifications to this picture so as to include the possibility of events, or operations, that are not in a fixed order. 

One of the main motivations arises from the consideration that, in a theory of quantum gravity, spacetime may lose its classical properties, possibly leading to indefiniteness in the causal relations~\citep{butterfield_isham_2001}. More specifically, \citet{Hardy2005, Hardy_2007, Hardy2008} has proposed that in order to combine the key elements of general relativity (a deterministic theory with a non-fixed causal structure) and quantum theory (a probabilistic theory with a fixed causal structure), a framework for probabilistic theories with a non-fixed causal structure must be adopted.

\begin{figure}[t] 
	\begin{center}
		\includegraphics[width=\columnwidth]{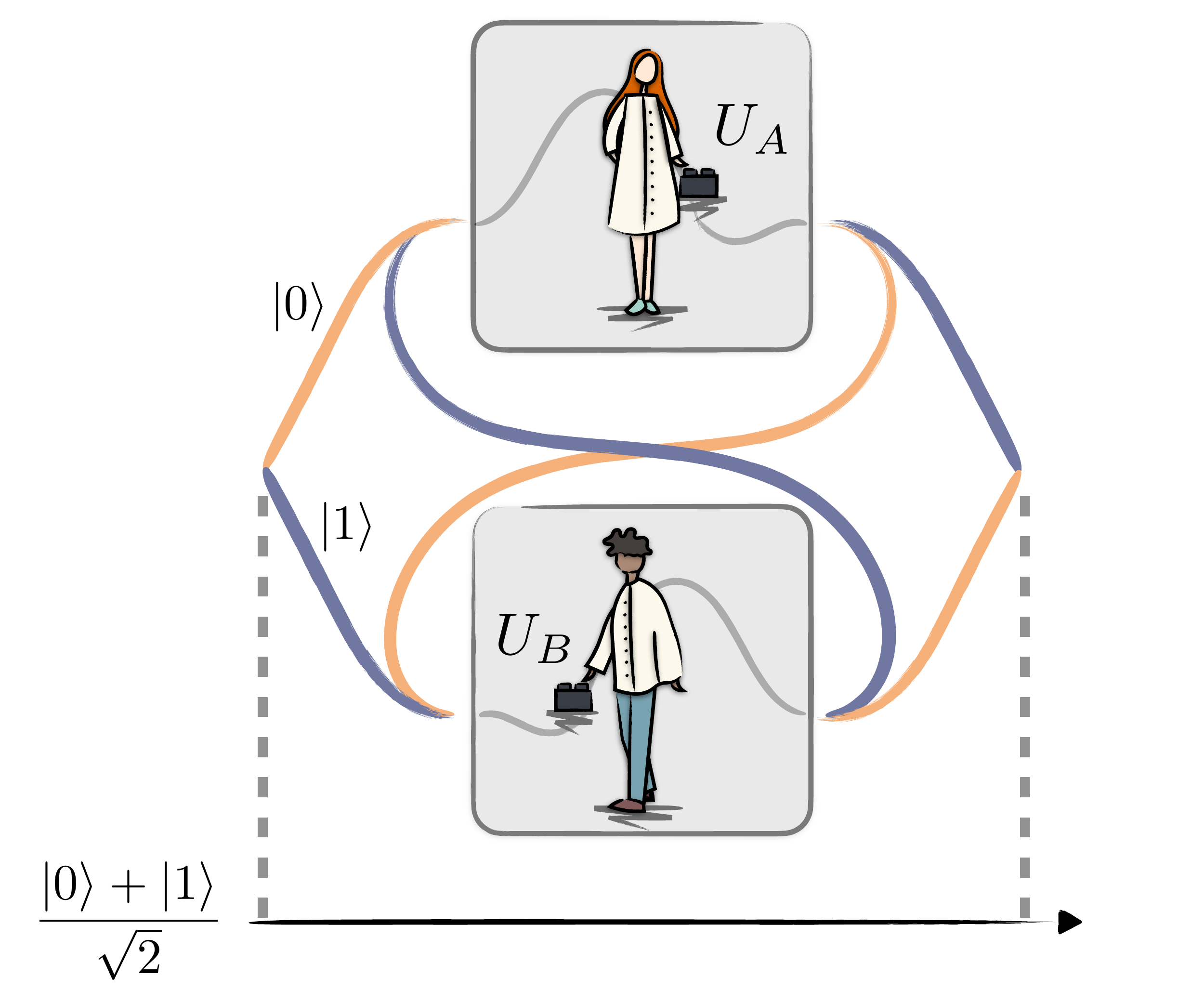} 
	\end{center}
\caption{Scheme of a bipartite quantum switch. Two parties, Alice and Bob, perform two operations $U_A$ and $U_B$, respectively, on a target system $\ket{\psi}$. A second quantum system controls the order in which the two parties perform their unitary operations. For example, when the ``control system'' is in a state $\ket{0}$, the target system could be sent first to Alice and then to Bob, and vice versa when the control system is in a state $\ket{1}$. Placing the control system in a quantum superposition state, $\bigl(\ket{0}+\ket{1}\bigr)/\sqrt{2}$, makes the order of operations $U_A$ and $U_B$ indefinite, resulting in the final state $(\ket{0}\otimes U_BU_A\ket{\psi}+\ket{1}\otimes U_AU_B\ket{\psi})/\sqrt{2}$.}
\label{fig:switch}
\end{figure}

Coming from a different perspective, and inspired by alternative models of computation such as Church's lambda calculus, \citet{Chiribella2008supermap} considered an abstract model of ``higher-order transformations'' \citep{taranto2025higher}. Rather than focusing only on transformations of states, this model examines maps between transformations, generating a hierarchy of structures that are not necessarily compatible with a causal concatenation of transformations as represented by a quantum circuit.  A concrete example of higher-order computation beyond quantum circuits was given in \citep{Chiribella2009_arxiv}, and later published in \citep{Chiribella2013}, based on the intuition that the wires connecting the boxes in a circuit can be moved and possibly put in superposition. This led to the introduction of the ``quantum switch'', illustrated in Fig.~\ref{fig:switch}: a process where the order in which two operations are applied to a ``target'' system depends on the state of a ``control'' system. 

From a foundational perspective, axiomatic reconstructions of quantum theory typically assume a global causal order of events \citep{hardy01, Dakic2010, Chiribella2011, Selby2021reconstructing, Masanes_2011}.  It is interesting to understand whether such an assumption is necessary or can be relaxed. This led \citet{OCB_2012} to formalize a framework for quantum theory without a background causal structure, connecting local quantum operations in the most general and logically consistent way. Within this framework, it was possible to conclude that a global causal order is not necessary for the local validity of quantum theory.

The above approaches have led to a large body of literature on the study of ``indefinite causal order'' \footnote{The terms ``indefinite causal order'', ``indefinite causal structure'',  ``indefinite causality'', and ``causal indefiniteness'' all indicate a lack of well-defined causal order and are often used interchangeably in the literature, as they will be throughout this review.} over the past few years. Among the reasons for the growing interest is that processes with indefinite causal order may offer advantages over causally ordered quantum circuits in a range of computational and information-theoretic tasks. This interest is further fueled by a growing number of experiments reporting the implementation of indefinite causal order, suggesting that the predicted advantages may be achievable in practice. Furthermore, this line of research has led to a more profound understanding of the nature of events, and of their causal relations, within the quantum framework, especially in scenarios beyond classical spacetime.

This review aims to provide a pedagogical introduction to the foundations and an overview of the main results within one of the most prominent frameworks---often referred to as the ``process matrix formalism''---that has emerged for the study of indefinite causal order and which has led to numerous tools, techniques, and conceptual insights. Given the rapid proliferation of publications, we focus on the seminal contributions that shaped the field, subsequent works that have made a lasting impact, and the most promising recent developments.

\subsection{Outline}

The outline of this review is set out below.

Sec.~\ref{firststeps} provides a concise, self-contained overview that lays the groundwork for understanding the following sections. It begins by outlining the basics of quantum systems, channels and operations, and then moves on to introduce the process matrix formalism. This tutorial section is designed to provide basic concepts and definitions while minimising technical complexity.

In Sec.~\ref{indefiniteorder}, the discussion delves into the details and core elements of the formalism. This part covers the axiomatic derivations of the formalism and introduces the notion of causal separability as a formal definition of definite causal order. It reviews causal witnesses, which are central to experimentally accessible demonstrations of indefinite causal order.  It presents the generalization of the process matrix formalism to multi-party scenarios, highlights aspects such as unitarity, classicality, and extensions to infinite-dimensional systems.

Sec.~\ref{switchsection} turns the focus to the quantum switch, the most commonly studied process with indefinite causal order. This section discusses various versions and generalizations of the quantum switch, as well as some causally ordered quantum circuits that reproduce its behavior.

Sec.~\ref{sec:causal_ineqs} considers the task of certifying indefinite causality from the statistics collected from an experiment under basic causality assumptions and without relying on the full quantum description of all of the devices involved in the experiment. This leads to a framework to analyze causality beyond a fully-quantum description, which includes device independent tests---based on, e.g., ``causal inequalities''---as well as semi-device-independent and theory-independent certifications of indefinite causality.

In Sec.~\ref{sec:applications}, the review examines scenarios where indefinite causality has been shown to offer potential advantages, such as in channel discrimination tasks, quantum computation, or quantum communication complexity. It also provides a clarifying perspective on the role that indefinite causality plays in other proposed advantages within the field.

Sec.~\ref{sec:Experim} provides an overview of the experimental investigations of indefinite causality conducted to date, emphasising the details of the experimental executions rather than the underlying theoretical frameworks, which are presented in the respective theoretical sections.

Sec.~\ref{sec:Gravity} offers an overview of works relating indefinite causal order to gravity and relativity. The main focus is on the quantum-gravitational realization and certification of processes with indefinite causal order; other topics include fixed-spacetime realization of the switch using relativistic time dilation, spacetime uncertainty induced by quantum clocks, quantum coordinate transformations, and preliminary work on the relation with the notion of background independence.

The interpretation of physical realizations of indefinite causal order, especially in table-top experiments, is summarized in Sec.~\ref{interpretationsection}. Special emphasis is given to the notion that events can be delocalized in time, opening up the possibility of indefinite causal relations on a classical background spacetime.

Lastly, Sec.~\ref{sec:AdjRes} discusses various studies related to indefinite causality in quantum systems. 
It first examines the effects of transformations on quantum processes, showing how continuous, reversible transformations cannot transform definite causal order into indefinite causal order. It then summarizes relevant works in the foundations of quantum mechanics: the reconstruction of quantum theory from operational models without definite causal order and the notion of non-contextuality without fixed causal structure. Finally, the section concludes with a discussion of quantum causal models, exploring the extension of classical causal models to quantum contexts and the implications of quantum correlations for these models.

We conclude with an outlook in Sec.~\ref{sec:Concl}, highlighting some of the key technical and conceptual challenges that remain open in the study of indefinite causality.

\section{Quantum theory without a background causal structure}
\label{firststeps}

Quantum theory, in its canonical formulation, makes a direct reference to the time ordering of events. 
A quantum state describes a system \emph{at a given time}. If we perform a sequence of measurements on the system, the formalism requires that we specify in advance their relative order: After the first measurement, we have to ``collapse'' the state before we can consider a second measurement.

To be concrete, a pure state is given by a (normalized) vector $\ket{\psi}$ in a Hilbert space $\mathcal{H}$. Say we want to perform two projective measurements, labeled $A$ and $B$, and calculate the joint probability $P(a,b)$ that measurement $A$ yields outcome $a$ and measurement $B$ yields $b$. If $A$ and $B$ are time-like separated events, with $A$ at an earlier time than $B$, we have to first evaluate the probability for $a$ to occur as $P(a) = \bra{\psi}\Pi^A_a\ket{\psi}$, where $\Pi^A_a$ is a projection operator corresponding to $A$'s outcome $a$. Next, we update the original state with the nonlinear transformation $\ket{\psi} \mapsto \ket{\Tilde{\psi}^A_a}$ = $\frac{\Pi^A_a\ket{\psi}}{\|\Pi^A_a\ket{\psi}\|}$  and calculate the conditional probability to find $B$'s outcome $b$ given $a$: $P(b|a)=\bra{\Tilde{\psi}^A_a}\Pi^B_b\ket{\Tilde{\psi}^A_a}$, where $\Pi^B_b$ is the projector corresponding to $B$'s outcome $b$. The joint probability then follows from Bayes' rule, $P(a,b) = P(b|a)P(a) = \bra{\Tilde{\psi}^A_a}\Pi^B_b\ket{\Tilde{\psi}^A_a} \bra{\psi}\Pi^A_a\ket{\psi}$.   
Clearly, knowing which measurement comes first is crucial to apply this procedure. 

In addition, the standard formalism treats space-like events in a radically different way from time-like ones. 
For space-like separation, we associate two separate Hilbert spaces, $\H^A$, $\H^B$, to the two potential events, with the state living in their tensor product, $\ket{\psi}^{AB}\in \H^{AB} \equiv \H^A\otimes\H^B$ (throughout the review we will use a similar short-hand notation for general tensor products). We then calculate the joint probabilities for outcomes $a, b$ with the formula $P(a,b) {=} \bra{\psi}^{AB} (\Pi^{A}_a\otimes \Pi^{B}_b)\ket{\psi}^{AB}{=}\tr [(\Pi^{A}_a\otimes \Pi^{B}_b) \ketbra{\psi}{\psi}^{AB}]$. 
Without fixing in advance whether the measurements are space-like or time-like, we do not know which of the two procedures above to apply. In other words, the ordinary formalism does not allow us to calculate probabilities for such events unless we already know their causal relations.

In order to formalize indefinite causal structures, it is necessary to first reformulate quantum theory in a way that does not require prior knowledge of the causal relations between events. We should be able to write down a general formula for the joint probability of events without specifying in advance whether they are space-like or, if time-like, their order. Such a formula should then allow us to reconstruct causal relations \textit{a posteriori}, based on measured statistics, similarly to how we can make inference about an unknown state.

As we will show, such a formula can indeed be written, and it takes the form of a generalized Born rule:
\begin{equation}\label{processborn}
    P(a,b) = \tr \left[\left(M^A_a\otimes N^B_b\right)^T\, W^{AB}\right],
\end{equation}
where $M^A_a$, $N^B_b$ are operators that represent the events (such as measurements or operations, along with the corresponding system updates) happening at two ``sites'' ($A$ and $B$) and $W^{AB}$ is the \textit{process matrix}, which encodes all the relevant information about how the two sites are connected, including information about the causal relations between the operations. 
Here, ${}^T$ denotes operator transposition---its presence is purely conventional and other versions of the formula exist in the literature, differing in the presence or positioning of the transposition.

Remarkably, the generalized probability rule \eqref{processborn} appears formally identical to the ordinary Born rule for space-like events.
This might seem surprising at first, because the ordinary Born rule does not allow causal influence between $A$ and $B$ (i.e., there is no possibility of exchanging signals), whereas signaling is possible with Eq.~\eqref{processborn}. As we will see, this is because the operators appearing in the formula, $M^A_a$, $N^B_b$, and $W^{AB}$ have a different physical meaning compared to the corresponding operators in the ordinary Born rule and, accordingly, satisfy different mathematical conditions. 

The remainder of this section is devoted to introduce, justify, and discuss Eq.~\eqref{processborn}. In order to do that, we will need to first review some basic elements of the ordinary quantum formalism. We will then present a new approach to quantum processes and measurements that unifies different causal scenarios. The new perspective will serve as a springboard to get to the generalized Born rule and to formalize processes with indefinite causal order.

\subsection{States, channels, and instruments}
\label{quantumreview}

One of the central features we want from our formalism is that it should allow us to distinguish rigorously scenarios with indefinite causal order---possibly of a quantum origin---from situations in which events do have a definite order, but we have classical uncertainty of what it is. To this end, it is more convenient to work with the ``mixed state'' (or ``open system'') formulation of quantum theory, namely with density matrices, quantum channels, etc. Below we briefly review the main tools we need from this formalism.

\subsubsection{Quantum states and quantum measurements}

Let ${\cal H}$ be a finite-dimensional\footnote{We will mostly use finite-dimensional systems for simplicity. For an introduction to the mathematical formalism of quantum theory, including infinite-dimensional systems, see, e.g., \citep{heinosaari}. An approach to indefinite causal structures in infinite dimensions is discussed in Sec.~\ref{subsubsec:infinite_dim}.} linear (Hilbert) space over the complex field and $\mathcal{L}(\H)$ be the space of linear operators acting on $\H$. A general quantum state, not necessarily pure, is defined as a density operator, namely as a positive semidefinite operator $\rho\in\L(\H), \; \rho\geq0$ with unit trace $\tr(\rho)=1$. Quantum states can describe probabilistic mixtures of pure states: an ensemble of states $\{\ketbra{\psi_a}{\psi_a}\}_a$ occurring with probabilities $\{p_a\}_a$ is described by the density operator $\rho = \sum_ap_a \ketbra{\psi_a}{\psi_a}$. Mixed states can also arise from ignoring a part of a composite system, as formalized by the definition of reduced states: $\rho^A = \tr_B\ketbra{\psi}{\psi}^{AB}$, where $\ket{\psi}^{AB} \in {\cal H}^{AB}$  and $\tr_B$ denotes the partial trace over $\H^{B}$.
Throughout this review, we use superscripts to denote the Hilbert space to which an object (vector, operator, etc.) belongs. This notation implies an implicit reordering of operators in certain expressions. For example, $\rho^{ABC}=\sigma^{AC}\otimes \tau^B$ implies that the right-hand side is permuted to preserve the reference order of the tensor factors (here, $\H^{ABC} \equiv \H^A\otimes\H^B\otimes\H^C$). We may drop the superscripts when the Hilbert spaces are clear from the context.

A general quantum measurement is described by a Positive Operator Valued Measure (POVM, of which a projective measurement is a particular case), which is a set of positive semidefinite operators $\{E_a\}_a$, $E_a\in\L(\H)$, $E_a\geq0$ (also called ``effect operators''), which sum to the identity:%
\footnote{A generalization to continuous outcomes is rather straightforward, although it requires measure-theoretic notions that are not particularly insightful for our purposes. We will thus limit ourselves to finite-outcome sets.}
$\sum_a E_a=\id$. The probability of obtaining the outcome $a$ when performing a measurement $\{E_a\}_a$ on a state $\rho$ is given by the Born rule:%
\footnote{Probabilities are also conditioned on the state, $P(a|\rho,\{E_a\}_a)=\tr(E_a\,\rho)$, but such a dependence is implicit in Eq.~\eqref{born} as the state is understood to be fixed. Other choices for writing such probabilities will depend on the context: Dependence on $\{E_a\}_a$ may be left implicit if the POVM is fixed, $P(a)=\tr(E_a\,\rho)$, or it may be encoded in a classical variable $x$, $P(a|x)=\tr(E_{a|x}\,\rho)$, where $\{E_{a|x}\}_{a}$ is a POVM for every $x$. The same considerations will apply below to the dependence on the process matrix (that will be left implicit) and on the quantum instruments.}
\begin{equation}\label{born}
    P(a|\{E_a\}_a)=\tr(E_a\,\rho).
\end{equation}
As we will see below, POVMs only characterize the outcome probabilities of measurements; they are not sufficient to describe how the system transforms upon measurement.

\subsubsection{Quantum channels and quantum instruments} \label{operations}
The evolution of a state from an initial to a final time is described by a \textit{quantum channel}. At the abstract level, a quantum channel is the most general deterministic transformation that sends states to states, where ``deterministic'' means it happens with unit probability (it either does not involve any measurement or, if it does, the measurement outcome is ignored). It is also required that a channel sends states to states when acting on a subsystem of a larger system, leading to the definition of a channel as a linear map $\mathcal{C}: \L(\H^X)\to \L(\H^Y)$ that is completely positive%
\footnote{A linear map $\mathcal{C}: \L(\H^X)\to \L(\H^Y)$ is positive if for every positive semidefinite operator $\rho\in \L(\H^X)$ we have that $\mathcal{C}(\rho)\geq0$. A map is completely positive if, for every linear space $\H^Z$ and every positive semidefinite operator $\sigma\in \L(\H^X\otimes\H^Z)$, we have that $\mathcal{C}\otimes \mathcal{I}^{Z}(\sigma)\geq0$, where $\mathcal{I}^{Z}$ is the identity map on $\L(\H^Z)$. 
Moreover, a linear map $\mathcal{C}: \L(\H^X)\to \L(\H^Y)$ is trace preserving if $\tr\bigl[\mathcal{C}(\rho)\bigr]=\tr(\rho) $ for every linear operator $\rho\in\L(\H^X)$.}
(CP) and trace preserving (TP). Note that, in general, the input ($\H^X$) and output ($\H^Y$) spaces of the channel can be different. Even for cases where we think of the input and output as being the same system (e.g., a channel that rotates a spin-$\frac{1}{2}$ particle), it will be convenient to explicitly assign different indices to the input and output (then isomorphic) spaces. It should be noted that all channels can be modeled in terms of a unitary interaction between the system and an environment, as ensured by Stinespring's dilation theorem \citep{Stinespring55,NielsenChuangBook}.

We will also need to consider probabilistic transformations, namely, those that act on a system when a measurement is made and a particular outcome $a$ is recorded. Such transformations are described by CP maps $\mathcal{M}_a$ that are not necessarily TP. This means that the state $\mathcal{M}_a(\rho)$, obtained from the application of a CP map to a state $\rho$, is not necessarily a normalized density operator. The interpretation is that this is one out of a collection of possible states, $\{\mathcal{M}_a(\rho)\}_a$,  where each state $\mathcal{M}_a(\rho)$---and correspondingly, each outcome $a$---occurs with a probability given by
\begin{equation}
   P(a|\{\mathcal{M}_a\}_a) = \tr\bigl[\mathcal{M}_a(\rho)\bigr].
\end{equation}
The collection of all possible maps $\{\mathcal{M}_a\}_a$, for all possible measurement outcomes, constitutes a \textit{quantum instrument} \citep{davies70}, defined by the property that the sum of all maps therein, $\sum_a \mathcal{M}_a$, must be CPTP (i.e., a quantum channel). As an example, the CP maps associated with a projective measurement over a basis $\{\ket{\psi_a}\}_a$ are $\mathcal{M}_a: \rho \mapsto \mathcal{M}_a(\rho) = \ketbra{\psi_a}{\psi_a}\rho\ketbra{\psi_a}{\psi_a}$. Similarly to channels, every instrument has a unitary dilation where, instead of being traced out, the auxiliary system is measured \citep{kraus1983states,watrous_2018}.

Many presentations avoid subnormalized states by introducing the nonlinear update map
\begin{equation} \label{eq:update}
    \rho\mapsto\frac{\mathcal{M}_a(\rho)}{\tr[\mathcal{M}_a(\rho)]} = \frac{\mathcal{M}_a(\rho)}{P(a|\{\mathcal{M}_a\}_a)}.
\end{equation}
This will not be necessary in our treatment: we will use a formulation of quantum theory that never needs to appeal to nonlinear transformations and in which, instead, the norm of subnormalized states encodes their probability of occurrence. We will come back to this point in Sec.~\ref{orderedprocesses}.

\subsection{Operation-state duality} 
\begin{figure}[hbt] 
	\begin{center}
		\includegraphics[width=.6\columnwidth]{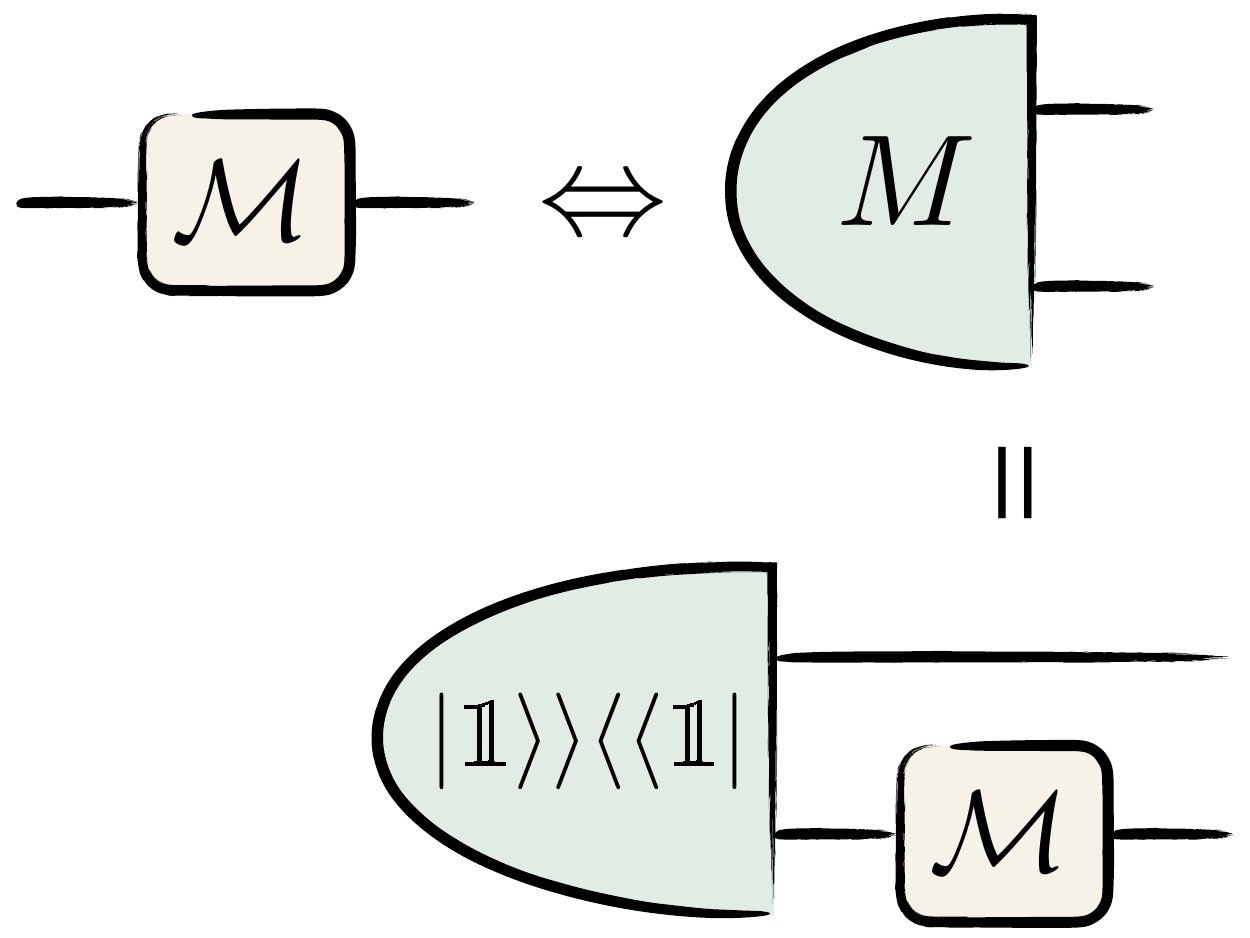} 
	\end{center}
\caption{Pictorial representation of the operation-state duality: A CP map $\cal{M}$ (upper left) is isomorphic to the bipartite state (upper right) $M$ obtained by applying the map $\cal{M}$ to a maximally entangled state [represented as in Eq.~\eqref{identitystate}] (lower right).}
\label{fig:op_state_dual}
\end{figure}

When analysing quantum states, channels, instruments, and more general objects that we will introduce later, it is very convenient to have a common representation in the same mathematical space. As each of these objects lives naturally in a linear space, one can always define isomorphisms that map all of them to a common space. A useful and physically motivated correspondence is between states and state transformations. Intuitively, given a specific CP map, one can apply it to one side of a bipartite state and obtain a new quantum state, thus giving a natural way to convert maps into states. The converse is also true: Given any bipartite state, one can identify a CP map that returns this state when applied to one side of an appropriate entangled state. As the correspondence is one-to-one, it provides a natural isomorphism between states and operations (see Fig.~\ref{fig:op_state_dual}).

\subsubsection{The Choi isomorphism} \label{Choiiso}
In order to represent operations acting on a space ${\cal H}^X$, we first fix a basis $\{\ket{i}^X\}_i$ of ${\cal H}^X$---its \emph{computational} basis\footnote{In the following, all basis-dependent operations (transposition ${}^T$, complex conjugation ${}^*$, link product $*$, etc.) will be defined by default in the computational basis.}---and introduce the (non-normalized) maximally entangled state 
\begin{equation} \label{identitystate}
    \kket{\id}^{XX'} \coloneqq \sum_i  \ket{i}^X\otimes\ket{i}^{X'} \ \in {\cal H}^X\otimes {\cal H}^{X'},
\end{equation}
where $\mathcal{H}^{X'}$ is a copy of the linear space $\mathcal{H}^X$.
By acting on one side of it with an arbitrary linear operator $V: {\cal H}^{X'} \to {\cal H}^Y$, we obtain the (in general non normalized) state
    \begin{align}
    \kket{V}^{XY} \coloneqq & \id\otimes V \kket{\id}^{XX'} \ \in {\cal H}^X\otimes {\cal H}^Y, \notag \\
    = & \sum_i \ket{i}^X \otimes(V\ket{i})^Y,
    \label{eq:def:dketV}
\end{align}
where $\id$ is the identity operator.  In the case where $V$ is a unitary (or, more generally, an isometry), then $\kket{V}$ is maximally entangled. For $V=\ketbra{\phi}{\psi}$, Eq.~\eqref{eq:def:dketV} implies that $\kket{V}=\bra{\psi}^T\otimes\ket{\phi}=\ket{\psi}^*\otimes\ket{\phi}$.
Thus, the isomorphism may be viewed as changing a bra into a complex-conjugated ket (and swapping the order).
Note that the mapping $V\mapsto \kket{V}$ is one-to-one, so that each operator corresponds to a unique ``double-ket'' vector, and vice versa. The inverse mapping is given by 
\begin{align}
    \kket{V}^{XY} \mapsto V =& \kket{V}^{T_X}, 
\end{align}
where $^{T_X}$ denotes partial transposition on $\mathcal{H}^X$. For instance, if $\kket{V}^{XY}\coloneqq \ket{\psi}^*\otimes\ket{\phi}\in\mathcal{H}^X\otimes\mathcal{H}^Y$, then $V=(\ket{\psi}^*)^T\otimes\ket{\phi}=\ket{\phi}\!{\bra{\psi}}: {\cal H}^X \to {\cal H}^Y$. Furthermore, if $\kket{V}$ is maximally entangled, then $V$ is necessarily an isometry, and a unitary if ${\cal H}^X$ and ${\cal H}^Y$ have the same (finite) dimension.

The operation-state correspondence is most commonly used for CP maps and density operators, in which case it takes the name of \textit{Choi isomorphism}, or \textit{Choi-Jamio{\l}kowski isomorphism} \citep{dePillis1967,Jamiolkowski1972,Choi1975}. The Choi isomorphism allows one to represent linear maps as linear operators in a bipartite space. The Choi operator of a linear map ${\cal M}: \L({\cal H}^X) \to \L({\cal H}^Y)$ is defined (still with respect to the computational basis of ${\cal H}^X$) as
\begin{align}
	M^{XY}\coloneqq& (\mathcal{I}^X\otimes \mathcal{M})(\kketbbra{\id}{\id}^{XX}) \notag \\
	  =& \sum_{ij} \ketbra{i}{j}^X\otimes \mathcal{M} ( \ketbra{i}{j})^Y,
\end{align}
where $\mathcal{I}$ is the identity map. As before, the mapping is one-to-one, with the inverse map given by 
\begin{equation}
    M \mapsto {\cal M}, \ {\cal M}(\rho) = \tr_X[(\rho^T \otimes \id^Y) M^{XY}] 
    \label{eq:inverse_Choi}
\end{equation}
$\forall \rho \in\L(\H^X)$. The important feature of this isomorphism is that a linear map $\mathcal{M}$ is completely positive if and only if its Choi operator $M$ is positive semidefinite, $M\geq~0$. In other words, $\mathcal{M}$ is a physical operation if and only if $M$ is (proportional to) a physical state. Additionally, direct calculation \citep{watrous_2018,wilde2011book} shows that a linear map $\mathcal{M}$ is trace preserving if and only if its Choi operator satisfies
\begin{equation}
\tr_{Y} M^{XY} =\id^X.
\end{equation}
Hence, a linear map $\mathcal{C}: \L(\H^X) \to \L(\H^Y)$ represents a quantum channel if and only if its Choi operator $C^{XY}\in \L(\H^{XY})$ respects $C^{XY}\geq0$ and $\tr_Y(C^{XY})=\id^X$. Similarly, a set of linear maps $\{\mathcal{M}_a\}_a$ [with each $\mathcal{M}_a: \L(\H^X) \to \L(\H^Y)$] is a quantum instrument if and only if their Choi operators respect $M_a^{XY}\geq0\ \forall a$ and $\sum_a \tr_Y( M^{XY}_a)=\id^X$.

\begin{figure}[!tbp] 
	\begin{center}
		\includegraphics[width=\columnwidth]{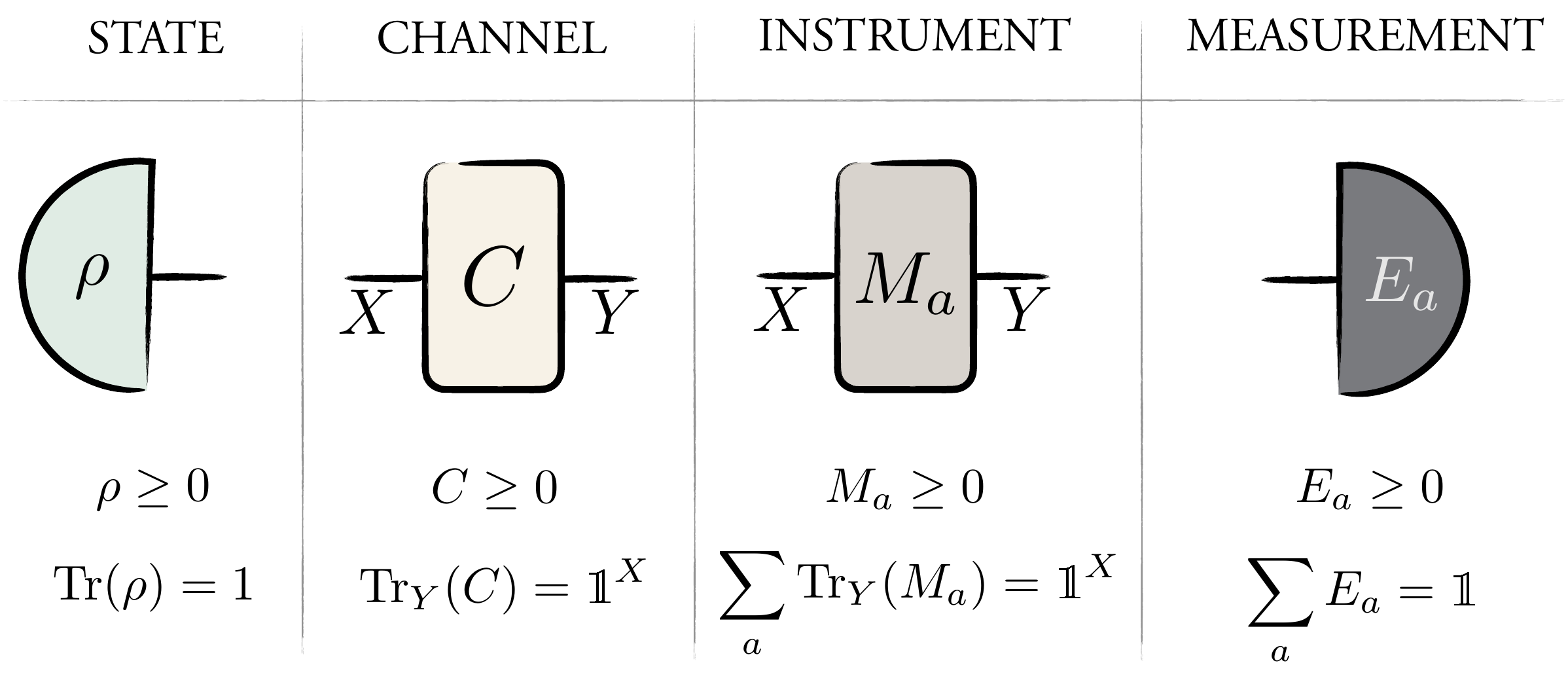} 
	\end{center}
\caption{Quantum circuit and Choi representation of quantum states, channels, instruments and measurements. For instruments and POVM measurements, only one element is depicted, rather than the full sets $\{M_a\}_a$, $\{E_a\}_a$. According to our definitions, the Choi representation of a POVM element $E_a$ is its transpose $E_a^T$.}
\label{fig:state_channel_instrument}
\end{figure}

When $\mathcal{C}$ is a unitary channel, that is, when there exists a unitary operator $U$ such that $\mathcal{C}(\rho)=U\rho\, U^\dagger$, its Choi operator may be written as $C=\kketbbra{U}{U}$ where $\kket{U}$ is the Choi vector of $U$ [as defined in Eq.~\eqref{eq:def:dketV}] and $\bbra{U}=\kket{U}^\dagger$.
If $\mathcal{C}$ is a channel with Kraus operators $\{K_i\}_i$, i.e., such that $\mathcal{C}(\rho)=\sum_i K_i\rho K_i^\dagger$ \citep{NielsenChuangBook,watrous_2018}, its Choi operator may similarly be written as $C=\sum_i\kketbbra{K_i}{K_i}$. 

Note that a quantum state $\rho\in\mathcal{L}(\H)$ may be viewed as a quantum channel by considering a map that transforms scalars into operators, that is, $\mathcal{C}:\mathcal{L}(\mathbb{C})\to\mathcal{L}(\H)$, $\mathcal{C}(1)\coloneqq\rho$. In this case, the Choi operator of this channel is simply given by $C=\rho$. Similarly, a quantum measurement with POVM elements given by $\{E_a\}_a$, $E_a\in\mathcal{L}(\H)$ can be viewed as a quantum instrument which transforms operators into scalars, that is, $\mathcal{M}_a:\mathcal{L}(\H)\to\mathcal{L}(\mathbb{C})$, $\mathcal{M}_a(\rho)\coloneqq\tr(E_a\,\rho)$. In this case, the Choi operators of this instrument are given by the transpose of its POVM elements, $M_a=E_a^T$.

Fig.~\ref{fig:state_channel_instrument} displays a summary and circuit representation of the basic quantum objects introduced so far. As customary in the literature, we may equivalently refer to quantum operations as CP maps or by their Choi operators, depending on the convenience of each situation.

\subsubsection{The link product}
\label{subsubsec:link_prod}

It can be convenient to know how to represent map composition at the level of Choi matrices. In other words, given two maps ${\cal M}_1: \L({\cal H}^X) \to \L({\cal H}^Y)$ and ${\cal M}_2: \L({\cal H}^Y) \to \L({\cal H}^Z)$, we want to connect the Choi representation of ${\cal M}_2\circ {\cal M}_1: \L({\cal H}^X) \to \L({\cal H}^Z)$ to that of ${\cal M}_1$ and ${\cal M}_2$. This is achieved by the \textit{link product} \citep{Chiribella2009}. In full generality, for two linear operators  $M_1 \in \L({\cal H}^{XY})$ and $M_2 \in \L({\cal H}^{YZ})$, it is defined as
\begin{align}
	 M_1*M_2 \coloneqq & (\id^{XZ} \otimes \bbra{\id}^{YY}) (M_1 \otimes M_2) (\id^{XZ} \otimes \kket{\id}^{YY}) \notag \\
	 = & \tr_Y \left[ (M_1^{XY}\otimes\id^{Z})^{T_Y} (\id^{X} \otimes M_2^{YZ})\right].
  \label{linkdefinition}
\end{align}
Indeed, if $M_1$ and $M_2$ are the Choi representations of ${\cal M}_1$ and ${\cal M}_2$, respectively, then $M_1*M_2$ is the Choi representation of ${\cal M}_2\circ {\cal M}_1$. Note that the link product is commutative, up to a reordering of the Hilbert spaces in the tensor products (as implied by the superscript notation).

The definition of the link product, Eq.~\eqref{linkdefinition}, also applies to the case where the output of a map does not (fully) overlap with the input of the other. In this case, the label $Y$ denotes the overlapping spaces between the two maps, while $X$ and $Z$ denote all the non-overlapping spaces for $M_1$ and $M_2$, respectively, where both $X$ and $Z$ might comprise both input and output systems. With this extension,
the link product unifies in an elegant way map composition and tensor product. Indeed, for the particular case where ${\cal H}^Y$ is trivial (i.e., 1-dimensional), so that there is no overlap between the two maps, the link product simplifies to $M_1*M_2=M_1 \otimes M_2$, which of course is the Choi representation of ${\cal M}_1 \otimes {\cal M}_2$. On the other hand, if ${\cal H}^X$ and ${\cal H}^Z$ are trivial, the link product simplifies to the trace: $M_1*M_2=\tr(M_1^T M_2)$. For example, the Born rule, Eq.~\eqref{born}, can be expressed as $E_a^T * \rho=\tr(E_a\,\rho)$. Notice also that the inverse Choi isomorphism, Eq.~\eqref{eq:inverse_Choi}, can be written as the application of a map to a state in their Choi representation: $\mathcal{M}(\rho)=M*\rho$.

If the Choi operators of both maps have rank 1---as for unitaries, isometries, and pure states---their link product also has rank 1. Correspondingly, the link product can be defined at the level of vectors $\kket{V_1} \in {\cal H}^{XY}$, $\kket{V_2} \in {\cal H}^{YZ}$ as \citep{Wechs2021}
\begin{align}
 \kket{V_1} * \kket{V_2} \coloneqq & (\id^{XZ} \otimes \bbra{\id}^{YY}) (\kket{V_1} \otimes \kket{V_2}) \nonumber \\
 = & (\kket{V_1}^{T_Y} \otimes \id^Z) (\id^X \otimes \kket{V_2}).
\end{align}
In this way, it holds that $ \kket{V_1}^{XY} * \kket{V_2}^{YZ} = \kket{V_2 V_1}^{XZ}$ and that  $\kket{V}^{XY}*\ket{\psi}^X=(V\ket{\psi})^Y$.

\subsection{Causally ordered processes} \label{orderedprocesses}

\subsubsection{Sequences of quantum measurements}

We now have all the mathematical tools to introduce a new perspective on quantum processes and measurements. We begin with a familiar scenario in which we have a system in the state $\rho$ subject to a sequence of two measurements (described as quantum instruments, see Sec.~\ref{operations}),  together with an evolution between the two given by a channel $\mathcal{C}$. We will say that the first measurement is performed by Alice, with associated label $A$, and the second by Bob, $B$. Let $\{{\cal M}_a\}_a$ and $\{{\cal N}_b\}_b$ be Alice's and Bob's instruments, respectively. In the standard approach, we first calculate the probabilities for Alice's measurement, $P(a) = \tr[\mathcal{M}_a(\rho)]$, and the conditional state associated with outcome $a$, $\Tilde{\rho}_a = \frac{\mathcal{M}_a(\rho)}{P(a)}$, as in Eq.~\eqref{eq:update}. Then, we evolve the new state with the map $\mathcal{C}$ and calculate the probabilities for Bob's measurement conditioned on Alice's outcome, $P(b|a) = \tr\left[{\mathcal{N}_b\circ \mathcal{C}(\Tilde{\rho}_a)}\right]$.

The conditional probability involved in the above procedure introduces a somewhat artificial asymmetry between the first and second measurement. A more temporally-agnostic quantity is the \textit{joint} probability, which we can obtain using Bayes' rule, $P(a,b)=P(b|a)P(a)$. Putting the above expressions together, the $P(a)$ factor simplifies and we find the more elegant form
\begin{equation} \label{jointprob}
    P(a,b) = \tr\left[\mathcal{N}_b\circ\mathcal{C}\circ\mathcal{M}_a(\rho)\right].
\end{equation}
We thus see that the peculiar nonlinearity of quantum state `collapse', Eq.~\eqref{eq:update}, is entirely due to the choice of calculating conditional rather than joint probabilities.

\subsubsection{Choi representation of causally ordered processes}
\label{subsubsec_process_matrix}

An important insight from Eq.~\eqref{jointprob} is that joint probabilities for quantum measurements are some bilinear function $\omega$ of the corresponding CP maps, $P(a,b) = \omega(\mathcal{M}_a, \mathcal{N}_b)$.
Crucially, this is also true if Alice and Bob, rather than measuring the same system at different times, measure two different parts of a joint state $\rho^{AB}$, in which case 
\begin{equation} \label{jointstate}
P(a,b) =  \omega(\mathcal{M}_a, \mathcal{N}_b) = \tr \left[(\mathcal{M}_a\otimes \mathcal{N}_b)(\rho^{AB})\right].
\end{equation}
This suggests a unification of all measurement scenarios discussed so far, regardless of the causal structure: any causally ordered scenario can be encoded in some function $\omega$ s.~t.\ $(\mathcal{M}_a,\mathcal{N}_b) \mapsto\omega(\mathcal{M}_a, \mathcal{N}_b) =P(a,b)$. We will call such a function a \textit{process}, and it can be understood in analogy to the definition of a \textit{state} as a linear function on observables: $E_a \mapsto \tr\left(E_a\,\rho\right) = P(a)$. 

The Choi isomorphism provides a particularly convenient way to represent such processes.
Let us label with $A_I$, $A_O$, $B_I$, $B_O$ the input and output spaces of Alice's and Bob's maps, respectively. When convenient, we will simplify the labels to $A\equiv A_IA_O$, $B\equiv B_IB_O$ for the tensor products of input and output spaces. A simple calculation, using the linearity of Alice and Bob's maps, shows that the joint probabilities in Eq.~\eqref{jointprob}, where Alice acts before Bob, can be expressed as
\begin{equation} \label{simpleBorn}
  \hspace{-0.3cm}
  P(a,b) = \tr \left[\left(M^{A}_{a}\otimes N^{B}_{b}\right)^T\, W^{AB}\right],
\end{equation}
where $M_{a}$, $N_{b}$ are the Choi representations of $\mathcal{M}_a$ and  $\mathcal{N}_b$, respectively, and we have introduced the \textit{process matrix}
\begin{equation} \label{Markovprocess}
    W^{AB} \coloneqq \rho^{A_I}\otimes C^{A_OB_I}\otimes\id^{B_O},
\end{equation}
where $C$ is the Choi representation of the channel $\mathcal{C}$.

We see that the process matrix encodes, in a single object, the initial state and the time evolution between Alice and Bob's operations. We also see that the same formula, Eq.~\eqref{simpleBorn}, remains valid for different causal scenarios. For example, in the joint-state scenario, Eq.~\eqref{jointstate} can be written in exactly the same form as Eq.~\eqref{simpleBorn}, with the process matrix
\begin{equation} \label{stateprocess}
    W^{AB} = \rho^{A_IB_I} \otimes \id^{A_OB_O}.
\end{equation}
In summary, Eq.~\eqref{simpleBorn} gives a probability rule that is agnostic about the causal relations between the measurements. In the literature, the words \textit{process} and \textit{process matrix} are often used interchangeably, similarly to how state and density matrix are often used as synonyms.

\subsubsection{Channels with memory} \label{subsubsec:channels_w_memory}
 
\begin{figure}[tbh] 
\centering
	\begin{center}
		\includegraphics[width=.65\columnwidth]{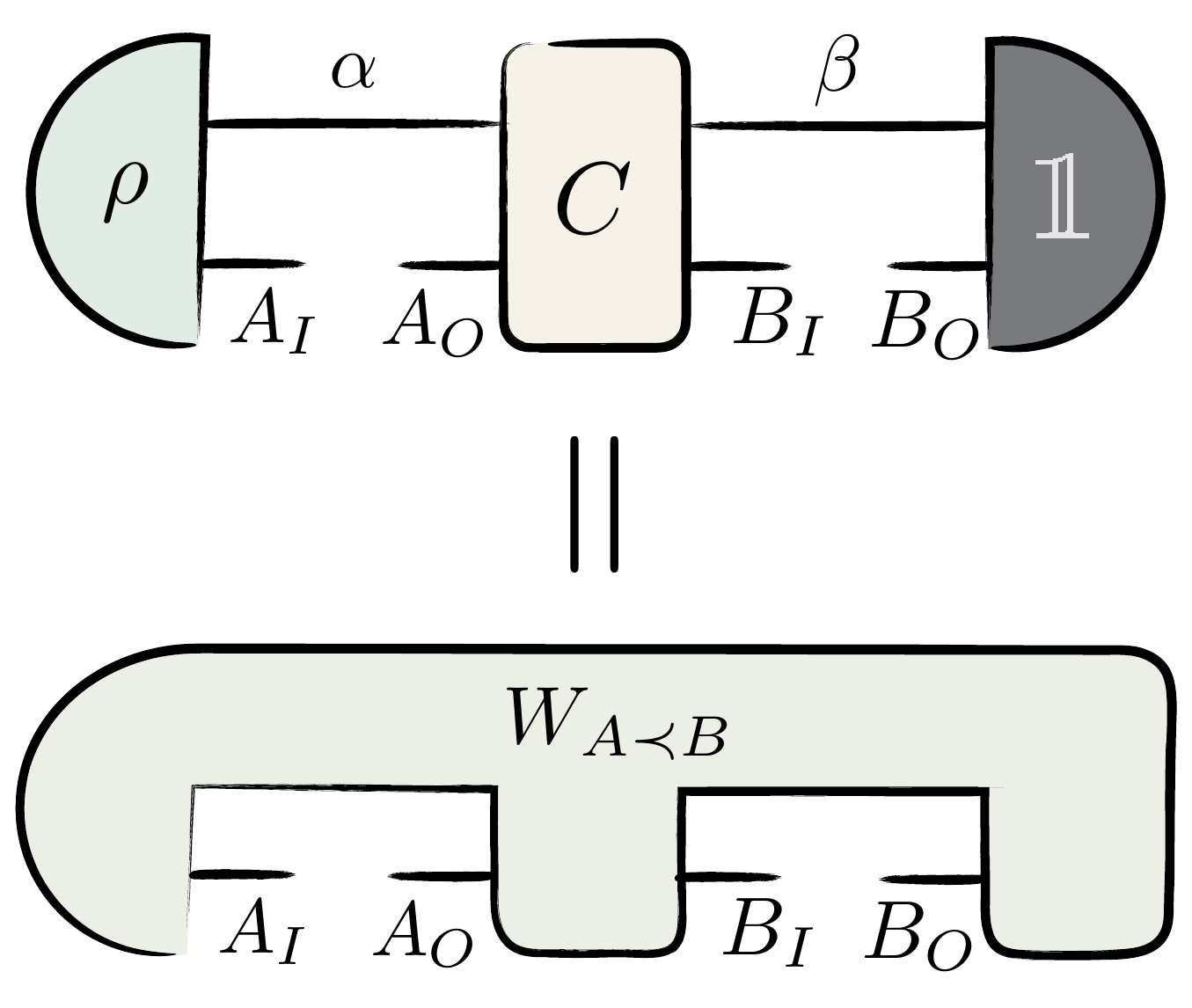} 
	\end{center}
\caption{Circuit illustration of a bipartite ordered quantum process, which also represents a channel with memory, as in Eq.~\eqref{eq:channelwithmemory}. } \label{fig:bipartite_ordered}
 \end{figure}

To conclude this overview of causally ordered processes, we should note that the process in Eq.~\eqref{Markovprocess} does not represent the most general scenario of two measurements separated in time. Alice might receive part of a state $\rho\in\L(\H^{\alpha}\otimes\H^{A_I})$ that is correlated with some other system in an auxiliary space $\H^{\alpha}$, representing an external environment. The channel connecting Alice's output to Bob's input can interact with the environment, 
$\mathcal{C}:\L(\H^{\alpha}\otimes\H^{A_O})\to\L(\H^{\beta}\otimes\H^{B_I})$, where $\H^{\beta}$ is another auxiliary space for the environment (that may in general differ from $\H^{\alpha}$); see Fig.\,\ref{fig:bipartite_ordered}. The process matrix representing this scenario is%
\footnote{When convenient, and where there is no risk of ambiguity, we omit identity operators. For instance, in the first line of Eq.~\eqref{eq:channelwithmemory}, $\rho^{\alpha A_I} \equiv \rho^{\alpha A_I} \otimes \id^{A_O \beta B_I}$ and $C^{\alpha  A_O \beta B_I}\equiv \id^{A_I}\otimes C^{\alpha  A_O \beta B_I}$.}
 	\begin{multline} \label{eq:channelwithmemory}
		W_{A\prec B}= \tr_{\alpha \beta} \left[\left(\rho^{\alpha A_I}\right)^{T_{\alpha }} C^{\alpha A_O \beta B_I}\right]\otimes\id^{B_O}\\
		= \rho^{\alpha A_I} * C^{\alpha  A_O\beta B_I} * \id^{\beta B_O}.
 	\end{multline}
  
Processes of this type are known as channels with memory  \citep{bowen2004,Kretschmann2005,caruso2012channel}, quantum strategies \citep{Gutoski2006}, or quantum combs  \citep{Chiribella2009}, see also Sec.~\ref{otherorderedprocesses}. They have a simple characterization as process matrices of the form 
\begin{subequations}
\begin{align}
\label{BOlast}
 W_{A\prec B} &= W^{A_IA_OB_I}\otimes\id^{B_O} \geq 0, \\
 \label{eq:TrBI_AOlast} 
 \tr_{B_I}W^{A_IA_OB_I} &= \rho^{A_I}\otimes\id^{A_O},\\
 \label{firstastate}
 \tr\rho^{A_I} &= 1.
\end{align}
\end{subequations}
As it turns out, these are the most general processes that can be realized when $A$ causally precedes $B$ and they include, as particular cases, channels without memory, Eq.~\eqref{Markovprocess}, and states, Eq.~\eqref{stateprocess}. Channels with memory are particularly useful in the study of open quantum systems with non-Markovian dynamics and initial system-environment correlations \citep{modioperational2012, Milz2017, Pollock2018}.

\subsection{Signaling and causal relations}\label{signaling section}

In the above overview we have obliquely referred to causal relations, but we have not explicitly discussed how to define them and how they fit into the formalism. In fact, causality and related notions have markedly different incarnations in various areas of physics, statistics, and philosophy \citep{Cox1992, woodward2003making, Pearl2009, Frisch2022}.
A minimal notion of causal relation, common in information theory, is that of \textit{signaling} \citep{Tsirelson1993, Popescu1994, Werner2001}. 
Consider a scenario where two parties, Alice and Bob, perform local operations according to some freely chosen classical variables (often called \textit{settings}) $x$ and $y$, respectively. Additionally, their operations involve measurements, whose outcomes yield classical variables $a$ and $b$. 
Such a scenario is characterized by a conditional probability distribution
$P(a,b|x,y)$. We say that Bob \textit{does not signal} to Alice (or that the probabilities are no-signaling from $B$ to $A$) if
\begin{align}
    P_A(a|x,y) = P_A(a|x,y'), \quad \forall x,y,y', \label{eq:nosig_B_to_A}
\end{align}
where Alice's marginal probability distribution is given by $P_A(a|x,y)\coloneqq\sum_b P(a,b|x,y)$. If the reverse is true, that is, if Bob can choose two variables $y\neq y'$ such that, at least for one choice of $x$, Alice can detect a difference in her outcome probabilities, then we say that Bob \textit{signals} to Alice. A conditional probability distribution is called \emph{one-way signaling} if it allows signaling in only one direction and \emph{two-way signaling} if signaling is possible in both directions, from $A$ to $B$ and vice versa, while \textit{no-signaling} means that no signaling is allowed in either direction.

In the presence of signaling from Bob to Alice, we say that Bob's local operation has a causal influence over Alice's, because a controllable change in the former produces an observable change in the latter. If Bob does not signal to Alice, we say that the probability distribution is \textit{compatible with $A \prec B$}, where the notation $A \prec B$ reads \emph{$A$ causally preceding $B$}. 
Note that this terminology does not imply any fundamental notion of cause and effect relations, but only deals with their operational manifestation: whether signaling is possible or not. 
In particular, a given probability distribution can be compatible with multiple causal relations: for example, a no-signaling distribution is compatible with both $A\prec B$ and $B\prec A$. 

When working at the level of probability distributions, we cannot distinguish whether the lack of signaling is due to the physical scenario or to the parties' choice of operations (e.g., Bob could perform the same operation for all values of $y$, making no difference to Alice's outcomes regardless of the available resources). For a definition that applies to the physical resource, independently of the particular operations, we say that a process is compatible with $A\prec B$ if, for any choice of operations, the resulting probabilities are no-signaling from $B$ to $A$.

It is instructive to verify that the causally ordered quantum processes introduced in Sec.~\ref{orderedprocesses} agree with this definition.
In a quantum communication scenario, the variables $x, y$ label choices of instruments, while $a, b$ denote the outcomes of the instruments. Accordingly, we write ${\cal M}^A_{a|x}$ for a CP map with setting $x$ and outcome $a$, similarly for ${\cal N}^B_{b|y}$, and the Born rule [Eq.~\eqref{processborn}] provides conditional probabilities $P(a,b|x,y)$. 
Consider a process matrix of the form $W_{A\prec B} = W^{A_IA_OB_I}\otimes\id^{B_O}$, describing a general channel with memory as in Sec.~\ref{subsubsec:channels_w_memory}. 
Alice's marginal probabilities are then given by
\begin{multline} \label{eq:1way_signaling}
P_A(a|x,y) \coloneqq \sum_b P(a,b|x,y)\\
= \sum_b \tr\left[\left(M^{A}_{a|x}\otimes N^{B}_{b|y}\right)^T \left(
W^{A_IA_OB_I}\otimes\id^{B_O}\right)\right] \\
= \tr\left[\left(M^{A}_{a|x}\otimes  \sum_b  \tr_{B_O}N^{B}_{b|y}\right)^T\, W^{A_IA_OB_I}\right] \\
= \tr\left[\left(M^{A}_{a|x}\right)^T \tr_{B_I} W^{A_IA_OB_I}\right].
\end{multline}
In the last step, we used the fact that the CP maps in Bob's instrument sum up to a CPTP map, and $\sum_b\tr_{B_O} N^{B}_{b|y} = \id^{B_I}$ is the Choi representation for the trace-preserving condition. We see that the dependency on $y$ disappears from the final expression, so that the probability is no-signaling from Bob to Alice, as expected.%
\footnote{The final expression can be further simplified: using Eq.~\eqref{eq:TrBI_AOlast}, and the inverse Choi isomorphism, we get $P_A(a|x,y) = \tr[{\cal M}^{A}_{a|x}(\rho^{A_I})]$, where $\rho^{A_I}$ is the reduced state that Alice receives (which, as expected, is independent of Bob's operation).}
From this exercise, we learn that a process matrix that has the identity on the output space of a party leads to probabilities that are no-signaling from that party, explaining the form of causally ordered processes in Eq.~\eqref{BOlast}. 

Just like probability distributions, processes can be compatible with multiple causal relations. In particular, a no-signaling process (of the form $W = \rho^{A_IB_I}\otimes\id^{A_OB_O}$, where $\rho$ is a shared state) can be realized both in a scenario where $A$ causally precedes $B$, or where $B$ causally precedes $A$. 

\subsection{Beyond causally ordered processes}
\label{subsec:W_indefinite_order}

\subsubsection{The process matrix}

We have seen that the process matrix encodes causal relations between parties, with a single probability rule that applies to all causally ordered scenarios,  Eq.~\eqref{simpleBorn}. This opens the possibility to consider more general scenarios, that are not compatible with any given order of events.

A first, natural generalization is a process with classical uncertainty on the causal order. 
Consider a situation where we are uncertain about the process connecting $A$ and $B$: it can be $W_1$, with probability $q_1$, or $W_2$, with probability $q_2$ (with $q_j\ge 0$, $q_1+q_2=1$). 
The probabilities for outcomes $a$, $b$ must then be
\begin{equation} \label{mixedprobabilities}
    P(a,b) = q_1\,P_1(a,b) + q_2\,P_2(a,b),
\end{equation}
where $P_j(a,b) = \tr \left[\left(M^{A}_{a}\otimes N^{B}_{b}\right)^T\, W_j^{AB}\right]$ (for $j=1,2$). Because this expression is linear in $W_j$, we can re-write it using our probability rule, Eq.~\eqref{simpleBorn}, and the single process matrix $W=q_1\,W_1 + q_2\,W_2$. We see that---just as for density matrices or channels---a convex combination of process matrices corresponds to a probabilistic mixture and, therefore, represents a meaningful physical scenario. In particular, this works if $W_1$ and $W_2$ encode different causal relations, in which case $W$ represents a situation where the order of events is well defined, but not known with certainty.

The central questions in the study of indefinite causal order are whether more general scenarios exist, how to characterize them, and what their physical significance is.
A challenge in this search is that such scenarios might involve genuinely new physics, which is hard to characterize without reference to a specific model. The main approach we follow, introduced by \citet{OCB_2012}, is to consider a formal, ``minimal'' extension of quantum theory, namely one that takes at face value the generalized probability rule, Eq.~\eqref{simpleBorn}, and considers the most general operators $W$ compatible with that rule. This results in the following:
\begin{definition} \label{def:process}
Given a set of parties $A, B,\dots$, with associated input and output Hilbert spaces,  a \emph{process matrix} is an operator $W^{AB\dots}\in \L(\H^{A_I}\otimes \H^{A_O}\otimes \H^{B_I}\otimes \H^{B_O}\otimes...)$ such that 
\begin{subequations}
\begin{align} \label{positive}
   W&^{A B\dots}\geq 0, \\ 
    \tr&[(M^A \otimes N^B\otimes\cdots)^T \, W^{AB\dots}] = 1, \notag \\
   & \quad \forall\; M^A \in \L(\H^A), \ N^B \in \L(\H^B),\dots \notag \\ \label{normalised_CPTPChoi}
   & \quad \quad \text{s.t.} \ \tr_{A_O} M^A = \id^{A_I}, \ \tr_{B_O} N^B = \id^{B_I},\dots
\end{align}
\end{subequations}
\end{definition}

\begin{figure}[hbt] 
	\begin{center}
		\includegraphics[width=.6\columnwidth]{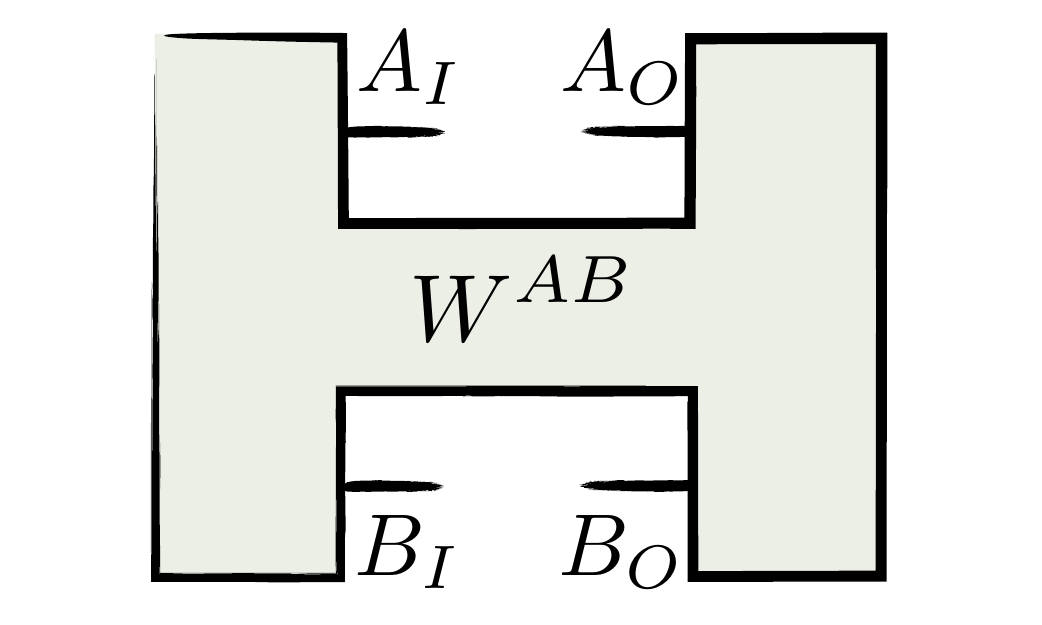} 
	\end{center}
\caption{Pictorial representation of a bipartite process matrix, where $A_I$ and $A_O$ label Alice's input and output spaces and $B_I$ and $B_O$ label Bob's input and output spaces, respectively. Alice and Bob's operations are then to be ``plugged'' into the corresponding open slots.}
\label{fig:ProcessMatrix}
\end{figure}

This definition can be derived from physically-motivated axioms, as we will discuss in Sec.~\ref{processmatrixsection}. For now, let us unpack its meaning. The operational interpretation of the process matrix is the same as before: when probed with instruments $\{M_a^A\}_a, \{N_b^B\}_b,\dots$---or, with reference to the pictorial representation of Fig.~\ref{fig:ProcessMatrix}, when these instruments are ``plugged'' into its empty slots---it returns joint probabilities $P(a,b,\dots)$ according to the generalized Born rule, Eq.~\eqref{simpleBorn} (extended to an arbitrary number of events). Condition \eqref{positive} is analogous to the positivity of density matrices: it ensures that probabilities are nonnegative (as we will see, it ensures a stronger condition, analogous to complete positivity for quantum operations). On the other hand, condition \eqref{normalised_CPTPChoi} states that the probability equals $1$ when all maps are trace preserving, namely, if each party applies a deterministic operation. This ensures normalization of probabilities: $\sum_{ab\dots}P(a,b,\dots)=1$ when the sum extends to all the outcomes of all the instruments. Note that normalization only needs to hold for $M^A, N^B \geq 0$ (namely for the Choi operators of CPTP maps); however, by linear extension, this is equivalent to requiring that it must hold for non positive semidefinite operators as well.

\subsubsection{Causally separable and nonseparable processes}

As we have seen, all the causally ordered processes of Sec.~\ref{orderedprocesses} are included as particular cases of process matrices. Furthermore, Definition~\ref{def:process} also includes processes with definite, but not fixed, causal order. For the bipartite case, these coincide with the probabilistic mixtures mentioned above:

\begin{definition} \label{def:Wcsep_bipartite}
A bipartite process matrix $W_\mathrm{c\mhyphen sep}$ for parties $A$, $B$  is said to be \emph{causally separable} if and only if it can be written as
\begin{align}
    W_\mathrm{c\mhyphen sep} = q \, W_{A\prec B} + (1{-}q) \, W_{B\prec A} \quad \text{with} \ q \in [0,1] \label{eq:def_csep},
\end{align}
where $W_{A\prec B}$ is a bipartite process matrix compatible with the order $A \prec B$ (i.e., it satisfies conditions \eqref{BOlast}--\eqref{firstastate}) and $W_{B\prec A}$ is a bipartite process matrix compatible with $B \prec A$ (i.e., it satisfies the same conditions, with $A$ and $B$ swapped).
\end{definition}

We now have a well-defined mathematical question: are there operators that satisfy the general definition above, [i.e., Eqs.~\eqref{positive}--\eqref{normalised_CPTPChoi}], but that cannot be decomposed as convex combinations of causally ordered process matrices? In other words, are there process matrices that are \emph{not} causally separable? This question was first posed by \citep{OCB_2012} and answered in the positive, by exhibiting the following example (with 2-dimensional input and output spaces):
\begin{equation}
   \hspace{-0.6cm} W_\mathrm{OCB} \!= \!\frac{1}{4} \!\left[ \id^{AB} \!+\! \frac{\id^{A_I}\sigma_z^{A_O}\sigma_z^{B_I}\id^{B_O} \!+\! \sigma_z^{A_I}\id^{A_O}\sigma_x^{B_I}\sigma_z^{B_O}}{\sqrt{2}} \right]\!, \label{eq:W_OCB}
\end{equation}
where $\sigma_z$ and $\sigma_x$ (and, later, $\sigma_y$) are the Pauli matrices and we left the tensor products implicit. 

This is indeed a positive semidefinite matrix and, using the characterization that we will introduce in Sec.~\ref{Wcharacterization}, it is also easy to see that is satisfies the normalization conditions, Eq.~\eqref{normalised_CPTPChoi}. $W_\mathrm{OCB}$ is not causally ordered, because it does not have an identity operator that factorizes on any output space [in other words, it does not satisfy condition \eqref{BOlast} for a channel with memory]. Instead, it contains two traceless terms, $\sigma_z^{A_O}\sigma_z^{B_I}$ and $\sigma_z^{A_I}\sigma_x^{B_I}\sigma_z^{B_O}$, each of which could be used to define a causally ordered process matrix: $W_{A\prec B}= \frac{1}{4} \left[ \id^{AB} + \sigma_z^{A_O}\sigma_z^{B_I} \right]$ and $W_{B\prec A}= \frac{1}{4} \left[ \id^{AB} +\sigma_z^{A_I}\sigma_x^{B_I}\sigma_z^{B_O} \right]$ (with implicit identity operators). An equal mixture $\frac{1}{2}\left( W_{A\prec B} + W_{B\prec A} \right)$ \textit{almost} gives $W_\mathrm{OCB}$, except that the factor attached to the traceless term would be $\frac{1}{2}$, instead of $\frac{1}{\sqrt{2}}$. Although this observation does not yet prove that $W_\mathrm{OCB}$ is causally nonseparable (one could try to add and subtract other traceless terms to $W_{A\prec B}$ and $W_{B\prec A}$, in order to get $W_\mathrm{OCB}$ as a convex combination), it provides a good intuition for it. Methods to formally prove the nonseparability of $W_\mathrm{OCB}$ will be presented in Secs.~\ref{nonsepsection}--\ref{subsec:causal_witness}. In Sec.~\ref{sec:causal_ineqs}, we will further show that $W_\mathrm{OCB}$ enables the violation of a \textit{causal inequality}, demonstrating an even stronger form of indefinite causal order. 

Although $W_\mathrm{OCB}$ is mathematically a valid example of indefinite causal order, it does not seem to have a natural physical interpretation. On the other hand, a physically motivated example is the \textit{quantum switch}, introduced by \citep{Chiribella2013}, which can be understood as a process implementing a quantum control of causal order, recall Fig.~\ref{fig:switch}.
A proper analysis of the switch in the process matrix formalism requires the introduction of multipartite processes, so we postpone it to Sec.~\ref{switchsection}. As we will see, not only the quantum switch is a causally nonseparable process, but it also has a physical interpretation, both in table-top quantum experiments and in scenarios involving quantum superpositions of spacetime metrics. 

\subsection{Defining and locating events}
\label{locatingevents}

We have introduced processes with indefinite causal order in a rather abstract way, through a formal extension of the Born rule for evaluating probabilities. This may leave one wondering about the general interpretation of the formalism: how can we identify operations without reference to their spacetime location? What do the labels $A, B,\dots$ represent? 

In familiar scenarios, where operations take place at fixed locations relative to a global spacetime coordinate system, the labels $A,B\dots$ can be directly associated with particular coordinate values, defined operationally relative to physical reference frames, such as `clocks' and `rulers'. It is clear that only causally ordered processes are compatible with this scenario.

The association of labels with physical reference frames naturally extends to scenarios where the location of the operations cannot be identified with specific points within a single, global coordinate system. For example, operations $A$ and $B$ may take place at particular times, as read by two distinct clocks (say, ``2pm'') but each clock might have some unknown shift as compared to a global reference time. This is a natural scenario leading to causally separable processes, such as probabilistic mixtures of causally ordered ones. 

Taking this interpretation further, one can consider reference systems (``clocks and rulers'') in quantum states, leading to events with indefinite spacetime location. Indeed, we will see that this is the idea behind some of the physical implementations of indefinite causal order, such as the quantum switch. 
Even in scenarios where there is no classical background spacetime, one can imagine to locate events in a similar way, labeling them relative to physical systems that act as reference frames. Most generally, we understand an ``event'' as the result of an intervention---a CP map in the quantum formalism---while the labels $A, B,\dots$ identify generalized, operationally defined, locations where interventions can take place.

To fix some terminology, a locus of intervention will be called a \textit{site}.\footnote{This terminology was introduced in \citep{MaguireThesis} and has been used, for example, in \citep{Costa2024}.} Formally, a site corresponds to a tensor factor $\mathcal{H}^{A_I}\otimes\mathcal{H}^{A_O}$ in the total Hilbert space $\left(\mathcal{H}^{A_I}\otimes\mathcal{H}^{A_O}\right)\otimes\left(\mathcal{H}^{B_I}\otimes\mathcal{H}^{B_O}\right)\otimes \, \ldots$  and plays a similar role to a subsystem in ordinary quantum theory. Expressions such as ``laboratories'', ``regions'', ``slots'', and ``parties''  have been used in the literature to refer to sites. Here, we will also commonly use the terminology ``party'' to refer to an idealized agent performing an operation at a site.

According to the interpretation described here, the generalized ``clocks and rulers'' that identify the sites are not part of the degrees of freedom that make up the process, even if they may be in quantum states, not localized in spacetime. This is unlike general relativity, which makes no reference to any background reference system. The incorporation of background independence into the process matrix formalism has been discussed in Ref. \citep{parker2021background}, see Sec.~\ref{quantumcoordinates} below.

\section{Quantum processes with indefinite causal order}
\label{indefiniteorder}

In this section, we will provide a detailed review of the process matrix formalism, briefly introduced earlier, along with the methods for characterizing indefinite causal order, with some intentional overlap with the initial discussion for clarity. We will start the presentation with processes involving only two sites, which already exemplify some of the main new features of the framework, and then generalize the formalism to arbitrary multipartite processes. This sounds like the multipartite case will come in another section; do we need this last sentence? A concise ``map'' of indefinite causal order, providing a short overview, can be found in Ref. \citep{EscandonMonardes2025}.

\subsection{The process matrix formalism}
\label{processmatrixsection}

\subsubsection{Axiomatic derivation} \label{indefiniteaxiomatic} 

In Sec.~\ref{firststeps}, we have introduced process matrices as a formal generalization of (causally ordered) quantum channels and states. Here, we discuss a principle-based derivation of both the generalized Born rule, Eq.~\eqref{processborn}, which we rewrite here for convenience:
\begin{equation} \label{simpleBorn2}
  \hspace{-0.4cm}
   P(a,b\mid\! \{\mathcal{M}_a^A\}_a, \{\mathcal{N}_b^B\}_b) \! = \! \tr \left[\left(M^{A}_{a}\otimes N^{B}_{b}\right)^T W^{AB}\right],
\end{equation}
and of bipartite process matrices, which, we recall, are defined by the conditions
\begin{subequations}
\begin{align} \label{positive2}
   W&^{A B}\geq 0, \\ 
    \tr&[(M^A \otimes N^B)^T\, W^{AB}] = 1, \notag \\
   & \quad \forall\; M^A \in \L(\H^A), \ N^B \in \L(\H^B) \notag \\ \label{normalised_CPTPChoi2}
   & \quad \quad \text{s.t.} \ \tr_{A_O} M^A = \id^{A_I}, \ \tr_{B_O} N^B = \id^{B_I}
\end{align}
\end{subequations}
(see Definition \ref{def:process}). The discussion below extends to general multipartite processes in a direct way.

The core idea is to consider a general framework where quantum theory applies locally in a set of \emph{sites} (or generalized event locations), while making no assumption about the causal structure connecting the sites. This means that observable events taking place at each site are described by ordinary quantum operations. 

Concretely, taking two sites labeled $A$ and $B$, we associate to each of them an input and an output Hilbert space, ${\cal H}^{A_I}$, ${\cal H}^{A_O}$, ${\cal H}^{B_I}$, and ${\cal H}^{B_O}$. We denote by $d^X$ the dimension of a Hilbert space ${\cal H}^{X}$ (which, for now, is assumed to be finite). Local events, as in ordinary quantum theory, are associated with local operations, namely with the outcomes $a$, $b$ of local instruments $\{\mathcal{M}_a^A\}_a$, $\{\mathcal{N}_b^B\}_b$. The purpose of the framework is to provide a way to evaluate joint probabilities for the events at the various sites, conditioned on the choice of instruments: $P(a, b\mid \{\mathcal{M}_a^A\}_a, \{\mathcal{N}_b^B\}_b)$. 

There are different ways to formalize the assumption of local validity of quantum theory, in terms of properties of probabilities predicted by the framework. Let us first state the relevant properties and then discuss their motivation and their role in the derivation of the process matrix formalism.

\textbf{Probabilities.\footnote{Although this is not a specific assumption about quantum theory---it is simply a statement that observable statistics are described by ordinary probability theory---it plays an important role in the derivation.}} Probability distributions are normalized, $\sum_{a b} P(a, b {\mid} \{\mathcal{M}_a^A\}_a, \{\mathcal{N}_b^B\}_b) {=} 1$, and non-negative, $P(a, b {\mid} \{\mathcal{M}_a^A\}_a, \{\mathcal{N}_b^B\}_b) \geq 0$. 

\textbf{Instrument non-contextuality.} Probabilities do not depend on the entire instruments: they are functions of the instrument elements alone.%
\footnote{Strictly speaking, we already make a (weaker) type of non-contextuality assumption when writing probabilities in the form ${P(a, b\mid \{\mathcal{M}_a^A\}_a, \{\mathcal{N}_b^B\}_b)}$, as we are assuming that the probabilities for observed events do not depend on the implementation of the instruments, e.g., on the particular interaction between system and measurement device. Non-contextuality assumptions are sometimes left implicit in the literature, by writing directly probabilities in the form $P\left(\mathcal{M}_a^A, \mathcal{N}_b^B \right)$, even though such an expression cannot be interpreted as a normalized probability measure over the space of CP maps.}
In other words, there exists a function $\omega$, called \textit{frame function} \citep{gleason57, caves2004}, such that
\begin{equation}
    P(a, b\mid \{\mathcal{M}_a^A\}_a, \{\mathcal{N}_b^B\}_b) = \omega\left(\mathcal{M}_a^A, \mathcal{N}_b^B \right).
\end{equation}
Note that we already encountered the function $\omega$ in Sec.~\ref{subsubsec_process_matrix} and called it a ``process'' there; indeed a process is assumed by definition to be linear in all arguments, while, for a frame function, linearity still has to be proven.

\textbf{Probabilistic mixtures.}
Each party can implement an arbitrary probabilistic mixture of any two instruments with matching sets of outcomes and compatible with the party's input and output spaces. The mixture of instruments $\{\mathcal{M}_a\}_a, \{\mathcal{M}'_a\}_a$, with probabilities $q$, $1-q$, respectively ($0\leq q \leq 1$), is represented by an instrument $\{\widetilde{\mathcal{M}}_a\}_a$, in which each map is the convex combination
\begin{equation}
    \widetilde{\mathcal{M}}_a = q\, \mathcal{M}_a + (1-q)\, \mathcal{M}'_a.
\end{equation}
Combined with \textbf{Instrument non-contextuality}, this assumption implies that the frame function is convex-linear in each of its variables:
\begin{multline}\label{convexlinear}
    \omega( q \, \mathcal{M} + (1-q) \, \mathcal{M}', \mathcal{N}) \\
    =  q \, \omega(\mathcal{M}, \mathcal{N}) + (1-q) \, \omega(\mathcal{M}', \mathcal{N})
\end{multline}
for all CP maps $\mathcal{M}$, $\mathcal{M}'$, $\mathcal{N}$ and all $0\leq q \leq 1$ (with the corresponding property for the second argument). 

\textbf{Coarse-graining.}
It is possible to coarse-grain two or more outcomes of an instrument, identifying them as a single outcome. As in ordinary quantum mechanics, coarse-graining is represented by a sum of maps. E.g., given an instrument $\{\mathcal{M}_a\}_{a=1}^n$, with $n\geq 2$, a coarse-graining of the last two outcomes is represented by an instrument with the $n-1$ elements 
\begin{equation}
    \left\{ \mathcal{M}_1,\dots, \mathcal{M}_{n-2}, \mathcal{M}_{n-1} + \mathcal{M}_n \right\}.
\end{equation}
Together with \textbf{Instrument non-contextuality}, coarse-graining implies that the frame function is additive:
\begin{equation}\label{additive}
    \omega(\mathcal{M} + \mathcal{M}', \mathcal{N}) \\
    =  \omega(\mathcal{M}, \mathcal{N}) + \omega(\mathcal{M}', \mathcal{N})
\end{equation}
for all CP maps $\mathcal{M}$, $\mathcal{M}'$, $\mathcal{N}$ (with the corresponding property for the second argument).

\textbf{Linearity.} (Assuming \textbf{Instrument non-contextuality}) the frame function $\omega$ is multi-linear in its arguments: 
\begin{equation}
    \omega\Big(\sum_{a}\alpha_a\,\mathcal{M}_a, \sum_{b} \beta_b\,\mathcal{N}_b\Big)
    = \sum_{ab}\alpha_a\,\beta_b\,\omega(\mathcal{M}_a, \mathcal{N}_b) 
\end{equation}
for all $\alpha_a, \beta_b \in \mathbb{C}$ and all CP maps $\mathcal{M}_a$, $\mathcal{N}_b$.

\textbf{Local extendibility.} The Local operations can be extended to act on additional systems prepared in an arbitrary state (uncorrelated with the systems involved in the process). This means that we can attach to Alice's and Bob's sites additional input systems, with Hilbert spaces $\H^{A'_I}$ and $\H^{B'_I}$, respectively, which can be prepared in an arbitrary (possibly entangled) state $\rho^{A_I'B_I'}$. Extended local operations $\mathcal{M}^{A'_IA}$, $\mathcal{N}^{B'_IB}$ should still give valid probability distributions when applied to the original process extended by the state $\rho^{A_I'B_I'}$.  Assuming \textbf{instrument non-contextuality} and \textbf{linearity}, the extended process is defined by a frame function denoted 
$\Tilde{\omega} = \omega_\rho \otimes\omega$, with $\omega_\rho(\cdot):=\tr(\rho\,\cdot)$, which acts on local operations of the form $E^{A'_I}\otimes\mathcal{M}^A$, $F^{B'_I}\otimes\mathcal{N}^B$ (where
$E^{A'_I}$, $F^{B'_I}$ are POVM elements) as
\begin{multline}
\Tilde{\omega}(E^{A'_I}\otimes\mathcal{M}^A, F^{B'_I}\otimes\mathcal{N}^B) \\ 
= \tr[(E^{A'_I}\otimes F^{B'_I})\rho^{A'_IB'_I}] \, \omega(\mathcal{M}^A, \mathcal{N}^B),
\label{extendibility}
\end{multline}
and whose action on arbitrary maps $\mathcal{M}^{A'_IA}$, $\mathcal{N}^{B'_IB}$ is then defined from this by linearity.

The properties above make no direct reference to any causal structure, and they are all satisfied by probability distributions arising from causally ordered processes. Therefore, they are all reasonable assumptions in the construction of a `minimal' extension of quantum theory, where only causal assumptions are lifted.

At the technical level, the formalism follows from \textbf{Probabilities}, \textbf{Linearity} (which subsumes \textbf{Instrument non-contextuality}), and \textbf{Local extendibility}. 
Indeed, due to the Riesz representation lemma on linear functionals \citep{ReedSimon}, \textbf{Linearity} implies that, for each $\omega$, there exists a unique operator $W = W^{AB}\in \mathcal{L}(\mathcal{H}^{A_I}\otimes\mathcal{H}^{A_O}\otimes\mathcal{H}^{B_I}\otimes\mathcal{H}^{B_O})$ such that $\omega\left(\mathcal{M}^A, \mathcal{N}^B \right) = \tr[(M^A\otimes N^B)^T\,W]$ for all CP maps $\mathcal{M}^A, \mathcal{N}^B$ with Choi matrices $M^A, N^B$, yielding the form~\eqref{simpleBorn2} of the generalized Born rule.  Normalization of \textbf{Probabilities} implies that the probability for any pair of CPTP maps must be 1, giving the constraints~\eqref{normalised_CPTPChoi2}. Such constraints are also sufficient to ensure normalization for arbitrary instruments because, using \textbf{Linearity} and the fact that the sum of all CP maps in an instrument is CPTP, we get
\begin{equation}
    \sum_{ab}\omega\left(\mathcal{M}_a^A, \mathcal{N}_b^B \right) = \omega\left(\sum_{a} \mathcal{M}_a^A, \sum_{b}\mathcal{N}_b^B \right) = 1.
\end{equation}
A slightly more technical argument, following \citep{barnum05}, shows that \textbf{Local extendibility} (together with \textbf{Probabilities}) implies $W\geq 0$\, as in Eq.~\eqref{positive2}.

It is interesting that positivity of probabilities, alone, does not imply $W\geq 0$. Indeed, there exist operators $W$, not positive semidefinite, for which $\tr[(M^A\otimes N^B)^T\,W] \geq 0$ for every $M^A\geq 0$, $N^B\geq 0$. Hence, \textbf{Local extendibility} implies a strictly stronger condition, in analogy with complete positivity for quantum channels. An alternative way to postulate this condition is to require that a process can be composed with local operations involving other quantum systems, with the composition resulting in a valid quantum operation~\citep{Araujo2017}.

The assumption of \textbf{Linearity} can be derived from other properties with a more direct physical interpretation. \citet{OCB_2012} derived it from \textbf{Probabilistic mixtures} and \textbf{Coarse-graining} (with the implicit assumption of \textbf{Instrument non-contextuality}), which can be motivated by requiring that the theory retains the quantum description of local operations, including the description of probabilistic post-processing. The result is a straightforward consequence of Eqs.~\eqref{convexlinear} and~\eqref{additive}. Later, \citet{Shrapnel2017} showed that \textbf{Instrument non-contextuality} alone is sufficient to derive \textbf{Linearity} [as well as the form of the trace rule in the definition of local extendibility, Eq.~\eqref{extendibility}], with an argument based on analogous results for the ordinary Born rule \citep{Busch2003, caves2004}, which are themselves versions of Gleason's theorem \citep{gleason57}. Non-contextuality is a core property of quantum theory, which can be interpreted as the requirement that \textit{only events that happen matter} for the calculation of probabilities. Further implications of instrument non-contextuality are discussed in Sec.~\ref {subsubsec:noncontextuality}.

Remarkably, it turns out that causally ordered quantum theory follows if, on top of the assumptions above, we add the assumption of causal order. In other words, if we impose that the probability distributions $P(a, b| \{\mathcal{M}_a^A\}_a, \{\mathcal{N}_b^B\}_b)$ do not allow for signaling from $B$ to $A$, we find that bipartite process matrices have to satisfy the additional constraints \eqref{BOlast}--\eqref{firstastate}, restricting them to quantum channels with memory \citep{Chiribella2009}. All the above results extend to arbitrary multipartite processes.

\subsubsection{Characterization}\label{Wcharacterization}

Having introduced process matrices, we now want to find some convenient characterization. The positivity condition, $W\geq 0$, is just like for density matrices. What makes things more complicated is the normalization condition, Eq.~\eqref{normalised_CPTPChoi2}. At first sight, this condition might seem hard to verify, as it needs to hold for the infinite set of CPTP maps. However, things become much simpler once we realize that Eq.~\eqref{normalised_CPTPChoi2} just implies some linear-affine constraints on the set of operators $W$.

The affine part arises from applying the constraint to CPTP maps whose Choi matrix is proportional to the identity: ${M}^{A}=\id^{A}/d^{A_O}$ and similarly for $N^B$.
(This represents a completely depolarizing operation, which transforms every state into the maximally mixed one.) Eq.~\eqref{normalised_CPTPChoi2} then gives 
\begin{equation}\label{affine}
    \tr W = d^O,
\end{equation}
where  $d^O \coloneqq  d^{A_O}d^{B_O}$  is the product of Alice and Bob's output space dimensions. This condition is analogous (apart from a multiplicative factor) to the normalization of density matrices, $\tr\rho = 1$. Enforcing normalization for all other CPTP maps turns out to impose additional linear constraints, that is, it forces process matrices to live in a strict subspace of operators, which we denote $\L_\mathrm{V}$ and that we will characterize shortly. We refer to  $\L_\mathrm{V}$ as the linear subspace spanned by ``valid'' process matrices (or, sometimes, simply the space of valid process matrices). It should always be understood that, for an operator $W$ to be a process matrix, it also has to satisfy the positivity condition, $W\geq 0$, and the affine constraint\footnote{Note that any positive semidefinite $W\neq 0$ in $\L_\mathrm{V}$ can always be ``renormalized'' to give a process matrix, $W\mapsto d^O\, W/\tr W$.
} $\tr W = d^O$. 

We will now present two approaches to the explicit characterization of $\L_\mathrm{V}$, first introduced in \citep{OCB_2012} and \citep{Araujo_2015}, respectively.

\paragraph{Characterization using a Hilbert-Schmidt basis.}\label{para:HS_terms}

To characterize the space $\L_\mathrm{V}$, it is useful to treat the set of linear operators $\L(\H)$ as a linear space endowed with a scalar product, called the Hilbert-Schmidt (HS) product:
\begin{equation}
    (A,B)_{\mathrm{HS}} \coloneqq \tr(A^{\dagger}B).
\end{equation}
For a Hilbert space $\H$ of dimension $d$, it is always possible to find a set of self-adjoint operators $\left\{\sigma_{i}\right\}_{i=0}^{d^2-1}$ that satisfies 
\begin{equation}
    \tr \left( \sigma_{i}\, \sigma_{j} \right)= d\, \delta_{i,j}, 
\end{equation}
where $\delta_{i,j}$ is the Kronecker delta.
This set constitutes a basis of $\L(\H)$ that is orthogonal with respect to the HS scalar product, called an HS basis. By convention, we fix $\sigma_{0}=\id$, which implies that all other basis elements are traceless. Every operator $A$ can therefore be represented in this basis as $A =\frac{1}{d} \sum_{i} \tr (\sigma_{i} \,A) \,\sigma_{i}$.
The Pauli matrices (with the identity operator) are an example of an HS basis for a two-dimensional Hilbert space. See, e.g., \citep{bertlmann08} for explicit constructions in arbitrary dimensions.

For bipartite process matrices, we can choose a product HS basis
\begin{equation}
\sigma_{ijkl} \coloneqq \sigma^{A_I}_{i}\otimes\sigma^{A_O}_{j}\otimes\sigma^{B_I}_{k}\otimes\sigma^{B_O}_{l},
\end{equation}
where each $\left\{\sigma_{i}^X\right\}_{i}$ is an HS basis of the corresponding space $\L(\H^X)$. It is then useful to introduce \textit{term types}: a basis element $\sigma_{j0 0 0 }$, with $j>0$, is of type $A_I$, as it acts nontrivially on $\H^{A_I}$ only. $\sigma_{j k 0 0 }$ with $j,k>0$, is of type $A_IA_O$, and so on. (Note that the label order is irrelevant in the definition of ``term types'': $A_IB_O$ is the same as type $B_OA_I$, etc.)

The constraint $\tr W=d^O$ fixes the coefficient of the $\sigma_{0 0 0 0 } = \id$ term, which motivates us to write a generic process matrix as
\begin{equation}\label{tracelessexpansion}
    W = \frac{1}{d^I}\big(\id + \textrm{ traceless terms}\big),
\end{equation}
where $d^I \coloneqq d^{A_I}d^{B_I}$ is now the product of Alice and Bob's input space dimensions. Restricting the traceless terms to (linear combinations of) specific types fully characterizes the space of valid process matrices $\L_\mathrm{V}$. We summarize ``valid'' and ``non-valid'' term types in Table \ref{terms} and refer to \citep{OCB_2012} for a derivation. Note that the traceless terms that can appear in $W$ are exactly those that \emph{do not} appear in the decomposition of $M^A \otimes N^B$, where $M^A$ and $N^B$ are Choi operators of quantum channels.

Given an operator of the form \eqref{tracelessexpansion}, written in an HS basis, it is now straightforward to verify whether it is a valid process matrix. For example, $W_\mathrm{OCB}$ from Eq.~\eqref{eq:W_OCB} only contains terms of valid types. Having verified that $W_\mathrm{OCB}\geq 0$ (e.g., by checking that all eigenvalues are non-negative) and that the coefficient that multiplies the identity term is $\frac{1}{d^I}$, it follows that it is indeed a valid process matrix.

\begin{table}[t]
    \centering
    \renewcommand{\arraystretch}{1.25}
    \setlength{\tabcolsep}{8pt}
    \begin{tabular}{|c|c|c|}
    \hline
    \multirow{3}{*}{\begin{tabular}{@{}c@{}}Valid\\ types\end{tabular}}
        & $\displaystyle A\vert\vert B$
        & \makebox[4.7cm][l]{$A_I$, $B_I$, $A_IB_I$} \\
    \cline{2-3}
        & $\displaystyle A\prec B$
        & \makebox[4.7cm][l]{$A_OB_I$, $A_IA_OB_I$} \\
    \cline{2-3}
        & $\displaystyle B\prec A$
        & \makebox[4.7cm][l]{$B_OA_I$, $B_IB_OA_I$} \\
    \hline
    \multicolumn{2}{|c|}{Non valid types}
        &
        \begin{tabular}{@{}l@{}}
        $A_O$, $B_O$, $A_OB_O$, $A_IA_O$, $B_IB_O$,\\
        $A_IA_OB_O$, $B_IB_OA_O$, $A_IA_OB_IB_O$
        \end{tabular}
        \\
    \hline
    \end{tabular}
    \caption{\label{terms}
    \textbf{Traceless term types.} A process can be decomposed in an operator basis (HS basis), consisting of the identity and a set of traceless terms, see Eq.~\eqref{tracelessexpansion}. Only traceless terms of the ``valid types'' can appear in a process matrix. In the bipartite case considered here, terms of type $A||B$ can only generate no-signaling correlations, while types indicated with $A\prec B$ (resp., $B\prec A$) can generate correlations where $A$ (resp., $B$) can signal to $B$ (resp., $A$). ``Non valid types'' cannot appear in a (standard) process matrix, but are meaningful in more general objects, such as conditional and postselected processes (see Secs.~\ref{conditionalprocesses} and \ref{postselectedprocesses} below).}
\end{table}

The HS term types have an intuitive interpretation in terms of the correlations they produce. First, note that the identity term does not produce any (non-trivial) correlation: $\tr[(M^A_a\otimes N^B_b)\frac{\id}{d^I}] = \tr(M^A_a)/d^{A_I}\tr(N^B_b)/d^{B_I}$, which makes Alice's and Bob's outcomes independent of each other and of the other party's choice of operation.%
\footnote{$P(a,b) = \tr(M^A_a)/d^{A_I}\tr(N^B_b)/d^{B_I}$ is the statistics obtained if each party implements their instrument on (independent) maximally mixed states.}
Second, note that any non-trivial correlation must arise from matching terms of the same type in $M^A_a\otimes N^B_b$ and in $W$, because the product of two traceless terms of different types is traceless (hence it gives a null contribution when calculating probabilities via the Born rule).

Now consider, as an example, a protocol where Alice ignores her input and prepares a state $\rho_{|x}$, conditioned on some classical variable $x$, while Bob measures a POVM $\{E_b\}_b$ and prepares the maximally mixed state. The corresponding instruments (in Choi representation) are $\{M^A_{|x}\}{=}\{\id^{A_I}\otimes\rho^{A_O}_{|x}\}$ (a single CPTP map for each choice of $x$) and $\{N^B_b\}_b=\{{E^{B_I}_b}^T\otimes\frac{\id^{B_O}}{d^{B_O}}\}_b$.
We can ask: what term types in the process matrix can contribute to signaling from $A$ to $B$---i.e., to correlations between $x$ and $b$---in this protocol? Expanding each local operation in an HS basis, we have $M^A_{|x} = \frac{1}{d^{A_O}}(\id^{A_IA_O} + \sigma_{|x}^{A_O})$, $N^B_b = c_b\id^{B_IB_O} + \tau_b^{B_I}$ for some traceless $\sigma_{|x}$ and $\tau_b$ (with $\sum_b \tau_b =0$) and some coefficients $c_b$ (with $\sum_b c_b =1$). We see that the only terms in $M^A_{|x}\otimes N^B_b$ that depend on both $x$ and $b$ are of type $A_O$ and $A_OB_I$: $c_b \, \sigma^{A_O}_{|x}\otimes \id^{A_IB_IB_O}$ and $\sigma^{A_O}_{|x}\otimes\tau_b^{B_I}\otimes \id^{A_IB_O}$, respectively. As terms of type $A_O$ do not appear in the process matrix, we conclude that the terms responsible for signaling in this protocol must be of type $A_OB_I$.

A similar analysis holds for other types: $A_IB_I$ can only generate no-signaling correlations, while terms of type $A_IA_OB_I$ can yield signaling if Alice performs operations that jointly involve her input and output systems nontrivially (e.g., a unitary operation has terms of type $A_IA_O$; combined with Bob's POVM, it gives terms of type $A_IA_OB_I$ in Alice and Bob's joint instrument, meaning that Alice can signal to Bob by changing the choice of unitary).

It is also interesting to look at the term types arising in causally ordered processes. no-signaling processes, $W=\rho^{A_IB_I}\otimes\id^{A_OB_O}$, contain (in addition to $\id^{AB}$) terms of types $A_I$, $B_I$ and $A_IB_I$---i.e., all the possible no-signaling terms. These terms can also appear in a channel without memory, $W= \rho^{A_I}\otimes C^{A_OB_I}\otimes \id^{B_O}$,  which may also include signaling terms of types $A_OB_I$ and $A_IA_OB_I$ (a channel $C^{A_OB_I}$ can only contain types $B_I$ and $A_OB_I$, because terms of type $A_O$ are forbidden by the trace-preserving condition $\tr_{B_I} C^{A_OB_I} = \id^{A_O}$). General channels with memory do not introduce any additional term types; therefore, we find that all the valid term types can arise from processes compatible with either $A\prec B$ or $B\prec A$. A consequence of this observation is that term types do not distinguish processes with definite or indefinite causal order, a point we will come back to in Sec.~\ref{subsec:characterization_sep}.

For general multipartite processes, which we discuss in Sec.~\ref{multipartite}, the identification of term types with possible correlations can be more involved, but the general structure remains: any term of a valid type in the process contributes to the probabilities only if it matches terms of the same type in the local operations. Furthermore, it remains true that the valid term types coincide with those arising from (arbitrarily) causally ordered process matrices.

Let us finally note that, despite the fact that they are forbidden in ``standard'' process matrices as introduced above, non-valid term types can have an interpretation in conditional processes, where we fix a measurement outcome at one or more sites (see Sec.~\ref{conditionalprocesses}), or in extensions of the process matrix formalism, including probabilistically postselected processes or processes arising from certain models of closed timelike curves (Sec.~\ref{postselectedprocesses}).

\paragraph{Basis-independent characterization.} \label{para:HSbasis_indep_charact}

The set of valid process matrices may also be characterized without referring to a particular operator basis. \citet{Araujo_2015} presents a method to characterize process matrices solely in terms of partial traces and a positive semidefinite constraint. More precisely, a linear operator $W\in\mathcal{L}(\mathcal{H}^{A_I}\otimes \mathcal{H}^{A_O}\otimes \mathcal{H}^{B_I}\otimes \mathcal{H}^{B_O})$ is a bipartite process matrix if and only if it respects
\begin{subequations}
\begin{align}
    & W\geq0,\label{eq:pos_constraint}\\
    & \tr(W)=d^O, \label{eq:trace_condition} \\
    & _{[1-A_O]}\tr_B W=0, \ _{[1-B_O]}\tr_A W=0, \label{eq:lin_constraint1} \\
    & _{[1-A_O][1-B_O]}W = 0, \label{eq:lin_constraint2}
\end{align}
\end{subequations}
where $_XW\coloneqq\tr_X(W)\otimes\frac{\id^X}{d^X}$ is the \textit{trace-and-replace} map, which consists in tracing out the system in space $\mathcal{H}^X$ and replacing it by the normalized identity operator, and with $_{[1-X]}W\coloneqq W-{}_XW$.%
\footnote{The trace-and-replace notation naturally involves a re-ordering of the Hilbert spaces. It is associative and commutative, so that for instance ${}_{XY}W = {}_X({}_YW) = {}_Y({}_XW)$, ${}_{[1-X]Y}W = {}_{[1-X]}({}_YW)$, $_{[1-X][1-Y]}W=_{[1-X]}(_{[1-Y]}W) = W - {}_XW - {}_YW + {}_{XY}W$. Note also that the map $W\mapsto {}_XW$ is trace preserving; that it is self-adjoint, in the sense that $\tr[S ({}_XW)] = \tr[({}_XS) W]$; and that it acts as a projector, so that ${}_X({}_XW) = {}_XW$ (as well as $_{[1-X][1-X]}W={}_{[1-X]}W$).}
The constraints of this characterization may be split in three natural parts: a positivity constraint given by Eq.~\eqref{eq:pos_constraint}, an affine trace normalization condition of Eq.~\eqref{eq:trace_condition}, and linear constraints 
given by Eqs.~\eqref{eq:lin_constraint1}--\eqref{eq:lin_constraint2}.%
\footnote{One may note that Eq.~\eqref{eq:lin_constraint1} can also be written as ${}_{[1-A_O]B}W = {}_{[1-B_O]A}W = 0$; Eqs.~\eqref{eq:lin_constraint1}--\eqref{eq:lin_constraint2} can then be combined together as a single linear constraint in the (equivalent) form ${}_{[1-A_O][1-B_O]}W + {}_{[1-A_O]B}W + {}_{[1-B_O]A}W = 0$ \citep{Araujo_2015}.}
The set of linear constraints Eqs.~\eqref{eq:lin_constraint1}--\eqref{eq:lin_constraint2} defines the subspace $\L_\mathrm{V}$ of valid process matrices introduced previously. It can be verified \citep{Araujo_2015} that the ``allowed terms'' considered in the explicit HS decompositions above are precisely those that satisfy these linear constraints.

\paragraph{Single-site processes.}\label{singleparty}

A particular type of process is one with a single site, $W^{A_IA_O}$. For such a process, the affine constraint reads $\tr[( M^{A_IA_O})^T\, W^{A_IA_O}]{=}1$ 
for every $M$ such that $\tr_{A_O}M^{A_IA_O}{=}\id^{A_I}$. This readily translates to $\tr W = d^{A_O}$ and $_{[1-A_O]} W{=}0$ or, equivalently, into the condition that $W^{A_IA_O}$ can only contain the (properly normalized) identity and HS terms of type $A_I$. Adding positivity, we find that single-party processes have the form
\begin{equation}\label{singleW}
    W^{A_IA_O} = \rho^{A_I}\otimes \id^{A_O}
\end{equation}
for some density matrix $\rho$ \citep{OCB_2012, morimae2014process}.

A consequence is that a single party, in isolation, cannot detect an indefinite causal structure, because all they can measure is a state, just as they would in a causally ordered scenario. Indefinite causal structure---in fact, any non-trivial causal structure---only appears when considering statistics gathered across two or more sites.

\paragraph{Dual affine of no-signaling channels.} \label{dualaffine}

Process matrices may also be defined and characterized as belonging to the dual affine set of no-signaling channels \citep{Chiribella2016optimal_networks}.
A multipartite quantum channel is said to be no-signaling when it cannot be used to send information from different sites \citep{Beckman2001,Eggeling2002,Piani2006}.
E.g., a bipartite quantum channel $\mathcal{C}:\mathcal{L}(\mathcal{H}^{A_I}\otimes \mathcal{H}^{B_I}) \to \mathcal{L}(\mathcal{H}^{A_O}\otimes \mathcal{H}^{B_O}) $ is no-signaling if and only if Alice's marginal output state ($\tr_{B_O}\mathcal{C}(\rho{^{A_IB_I}})$) only depends on her own part of the input state ($\tr_{B_I}\rho{^{A_IB_I}}$) and not on Bob's part, and vice versa. The definition for the multipartite case follows analogously.
It can be shown that a multipartite channel is no-signaling if and only if it can be written as a linear combination of independent (i.e., product) channels \citep{Chiribella2013,Gutoski2009}. 
That is, a Choi operator $C\in\L(\H^{A_I}\otimes \H^{A_O}\otimes \H^{B_I}\otimes \H^{B_O} \otimes \cdots)$ is that of a no-signaling channel if and only if $C\geq0$ and it admits some affine decomposition $C=\sum_i \gamma_i \,M^A_i \otimes N^B_i \otimes \cdots$, where $M^A_i, N^B_i, \ldots$ are Choi operators of quantum channels
and the $\gamma_i$'s are real (but not necessarily positive) numbers respecting $\sum_i \gamma_i = 1$. 

The dual affine of a set of linear operators $\mathcal{S}\subseteq\L(\H)$ is the set of all operators $W\in\L(\H)$ that satisfy $\tr(S^T\, W)=1$ for any $S\in\mathcal{S}$. 
In Definition~\ref{def:process}, process matrices are defined as positive semidefinite operators $W\geq0$ such that, for any set of independent channels with Choi operators $M^A$, $N^B$, \ldots, we have that $\tr[(M^A \otimes N^B\otimes\cdots)^T\, W^{AB\dots}] = 1$. 
Due to linearity and to the observation above, Definition~\ref{def:process} is equivalent to imposing that $\tr[(C^{AB\dots}_\mathrm{nsig})^T\, W^{AB\dots}] = 1$, for all no-signaling channels with Choi operator $C^{AB\dots}_\mathrm{nsig}$.
Hence, the set of valid process matrices is the set of positive semidefinite operators that are in the dual affine of independent channels, or of no-signaling channels, equivalently.
This dual affine formulation provides a different interpretation of processes matrices and allows for a simple connection between primal and dual formulations for some optimization problems \citep{Chiribella2016optimal_networks,Bavaresco2020,Bavaresco2020b}. 

\paragraph{Process matrices and semidefinite programming.} \label{para:SDP} 

As presented in this section, the set of valid process matrices is characterized by positive semidefinite, affine and linear constraints. Sets of this form may be analyzed via semidefinite programming (SDP), a well-studied class of convex optimization problems that admits efficient and robust numerical approaches \citep{boyd,watrous_2018,Skrzypczyk2023}. SDP methods have found applications in several branches of quantum information science, and investigating indefinite causality is one of them.  Further below, we will in particular discuss how to employ SDP methods for detecting and quantifying indefinite causality \citep{Araujo_2015,Branciard_2016,Bavaresco_2019,Dourdent2022}, characterizing processes admitting a classical or quantum control circuit implementation \citep{Wechs2021}, and for analyzing the performance of process matrices in several other tasks such as maximizing the violation of causal inequalities \citep{Branciard2016}, transforming unitary operations \citep{Chiribella2016optimal_networks,Quintino2019,Quintino2019b}, channel discrimination \citep{Bavaresco2020,Bavaresco2020b}, comparing unknown unitary channels \citep{hashimoto22}, quantum metrology \citep{Liu2023a,mothe2023reassesing}, evaluating the quantum query complexity \citep{abbott2024query}, and for finding optimal ordered simulations of the quantum switch \citep{bavaresco2024simulated}. These and other results are discussed in Sec.~\ref{sec:applications}.

\subsubsection{Probing quantum processes and generalized observables}
\label{observables}

\begin{figure}[htb] 
	\begin{center}
		\includegraphics[width=.6\columnwidth]{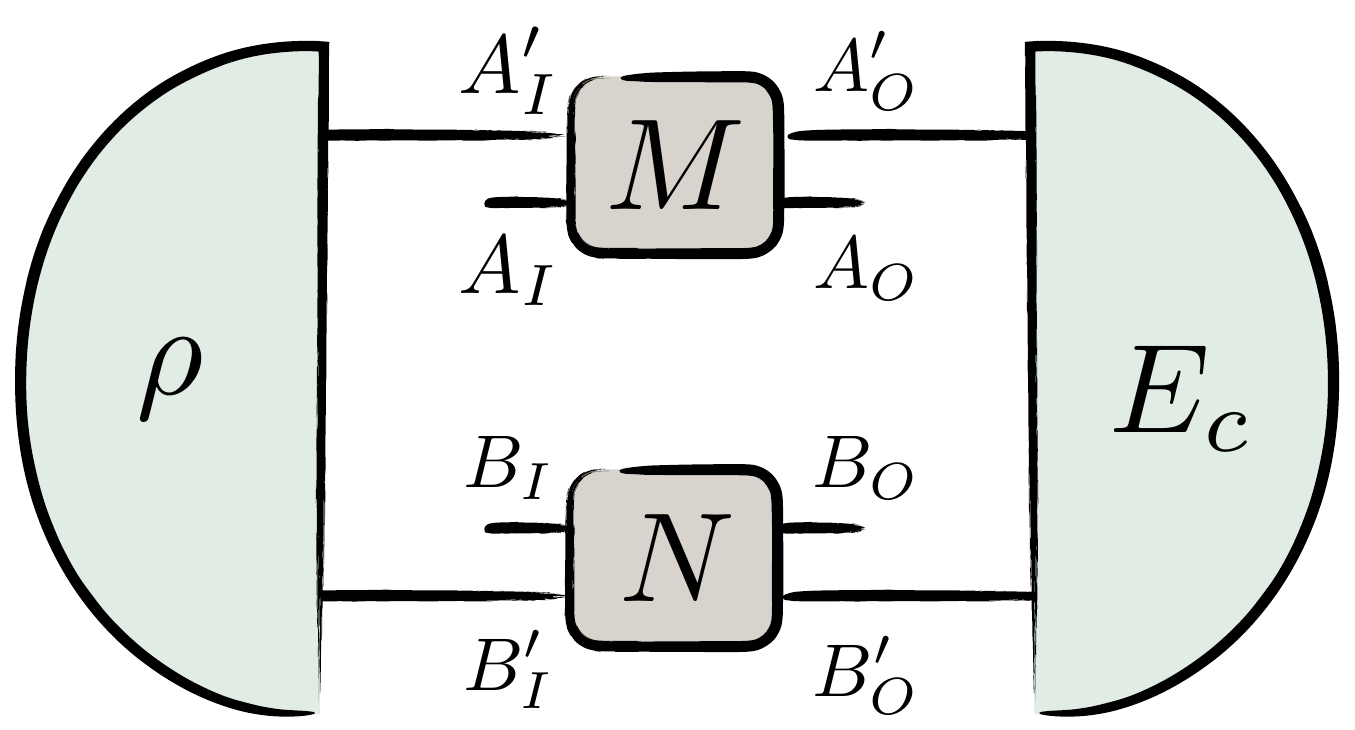} 
	\end{center}
\caption{Pictorial representation of the most general  measurement on a bipartite process matrix realizable through local operations, Eq.~\eqref{sandwhichoperation}, where $A_I$ ($B_I$) and $A_O$ ($B_O$) label Alice’s (Bob's) input and output spaces and $A_I'$, $B_I'$, $A_O'$, $B_O'$ are the corresponding arbitrary external auxiliary systems.  This operation can be ``plugged'' into the two slots of a bipartite process, Fig.~\ref{fig:ProcessMatrix}.} 
\label{fig:probing_qproc}
\end{figure}

One may wonder how one can, in general, probe and extract information from quantum processes, i.e., apply some operations that return some outcome. So far, we have mostly discussed operations consisting of local instruments that only act on the input and output space of the respective sites.
More generally, we can allow each party to include arbitrary external auxiliary systems in their operations: for example, Alice can perform CP maps $M^{A_IA_I'A_OA_O'}$, where the primed labels denote additional systems that can be prepared and measured independently of the probed process. As these auxiliary systems can be prepared in an arbitrary, possibly entangled, state, we call these general operations \textit{entanglement assisted local instruments}---with the understanding that arbitrary joint measurements of the auxiliary systems are also allowed. Recalling that every local instrument can be implemented by letting the system interact with a respective auxiliary system through a channel, and then measuring the auxiliary system, see e.g.,\ \citep{wilde2011book}, we do not lose in generality if we assume that the local operations are CPTP. Therefore, the most general measurement that can be performed through local operations on a bipartite process takes the form%
\footnote{Note that the most general operations that produce well-defined probability distributions, $P(c) = \tr [M_c^T\, W]$, are given by sets of positive semidefinite operators $\{M_c\}_c$, such that $\sum_c M_c$ is a no-signaling channel, see also Sec.~\ref{dualaffine}. However, such operators may not be realizable using local operations and shared no-signaling resources \citep{Beckman2001}, and therefore are outside the operational framework considered here.}
(in terms of the link product, recall Sec.~\ref{subsubsec:link_prod})
\begin{multline} \label{sandwhichoperation}
    M_c^{A_IA_OB_IB_O} = \\
    \! \! \rho^{A_I'B_I'} * \left(M^{A_IA_I'A_OA_O'} \otimes N^{B_IB_I'B_OB_O'}\right) * (E_c^{A_O'B_O'})^T,
\end{multline}
where $\rho$ is an arbitary joint state, $M$ and $N$ are CPTP maps, and $\{E_c\}_c$ is a POVM, see Fig.~\ref{fig:probing_qproc}.

It is also useful to make a connection with the notion of observables in ordinary quantum theory, defined as operators $O$ whose expectation values on a state $\rho$ are given by $\langle O\rangle = \tr [O\rho]$.  Generalizing to arbitrary processes, an operator $S$ living on the product of all input and output spaces can be associated with an operational procedure, producing an expectation value $\langle S \rangle =\tr[S^T \, W]$ (where the transpose is introduced for consistency with the way we write the generalized Born rule, see Eq.~\eqref{processborn} and the comment that follows it). Concretely, it is always possible to decompose any given operator $S$ in the form
\begin{equation}
    S = \sum_{x,y,a,b} \gamma_{x,y,a,b} \, M_{a|x}^A \otimes N_{b|y}^B \label{eq:decomp_S}
\end{equation}
for some positive semidefinite matrices $M_{a|x}^A$, $N_{b|y}^B$, and for some coefficients $\gamma_{x,y,a,b}$.
Moreover, the matrices $M_{a|x}^A$, $N_{b|y}^B$ can be interpreted as the Choi matrices of local CP maps, which, without loss of generality (up to a rescaling of the coefficients $\gamma_{x,y,a,b}$) can be assumed to be trace-nonincreasing, and which can thus be taken to be part of some quantum instruments $\{M_{a|x}^A\}_a$ and $\{N_{b|y}^B\}_b$ labeled by $x$ and $y$ \citep{Araujo_2015}. 
Such instruments can be implemented on an arbitrary process $W$, so that the expectation value $\langle S \rangle$ can be obtained by combining the probabilities for the individual CP maps:
\begin{align}
   \langle S \rangle = \tr[S^T \, W] & = \sum_{x,y,a,b} \gamma_{x,y,a,b} \, \tr[(M_{a|x}^A \otimes N_{b|y}^B)^T \, W] \notag \\
    & = \sum_{x,y,a,b} \gamma_{x,y,a,b} \, P(a,b|x,y). \label{eq:reconstruct_S}
\end{align}
Notice that some of the coefficients $\gamma_{x,y,a,b}$ may vanish, meaning that not all the outcomes in an instrument necessarily contribute to the expectation value. Furthermore, just as for observables on states, the decomposition \eqref{eq:decomp_S} is not unique, so that there can be different procedures associated with the same observable $S$. An even larger variety of equivalent procedures can be found by considering decompositions of the form $S = \sum_{z,c} \gamma_{z,c} \, M_{c|z}$, where $M_{c|z}$ are given by entanglement-assisted local instruments, as in Eq.~\eqref{sandwhichoperation}.

Note that, in ordinary quantum mechanics, observables are typically associated with self-adjoint operators, for which the spectral decomposition provides a privileged projective measurement realization. 
Similarly, one can (and one typically does) restrict generalized observables to be self-adjoint. The only caveat is that the spectral decomposition might not readily provide an instrument measurable through local operations. This is because, given a spectral decomposition $S=\sum_c \lambda_c \,M^{AB}_c$, where the $M^{AB}_c$'s are proportional to orthogonal projectors and $\sum_c M^{AB}_c = \frac{1}{d^O} \id^{AB}$  (in which the normalization ensures that $\{M^{AB}_c\}_c$ is a valid instrument), it might not always be possible to express the whole set $\{M^{AB}_c\}_c$ in the form \eqref{sandwhichoperation} for a complete POVM $\{E_c\}_c$ and a fixed choice of $M,N,\rho$. In practice, it is typically more convenient to combine measurements from several local instruments, as in the decomposition \eqref{eq:decomp_S}. 

\subsubsection{Process matrices as constrained states or channels}
\label{processstatechannel}

\paragraph{Analogies and disanalogies with states.}\label{processasstate}

\begin{figure}[hbt] 
	\begin{center}
		\includegraphics[width=.65\columnwidth]{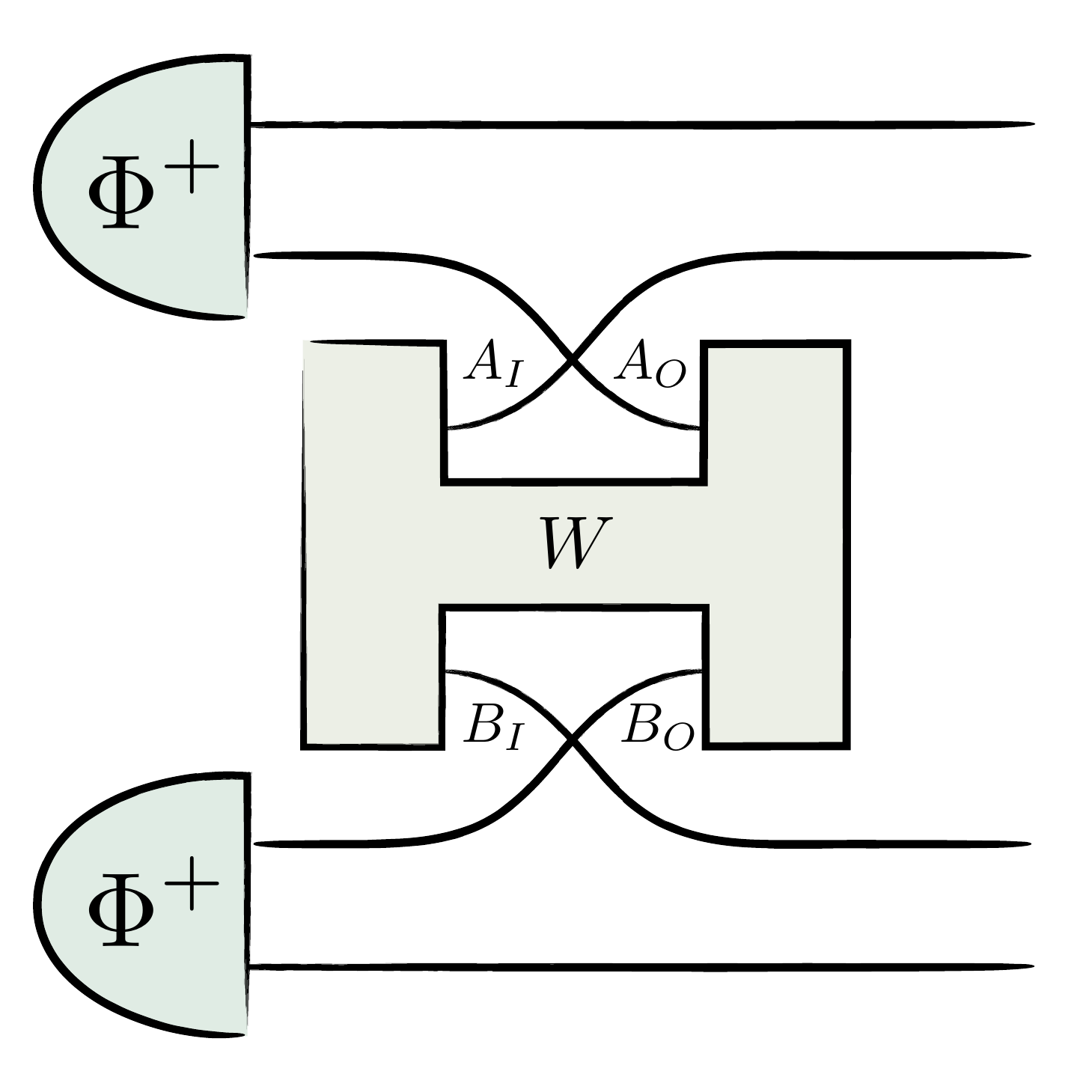} 
	\end{center}
\caption{Mapping a process to a state. The state (density matrix) obtained by plugging half of a maximally entangled state into each local output is proportional to the process matrix.}
\label{fig:ProcAsSt}
\end{figure}

The Born rule for processes, Eq.~\eqref{processborn}, can be seen as a generalization of the ordinary Born rule for quantum states, where POVMs are replaced by instruments and density matrices are replaced by process matrices. Indeed, since the process matrix is positive semidefinite and it has a fixed trace, it is formally proportional to a density matrix. It is in fact always possible to transform a process into a state: by feeding half of a maximally entangled state $\ket{\Phi^+}\propto\kket{\id}$ into the local output space at each site, and combining the other halves with the systems from the local input spaces, one obtains a state whose density matrix is equal to the process matrix, up to normalization and ordering of the tensor factors: see Fig.~\ref{fig:ProcAsSt}.  Thanks to their formal and physical relation, several properties and techniques associated with states can be transferred to processes. However, there are also important differences.

First, we should emphasize the different physical meaning of the Hilbert spaces in which states and processes are defined. For states of composite quantum systems, a decomposition into tensor factors represents a division into subsystems, where all subsystems are usually considered to coexist ``at a given time''. By contrast, even in the case of causally ordered processes, different sites (or parties) associated with corresponding tensor factors may represent a single system at different times.%
\footnote{More generally, the traditional notion of ``system'' does not always apply to a process with indefinite causal order, in the sense that we may not be able to interpret experiments in terms of evolving systems that are exchanged between laboratories \citep{GRINBAUM201722}.}
Furthermore, each site in a process is generally associated with two tensor factors (one input and one output space---although one of them may be trivial), while states have a single tensor factor for each subsystem: an input space. The physical difference between input and output spaces is that the output is responsible for the causal influence (i.e., signaling), out of a site. As we will discuss in more details in Sec.~\ref{processcomposition} below, processes built on tensor products of sites that contain non-trivial input and output spaces generally cannot be interpreted as being attached to a ``larger site'', while this is always possible for states. 

Another important difference, on the mathematical level, is that process matrices are subject to additional linear constraints, Eqs.~\eqref{eq:lin_constraint1}--\eqref{eq:lin_constraint2}. This means that, even though all processes can be mapped to states, not all states correspond to valid processes, for a given decomposition of the Hilbert space into different parties and into their input and output systems. A notable consequence is that, depending on the scenario, pure states may not have a corresponding process (i.e., rank-one projectors can violate the linear validity constraints of process matrices). This is true in particular for the bipartite cases we have considered so far: if the input and output spaces have the same (nontrivial) dimension, then the process matrix cannot have rank one.%
\footnote{Even in a multipartite case, if we plug unitaries in all but one sites of a rank-one operator, we obtain a rank-one operator for the remaining site. This cannot be a valid process matrix, because single-party process matrices must have the identity on the output space, see Sec.~\ref{singleparty}.
As discussed in Sec.~\ref{subsec:unitary_processes} below, rank-one processes exist if at least one party has a trivial output space. It is an open question 
whether (allowing for different local input and output dimensions) there exist rank-one process matrices where all outputs are non-trivial.}
This in fact opens some potential ambiguity in the notion of ``pure process'': while pure (i.e., rank-one) states can be equivalently characterized as extremal (namely as states that cannot be written as non-trivial convex combinations of other states), a similar identification does not work for processes---indeed, not all process matrices can be decomposed as convex combinations of valid rank-one processes. Here we will keep the identification of purity with rank one, and accept that not all extremal processes are pure; see also Sec.~\ref{subsec:unitary_processes} for a different notion of ``purity''.  

Yet another difference arises from the probabilities that processes and states can generate. Processes can be probed (at each site) with instruments, which (in the Choi form) satisfy $\sum_a \tr_{A_O}M_a^{A_IA_O} = \id^{A_I}$, while the POVMs used to probe states satisfy the stronger condition $\sum_a E_a = \id$. Thus, all POVMs (with appropriate renormalization and factorizing the Hilbert space so as to interpret it as an input-output pair) can be mapped to instruments, but the opposite is not true \citep{Costa2018}. This is the reason that process matrices allow signaling, despite the formal analogy with states.

Finally, states and processes differ in the way one restricts from a set of parties to a subset. For states, one can ``ignore'' a party by tracing out the corresponding subsystem: the \textit{reduced state} $\rho^A=\tr_{B}\rho^{AB}$ provides all the information relevant to observables on site $A$ alone. One can also define (up to normalization) the \textit{conditional state} $\rho^A_{E_b}=\tr_{B}\left[(\id^A\otimes E_b^B)\rho^{AB}\right]$ to reproduce the observations at $A$ conditioned on observing a POVM element $E_b$ at $B$. 
In the case of process matrices, it is not always possible to ``ignore a party'' without specifying (or making a guess about) the operation it performs. Indeed, signaling from $B$ to $A$ is precisely the property that a choice of CPTP map at $B$ changes the marginal statistics at $A$. 
Furthermore, given a valid process matrix $W^{AB}$, the operator $W^A_{M_b}=\tr_{B}\left[(\id^A\otimes M_b^B)^T\, W^{AB}\right]$ obtained by fixing a CP map $M_b$ at $B$ is not necessarily a valid process matrix, even up to normalization. In Sec.~\ref{conditionalprocesses}, we will formally define such operators as ``conditional process matrices'', in analogy to conditional states,  and discuss their properties in more details.

\paragraph{Processes as channels.} \label{processaschannel}

\begin{figure}[!tbp] 
	\begin{center}
		\includegraphics[width=\columnwidth]{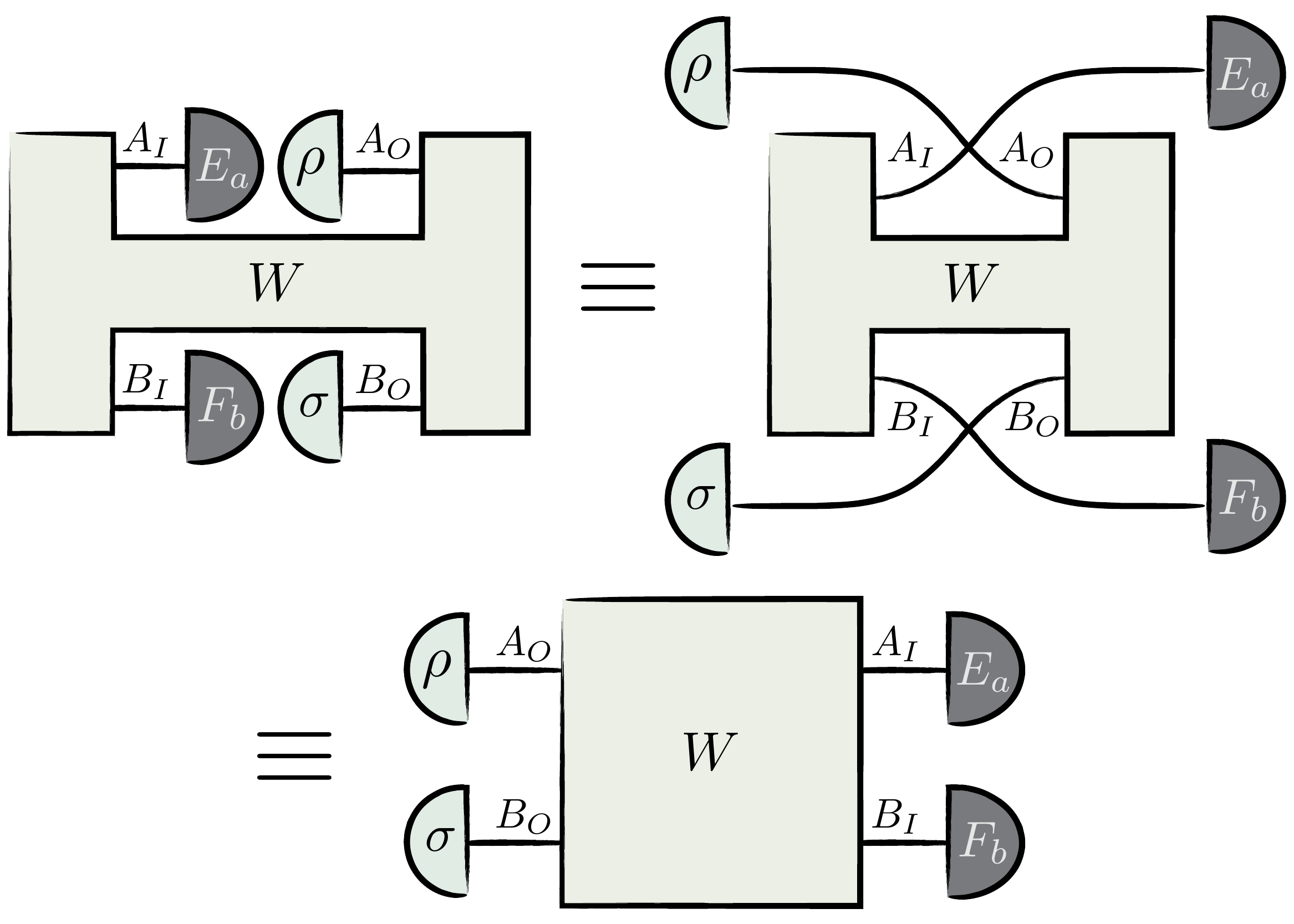} 
	\end{center}
\caption{Pictorial representation of the correspondence between processes and channels. In the top left representation, Alice (Bob) measures the input system with a POVM element $E_a$ ($F_b$) and, independently of the measurement outcome, reprepares a state $\rho$ ($\sigma$), yielding the probabilities given in Eq.~\eqref{eq:process2cahnnel}. In the top right, measurements and preparations on the process are ``pulled out'' for visual convenience; an equivalent interpretation is a scenario where the local systems are swapped with external auxiliary systems. One can thus see $W$ as a channel from $A_OB_O$ to $A_IB_I$, as illustrated at the bottom.}
\label{fig:ProcAsChan}
\end{figure}

Process matrices can also be put in relationship with quantum channels. This can be very useful in addressing several questions, as it allows applying to processes a broad range of existing tools and results for channels. A direct way to see the relationship is that a process can directly be used as a channel from all output spaces to all input spaces. This is the case when Alice (resp., Bob) prepares a state $\rho$ (resp., $\sigma$) after having measured a POVM element $E_a$ (resp., $F_b$), where the states are prepared independently of the measurement outcomes, as illustrated in the top left of Fig.~\ref{fig:ProcAsChan}. Plugging the corresponding instruments into the Born rule for processes, we get the probabilities
\begin{multline} \label{eq:process2cahnnel}
   \hspace{-3mm} P(a,b) = \tr \! \left[\left(\!\big(E^{A_I}_{a}\big)^{\!T} \!\otimes\rho^{A_O}\otimes \big(F^{B_I}_{b}\big)^{\!T} \!\otimes \sigma^{B_O}\! \right)^{\!T} W^{AB}\right] \\
    = \tr \left[\left(E^{A_I}_{a} \otimes F^{B_I}_b\right) \mathcal{W}(\rho^{A_O}\otimes\sigma^{B_O})^{A_IB_I}\right],
\end{multline}
where the map $\mathcal{W}: \mathcal{L}(\mathcal{H}^{A_OB_O}) \to \mathcal{L}(\mathcal{H}^{A_IB_I})$ is defined by 
\begin{equation}
    \mathcal{W}(\rho^{A_O}\otimes\sigma^{B_O}) \coloneqq  \tr_{A_OB_O} \left[\left(\rho^{A_O}\otimes \sigma^{B_O}\right)^T  W^{AB}  \right].
\end{equation}
This is precisely the expression of the inverse Choi isomorphism from Eq.~\eqref{eq:inverse_Choi}, applied to $W^{AB}$, so effectively Alice and Bob receive in their input space the state obtained by applying $\mathcal{W}$ to the states they prepare and send out into their output spaces (see the bottom of Fig.~\ref{fig:ProcAsChan}). Since the affine constraints on $W^{AB}$ imply%
\footnote{This can for instance be seen, using Eqs.~\eqref{eq:lin_constraint1}--\eqref{eq:lin_constraint2}, by writing ${}_{[1-A_OB_O]}({}_{A_IB_I}W) = {}_{A_IB_I}W - {}_{A_I}({}_{A_OB_IB_O}W) - {}_{B_I}({}_{A_IA_OB_O}W) + {}_{A_IB_IA_OB_O}W = {}_{A_IB_I}W - {}_{A_I}({}_{B_IB_O}W) - {}_{B_I}({}_{A_IA_O}W) + {}_{A_IB_IA_OB_O}W = {}_{A_IB_I[1-A_O][1-B_O]}W = 0$, which means that $\tr_{A_IB_I}W \propto \id^{A_OB_O}$. The normalization condition of Eq.~\eqref{eq:trace_condition} then gives the exact proportionality constant.}
$\tr_{A_IB_I}W = \id^{A_OB_O}$, and since $W\geq 0$, it follows that $\mathcal{W}$ is a CPTP map, of which the process matrix $W$ is precisely the Choi representation.
An alternative way to implement the channel $\mathcal{W}$ is to let each party perform a swap operation with external auxiliary systems of appropriate dimension, and perform preparations and measurements on the auxiliary systems (top right of Fig.~\ref{fig:ProcAsChan}). 

When interpreting the process as a channel $\mathcal{W}$ from outputs to inputs (with each input preceding the output according to the local time direction at each site), the latter might be seen as somehow working ``backwards in time''. When one plugs in (arbitrary) local operations, the channel and local operations might then be seen as being composed ``in a loop''.
As the linear constraints for processes are strictly stronger than the trace-preserving condition for channels, not all channels correspond to valid processes, and the additional constraints may be interpreted as preventing any paradoxes arising from such ``loops''. More specifically, the channels $\mathcal{W}: \L(\H^{A_OB_O})\rightarrow \L(\H^{A_IB_I}), \rho\mapsto \mathcal{W}(\rho)$ that do correspond to valid processes are those that are no-signaling from $A_O$ to $A_I$ (i.e. such that $\tr_{B_I}\mathcal{W}(\rho)$ can only depend on $\tr_{A_O}(\rho)$, which translates the constraint $_{[1-A_O]B_I}W=0$ that follows from Eqs.~\eqref{eq:lin_constraint1}--\eqref{eq:lin_constraint2}), no-signaling from $B_O$ to $B_I$, and that further satisfy $\mathcal{W}(\rho) = \mathcal{W}({}_{A_O}\rho+{}_{B_O}\rho-{}_{A_OB_O}\rho)$, for any  $\rho\in\L(\H^{A_OB_O})$ (which translates the constraint $_{[1-A_O][1-B_O]}W=0$). Another equivalent way to translate the latter constraint is as follows: suppose the channel $\mathcal{W}$ is fed with some states $\rho_{x,y}^{A_OB_O}=\rho_x^{A_O}\otimes\rho_y^{B_O}$, with two possible local input states on each side ($x,y=0,1$) chosen at random; then the average output state does not depend on the parity of the inputs $x\oplus y$ (i.e., $\mathcal{W}(\rho_{0,0})+\mathcal{W}(\rho_{1,1})=\mathcal{W}(\rho_{0,1})+\mathcal{W}(\rho_{1,0})$). This property was called ``parity erasure'' in \citep{liu2025}; see also the discussion in the multipartite case in Sec.~\ref{para:multipartite_charact}.

\paragraph{Tensor product of processes.}
\label{processcomposition}

Given a set of states $\{\rho_j\}_j$, their tensor product $\bigotimes_j \rho_j$ can be interpreted as a single-party state on which arbitrary measurements can be made. The same is true for channels: $
\bigotimes_j\mathcal{M}_j$ can be seen as a channel from a single party, who inputs states from the product input space, to another single party, who can make arbitrary measurements on the product output space. 
However, as discussed in \citep{Chiribella2009}, the tensor product of causally ordered process matrices (there named `quantum combs') is not uniquely defined, and may not lead to a valid ordered process. Also, \citet{Jia2018} presents examples where the tensor product of process matrices is not a valid process matrix, marking another way in which processes differ from states and channels.  

To see why this is the case, it is sufficient to consider two processes $W_{A\prec B}$, $W_{B\prec A}$ corresponding to channels in the opposite direction, say the identity channel from $A$ to $B$ and from $B$ to $A$ (with some arbitrary input state on $A_I$ and $B_I$) respectively. If we try to interpret the product process $W_{A\prec B}\otimes W_{B\prec A}$ as a bipartite process, where the local spaces are the products of the individual spaces, it means we should be able to perform arbitrary joint operations at each site. We can however quickly show how this creates ``causal loops'', leading to non-normalized probabilities. For example, Alice can take the input from $W_{B\prec A}$ (coming from Bob) and feed it, untouched, into $W_{A\prec B}$, which is sent to Bob. Bob can then take the input from $W_{A\prec B}$ and feed it to $W_{B\prec A}$ after performing a basis permutation---say $\sigma_x$ in the case of qubits. After such a ``loop'', Alice receives the opposite of the bit she sent out, in contradiction with the fact that she performs the identity operation.  (This is akin to the ``grandfather paradox'' for time traveling, in which an agent travels to a time before their own grandfather had children and kills him, thereby preventing their own birth \citep{schachner33,barjavel43}). At the formal level, applying the Born rule to this situation would give a factor $(\tr \sigma_x)^2 = 0$, meaning that there is $0$ probability for the protocol to succeed, despite the fact that Alice and Bob perform deterministic operations. In other words, $W_{A\prec B}\otimes W_{B\prec A}$ is not a valid bipartite process matrix. It was shown by \citep{Guerin2019} that, under a few assumptions, there is no workaround: there is no consistent composition rule for processes if one tries to combine the sites from different processes into single sites.

A way to interpret this result is that, when taking tensor products, there is no ground to identify sites from different processes and join them into single sites (in fact, this is clearly ill-defined for products of processes with different numbers of sites). Instead, a product of processes should be interpreted as a process with a larger number of sites, given by the union of sites in the factor processes, where only product operations are allowed. It is easy to see that the resulting operator satisfies the normalization conditions and is indeed a valid (multipartite, see Sec.~\ref{multipartite}) process matrix.

The composition of process matrices, which may not be restricted to tensor product, and conditions for being valid and consistent are characterized in \citep{Kissinger2019,Simmons2022Higer,Simmons2024,Hoffreumon2022_proj,perinotti16higher,bisio19higher} from a category and type theory perspective, and in \citep{milz2023_transformations} from a linear algebra one, see also Subsec.~\ref{subsec:TransfPM}.

\subsection{Causal (non)separability}\label{nonsepsection}

Now that we have defined and characterized process matrices in general, let us turn to causally separable ones.
As introduced and motivated in Sec.~\ref{subsec:W_indefinite_order} already, these are the processes that are compatible with a definite (although not necessarily fixed) causal order. In the bipartite case, causally separable process matrices are simply probabilistic mixtures of causally ordered processes, compatible with $A\prec B$ or $B\prec A$ (see Definition~\ref{def:Wcsep_bipartite}).

\subsubsection{Characterization} \label{subsec:characterization_sep}

It is convenient, for their characterization, to incorporate the values $q$ and $1-q$ defining the weights of the convex combination into the (then no longer normalized) process matrices that are being mixed. Combining Definition~\ref{def:Wcsep_bipartite} with the characterization of Eqs.~\eqref{BOlast}--\eqref{firstastate} for processes compatible with the order $A\prec B$ (and the analogous characterization for $B\prec A$), we obtain that a bipartite causally separable process matrix is one that can be decomposed as 
\begin{subequations}
\begin{align}
 W_\mathrm{c\mhyphen sep} & = W_{(A,B)}^{A_IA_OB_I}\otimes\id^{B_O} + W_{(B,A)}^{B_IB_OA_I}\otimes\id^{A_O} \label{eq:charact_csep_bi1} \\[1mm]
 \text{with } \ & W_{(A,B)}^{A_IA_OB_I} \geq 0, \ W_{(B,A)}^{B_IB_OA_I} \geq 0,\label{eq:charact_csep_bi2} \\
 & \tr_{B_I}W_{(A,B)}^{A_IA_OB_I} = \rho_{(A)}^{A_I}\otimes\id^{A_O}, \label{eq:charact_csep_bi3} \\
 & \tr_{A_I}W_{(B,A)}^{B_IB_OA_I} = \rho_{(B)}^{B_I}\otimes\id^{B_O}, \label{eq:charact_csep_bi4}  \\
 & \tr \rho_{(A)}^{A_I} + \tr \rho_{(B)}^{B_I} = 1. \label{eq:charact_csep_bi5}
\end{align}
\end{subequations}
for some (positive semidefinite) operators $W_{(A,B)}^{A_IA_OB_I} \in \L(\H^{A_IA_OB_I})$, $W_{(B,A)}^{B_IB_OA_I} \in \L(\H^{B_IB_OA_I})$, $\rho_{(A)}^{A_I} \in \L(\H^{A_I})$ and $\rho_{(B)}^{B_I} \in \L(\H^{B_I})$.

One may note that the characterization above implies the validity of the process matrix $W_\mathrm{c\mhyphen sep}$. Each individual term $W_{(A,B)}^{A_IA_OB_I}\otimes\id^{B_O}$, $W_{(B,A)}^{B_IB_OA_I}\otimes\id^{A_O}$ can be characterized according to the ``terms allowed'' in an HS basis (see Sec.~\ref{para:HS_terms}), but their combination in $W_\mathrm{c\mhyphen sep}$ does not exclude any more terms than the ``nonvalid'' ones. 
In other words, the linear span of causally separable processes coincides with the whole space $\L_\mathrm{V}$ of valid process matrices (contrarily to the span of processes compatible with a given fixed order: e.g., processes compatible with $A\prec B$, of the form $W_{(A,B)}^{A_IA_OB_I}\otimes\id^{B_O}$, cannot contain terms of (valid) types $B_OA_I$ or $B_IB_OA_I$).

\subsubsection{Robustness of causal nonseparability}
\label{subsubsec:robustness}

Since the constraints from Eqs.~\eqref{eq:charact_csep_bi1}--\eqref{eq:charact_csep_bi5} are affine or positive semidefinite,  determining whether a given process matrix is causally separable or not is an SDP problem (see Sec.~\ref{para:SDP}). 
A convenient way to formulate this as an optimization problem is to look at how much noise can be added to a given process matrix $W$, before it becomes causally separable. Specifically, consider mixing $W$ with another fixed process matrix $W^\circ$ (typically, the ``fully mixed'' process matrix $W^\circ = \id^\circ \coloneqq  \id^{AB}/d^I$). 
One may then look at the minimal value of $r \ge 0$ such that the mixture $W(r) = \frac{1}{1+r}(W + r W^\circ)$ is causally separable.

It is convenient here to introduce the set ${\cal W}_\mathrm{c\mhyphen sep}$ of non-normalized causally separable process matrices, which will allow us to ignore the $\frac{1}{1+r}$ factor above. 
This set---a closed convex cone---is obtained as the Minkowski sum%
\footnote{A convex cone ${\cal W}$ is a subset of a vector space such that for any $W_1, W_2\in{\cal W}$ and any scalars $q_1, q_2\ge 0$, one has $q_1\,W_1 + q_2\,W_2\in{\cal W}$. The so-called Minkowski sum of two subsets ${\cal W}_1, {\cal W}_2$ is simply defined as ${\cal W}_1 + {\cal W}_2 \coloneqq \{W_1+W_2|W_1\in{\cal W}_1,W_2\in{\cal W}_2\}$.}
\begin{align}
    & {\cal W}_\mathrm{c\mhyphen sep} = {\cal W}_{A\prec B} + {\cal W}_{B\prec A} \label{eq:Wcsep_cone_Minkowski_sum}
\end{align}
of the cones ${\cal W}_{A\prec B}$ and ${\cal W}_{B\prec A}$ of non-normalized process matrices compatible with the orders $A\prec B$ and $B\prec A$, resp., and can be characterized more explicitly, according to Eqs.~\eqref{eq:charact_csep_bi1}--\eqref{eq:charact_csep_bi4} [while ignoring Eq.~\eqref{eq:charact_csep_bi5}], as
\begin{align}
    & {\cal W}_\mathrm{c\mhyphen sep} \notag \\
    & = \big\{W \, \big| \, \exists\, W_{(A,B)}^{A_IA_OB_I} \!\ge\! 0, W_{(B,A)}^{B_IB_OA_I} \!\ge\! 0, \rho_{(A)}^{A_I} \!\ge\! 0, \rho_{(B)}^{B_I} \!\ge\! 0, \notag \\
    & \qquad \qquad W = W_{(A,B)}^{A_IA_OB_I}\otimes\id^{B_O} + W_{(B,A)}^{B_IB_OA_I}\otimes\id^{A_O}, \notag \\
    & \qquad \qquad \tr_{B_I}W_{(A,B)}^{A_IA_OB_I} = \rho_{(A)}^{A_I}\otimes\id^{A_O}, \notag \\
    & \qquad \qquad \tr_{A_I}W_{(B,A)}^{B_IB_OA_I} = \rho_{(B)}^{B_I}\otimes\id^{B_O} \big\}. \label{eq:Wcsep_cone}
\end{align}
With this, the optimization problem described above can be expressed as
\begin{align}
    \min\ & r \notag \\
    \text{s.t.}\ & W + r \, W^\circ \in {\cal W}_\mathrm{c\mhyphen sep}, \; r \ge 0. \label{eq:primal}
\end{align}
The value $r^\mathrm{opt}$ of the optimal solution to this SDP problem defines the \emph{robustness} of (the causal nonseparability of) $W$ with respect to $W^\circ$. If $r^\mathrm{opt} > 0$, then this implies that the process matrix $W$ is causally nonseparable (since in that case, $W \notin {\cal W}_\mathrm{c\mhyphen sep}$); on the other hand, if $r^\mathrm{opt} = 0$, then $W$ is causally separable.

If $W^\circ = \id^\circ$, then we refer to $r^\mathrm{opt}$ as the \emph{random robustness}.
A slightly different approach is, instead of fixing $W^\circ$ above, to also optimize over it (in the whole set of valid process matrices).
The corresponding optimization problem is then again an SDP problem,%
\footnote{Explicitly, the optimization problem for the generalized robustness is
\begin{align}
    \min\ r \quad \ \text{s.t.}\ & W + r \, W^\circ \in {\cal W}_\mathrm{c\mhyphen sep}, \; r \ge 0, \notag \\
    & W^\circ \ge 0, W^\circ \in \L_\mathrm{V}, \tr W^\circ = d^O. \notag
\end{align}
To express it as an SDP, one may incorporate $r$ into $W^\circ$ and write \citep{Araujo_2015}
\begin{align}
    \min \tr W^\circ/d^O \quad \ & \text{s.t. } W + W^\circ \in {\cal W}_\mathrm{c\mhyphen sep}, W^\circ \in {\cal W}_\mathrm{V}, \notag
\end{align}
where ${\cal W}_\mathrm{V}$ is the cone of nonnormalized valid process matrices (i.e., positive semidefinite matrices in $\L_\mathrm{V}$). \label{ftn:primal_gen_robustness}}
and gives, in that case, the so-called \emph{generalized robustness}.
In \citep{Araujo_2015}, it was shown that the generalized robustness satisfies some convexity and monotony properties that are desirable to define a proper \emph{measure of causal nonseparability}.

Solving the respective optimization problems numerically for the process matrix $W_\mathrm{OCB}$ of Eq.~\eqref{eq:W_OCB}, for instance, gives a random robustness of $\sim 0.414>0$ \citep{Araujo_2015,Branciard_2016}, and a generalized robustness of\footnote{Using the causal witness that we will present in Sec.~\ref{subsec:causal_witness}, we will find that the analytical values for these robustnesses are $\sqrt{2}-1$ and $(\sqrt{2}-1)^2$, respectively.} $\sim 0.172>0$.
This formally proves our claim in Sec.~\ref{subsec:W_indefinite_order}, that $W_\mathrm{OCB}$ is causally nonseparable.

\subsubsection{Analogies and disanalogies with entanglement}
\label{subsubsec:analogy_csep_entgmt}

The causal separability / nonseparability property of process matrices share some similarities with the separability / entanglement of quantum states. Let us start by pointing out that quantum states (in the form of density matrices) and process matrices both form convex sets of positive semidefinite operators. Also, similarly to separable states, the set of causally separable processes is a convex subset. These convex properties allow for similar approaches, justifying the use of a similar nomenclature. In particular, they allow us to analyze and quantify causal nonseparability in terms of robustness measures (as above) and witnesses (see below), similarly to robustness of entanglement \citep{vidal99} and entanglement witnesses \citep{horodecki96,terhal00}. Device-independent certifications of causal nonseparability can also be considered. Their analogy with the case of entanglement (which can be certified device-independently through the violation of Bell inequalities) will be discussed in Sec.~\ref{subsubsec:analogy_noncausal_nonlocal}. 

There are, however, some fundamental structural differences between causal nonseparability and entanglement. Most remarkably, as discussed in Sec~\ref{subsec:characterization_sep}, the set of causally separable process matrices admits an SDP characterization. Therefore, deciding whether a process matrix is causally nonseparable admits an efficient computational approach. This should be contrasted with the separability of quantum states. Apart from the special case of two qubits and qubit-qutrit \citep{horodecki96}, deciding whether a quantum state is entangled or not does not admit an SDP characterization, and it is expected to be a very hard computational problem. In precise terms, weak membership of the set of separable states is an NP-hard problem on the dimension of states \citep{gurvits02}.

\subsection{Causal witnesses} \label{subsec:causal_witness}

\subsubsection{Separating causally nonseparable from causally separable processes}

Another way to verify the causal nonseparability of a given process, originally inspired by the aforementioned analogy with entanglement witnesses \citep{horodecki96,terhal00}, is via the use of \emph{witnesses of causal nonseparability}---or \emph{causal witnesses}, for short \citep{Araujo_2015,Branciard_2016}.

\begin{figure}[htb] 
	\begin{center}
		\includegraphics[width=.9\columnwidth]{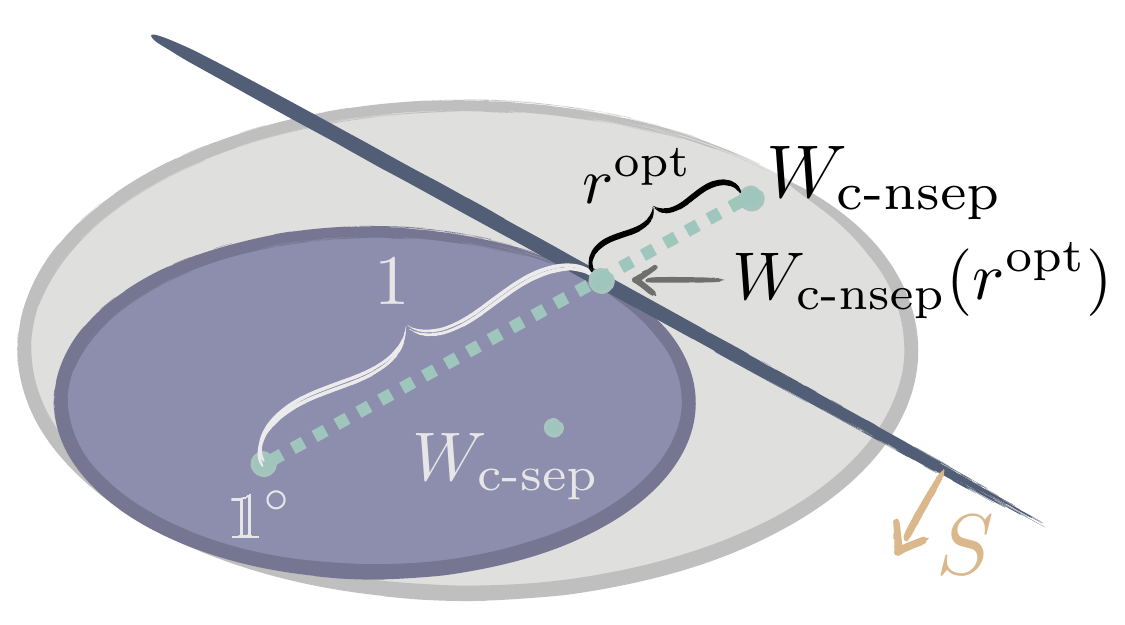} 
	\end{center}
\caption{Depiction of the convex set of causally separable processes (purple region), contained in the whole set of valid processes (gray). 
    The hyperplane $S$, separating a causally nonseparable process $W_\mathrm{c\mhyphen nsep}$ from the set of causally separable processes, represents a causal witness. $r^\mathrm{opt}$ is the random robustness of $W_\mathrm{c\mhyphen nsep}$, which corresponds to the distance between $W_\mathrm{c\mhyphen nsep}$ and the set of separable processes, along the line connecting it with $\id^{\circ}$ (and relative to the distance between $\id^{\circ}$ and the crossing point $W_\mathrm{c\mhyphen nsep}(r^\mathrm{opt})$); as depicted here, $S$ is tangent to the set of causally separable processes at that crossing point, and is thus optimal with respect to random robustness.
    }
\label{fig:witness}
\end{figure}

The basic idea comes from the geometric picture sketched in Fig.~\ref{fig:witness}, and is based on the separating hyperplane theorem \citep{rockafellar70}: since the set of causally separable process matrices is closed and convex, then any causally nonseparable process can be separated from the causally separable set by a hyperplane. 

To formalize this, we define a \emph{causal witness} as any linear operator $S\in\L(\H^{AB})$ satisfying
\begin{align} 
    \tr[S^T\, W_\mathrm{c\mhyphen sep}] \ge 0 \text{ for all causally separable } W_\mathrm{c\mhyphen sep}. \label{eq:def_witness}
\end{align}
Such operators can indeed (potentially) ``witness'' causal nonseparability, in the sense that obtaining $\tr[S^T\, W] < 0$, for a given $W$, guarantees that $W$ is causally nonseparable.
The separating hyperplane theorem then translates into the existence, for any causally nonseparable process $W_\mathrm{c\mhyphen nsep}$, of a causal witness that detects it, i.e., such that
\begin{align}
    \tr[S^T\, W_\mathrm{c\mhyphen nsep}] < 0.
\end{align}
As detailed in the following subsection~\ref{subsubsec:construct_witness}, for any causally nonseparable process matrix, one can efficiently construct a causal witness that certifies this property via SDP (in contrast with entanglement witnesses, see Sec.~\ref{subsubsec:analogy_csep_entgmt} above).
Furthermore, causal witnesses define observables that can be measured through an appropriate set of operations (see Sec.~\ref{observables}, and the more specific discussion below), making them a practical tool to detect causal non-separability in experiments, without requiring one to fully reconstruct the process matrix.

\subsubsection{Constructing a causal witness}
\label{subsubsec:construct_witness}

SDP optimization problems naturally come in two related forms, dual to each other. In particular, the ``primal form'' of Eq.~\eqref{eq:primal} is intimately linked to its ``dual form'', which reads (again, for any fixed process matrix $W^\circ$)%
\footnote{For the case of generalized robustness, the dual problem to that given in Footnote~\ref{ftn:primal_gen_robustness} writes
\begin{align}
    \min\, & \tr[S^T \, W] \notag \\
    \text{s.t. } & S \in ({\cal W}_\mathrm{c\mhyphen sep})^*,  S \in \L^\texttt{H}(\H^{AB}), \id/d^O - S \in ({\cal W}_\mathrm{V})^*, \notag
\end{align}
where $({\cal W}_\mathrm{V})^*$ is the dual cone to the cone ${\cal W}_\mathrm{V}$ of nonnormalized valid process matrices.}
\begin{align}
    \min\, & \tr[S^T \, W] \notag \\
    \text{s.t. } & \tr[S^T \, W^\circ] \leq 1, \notag \\
    & S \in ({\cal W}_\mathrm{c\mhyphen sep})^*,  S \in \L^\texttt{H}(\H^{AB}), \label{eq:dual}
\end{align}
where
\begin{align}
    ({\cal W}_\mathrm{c\mhyphen sep})^* \coloneqq \{S|\forall W\in {\cal W}_\mathrm{c\mhyphen sep}, \tr[S^T \, W] \ge 0\}
\end{align}
is the dual%
\footnote{We generically define the dual ${\cal W}^*$ of a cone of matrices ${\cal W}$ as ${\cal W}^*\coloneqq\{S|\forall\,W\in{\cal W}, \tr[S^T W]\ge 0\}$. 
Notice that for a linear subspace $\L$, the dual coincides with the orthogonal complement, which we denote with $^\perp$: $\L^*=\L^\perp=\{S|\forall\,W\in\L, \tr[S^T W] = 0\}$.
\\
We will make use below of the following duality relations for the Minkowski sum and intersection of two closed convex cones ${\cal W}_1, {\cal W}_2$:
\begin{align}
    ({\cal W}_1 + {\cal W}_2)^* = {\cal W}_1^* \cap {\cal W}_2^*, \quad ({\cal W}_1 \cap {\cal W}_2)^* = {\cal W}_1^* + {\cal W}_2^*. \notag
\end{align}
}
to the cone of (nonnormalized) causally separable process matrices ${\cal W}_\mathrm{c\mhyphen sep}$ described in Eq.~\eqref{eq:Wcsep_cone}, and $\L^\texttt{H}(\cdot)$ denotes the subset of Hermitian operators in $\L(\cdot)$.

According to our definition from Eq.~\eqref{eq:def_witness}, the dual cone $({\cal W}_\mathrm{c\mhyphen sep})^*$ describes the \emph{cone of causal witnesses}. It admits different (equivalent) explicit characterizations.%
\footnote{The first explicit characterization of $({\cal W}_\mathrm{c\mhyphen sep})^*$ was given in \citep{Araujo_2015,Branciard_2016}. Starting from the observation that one can write ${\cal W}_\mathrm{c\mhyphen sep} = ( {\cal P} \cap \L_{[1-B_O]} + {\cal P} \cap \L_{[1-A_O]} ) \cap \L_\mathrm{V}$, with ${\cal P}$ the cone of positive semidefinite matrices in $\L(\H^{AB})$, $\L_{[1-B_O]} \coloneqq \{W^{AB}|{}_{[1-B_O]}W = 0\}$ and $\L_{[1-A_O]} \coloneqq \{W^{AB}|{}_{[1-A_O]}W = 0\}$, the duality relations of the previous footnote give [noticing that ${\cal P}$ is self-dual, that $\L_{[1-B_O]}^\perp = \{S^{AB}|{}_{[1-B_O]}S = S\} = \{S^{AB}|{}_{B_O}S = 0\}$ and that ${\cal P} + \L_{[1-B_O]}^\perp = \{S^{AB} | {}_{B_O}S \ge 0\}$ (and similarly for ${\cal P} + \L_{[1-A_O]}^\perp$)]
\begin{align}
    ({\cal W}_\mathrm{c\mhyphen sep})^* & = ({\cal P} + \L_{[1-B_O]}^\perp) \cap ({\cal P} + \L_{[1-A_O]}^\perp) ) + \L_\mathrm{V}^\perp \notag \\
    & = \{S\, |\, \exists\, S_P, S^\perp
\in \L(\H^{AB}), \ S = S_P + S^\perp, \notag\\[-1mm]
    & \hspace{10mm} {}_{B_O}S_P \ge 0, {}_{A_O}S_P \ge 0, S^\perp \in \L_\mathrm{V}^\perp\} . \notag
\end{align}
The characterization of Eqs.~\eqref{eq:dual_Wcsep_cone_decomp}--\eqref{eq:dual_WAB_cone}, on the other hand, follows from writing ${\cal W}_\mathrm{c\mhyphen sep} = {\cal W}_{A\prec B} + {\cal W}_{B\prec A}$ with ${\cal W}_{A\prec B} = {\cal P} \cap \L_{[1-B_O]} \cap \L_{[1-A_O]B}$ (and similarly for ${\cal W}_{B\prec A}$), with now $\L_{[1-A_O]B} \coloneqq \{W^{AB}|{}_{[1-A_O]B}W = 0\}$, and using again the previous duality relations---noting here (in addition to $\L_{[1-B_O]}^\perp = \{S^{AB}|{}_{B_O}S = 0\}$) that $\L_{[1-A_O]B}^\perp = \{S^{AB}|{}_{[1-A_O]B}S = S\} = \{S^{AB}=S_{(A)}^A \otimes \id^B|{}_{A_O}S_{(A)} = 0\}$.
\\
Finally, one may note that Eq.~\eqref{eq:dual_WAB_cone} is also equivalent to the following simpler form: 
\begin{align}
    ({\cal W}_{A\prec B})^* = \big\{S \, \big| \, & \exists\, S_{(A)}'\in\L(\H^A), \notag \\
    & \tr_{B_O} S \ge S_{(A)}' \otimes \id^{B_I}, \tr_{A_O} S_{(A)}' = 0 \big\}. \notag
\end{align}
[Indeed, starting from a decomposition $S = S_+ + S_{(AB)} + S_{(A)}\otimes\id^{B}$ as in Eq.~\eqref{eq:dual_WAB_cone}, one gets the constraints above with $S_{(A)}' = d^{B_O} S_{(A)}$; conversely, starting from the constraints above, one gets a decomposition as in Eq.~\eqref{eq:dual_WAB_cone} with $S_{(A)} = {}_{[1-A_O]}S_{(A)}'/d^{B_O}$, $S_{(AB)} = {}_{[1-B_O]}S$ and $S_+ = S - S_{(AB)} - S_{(A)} \otimes \id^B = [\tr_{B_O}S - {}_{[1-A_O]}S_{(A)}' \otimes \id^{B_I}] \otimes \id^{B_O}/d^{B_O}
= [(\tr_{B_O}S - S_{(A)}' \otimes \id^{B_I}) + {}_{A_O}S_{(A)}' \otimes \id^{B_I}] \otimes \id^{B_O}/d^{B_O} \ge 0$.]}
E.g., noting that the constraint $\tr[S^T \, W] \ge 0$ for all $W\in {\cal W}_\mathrm{c\mhyphen sep}$ requires that both $\tr[S^T \, W_{A\prec B}] \ge 0$ for all process matrices compatible with the order $A\prec B$, and $\tr[S^T \, W_{B\prec A}] \ge 0$ for all those compatible with $B\prec A$, one can write
\begin{align}
    ({\cal W}_\mathrm{c\mhyphen sep})^* = ({\cal W}_{A\prec B})^* \cap ({\cal W}_{B\prec A})^*, \label{eq:dual_Wcsep_cone_decomp}
\end{align}
where $({\cal W}_{A\prec B})^*$ and $({\cal W}_{B\prec A})^*$ are the dual to the cones ${\cal W}_{A\prec B}$ and ${\cal W}_{B\prec A}$ introduced in Eq.~\eqref{eq:Wcsep_cone_Minkowski_sum}, which can themselves be characterized as \citep{Wechs_2019}
\begin{align}
    ({\cal W}_{A\prec B})^* = \big\{S \, \big| & \, \exists\, S_+, S_{(AB)} \in\L(\H^{AB}), S_{(A)}\in\L(\H^{A}), \notag \\
    & \!\! S = S_+ + S_{(AB)} + S_{(A)}\otimes\id^{B}, \notag \\
    & \!\!\!\!\! S_+\ge 0, {}_{B_O}S_{(AB)} = 0, {}_{A_O}S_{(A)} = 0 \big\}, \label{eq:dual_WAB_cone}
\end{align}
and similarly for $({\cal W}_{B\prec A})^*$.

Furthermore, one can show that ``strong duality'' holds for the above primal and dual SDP problems~\eqref{eq:primal} and~\eqref{eq:dual} whenever $W^\circ$ is in the interior of the set of causally separable process matrices%
\footnote{Strong duality for SDP problems is ensured in the case where the so-called Slater conditions are satisfied: namely, when the optimizations in both the primal and dual problems are over nonempty sets, and one of the constraints can be strictly satisfied \citep{watrous_2018,Skrzypczyk2023}.}%
---as is the case for \mbox{$W^\circ = \id^\circ$}---meaning that their optimal solutions $r^\mathrm{opt}$ (the robustness, recall Sec.~\ref{subsubsec:robustness}) and $S^\mathrm{opt}$ are then related by
\begin{align}
    \tr[(S^\mathrm{opt})^T \, W ] = -r^\mathrm{opt}. \label{eq:strong_duality}
\end{align}
With such a choice of $W^\circ$, any causally nonseparable process matrix will thus provide a solution $S^\mathrm{opt} \in ({\cal W}_\mathrm{c\mhyphen sep})^*$ to the dual SDP problem above that satisfies $\tr[(S^\mathrm{opt})^T \, W ] = -r^\mathrm{opt} < 0$. 
In other words, the above SDP approach indeed allows one, for any causally nonseparable process matrix, to construct a causal witness explicitly that detects its causal nonseparability.
Note that the witness $S^\mathrm{opt}$ thus obtained is optimal with respect to the noise model under consideration, in the sense that it detects the causal nonseparability of $W(r) = \frac{1}{1+r}(W + r W^\circ)$ for all $r > r^\mathrm{opt}$, i.e., all values of $r$ such that $W(r)$ is causally nonseparable (see Fig.~\ref{fig:witness}).

\medskip

The idea of causal witnesses was first introduced in \citep{Araujo_2015}, and further investigated in \citep{Branciard_2016}. An explicit witness was constructed by solving the dual problem of Eq.~\eqref{eq:dual} with $W^\circ = \id^\circ$ for the process matrix $W_\mathrm{OCB}$ of Eq.~\eqref{eq:W_OCB}, namely,
\begin{align}
    & S_\mathrm{OCB}^\mathrm{opt} = \frac{1}{4} \!\left[ \id^{AB} - \id^{A_I}\sigma_z^{A_O}\sigma_z^{B_I}\id^{B_O} - \sigma_z^{A_I}\id^{A_O}\sigma_x^{B_I}\sigma_z^{B_O} \right]\!. \label{eq:S_OCB}
\end{align}
One can verify that this indeed gives the random robustness $r^\mathrm{opt} = - \tr[(S_\mathrm{OCB}^\mathrm{opt})^T \, W_\mathrm{OCB}] = \sqrt{2}-1>0$, as already obtained (numerically) from the primal problem in Sec.~\ref{subsubsec:robustness} above. The same witness was found in \citep{Branciard_2016} to also detect the causal nonseparability of a larger family of processes considered by \citep{Brukner2015}, where the $\frac{1}{\sqrt{2}}$ weights in $W_\mathrm{OCB}$ are replaced by some continuous parameters.

\subsubsection{Measuring a causal witness}
\label{measuringwitnesses}

In order to make use of a causal witness and demonstrate causal nonseparability in practice, one needs to access the quantity $\tr[S^T \, W]$ experimentally. As discussed in Sec.~\ref{observables}, this quantity can be constructed from the statistics observed in a specific experiment, similarly to any textbook quantum observable: see Eqs.~\eqref{eq:decomp_S}--\eqref{eq:reconstruct_S}. 
Such measurements of causal witnesses have been implemented on the quantum switch (corresponding in fact to multipartite scenarios, see below) in the experiments of \citep{rubino2017,goswami2018}, 
see Secs.~\ref{subsubsec:ExpPathAndPol} and~\ref{subsubsec:OAM}.

Depending on the concrete laboratory setting, one may sometimes also have certain restrictions on the instruments that the parties can implement.
In such cases, not all causal witnesses can be measured (and not all causally nonseparable processes can then be certified). One can take these constraints into account when constructing witnesses to be measured, and restrict these to certain subspaces of $({\cal W}_\mathrm{c\mhyphen sep})^*$. For example, if a party, say $A$, can only perform CPTP operations (or even only unitaries) rather than instruments with multiple possible outcomes, then its Choi operators $M^A$ necessarily satisfy the TP condition $\tr_{A_O} M^A= \id^{A_I}$, which implies ${}_{A_IA_O}M^A = {}_{A_O}M^A$ (together with ${}_{A_IA_O}M^A = {}_{A_I}M^A$, for unitaries). Any witness $S$ that can be obtained as a combination of the form of Eq.~\eqref{eq:decomp_S} then necessarily also satisfies ${}_{A_IA_O}S = {}_{A_O}S$ (together with ${}_{A_IA_O}S = {}_{A_I}S$, if only unitaries can be implemented).
These constraints can directly be incorporated in the dual SDP problem~\eqref{eq:dual} [see \citep{Branciard_2016} for a more detailed analysis applied to the quantum switch].

Note, finally, that faithfully reconstructing $\tr[S^T\, W]$ as in Eq.~\eqref{eq:reconstruct_S} requires one to trust that the experimental setup indeed implements the prescribed instruments $\{M_{a|x}^A\}_a$ and $\{N_{b|y}^B\}_b$ that enter into the decomposition of Eq.~\eqref{eq:decomp_S}. Departures from the ideal operations---under some assumptions---can be incorporated in the definition of the witness [e.g., by introducing a correction to the causal separability bound, as was done in \citep{goswami2018}]. In any case, measuring a causal witness in such a way is \emph{device-dependent}. We will see, in Sec.~\ref{sec:causal_ineqs}, that there can also be device-independent, or semi-device-independent approaches to demonstrate causal nonseparability.

\subsection{Multipartite process matrix formalism}
\label{multipartite}

\subsubsection{Multipartite process matrices}

The process matrix formalism, formally introduced in the bipartite case in Sec.~\ref{processmatrixsection}, generalizes easily to the multipartite case. In particular, for a scenario with $N$ parties $A^1, \ldots, A^N$ with input and output Hilbert spaces $\H^{A_I^k}$, $\H^{A_O^k}$, and with instruments (in the Choi representation) $\{M^{A^k}_{a_k}\}_{a_k}$, for $k \in \N \coloneqq \{1, \ldots, N\}$,%
\footnote{We will generally identify parties $A^k$ with their label $k$, and the set of parties (or any subset of it) with the set of labels $\N$ (or with the corresponding subset).}
the generalized Born rule of Eq.~\eqref{simpleBorn2}
extends naturally to
\begin{align}\label{processborn_Npartite}
    & P(a_1,\ldots,a_N\mid \{\mathcal{M}_{a_1}^{A_1}\}_{a_1}, \ldots, \{\mathcal{M}_{a_N}^{A_N}\}_{a_N}) \notag \\
    & \hspace{8mm} = \tr \left[\left(M^{A^1}_{a_1}\otimes \cdots \otimes M^{A^N}_{a_N}\right)^T \, W^{A^1\cdots A^N}\right]\!,
\end{align}
where $W^{A^1\cdots A^N} \in \L(\H^{A^1\cdots A^N})$ (with each $\H^{A^k} = \H^{A_I^kA_O^k}$) is now an $N$-partite process matrix.

\paragraph{Characterization.}
\label{para:multipartite_charact}

Validity conditions for such $N$-partite process matrices generalize as follows: these must still be positive semidefinite, $W \ge 0$, and normalized in such a way that $\tr W = d^O$, as in Eq.~\eqref{affine}, where now $d^O : = \Pi_{k\in\N} \, d^{A_O^k}$ is the product of all output space dimensions.
The ``valid term types'' in an HS basis decomposition are the same as those that can appear in causally ordered processes (considering all possible fixed orders; see Sec.~\ref{para:HS_terms}). These are, in addition to the identity, those such that at least one party $A^k$ has a nontrivial operator $\sigma$ on $\H^{A_I^k}$ with the identity $\id$ on $\H^{A_O^k}$---i.e., terms of type $\cdots A_I^k\cdots$ for some $k$, where the dots do not include $A_O^k$ \citep{OreshkovGiarmatzi_2016}.
One can also provide an HS basis-independent characterization (see Sec.~\ref{para:HSbasis_indep_charact}) for the linear space $\L_\mathrm{V}^\N$ spanned by valid $N$-partite process matrices as follows \citep{Araujo_2015, Wechs_2019}:%
\footnote{Equivalently, in just one linear constraint \citep{Araujo_2015}:
\begin{align*}
    W \in \L_\mathrm{V}^{\N} \Leftrightarrow & \sum_{\emptyset\neq\X \subseteq \N}{}_{\prod_{k\in\X}[1-A_O^k]\prod_{k'\in\N\backslash\X}A_I^{k'}\!A_O^{k'}}W \notag \\
    & \ = {}_{\prod_{k\in\N}[1-A_O^k+A_I^kA_O^k]}W - {}_{\prod_{k\in\N}A_I^kA_O^k}W = 0,
\end{align*}
with a natural (associative and commutative) extension of the notation: ${}_{[1-A_O^k+A_I^kA_O^k]}W \coloneqq W - {}_{A_O^k}W + {}_{A_I^kA_O^k}W$. The second line in the equation above can be written as $\mathcal{P}_\mathrm{nsig}(W) - \frac{\tr W}{d^Id^O} \id = 0$, where $\mathcal{P}_\mathrm{nsig}$ is the projector onto the linear span of $k$-partite no-signaling channels, and is thus equivalent to $\mathcal{P}_\mathrm{nsig}(W)\propto \id$: see the next paragraph.}
\begin{subequations}
\begin{align}
    & W \in \L_\mathrm{V}^{\N} \notag \\
    & \Leftrightarrow \forall \X \subseteq \N, \X \neq \emptyset, {}_{\prod_{k\in\X}[1-A_O^k]}\tr_{\N\backslash\X}W = 0 \label{eq:charact_LVN_1} \\
    & \Leftrightarrow \forall \X \subsetneq \N, \X \neq \emptyset, \tr_{\N\backslash\X} W \in \L_\mathrm{V}^{\X} \notag \\
    & \hspace{10mm} \text{and } {}_{\prod_{k\in\N}[1-A_O^k]}W = 0, \label{eq:charact_LVN_2}
\end{align}
\end{subequations}
where, for any (nonempty) subset $\X$ of $\N$, $\L_\mathrm{V}^\X$ is the subspace of $\L(\bigotimes_{k \in \X}\H^{A^k})$ spanned by valid $|\X|$-partite process matrices, for the parties $A^k$ with $k \in \X$, and $\tr_{\N\backslash\X}$ is the partial trace over all other parties' input and output spaces (with $\N \backslash \mathcal{X}$ denoting the set difference of $\N$ and $\mathcal{X}$).

As discussed in Sec.~\ref{dualaffine}, the set of process matrices is the set of positive semidefinite matrices that are in the dual affine of no-signaling channels.
This dual affine approach allows for an alternative and simple characterization for multipartite process matrices. 
Note indeed that a linear operator $C\in \L\big(\bigotimes_{k=1}^N \mathcal{H}^{A_I^kA_O^k}\big)$ is the Choi matrix of an $N$-partite no-signaling channel if and only if $C\geq0$, $_{A^k_O}C= {}_{A^k_IA^k_O}C$ for every $k$, and $\tr(C)=d^I$ \citep{Chiribella2013}.
Defining the projectors \citep{milz2023_transformations,Hoffreumon2022_proj}, $\mathcal{P}_k(C)\coloneqq C- {}_{A^k_O}C + {}_{A^k_IA^k_O}C$, it is easily seen that for any $k$ and $l$, we have the commutation relation  $\mathcal{P}_k\circ \mathcal{P}_l=\mathcal{P}_l\circ \mathcal{P}_k$. Hence, we have that ${}_{A^k_O}C={}_{A^k_IA^k_O}C$ for every $k$ if and only if $C=\mathcal{P}_\mathrm{nsig}(C)$, where $\mathcal{P}_\mathrm{nsig}\coloneqq\mathcal{P}_N \circ \cdots \circ \mathcal{P}_2 \circ \mathcal{P}_1$ is the projector onto the linear space spanned by $k$-partite no-signaling channels.  
This provides a systematic characterization for arbitrary $N$-partite process matrices from the dual affine relationship between process matrices and no-signaling channels \citep{Chiribella2016optimal_networks}: namely, a positive semidefinite operator $W\geq0$ is an $N$-partite process matrix if and only if $\mathcal{P}_\mathrm{nsig}(W)=\frac{\id}{d^I}$ \citep{milz2023_transformations,Hoffreumon2022_proj}.

Furthermore, as in the bipartite case (recall Sec.~\ref{processaschannel}), multipartite process matrices can be seen to represent channels $\mathcal{W}: \mathcal{L}(\bigotimes_k \mathcal{H}^{A_O^k}) \to \mathcal{L}(\bigotimes_k \mathcal{H}^{A_I^k})$ from all the output spaces of the parties to all their input spaces. Again, not all channels correspond to valid processes, as the constraints of Eqs.~\eqref{eq:charact_LVN_1} or~\eqref{eq:charact_LVN_2} are stronger than just the trace-preserving condition. The additional constraints can be identified with a ``parity-erasure'' condition \citep{liu2025}: suppose that all parties $A_k$ can choose at random one out of two possible states $\rho_{x_k}$ ($x_k=0,1$) to send through the channel. Then for any nonempty subset of parties $\X \subseteq \N$, the average marginal output state of $\mathcal{W}$ for those parties does not depend on the parity $\bigoplus_{k\in\X}x_k$ of the input values $x_k$ for those parties: $\sum_{\vec x\text{ s.t. }\bigoplus_{k\in\X}x_k=0}\rho_\X^\text{out}(\vec x) = \sum_{\vec x\text{ s.t. }\bigoplus_{k\in\X}x_k=1}\rho_\X^\text{out}(\vec x)$, where $\rho_\X^\text{out}(\vec x)\coloneqq\tr_{A_I^{\N\backslash\X}}\mathcal{W}(\bigotimes_k \rho^{A_O^k}_{x_k})$ and $\vec x\coloneqq(x_1,\ldots,x_N)$.%
\footnote{This can be seen by first noting that the constraints of Eq.~\eqref{eq:charact_LVN_1}, altogether, are also equivalent to
${}_{\prod_{k\in\X}[1-A_O^k]}\tr_{A_I^{\N\backslash\X}}W = 0$ for all nonempty $\X \subseteq \N$ (where $\tr_{A_I^{\N\backslash\X}}$ denotes the partial trace over all spaces $\H^{A_I^\ell}$ with $\ell\in\N\backslash\X$).
This is equivalent to $\tr_{A_I^{\N\backslash\X}}\tr_{A_O^\N}\big[{}_{\prod_{k\in\X}[1-A_O^k]}W\cdot(\varrho_1^{A_O^1}\otimes\cdots\otimes\varrho_N^{A_O^N}\otimes\id^{A_I^\N})^T\big] = \tr_{A_I^{\N\backslash\X}}\tr_{A_O^\N}\big[W\cdot({}_{\prod_{k\in\X}[1-A_O^k]}(\varrho_1^{A_O^1}\otimes\cdots\otimes\varrho_N^{A_O^N})^T\otimes\id^{A_I^\N})\big] = \tr_{A_I^{\N\backslash\X}} \mathcal{W}\big[\bigotimes_{k\in\X}(\varrho_k^{A_O^k}-\id^{A_O^k}/d^{A_O^k})\bigotimes_{\ell\in\N\backslash\X}\varrho_\ell^{A_O^\ell}\big] = 0$ for all trace-1 $\varrho_1^{A_O^1},\ldots,\varrho_N^{A_O^N}$, which in turn is equivalent to $\tr_{A_I^{\N\backslash\X}} \mathcal{W}\big[\bigotimes_{k\in\X}(\rho_0^{A_O^k}-\rho_1^{A_O^k})\bigotimes_{\ell\in\N\backslash\X}(\rho_0^{A_O^\ell}+\rho_1^{A_O^\ell})\big] = 0$ for all states $\rho_{x_k}^{A_O^k}$ as we consider. Expanding the tensor products, we find that this last expression is equal to $\big(\sum_{\vec x\text{ s.t. }\bigoplus_{k\in\X}x_k=0}\rho_\X^\text{out}(\vec x)\big) - \big(\sum_{\vec x\text{ s.t. }\bigoplus_{k\in\X}x_k=1}\rho_\X^\text{out}(\vec x)\big) = 0$, which is nothing but the parity-erasure condition. Notice that for a singleton $\X=\{k\}$, the parity-erasure constraint reduces to the constraint that $A_O^k$ does not signal to $A_I^k$ (no-signaling to one's own past).}
\citet{liu2025} showed that this parity-erasure property must in fact be satisfied by processes in more general operational probabilistic theories as well, arguing that it follows from the local validity of causality at each party's site, and suggested this could provide an information-theoretic axiomatization of the process matrix formalism.

\paragraph{Conditional process matrices.} \label{conditionalprocesses}

In the context of multipartite scenarios, it is important to understand how to recover an effective description for a subset of parties, conditioned on the remaining parties performing some specified operations. To this end, conditioning on the party $A^k$ applying some CP map with Choi operator $M^{A^k}$, we introduce the \textit{conditional process matrix} \citep{OreshkovGiarmatzi_2016}
\begin{equation}
W_{M^{A^k}} \coloneqq \tr_k \big[ (\id^{A^{\N\backslash k}}\otimes M^{A^k})^T \, W \big],
\label{eq:def_conditional_WMk}
\end{equation}
with\footnote{We identify singletons with their element: e.g., $\N\backslash k = \N\backslash \{k\}$.} $\tr_k = \tr_{A_I^kA_O^k}$, $\id^{A^{\N\backslash k}} = \bigotimes_{j\in\N\backslash k}\id^{A^j}$. 
Conditioning on more parties' operations is a straightforward extension. Clearly, the conditional process matrix is the operator one has to use when applying the generalized Born rule of Eq.~\eqref{processborn_Npartite} to the other parties while fixing an outcome $a_k^*$, associated with a CP map $M_{a_k^*}^{A^k}$, for party $A^k$: $P(\vec a_{\N\backslash k},a_k^*) = P(a_1,\ldots,a_{k-1},a_k^*,a_{k+1},\ldots,a_N) = \tr[(\bigotimes_{j\in\N\backslash k} M^{A^j}_{a_j})^T\, W_{M^{A^k}_{a_k^*}}]$ (where $\vec a_\K$ denotes the list of outputs corresponding to a subset $\K$ of $\N$).

Note that we allow here for a minor terminology tension with ``conditional probabilities'', since those, as opposed to our definition of conditional process matrices, imply a ``renormalization'' $P(\vec a_{\N\backslash k}|a_k^*)=P(\vec a_{\N\backslash k},a_k^*)/P(a_k^*)$. Behind this choice is the fact, already mentioned in Sec.~\ref{processasstate}, that a conditional process matrix is not necessarily proportional to a valid process matrix (it may contain `non valid term types', cf Sec.~\ref{para:HS_terms}), so it cannot be renormalized to one.%
\footnote{For a simple example, take $W^{AB}=\KKet{\id}\BBra{\id}^{A_OB_I}$ (an identity channel from $A$ to $B$, with trivial $\H^{A_I}, \H^{B_O}$). If $B$ observes a POVM element $E$, $A$'s conditional processes matrix is $\tr_{B_I} \left[(E^{B_I})^T \KKet{\id}\BBra{\id}^{A_OB_I}\right] = E^{A_O}$. This is not proportional to a valid process matrix (unless $E\propto\id$), because a single-site process matrix must have the identity on the output space, see Sec.~\ref{singleparty}.}
This happens when $P(a_k^*)$ depends nontrivially on the operations of some of the other parties after ignoring their outcomes, i.e., when $P(a_k^*)$ depends on the CPTP map $\sum_{a_j}\mathcal{M}^{A^j}_{a_j}$ for at least one $j\in\N\backslash k$ (indicating signaling from $A^j$ to $A^k$). Indeed, to calculate the conditional probabilities $P(\vec a_{\N\backslash k}|a_k^*)$, we need to plug in the ``renormalized'' conditional process $W_{M^{A^k}_{a_k^*}}/P(a_k^*)$ into the Born rule. This introduces a nonlinearity into the expressions, because now, via $P(a_k^*) = \sum_{(a_
j)_{j\in\N\backslash k}} \tr[(\bigotimes_{j\in\N\backslash k} M^{A^j}_{a_j})^T\, W_{M^{A^k}_{a_k^*}}]$, the CP maps $M^{A^j}_{a_j}$
appear at the denominator. This implies that the probabilities cannot be re-written in the form of a generalized Born rule with a valid process matrix.

On the other hand, when $P(a_k^*)$ \textit{is} independent of any other party's operations, the conditional process matrix is proportional to a valid process matrix. In particular, this happens if we condition on $A^k$ performing a CPTP map $M^{A^k}$ (equivalently, if we ignore $A^k$'s measurement outcomes, $M^{A^k}=\sum_{a_k}M^{A^k}_{a_k}$).
In this case, $W_{M^{A^k}}$ is a valid process matrix, without any renormalization (because the marginal probability to observe a given CPTP map is always one). It can be seen as a process analogue of a reduced state, and it is sometimes called \textit{reduced process} in the literature \citep{Araujo_2015}.%
\footnote{\textit{Reduced process} is also used in some sources to denote a mere partial trace, $\tr_k W$ \citep{Wechs_2019}.}
Notice however that, unlike for reduced states, a reduced process $W_{M^{A^{k}}}$ depends in general on the choice of CPTP map (here, $M^{A^{k}}$) of the party that is traced out (when that party can signal to some other party). 

In the particular case of causally ordered processes (see Sec.~\ref{paramultiseparability_FO} below), processes conditioned on a measurement outcome observed by the last party have been known as elements of \textit{measuring strategies} \citep{Gutoski2009} or \textit{testers} \citep{Chiribella2009}. Since their only constraint is that they have to sum up to a causally ordered process, it follows that, up to normalization, any positive operator spanning any number of sites has an implementation as a causally ordered process subject to postselection on a particular final outcome.

\subsubsection{Multipartite causal (non)separability}
\label{sectionmultiseparability}

\paragraph{Compatibility with fixed causal orders.}\label{paramultiseparability_FO}

As already referred to above, the notions of signaling, and thereby of compatibility with some given causal order, extend naturally to multipartite scenarios. In particular, we will say that a process is compatible with a subset of parties $\K_1$
acting before another (non-overlapping) subset of parties $\K_2$
---denoted $\K_1\prec\K_2$---if there cannot be, whatever the quantum instruments being used, any signaling from any party in $\K_2$ to the parties in $\K_1$ \citep{OreshkovGiarmatzi_2016}.

Process matrices that are compatible with a given order can be characterized in a similar way as in the bipartite case. E.g., when $\K_1$ and $\K_2$ are complementary ($\K_1\cup\K_2=\N$), then any HS decomposition of a process matrix compatible with $\K_1\prec\K_2$ can only contain terms which, in addition to being of ``valid types'' for an $N$-partite process matrix (terms in $\L_\mathrm{V}^\N$), are such that their ``restriction''%
\footnote{The ``restriction'' of an HS term to a subset of parties is obtained by just ignoring the factors attributed to the other parties. E.g., the restriction of $\sigma^{A_I}_{i}\otimes\sigma^{A_O}_{j}\otimes\sigma^{B_I}_{k}\otimes\sigma^{B_O}_{l}\otimes\sigma^{C_I}_{m}\otimes\sigma^{C_O}_{n}$ to $B$ and $C$ is simply $\sigma^{B_I}_{k}\otimes\sigma^{B_O}_{l}\otimes\sigma^{C_I}_{m}\otimes\sigma^{C_O}_{n}$.}
to $\K_2$ are of a valid type for process matrices with parties in $\K_2$ (terms in $\L_\mathrm{V}^{\K_2}$) \citep{OreshkovGiarmatzi_2016}.
Equivalently, a basis-independent characterization can be given to the linear space $\L_{\K_1\prec\K_2}$ spanned by valid process matrices compatible with $\K_1\prec\K_2$ (still with%
\footnote{\citet{Wechs_2019} also provided the corresponding constraints for any two disjoint, but not necessarily complementary subsets of parties.}
$\K_1\cup\K_2=\N$) as follows \citep{Wechs_2019}:
\begin{subequations}
\begin{align}
    & W \in \L_{\K_1\prec\K_2} \notag \\
    & \Leftrightarrow \forall \X_1 \subseteq \K_1, \X_1 \neq \emptyset, {}_{\prod_{k\in\X_1}[1-A_O^k]}\tr_{\N\backslash\X_1}W = 0 \notag \\
    & \qquad \text{and} \ \forall \X_2 \subseteq \K_2, \X_2 \neq \emptyset, {}_{\prod_{k\in\X_2}[1-A_O^k]}\tr_{\K_2\backslash\X_2}W = 0 \label{eq:LV_K1_K2} \\
    & \Leftrightarrow \tr_{\K_2} W \in \L_\mathrm{V}^{\K_1} \notag \\[-1mm]
    & \qquad \text{and} \ \forall M^{A^{\K_1}} \in \L(\H^{A^{\K_1}}), W_{M^{A^{\K_1}}} \in \L_\mathrm{V}^{\K_2},
\end{align}
\end{subequations}
where $W_{M^{A^{\K_1}}}$ is a conditional process matrix defined as in Eq.~\eqref{eq:def_conditional_WMk}, conditioned here on a joint operator $M^{A^{\K_1}} \in \L(\H^{A^{\K_1}})$ for all parties in $\K_1$ (with $\H^{A^{\K_1}} = \bigotimes_{k \in \K_1}\H^{A^k}$).

One can then also consider several (disjoint) subsets of parties $\K_j$: a process is compatible with the order $\K_1 \prec \K_2 \prec \cdots \prec \K_m$ when there is no signaling from the ``future subsets'' to the ``past subsets'', i.e., when it is compatible with $\K_{(\leq j)}\prec\K_{(> j)}$ for any $j = 1, \ldots, m-1$, where $\K_{(\leq j)} = \bigcup_{i\leq j} \K_i$ and $\K_{(> j)} = \bigcup_{i'>j} \K_{i'}$. The corresponding subspace of process matrices can then readily be characterized as $\L_{\K_1 \prec \K_2 \prec \cdots \prec \K_K} = \bigcap_{j}\L_{\K_{(\leq j)}\prec\K_{(> j)}}$.
When the subsets under consideration are the $N$ singletons $\{A^k\}$, we obtain the conditions for compatibility with a fixed causal order between all parties, say $A_1 \prec A_2 \prec \cdots \prec A_N$.
From the characterization of Eq.~\eqref{eq:LV_K1_K2} above, after simplification we find
\begin{align}
    & W \in \L_{A_1\prec A_2\prec \cdots\prec A_N} \notag \\
    & \Leftrightarrow \forall k = 1, \ldots, N, {}_{[1-A_O^k]}\tr_{\{k+1,\ldots,N\}}W = 0,
\end{align}
which are indeed the constraints that characterize ``channels with memory'', ``quantum strategies'', or ``quantum combs'' \citep{Kretschmann2005, Gutoski2006,Chiribella2009}.

\paragraph{Different definitions of causal (non)separability.}

Beyond the compatibility with fixed causal orders, the definition of causal (non)separability becomes more subtle in the multipartite setting. First, note that only considering convex mixtures of different fixed orders (as causal separability reduces to in the bipartite case) would not encapsulate the idea that the causal order can be established dynamically: that is, that the causal order between future operations may depend (in a well-defined, classical manner) on the choice of past operations.

To allow for such ``dynamical'' causal orders (also called ``adaptive-definite'' by \citep{baumeler2014}, as opposed to ``convex-definite'' for mere convex mixtures of different orders), we resort to recursive definitions of causal separability for general $N$-partite scenarios. The first proposal, by \citet{OreshkovGiarmatzi_2016}, was as follows: an $N$-partite causally separable process matrix $W$ is one that can be decomposed as a convex mixture of process matrices $W^{(k)}$, where each $W^{(k)}$ is compatible with party $A^k$ acting first, and is such that whatever CP map $M^{A^k}$ that party applies, the $(N-1)$-partite conditional process matrix $W_{M^{A^k}}^{(k)}$ is itself causally separable. (For $N=1$, any process matrix is causally separable.)

As it turns out, this definition allows for a phenomenon called ``activation of non-causality'', in which a causally separable process becomes causally nonseparable when some auxiliary entangled state is attached to it \citep{OreshkovGiarmatzi_2016,Wechs_2019}. Processes that do not allow for such activation were then called ``extensibly causally separable'' \citep{OreshkovGiarmatzi_2016}. Arguing that the notion of causal nonseparability should be seen as already contained in the process, rather than as somehow originating from the attached auxiliary state, the following alternative definition was proposed in \citep{Wechs_2019}, which already encapsulates the possibility to attach auxiliary states (as the whole process matrix formalism allows for)---and which now seems to be more widely accepted than the previous one:
\begin{definition} \label{def:Wcsep_Npartite}
An $N$-partite process matrix $W_\mathrm{c\mhyphen sep}$ is said to be \emph{causally separable} if and only if for any auxiliary input Hilbert spaces $\H^{A_{I'}^k}$ attached to each party and any state $\rho \in \L(\bigotimes_{k}\H^{A_{I'}^k})$, the composed process $W_\mathrm{c\mhyphen sep}\otimes\rho$ can be decomposed as
\begin{align}
    W_\mathrm{c\mhyphen sep}\otimes\rho =  \sum_{k\in\N} q_k  \, W_\rho^{(k)} \ \ \ \text{with} \ q_k\ge 0, \ {\textstyle \sum_k} q_k = 1, \label{eq:def_csep_Npartite}
\end{align}
where each $W_\rho^{(k)}\in\L(\bigotimes_{k}\H^{A_I^kA_{I'}^kA_O^k})$ is an $N$-partite process matrix compatible with party $A^k$ acting first, and is such that whatever CP map $\M^{A^k}:\L(\H^{A_I^kA_{I'}^k})\to\L(\H^{A_O^k})$ that party applies, the $(N-1)$-partite conditional process matrix $(W_\rho^{(k)})_{M^{A^k}} = \tr_k \big[ (\id^{A^{\N\backslash k}}\otimes M^{A^k})^T \, W_\rho^{(k)} \big]$ is itself causally separable.%
\footnote{Here $\tr_k$ is understood as $\tr_{A_I^kA_{I'}^kA_O^k}$.}
(For $N=1$, any process matrix is causally separable.)
\end{definition}

Although phrased slightly differently, this definition is equivalent to that of ``extensible causal separability'' of \citep{OreshkovGiarmatzi_2016}. 
Similarly to the bipartite case, it allows for a characterization of causally separable processes in terms of SDP constraints. E.g., for $N=3$, with parties denoted $A, B, C$, a causally separable process is one that can be decomposed in the form (that the reader may compare to Eqs.~\eqref{eq:charact_csep_bi1}--\eqref{eq:charact_csep_bi5}, for the bipartite case)
\begin{subequations}
\begin{align}
 & W_\mathrm{c\mhyphen sep} = \ \  W_{(A,B,C)}^{ABC_I}\otimes\id^{C_O} + W_{(A,C,B)}^{ACB_I}\otimes\id^{B_O} \notag \\
 & \qquad \qquad + W_{(B,A,C)}^{BAC_I}\otimes\id^{C_O} + W_{(B,C,A)}^{BCA_I}\otimes\id^{A_O} \notag \\
 & \qquad \qquad + W_{(C,A,B)}^{CAB_I}\otimes\id^{B_O} + W_{(C,B,A)}^{CBA_I}\otimes\id^{A_O} \label{eq:charact_csep_tri1} \\[1mm]
 & \text{with } \ W_{(A,B,C)}^{ABC_I} \geq 0, \ \ldots \label{eq:charact_csep_tri2} \\
 & \quad \tr_{C_I}W_{(A,B,C)}^{ABC_I} = W_{(A,B)}^{AB_I}\otimes\id^{B_O}, \ \ldots 
 \label{eq:charact_csep_tri3} \\
 & \quad \tr_{B_I}W_{(A,B)}^{AB_I} + \tr_{C_I}W_{(A,C)}^{AC_I} = \rho_{(A)}^{A_I}\otimes\id^{A_O}, \ \ldots \label{eq:charact_csep_tri4}  \\
 & \quad \tr \rho_{(A)}^{A_I} + \tr \rho_{(B)}^{B_I} + \tr \rho_{(C)}^{C_I} = 1, \label{eq:charact_csep_tri5}
\end{align}
\end{subequations}
where the dots signify that the analogous constraints apply to the other terms, obtained by permutation of the parties.
Here the matrices $W_{(A,B,C)}^{ABC_I}$ etc. are not necessarily valid process matrices, even up to normalization, so that the above characterization indeed allows for dynamical causal orders (otherwise, if each of them was a valid process matrix, Eq.~\eqref{eq:charact_csep_tri1} would merely describe a convex mixture of processes compatible with a fixed order).

An interesting tripartite scenario, which is particularly relevant to the study of the quantum switch (see Sec.~\ref{switchsection}), is one where one party---say $C$---has no (or a trivial) output space. In that case, it can be seen that the above definition [in contrast to the former definition from \citep{OreshkovGiarmatzi_2016}] reduces to saying that a causally separable process is a mixture of processes compatible with the fixed orders $A\prec B\prec C$ and $B\prec A\prec C$. This recovers the definition of causal separability initially proposed by \citep{Araujo_2015} as a straightforward generalization of the bipartite case to that specific tripartite scenario, based on a rather intuitive consideration: if $C$ has no output space then it can always be considered to come last, so that the only relevant orders should just be the two orders above.

Beyond the case of $N=3$, note, however, that the natural generalization of Eqs.~\eqref{eq:charact_csep_tri1}--\eqref{eq:charact_csep_tri5} to $N\ge 4$---which turns out to characterize precisely the class of ``quantum circuits with classical control of causal order (QC-CCs)'', see Sec.~\ref{subsec:Gen}
\citep{Wechs2021}---only gives a sufficient condition for causal nonseparability; necessary conditions were also derived in \citep{Wechs_2019}, and it remains an open question to prove whether some of the conditions are both necessary and sufficient.
These necessary and sufficient conditions for causal separability can both be written as SDP problems, and the general technique of causal witnesses introduced in Sec.~\ref{subsec:causal_witness} above can be generalized to witness multipartite causal nonseparability (or to witness the non-inclusion in the QC-CC class).

\subsection{Processes with open past and future as higher order transformations}
\label{subsec:open_past_future}

In the previous sections, the process matrix formalism was introduced and employed as a method to describe the causal relations between different sites. In this section,
we show how to use it to describe transformations between quantum operations.

\begin{figure}[hbt] 
	\begin{center}
		\includegraphics[width=.7\columnwidth]{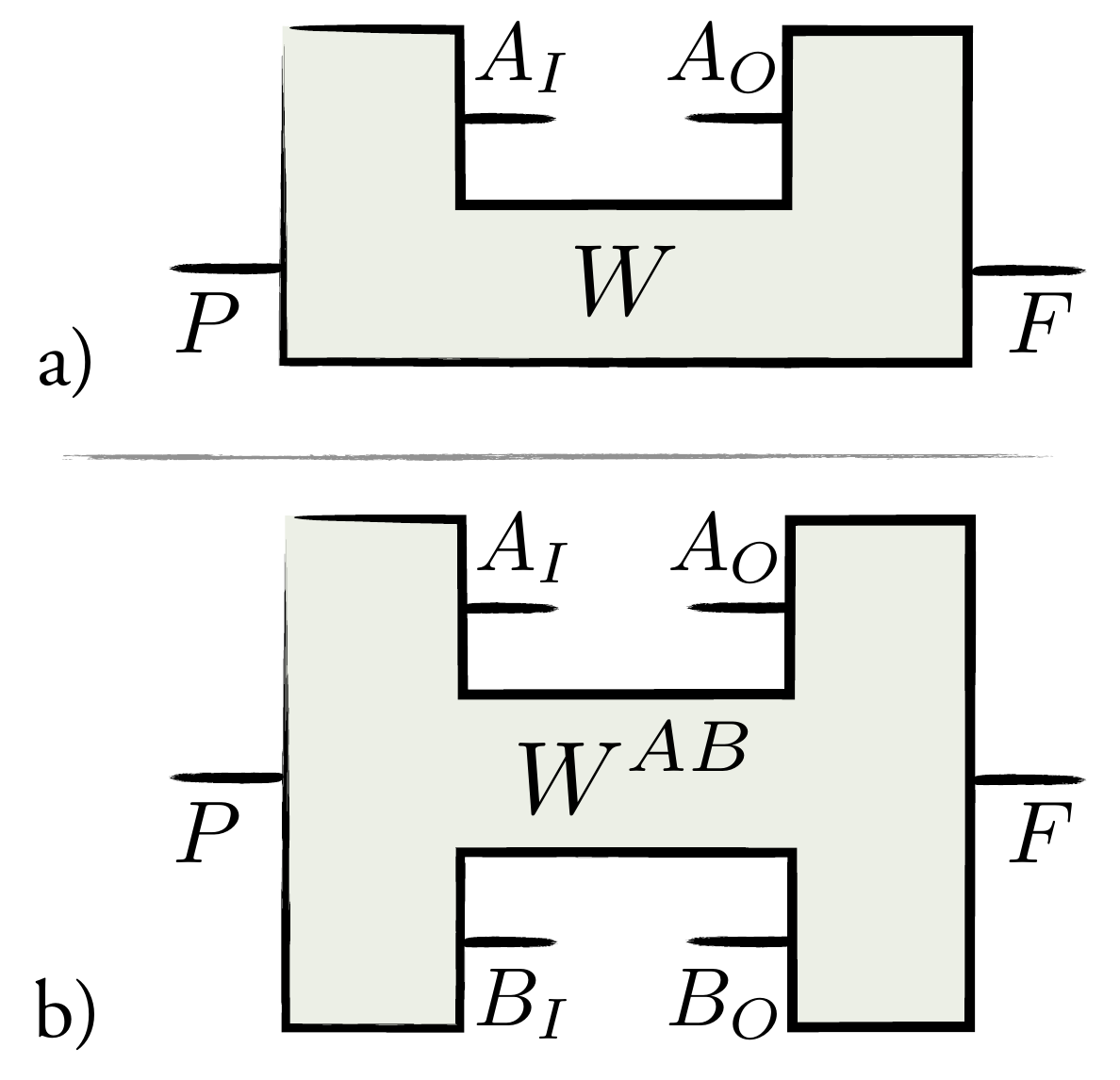} 
	\end{center}
\caption{\textbf{a)} Pictorial representation of a one-slot process which transforms a quantum channel from $A_I$ to $A_O$ into a quantum channel mapping $P$ (past) to $F$ (future).  \textbf{b)} Pictorial representation of a two-slot process which transforms a pair of quantum channels, one from $A_I$ to $A_O$, another from $B_I$ to $B_O$, into a quantum channel mapping $P$ to $F$.}
\label{fig:ProcessMatrix1}
\end{figure}

We start by describing how a tripartite process may used to transform quantum channels into quantum channels. Let us consider a process with an  ``open slot'', into which Alice may plug a quantum channel from $A_I$ to $A_O$, to obtain a quantum channel from site $P$ (past) to site $F$ (future), as in Fig.~\ref{fig:ProcessMatrix1}~\textbf{a)}.
Formally, such a one-slot process can be seen as a tripartite process, $W\in\mathcal{L}(\H^P\otimes\H^{A_IA_O}\otimes\H^F)$, where the linear space $\H^P$ represents a site which only has an output space and the linear space $\H^F$ represents a site which only has an input space.  In this way, if Alice plugs an arbitrary quantum channel with Choi operator $C^{A_IA_O}\in\mathcal{L}(\H^{A_IA_O})$  into the process $W$, the output is a quantum channel from $\L(\H^P)$ to $\L(\H^F)$ with its Choi operator given by \begin{equation}
C^{A_IA_O}*W=\tr_{A_IA_O}\big[(\id^P\otimes {C^{A_IA_O}}^T\otimes\id^F)\,W\big].
\end{equation}

One may thus also represent the action of such a one-slot process matrix with open past and future as a linear map, which transforms the Choi operator of a quantum channel into the Choi operator of {another} quantum channel. More precisely, via the Choi isomorphism (Sec.~\ref{Choiiso}), the action of any process matrix $W\in\mathcal{L}(\H^{PA_IA_OF})$ on a quantum channel (or any other linear operator) $C^{A_IA_O}\in\mathcal{L}(\H^{A_IA_O})$ is equivalently described by a map 
\begin{align}
\mathscr{W}:\L(\H^{A_IA_O}) \to & \, \L(\H^{PF}) \notag \\
C^{A_IA_O} \mapsto & \, \mathscr{W}(C^{A_IA_O})\coloneqq C^{A_IA_O}*W.
\end{align}

Additionally, one may also analyze and characterize process matrices without making any reference to the Choi isomorphism by means of \textit{linear supermaps} \citep{Chiribella2008supermap}. Let us now illustrate this for the case of a single party $A$, with open past and open future. In this approach, quantum channels are not described by their Choi operators but by a linear map $\mathcal{C}:\L(\H^{A_I})\to \L(\H^{A_O})$, hence, a process with open past and future is a supermap \citep{Chiribella2008supermap}, a linear transformation from maps to maps, mathematically described by
\begin{equation}
	\widetilde{\mathscr{W}}:[\L(\H^{A_I})\to \L(\H^{A_O})]\to [\L(\H^{P})\to \L(\H^{F})].
\end{equation}
In this approach, if $\mathcal{C}$  is a quantum channel mapping $\L(\H^{A_I})$ to $\L(\H^{A_O})$, $\widetilde{\mathscr{W}}(\mathcal{C})$ is a quantum channel mapping $\L(\H^{P})$ to $\L(\H^{F})$.

One-slot processes can always be implemented by causally ordered circuits by simply concatenating an encoding and a decoding quantum channels, which may also act on an auxiliary system which can eventually be traced out \citep{Chiribella2008supermap}. These are also known as deterministic quantum supermaps \citep{Chiribella2008supermap,Zyczkowski2008}, 1-turn quantum strategies \citep{Gutoski2006}, one-slot combs \citep{Chiribella2008architecture,Chiribella2009} and quantum superchannels \citep{gour18,Quintino2019b,Yokojima2020}.

In a similar vein, the process matrix formalism allows one to analyze two-slot quantum processes, which transform a pair of quantum channels into  an output channel. Formally, a two-slot process  can be seen as a four-partite process matrix $W\in\mathcal{L}(\H^P\otimes\H^{A_IA_O}\otimes\H^{B_IB_O}\otimes\H^F)$ where the past party $P$ has a trivial input space and the future party $F$ has a trivial output space, as illustrated in Fig.~\ref{fig:ProcessMatrix1}~\textbf{b)}. Differently from the one-slot case, two-slot processes may not respect a definite causal order.
A standard example of a two-slot process with indefinite causality is the quantum switch, already presented in Fig.~\ref{fig:switch} and which we will discuss in more details in Sec.~\ref{switchsection}. 
When acting on a pair of unitary channels described by the operators $U_A$ and $U_B$, the quantum switch transforms these into a unitary output channel with operator $U_\text{out}=\ketbra{0}{0}^\mathrm{c}\otimes U_BU_A + \ketbra{1}{1}^\mathrm{c}\otimes U_AU_B$, where the first system, with superscript $\mathrm{c}$, corresponds to the control system, and the second one, onto which $U_A$ and $U_B$ are applied, to the target system.

Analogously, we may also consider $k$-slot processes, which can be seen as $(k+2)$-partite processes matrices where $k$ parties have (in general) non-trivial input and output spaces, the past party $P$ has a trivial input space and the future party $F$ has a trivial output space.
Indeed, process matrices effectively transform quantum operations into quantum operations---in this context, process matrices are then the Choi operators of \textit{higher-order operations} \citep{perinotti16higher,bisio19higher,taranto2025higher}. This approach allows one to extend the range of problems and circumstances where the process matrix formalism can be used. In Sec.~\ref{sec:applications}, we present several applications that employ this open-past and open-future perspective. 

Note that the quantum processes discussed in the previous sections, which produce probabilities as a function of local maps, can  also be seen as higher-order transformations where the input and output spaces of the resulting operation are trivial (i.e, one-dimensional linear spaces).  Conversely, as presented in this section, any process with open past and future can be viewed as a particular type of process matrix where one site (the past) has trivial input space, while another (the future) has trivial output. Therefore, the two points of view are ultimately equivalent and the choice between the two is a matter of convenience depending on the context.

\subsection{Unitary processes}
\label{subsec:unitary_processes}

{Some of the distinctive features of quantum theory, such as superposition, are best understood by restricting the attention to pure states and unitary evolution, which represent the maximal possible knowledge of a closed system, so it is meaningful to look for similar notions for processes.
As quantum processes combine states and evolution in one object, different notions of ``purity'' and ``unitarity'' appear in the literature. As mentioned in Sec.~\ref{processasstate}, here we adopt the terminology first introduced in \citep{Araujo_2015}, where pure processes are identified with rank-1 process matrices, such that $W=\proj{V}$ for some vector $\ket{V}$. As we have seen in sec.~\ref{processaschannel}, a process can be viewed as a channel from the tensor product of all output spaces to that of all input spaces, so we can define a unitary process as one corresponding to a unitary channel,  such that $W=\kketbbra{U}{U}$, where $\kket{U}$ is the Choi vector of some unitary operator $U$. A prominent example of a unitary, causally nonseparable process is the quantum switch, which we will discuss in detail in Sec.~\ref{switchsection}.} 

Motivated by finding a principle that characterizes physically implementable processes with indefinite causality, \citet{Araujo2017purification} introduced a different definition (referred there as ``pure processes''), corresponding to the ``reversibility preserving processes'' of \citep{Yokojima2020}. These are processes with open past and open future that preserve the reversibility of quantum operations. That is, they transform unitary operations into a unitary operation, including when the input unitary operations act on a larger space containing external auxiliary systems. Remarkably, \citet{Araujo2017purification} showed that this definition is equivalent to that of a unitary process: a process $W$ is reversibility preserving if and only if there exists a unitary operator $U$ such that $W=\kketbbra{U}{U}$. We stress that, although all unitary processes are rank-1 process matrices, not all rank-1 (valid) process matrices are unitary processes (they may be non-unitary isometric channels). 

\citet{Barrett2020} and \citet{Yokojima2020} showed that, in the two-slot case, all unitary processes have a similar essence as the quantum switch, that is, they can always be written as coherent control of causally ordered process. More rigorously, if $W=\kketbbra{U}{U}$ is a two-slot unitary process, the unitary $U$ decomposes as a direct sum $U=U_{A\prec B}\oplus U_{B\prec A}$, where $W_{A\prec B}=\kketbbra{U_{A\prec B}}{U_{A\prec B}}$ and $W_{B\prec A}=\kketbbra{U_{B\prec A}}{U_{B\prec A}}$ are {unitary and} causally ordered processes and $\oplus$ denotes the direct sum (i.e., the two terms act on orthogonal subspaces). This characterization leads to a recipe on how to implement two-slot unitary processes in terms of coherent quantum control of causal order \citep{Wechs2021}.
Also, this characterization can be used to show that two-slot unitary processes cannot violate causal inequalities \citep{Yokojima2020,Branciard2016,OreshkovGiarmatzi_2016}, see Sec.~\ref{switchnonseparability}. A further consequence is that unitarily extendible bipartite processes (without open past and future) are necessarily causally separable.

Another implication of the above result is that two-slot unitary processes with different causal order cannot be directly superposed,  as independently proven in \citep{Costa2020}. More precisely, a non-trivial linear combination $\alpha  \kket{U_{A\prec B}} + \beta \kket{U_{B\prec A}}$ in general does not define a valid process--- unless the differently ordered components live in orthogonal subspaces. In other words, it is necessary to have some additional ``control'' degree of freedom in order to define a unitary two-slot process with indefinite causal order (as in the quantum switch). The result extends to more than two slots, although with additional assumptions: $\alpha  \kket{U_{A\prec B}} + \beta \kket{U_{B\prec A}}$ cannot be a valid process if $\kket{U_{A\prec B}}$, $\kket{U_{B\prec A}}$ are Markovian (i.e., they are products of independent unitaries) and the input and output dimensions of all sites are equal \citep{Costa2020}. It is currently unknown if the result holds under weaker conditions.

More generally, in multi-slot scenarios, unitary processes are known to have a more complex structure. For instance, there exist three-slot unitary processes which violate causal inequalities~\citep{Araujo2017purification,Baumeler_2016}; see Sec.~\ref{violationclassicalprocesses}.
\citet{Wilson2022Polycategorical} introduced the concept of polyslots, a strengthening of locally-applicable transformations, which provides a categorical approach that characterizes the unitary process matrices when applied to unitaries.
Moreover, \citet{Tselentis2023} showed that the causal structure of unitary processes (with or without indefinite causal order) must respect a graph-theoretic criterion, called ``siblings-on-cycles graph'' structure, and conjectured that there exists a valid unitary quantum process for every siblings-on-cycles graph structure.
It is an important open question, whether there exists a simple necessary and sufficient characterization of general multi-slot unitary processes.
    
\subsection{Classical processes} \label{classicalprocesses}

In the process matrix formalism, (local) classical operations can be described by transition probabilities $P(a, o|x,i)$, i.e., by the conditional probabilities that the measurement result $a$ is observed
and the classical output state $o$ is sent out when the measurement setting is $x$ and the input state is $i$. For each choice of setting variable $x$, classical operations can be expressed in the quantum formalism as instruments, more specifically, as sets of CP maps with the CJ operators that are diagonal in the computational basis, i.e. $M_{a|x} = \sum_{o,i} P(a,o|x,i) \ketbra{i}{i}\otimes\ketbra{o}{o}$. To express
correlations for such classical maps for an arbitrary number of observers, it is sufficient to consider process matrices that are diagonal both in the input and the output spaces---we call them ``classical process matrices''.  
Thus, both the classical process matrices and the local operations can be expressed in the diagonal form, and their CJ operators contain only the identity $\id$ and the Pauli-Z matrix $\sigma_z$ (where we identify the computational basis with the basis of the Pauli-Z operator). 

In their original work, \citet{OCB_2012} showed that, in the bipartite case, such process matrices are always compatible with a global causal order, i.e., they are causally ordered or causally separable. Relaxing the assumptions, \citet{BaumannBrukner} showed that, for bipartite process matrices, assuming that they are diagonal in the input spaces only is actually enough to imply compatibility with a global causal order. This is no longer true for classical process matrices with three or more parties,  which were furthermore shown to be able to also violate causal inequalities \citep{Baumeler2014PRA}, see Sec.~\ref{violationclassicalprocesses}.

\begin{figure}[hbt] 
    \begin{center}
    \includegraphics[width=.55\columnwidth]{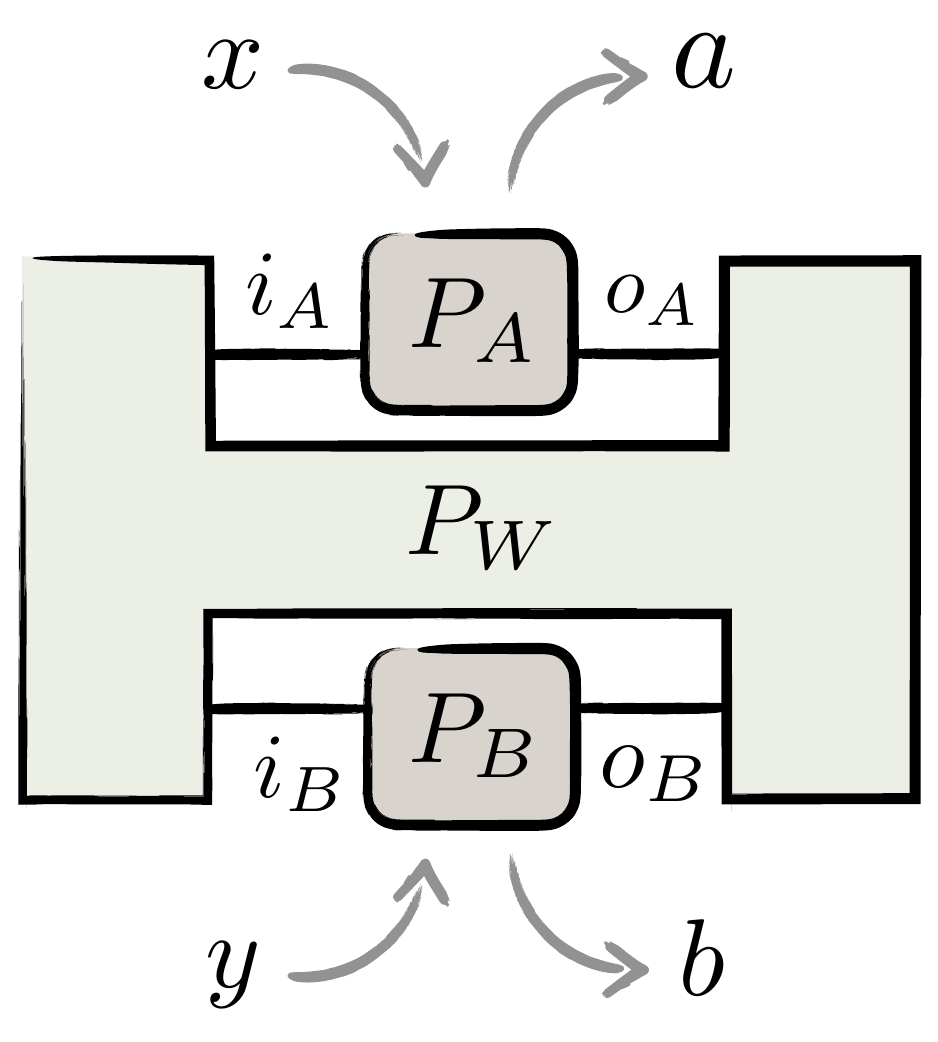} 
    \end{center}
\caption{A classical process is defined by transition probabilities from local outputs ($o_A, o_B$) to local inputs ($i_A, i_B$). It combines with the local operations (which map probabilistically local inputs to local outputs) to give the probabilities for the observed outcomes $a, b$, conditional on the settings $x,y$, according to the ``classical generalized Born rule'' of Eq.~\eqref{eq:classical_Born_rule}.}
\label{fig:classical_ProcessMatrix}
\end{figure}

\begin{figure}[htbp] 
	\begin{center}
		\includegraphics[width=\columnwidth]{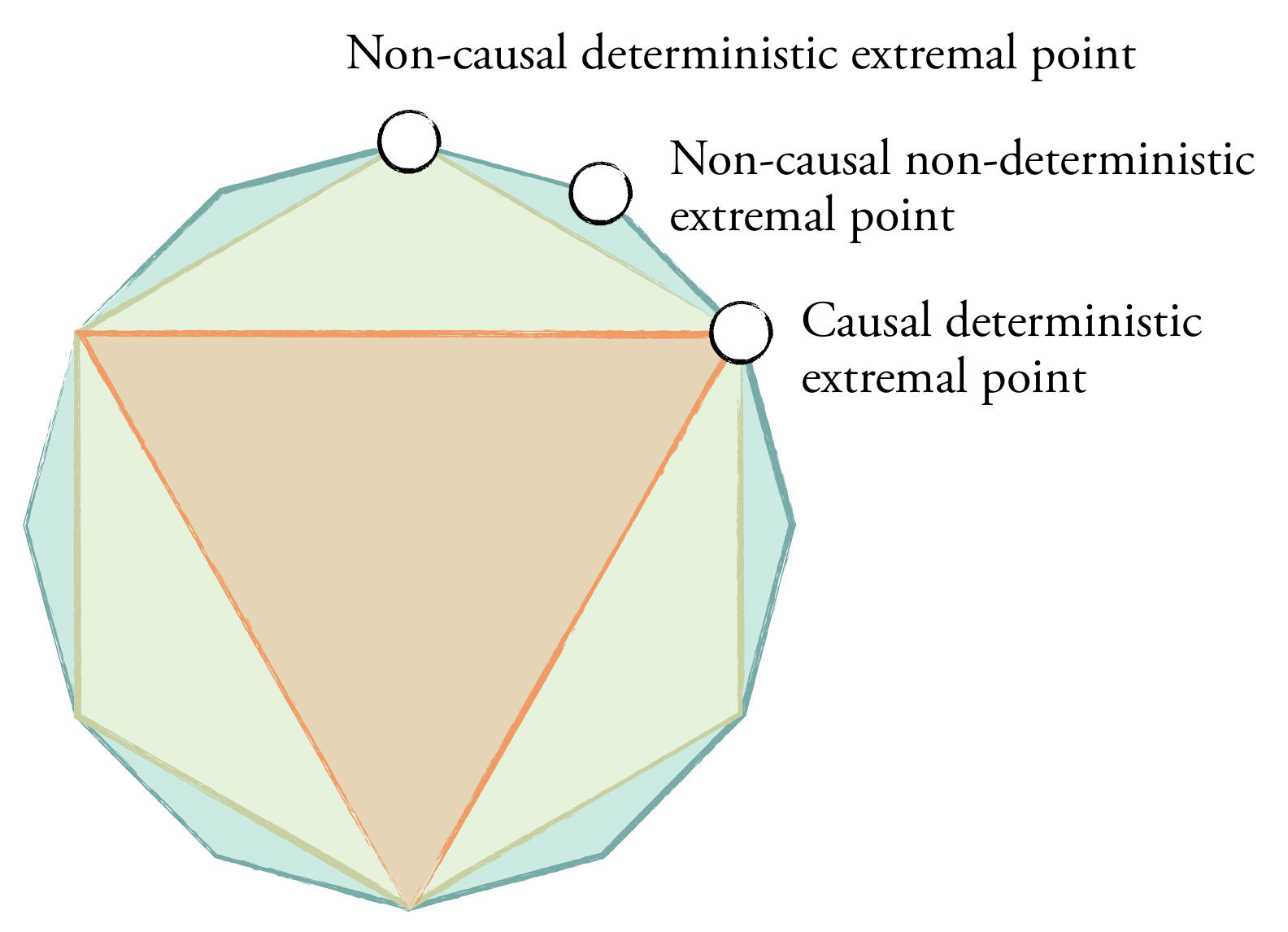} 
	\end{center}
\caption{Qualitative depiction of polytopes of classical processes. The inner polytope, colored in orange, represents the set of all processes achievable within a definite causal order. The outer polytope, in blue, illustrates the broader set of all valid classical processes, including those that lack a definite causal order. The middle polytope, in green, includes all processes that can be expressed as mixtures of deterministic ones, with or without definite causal order. Figure adapted from \citep{Baumeler_2016}.
}
\label{fig:polytopes}
\end{figure}

When restricted to classical processes and operations, the generalized Born rule can be written in a ``fully classical'' form: 
\begin{multline}
  P(a,b|x,y) = \sum_{o_A o_B i_A i_B} P_A(a,o_A|x,i_A) \, P_B(b,o_B|y,i_B) \times \\
  \times P_W(i_A,i_B|o_A, o_B) \label{eq:classical_Born_rule}
\end{multline}
for the bipartite case, see Fig.~\ref{fig:classical_ProcessMatrix}. Here, $P_A$ and $P_B$ are the transition probabilities defining the local operations, while $P_W(i_A,i_B|o_A, o_B)$ are the diagonal entries of the process matrix, which are normalized---and can be interpreted as---transition probabilities from outputs to inputs (similar to the interpretation of general quantum processes as channels, see Sec.~\ref{processaschannel}). This enables a treatment of classical processes within the framework of classical probabilities, where the conditions defining valid processes translate into constraints on the conditional probabilities $P_W$. This approach was used by \citet{Baumeler_2016} to characterize the set---a convex polytope, in that case---of all valid (``logically consistent'' in their terminology) classical processes, with or without definite causal order, for up to three parties, and with binary inputs and outputs. They also derived a smaller polytope, the intermediate one in Fig.~\ref{fig:polytopes}, whose extremal points are the valid deterministic processes, in which all inputs $i_A,i_B,\dots$ are given as a function of all outputs, $o_A,o_B,\dots$, called the ``process function''. As with the previous larger polytope, this includes processes with indefinite causal order. Notably however, there are valid classical processes that cannot be written as convex combinations of deterministic ones, hence which lie outside of this intermediate polytope. These processes can still be written as mixtures of deterministic functions, but where at least one of them is not a valid process function. 
\citet{Baumeler_2016} argued that such mixtures should be regarded as nonphysical because they are unnaturally fine-tuned, in the sense that a tiny variation in the mixture may turn valid into non-valid processes. 
\citet{kunjwal2024Generalizing} referred to the correlations arising from such processes as \textit{antinomic} and argue they should in fact not be considered classical by relating them to non-classical correlations in Bell scenarios, see Sec.~\ref{subsubsec:analogy_noncausal_nonlocal}.

\begin{figure}[htbp] 
	\begin{center}
		\includegraphics[width=\columnwidth]{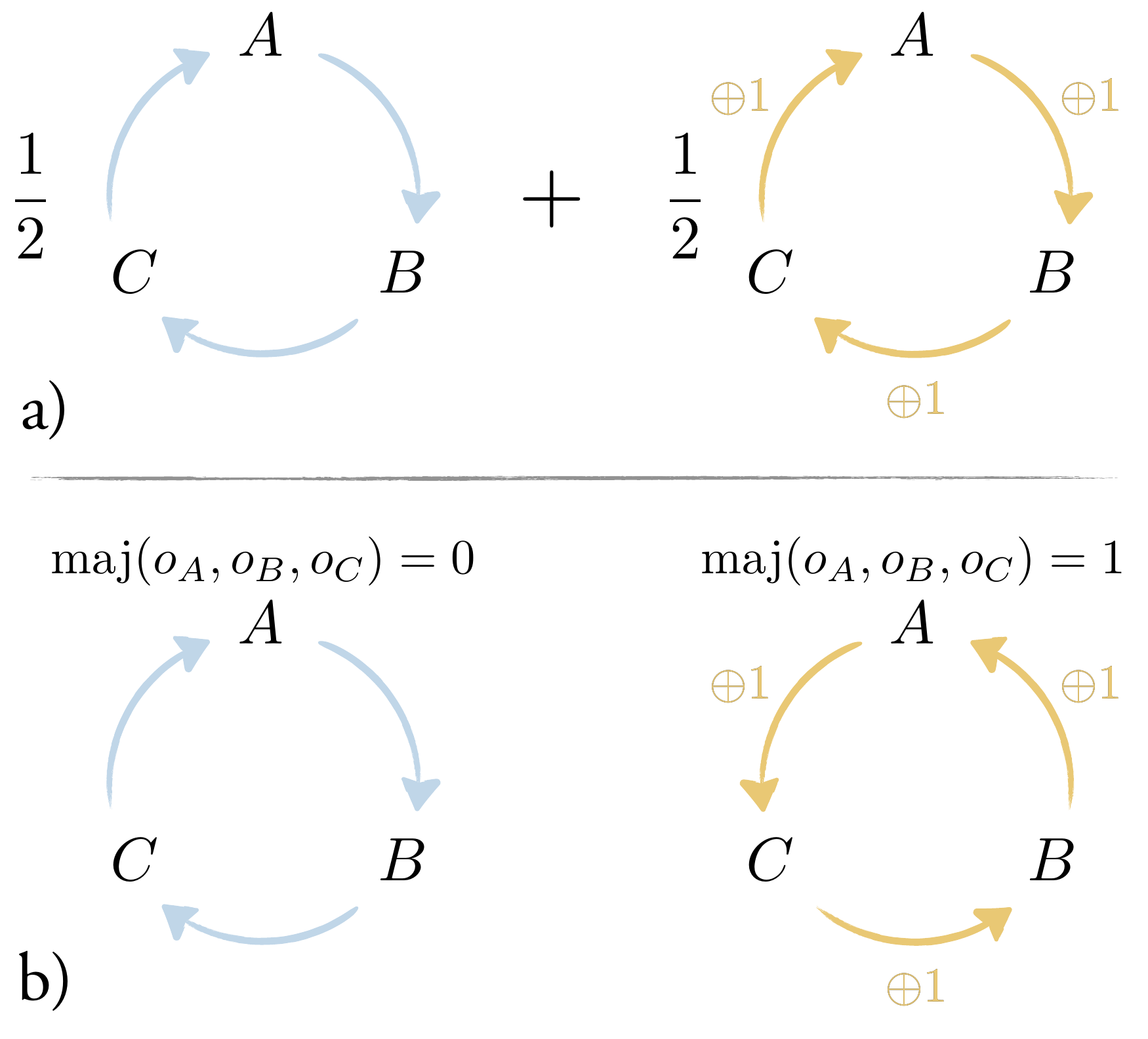} 
	\end{center}
\caption{\textbf{a)} Illustrative example of a mixture involving two non-valid process matrices: a circular identity channel combined with a circular bit-flip channel. \textbf{b)} Process that sends the output of each party to the next in a loop, with the direction determined by whether the majority of outputs is 0 or 1. Figure adapted from \citep{Baumeler_2016}.}
\label{fig:W_3}
\end{figure}

A qualitative representation of \textit{(i)} the polytope of classical processes without predefined causal order, \textit{(ii)} the polytope whose all extremal points are classical deterministic processes, and \textit{(iii)} the causal polytope with causal deterministic extremal points is given in Fig.~\ref{fig:polytopes}. An explicit and illustrative example of a valid process that is a mixture of two non-valid ones, first introduced by \citet{Baumeler2014PRA}, is given in Fig.~\ref{fig:W_3}a. It is a three-party process between, say Alice, Bob, and Charlie, of the form:
\begin{align}
    W_3^{ABC}=\frac{1}{8}(\id^{\otimes 6} & + \id \otimes \sigma_z \otimes \sigma_z\otimes \sigma_z\otimes \sigma_z \otimes \id  \nonumber \\
    & + \sigma_z \otimes \id \otimes \id \otimes \sigma_z \otimes \sigma_z  \otimes \sigma_z \nonumber \\
    & + \sigma_z \otimes \sigma_z \otimes \sigma_z \otimes  \id  \otimes 
    \id \otimes \sigma_z ).
\end{align}
The second summand in $W_3$ 
is responsible for signaling correlations from Alice and Bob to Charlie, and similarly for the next two summands.

An alternative way of understanding the process is to examine its action as a map from the parties' outputs to their inputs. Indeed, if the respective outputs of the parties are $o_A, o_B, o_C \in \{0,1\}$, the process maps them to the inputs $i_A, i_B, i_C \in \{0,1\}$ according to the following probability distribution
\begin{align} 
    & P(i_A,i_B,i_C|o_A,o_B,o_C) \notag \\ 
    & \quad = \begin{cases}
      1/2, & \text{if}\ i_A =o_C,i_B=o_A,i_C=o_B \\
      1/2, &  \text{if} \ i_A =\bar{o}_C,i_B=\bar{o}_A,i_C=\bar{o}_B  \\
      0, &  \text{otherwise,}
    \end{cases} \label{loop1}
\end{align}
where $\bar{o}_k= o_k \oplus 1$ ($k=A,B,C$). We see that process $W_3$ realizes a uniform mixture of two causal loops (in the same direction) where the output of one party is sent as an input to the next one in the loop, and where the output of one party is flipped and then sent as an input to the next one in the loop. Since in the ``flipped'' loop an even number of bit-flips are applied, any party can signal to their predecessor in the loop but not to themselves, in agreement with valid processes.  

In their work, \citet{Baumeler_2016} introduced a process, now standardly referred to as the ``Lugano process'', with the remarkable property of being fully deterministic and yet being able to violate causal inequalities.\footnote{This process was first discovered in Vienna by Ara{\'u}jo and Feix, who privately communicated it to Baumeler and Wolf, based in Lugano. Upon agreement, Baumeler and Wolf later published the results, leading to the process being referred to as the ``Lugano process''.}
Being deterministic, the process can be expressed as a \textit{process function} from outputs to inputs:
\begin{align}
    i_A =  \bar{o}_B o_C , \quad
    i_B =  \bar{o}_C o_A , \quad
    i_C = \bar{o}_A o_B. \label{loop2}
\end{align}
A possible way to interpret this process is that it sends the output of each party to the next one, in a loop, but the direction of the loop depends on whether the majority of outputs is $0$ or $1$, see Fig.~\ref{fig:W_3}b.
Alternatively, the process can be interpreted as implementing ``bipartite conditional one-way signaling'', in the sense that each pair of parties shares a noiseless one-way channel, with the channel direction determined by the output of the third party. For example, for $o_C=0$, $A$ signals to $B$, $i_B=o_A$, but $A$ receives the constant $i_A = 0$. For $o_C=1$, $B$ signals to $A$, $i_A=\bar{o}_B$, but $B$ receives the constant $i_B = 0$. The conditional one-way nature of the process is what prevents the information from traveling from a party's output to the same party's input, which would contradict the party's ability to prepare the output as an arbitrary function of the input. A complete characterization of valid process function was provided by \citep{dourdent2025}, improving on (and correcting) previous attempts in \citep{Baumeler_2019} and \citep{Tobar_2020}.

Another interesting aspect of the Lugano process is that, as all process functions, it can be extended to a reversible one (i.e., defined by an invertible function), by introducing global past and future variables (in the terminology of Sec.~\ref{subsec:open_past_future}, such an extension is a three-slot process with open past and future). Furthermore, every invertible function $o\mapsto i = f(o)$ defines a unitary channel from output to inputs, $U_f=\sum_{o}\ketbra{f(o)}{o}$, and it was proven in~\citep{Araujo2017purification} that the unitary extension of the Lugano process is a valid quantum process.\footnote{This process was first discovered in Lugano by Baumeler and Wolf, who privately communicated it to Ara{\'u}jo \emph{et al.}, based in Vienna, who later published it upon agreement.} More generally, \citet{Baumeler_2019} show that any deterministic process (a process function) can be made reversible, hence it can also be made unitary. 

In Sec.~\ref{violationclassicalprocesses}, we introduce causal games and the corresponding causal inequalities violated by both processes in Eq.~(\ref{loop1}) and~(\ref{loop2}).

The Lugano-type processes can also be understood as examples of deterministic reversible dynamics in the presence of closed time-like curves (CTCs), as introduced in \citep{Baumeler_2019, Tobar_2020}. The framework of \citep{Baumeler_2019} extends the usual concept of time evolution to that where, for a number of spacetime regions, the state in the past of each region is computed as a function of the state in the future of all regions. In addition, the framework allows for arbitrary operations to be performed in each region, and is thus compatible with ``freedom of choice''. This shows that it is possible to have deterministic reversible dynamics compatible with arbitrary local operations, where the state observed in each region depends non-trivially on the states prepared in all other regions, in contrast to standard approaches to CTCs \citep{deutsch1991quantum,Lloyd2011}, which are either in conflict with local physics or with the ``freedom of choice'' condition.

Interesting connections have also been found between classical processes with indefinite causal order and quantum information processing, in particular, by extending the set of local operations and classical communication (LOCC). \citet{Akibue2017} introduces the concept of LOCC$^*$, which extends LOCC by the parties to make use of general conditional probabilities, which are not required to be consistent classical process, and prove that in the bipartite case, LOCC$^*$ is the set of bipartite channels with separable Choi operators (SEP). Then, \citet{Kunjwal2023} extends LOCC by allowing the parties to have classical process, and show that this extension of LOCC enables the possibility of performing SHIFT measurements \citep{Bennett1999NLwithout}, what cannot be done by standard LOCC, a result that was then generalized in \citep{dourdent2025}. The connection between SHIFT measurements and the quantum switch is also discussed in \citep{Steffinlongo2026}.

\citet{Baumeler2016_FixedPoints, Baumeler_2019} showed that classical process necessarily have a unique fixed point. \citet{Baumeler2020grandfather} used this result to show that, in the framework of classical process matrices, the grandparent antinomy and the information antinomy are equivalent: if in some quasi process the parties can generate the grandparent antinomy, then there exists a choice of local interventions for the information antinomy, and vice versa.

\subsection{Infinite-dimensional process matrices}
\label{subsubsec:infinite_dim}

Most of the literature on indefinite causal order focuses on processes where local input and output Hilbert spaces at each site have finite dimensions. It is desirable to extend the process matrix formalism to include infinite-dimensional Hilbert spaces, which occur naturally in quantum theory. Even though the basic definitions generalize in a direct way---a quantum process is a multilinear functional of quantum operations---this is still a relatively little explored topic. One of the technical issues is the definition of the Choi isomorphism, which does not directly extend to the infinite-dimensional case because the ``identity vector'' $\kket{\id} = \sum_i\ket{i}\ket{i}$ is not well defined in an infinite-dimensional Hilbert space \citep{Holevo2011}.

Causally ordered process matrices with arbitrary dimension (and possibly infinity) were analyzed in \citep{accardi1982stochastic} via the $C^*$ algebra formalism under the name of ``Quantum Stochastic Processes'', and in \citep{Kretschmann2005} under the name of channels with memory. Additionally, quantum and post-quantum causal modeling with infinite dimension were also analyzed in \citep{milz2020Kolmogorov}.

For general process matrices, an explicit approach, based on phase-space methods, has been proposed by \citep{Giacomini2015}. In this approach, one models a quantum system in terms of (possibly multi-dimensional) generalized coordinates $x$ and their canonically conjugate momenta $p$, which we denote compactly as $\xi\equiv (x,p)$. Hilbert-space operators, such as a state $\rho$, are represented as phase-space functions, $\rho(\xi)$, such that $\tr \rho = \int d\xi \rho(\xi)$. A convenient phase-space representation is the Wigner function, which has the advantage that HS scalar products have the simple representation $\tr (E \rho) = \int d\xi E(\xi) \rho(\xi)$, where $E(\xi)$ and $\rho(\xi)$ are the Wigner-function representations of the operators $E$ and $\rho$, respectively. [See e.g.\ \citep{scully_zubairy_1997} for an introduction to phase-space methods.]

For a quantum process involving two sites $A, B$, we associate phase-space variables to each input and output space. 
The Born rule for processes then reads
\begin{equation} \label{BornWigner}
    P(a,b) = \int d\vec{\xi}\, M_a(\xi_{A_I}, \xi_{A_O}) N_b(\xi_{B_I}, \xi_{B_O}) W(\vec{\xi} \,),
\end{equation}
where the functions $M_a(\xi_{A_I}, \xi_{A_O})$ and $N_b(\xi_{B_I}, \xi_{B_O})$ are phase-space representations of the CP maps applied at sites $A$ and $B$, respectively, and $W(\vec{\xi})$, with $\vec{\xi}\coloneqq(\xi_{A_I}, \xi_{A_O}, \xi_{B_I},\xi_{B_O})$, is the phase-space representation of the process matrix. We see that the phase-space representation provides a continuous-variable version of the process matrix formalism that retains several of the properties arising from the Choi-Jamio{\l}kowski isomorphism. For example, it is easy to see that, if a process does not depend on the output variables of a party, such that $W(\vec{\xi}) = {W}(\xi_{A_I},\xi_{A_O},\xi_{B_I})$, then that party cannot signal to the others, in close analogy with the form $W^{A_IA_OB_I}\otimes\id^{B_O}$ of causally ordered processes, see Eq.~\eqref{BOlast}.

\section{The quantum switch}
\label{switchsection}

Up to this point, the discussion has centered on key aspects of the process matrix formalism, including its axiomatic derivation and the concept of causal separability as a formal definition of definite causal order. This section shifts the focus to the quantum switch, a widely studied example of a process exhibiting an indefinite causal order.

\subsection{The quantum switch process}
\label{subsec:W_QS}

\begin{figure}[htb] 
	\begin{center}
		\includegraphics[width=.7\columnwidth]{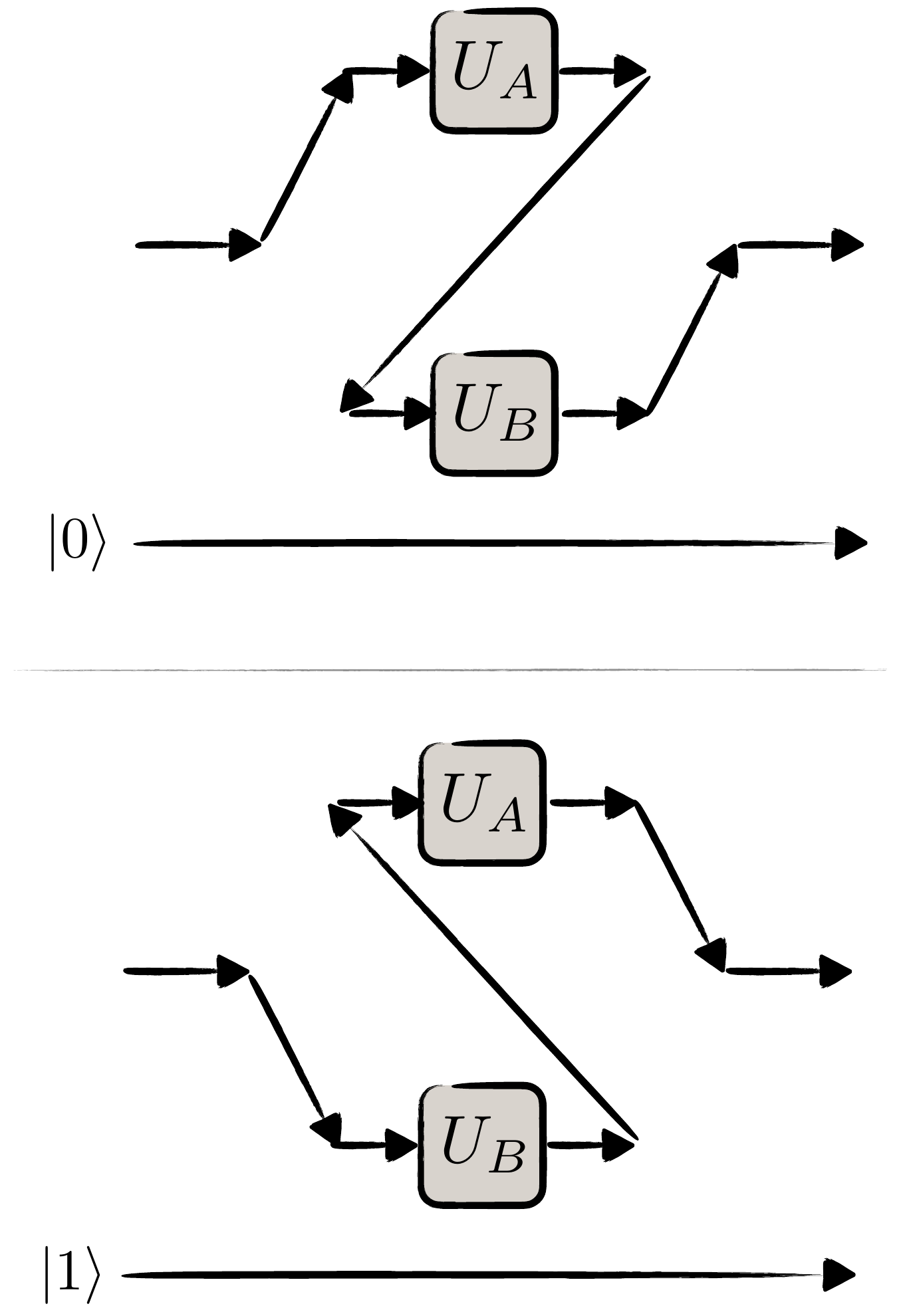} 
	\end{center}
\caption{The quantum switch. The connection between two quantum operations, $U_A$ and $U_B$, and hence the order of their applications, depends on the state of a control system. Figure adapted from \citep{Chiribella2013}.}
\label{fig:quantum_switch}
\end{figure}

The \textit{quantum switch}---sometimes just \textit{switch} for brevity---is the most studied example of a process with indefinite causal order, thanks to the simplicity of its definition and its natural physical interpretation, see Fig.~\ref{fig:quantum_switch}. It was originally introduced in \citep{Chiribella2009_arxiv}---a precursor to \citep{Chiribella2013}---as a quantum-coherent generalization of the \textit{classical switch}. The latter is a process where a classical variable decides the order in which two operations are applied on a target system. The original motivation was to show that the switch (both classical and quantum) cannot be implemented by a quantum circuit with fixed connections between operations (and with no ``backwards in time'' loops).  
The interest of most subsequent literature---as well as of this review---is rather in the incompatibility of the quantum switch with any definite causal order, including classical uncertainty or classical control, in agreement with the notion of causal non-separability outlined in Sec.~\ref{sectionmultiseparability}.

\subsubsection{Quantum switch with open past and future} \label{subsubsec:QS_with_open_PF}
There are several variants and generalizations of the quantum switch. A common approach is to define it as a \textit{unitary supermap}, i.e., as a higher-order transformation---in the sense of Sec.~\ref{subsec:open_past_future}---mapping unitaries to unitaries. Given two Hilbert spaces, a \textit{control},  $\mathcal{H}^{\textup{c}}\cong \mathbb{C}^2$, and a \textit{target}, $\mathcal{H}^{\textup{t}}$ (of arbitrary dimension), the quantum switch transforms a pair of unitaries acting on the target, $U_{A,B}:\mathcal{H}^{\textup{t}}\rightarrow\mathcal{H}^{\textup{t}}$, to a unitary acting on the joint control-target system, $S(U_{A},U_{B}): \mathcal{H}^{\textup{c}}\otimes\mathcal{H}^{\textup{t}}\rightarrow \mathcal{H}^{\textup{c}}\otimes\mathcal{H}^{\textup{t}}$, defined as
\begin{align} \label{unitaryswitch}
S(U_A,U_B) \coloneqq  \ketbra{0}{0} \otimes U_B U_A + \ketbra{1}{1} \otimes U_AU_B.
\end{align}
The interpretation is that the two unitaries act on the target in the order $U_A \prec U_B$ if the control is in state $\ket{0}$ and in the order $U_B \prec U_A$  for the initial control state $\ket{1}$. 

To define the action of the switch on arbitrary local operations, we can first decompose them into \textit{Kraus operators} \citep{NielsenChuangBook}: The action of any CP map $\mathcal{M}$ can be written as $\mathcal{M}(\rho) = \sum_j K_j \rho K_j^{\dagger}$ for a set of operators $\{K_j\}_j$. In turn, any set of operators defines a CP map in this way, with the trace-preserving condition being equivalent to $\sum_j K_j ^{\dagger} K_j = \id$. Given the Kraus operators $K_{A\,i}$, $K_{B\,j}$ for two local maps $\mathcal{M}_{A,B}$, we can extend the action of the switch, Eq.~\eqref{unitaryswitch}, by linearity and define the operators
\begin{multline}
W_{i j} \coloneqq  S(K_{A\,i},K_{B\,j}) \\
= \ketbra{0}{0} \otimes K_{B\,j}K_{A\,i} + \ketbra{1}{1} \otimes K_{A\,i}K_{B\,j}.
\end{multline}
These are the Kraus operators that define the CP map---acting on control and target---obtained by inserting the two local operations into the switch:
\begin{align} \label{switchsupermap}
\mathcal{S}\left(\mathcal{M}_{A},\mathcal{M}_{B}\right) (\rho) \coloneqq  \sum_{ij} W_{i j} \rho W_{i j}^{\dagger}. 
\end{align}
It can be readily verified that this map is trace preserving whenever both $\M_A$ and $\M_B$ are trace preserving.

Although it is not immediately obvious in this form, the definition \eqref{switchsupermap} does not depend on the choice of Kraus representation of the two local maps, hence it is a well-defined linear supermap---and therefore, it defines a valid two-slot process with open past and future (i.e., a four-partite process where a ``past'' site has trivial input and a ``future'' site has trivial output, see Sec.~\ref{subsec:open_past_future}). This can be verified by passing to the process matrix representation, which greatly facilitates the analysis. To this end, we introduce the local spaces $\mathcal{H}^{A_I}\cong \mathcal{H}^{A_O} \cong \mathcal{H}^{B_I} \cong \mathcal{H}^{B_O} \cong \mathcal{H}^{\textup{t}}$, and the past and future spaces, $\mathcal{H}^{P} \coloneqq \mathcal{H}^{P_c} \otimes \mathcal{H}^{P_t}$,  $\mathcal{H}^{F} \coloneqq \mathcal{H}^{F_c} \otimes \mathcal{H}^{P_t}$,  with $\mathcal{H}^{P_c} \cong \mathcal{H}^{F_c} \cong \mathcal{H}^{\textup{c}}$ and $\mathcal{H}^{P_t} \cong \mathcal{H}^{F_t} \cong \mathcal{H}^{\textup{t}}$. Since the switch is a unitary process (and thus ``pure'', in the language of Sec.~\ref{subsec:unitary_processes}), it has a rank-one process matrix,
\begin{multline} \label{unitaryswitchvector}
    W_\textup{QS} =  \ketbra{w_\textup{QS}}{w_\textup{QS}} , \quad \text{with} \\[1mm]
    \ket{w_\textup{QS}} \coloneqq \ket{0}^{P_\textup{c}} \kket{\id}^{P_\textup{t}A_I} \kket{\id}^{A_OB_I} \kket{\id}^{B_OF_\textup{t}} \ket{0}^{F_\textup{c}}
    \\
+ \ket{1}^{P_\textup{c}} \kket{\id}^{P_\textup{t}B_I} \kket{\id}^{B_OA_I} \kket{\id}^{A_OF_\textup{t}} \ket{1}^{F_\textup{c}} \\
\in {\cal H}^{PABF}.
\end{multline}
Recalling that $\kket{\id}^{XY}$ is the pure-process representation of the identity map from $X$ to $Y$, it is intuitive (and simple to verify), that composing $\ket{w_\textup{QS}}$ with local unitaries gives back Eq.~\eqref{unitaryswitch}, in the sense that
\begin{align}
\label{eq:Wswitch}
 \hspace{-0.22cm}   \left(\kket{U_A}^{A_IA_O}\otimes\kket{U_B}^{B_IB_O}\!\right)* \ket{w_\textup{QS}} = \kket{S(U_A,U_B)}^{PF}\!\!,
\end{align}
where we used the link product introduced in Sec.~\ref{subsubsec:link_prod}. 

For arbitrary local operations, with Choi representations $M_A$, $M_B$, the map $(M_A,M_B) \mapsto \left(M_A\otimes M_B\right)*W_\textup{QS}$ is linear by construction, and it is again straightforward (e.g., going through the Kraus representation) to verify that it produces the Choi matrix of the map $\mathcal{S}\left(\mathcal{M}_{A},\mathcal{M}_{B}\right)$, defined in Eq.~\eqref{switchsupermap}. This shows that, indeed, $W_\textup{QS}$ is the process matrix of the quantum switch.

Finally, as shown in Ref. \citep{dong2023uniquely}, the action of the quantum switch on local unitary operations presented in Eq.~\eqref{unitaryswitch}, combined with the fact the valid process matrices are positive semidefinite, uniquely defines the process matrix of the quantum switch. This ensures that the process matrix in Eq.~\eqref{unitaryswitchvector} is indeed the only one that agrees with~\eqref{unitaryswitch}.

\subsubsection{Quantum switch with fixed past} \label{subsubsec:QS_with_fixed_P}
As discussed, preparing the control system in one of the basis states $\ket{0}, \ket{1}$ results in the two operations being performed in one of the two possible orders. The interesting scenario is when the control is prepared in a superposition of these states. For this reason, one often considers a version of the switch where the past control-target system is initialized in state $\ket{+}^{P_\textup{c}}\otimes\ket{\phi}^{P_\textup{t}}$, where $\ket{+} \coloneqq  \frac{1}{\sqrt{2}}(\ket{0}+\ket{1})$ and $\ket{\phi}$ is an arbitrary state of the target's system. The resulting process matrix is
\begin{align}
W_\textup{QS}^{+,\phi} = & \ketbra{w_\textup{QS}^{+,\phi}}{w_\textup{QS}^{+,\phi}},  \quad \text{with} \notag \\[2mm]
\ket{w_\textup{QS}^{+,\phi}} \coloneqq & (\ket{+}^{P_\textup{c}}\otimes\ket{\phi}^{P_\textup{t}}) * \ket{w_\textup{QS}} \notag \\
= & \frac{1}{\sqrt{2}} \ket{\phi}^{A_I} \kket{\id}^{A_OB_I} \kket{\id}^{B_OF_\textup{t}} \ket{0}^{F_\textup{c}} \notag \\
& + \frac{1}{\sqrt{2}} \ket{\phi}^{B_I} \kket{\id}^{B_OA_I} \kket{\id}^{A_OF_\textup{t}} \ket{1}^{F_\textup{c}} \notag \\
& \hspace{15mm} \in {\cal H}^{ABF}. \label{eq:W_QS_v2}
\end{align}

This version of the switch can be interpreted directly as an equal superposition of two pure, causally ordered processes, corresponding to the orders $A\prec B$ and $B\prec A$, respectively (this is unlike the open-past switch, Eq.~\eqref{unitaryswitchvector}, because, in that case, the two components of the sum do not correspond to valid processes.) This interpretation gives rise to the common term ``superposition of causal orders'' for the switch. However, \citet{Costa2020} argues that this terminology overlooks the role of the control system, which is necessary to observe any indefinite causal order---hence it is more appropriate to interpret Eq.~\eqref{eq:W_QS_v2} as displaying \textit{entanglement} between causal order and the control. As discussed in Sec.~\ref{subsec:open_past_future}, there is no simple way to define a direct superposition of oppositely ordered processes.

\subsection{Causal non-separability of the switch}
\label{switchnonseparability}

If we trace out the final control system from the switch supermap, Eq.~\eqref{switchsupermap}, we obtain the \textit{classical switch} introduced in \citep{Chiribella2013}:
\begin{align}
&\mathcal{S}_{\textup{cl}}\left(\mathcal{M}_{A},\mathcal{M}_{B}\right) (\rho)\coloneqq \tr_{\textup{c}}\mathcal{S}\left(\mathcal{M}_{A},\mathcal{M}_{B}\right) (\rho) \notag \\
&= \mathcal{M}_{B}\circ\mathcal{M}_{A}\left(\bra{0}_{\textup{c}}\rho\ket{0}_{\textup{c}}\right) +\mathcal{M}_{A}\circ\mathcal{M}_{B}\left(\bra{1}_{\textup{c}}\rho\ket{1}_{\textup{c}}\right).
\end{align}
This supermap cannot be realized by a standard  quantum circuit with  fixed order if the operations $\mathcal{M}_{A,B}$ are treated as ``black boxes''---that is, if the maps are inserted in pre-assigned ``slots'' of the circuit and other parts of the circuit cannot depend on the choice of maps. 
However, the classical switch can still be realized as a process with definite causal order, where the initial state is measured on the control's computational basis and the order of gates is decided as a function of the classical outcome of the measurement. Intuitively, this is not the case for the quantum switch, because it retains coherence in the control's states. Nevertheless, to make the argument rigorous, we have to take into account the multipartite definition of causal separability, discussed in Sec.~\ref{sectionmultiseparability}, which includes the possibility of dynamical causal order. 

It is still quite straightforward to see that the fixed-past version of the quantum switch $W_\textup{QS}^{+,\phi}$, Eq.~\eqref{eq:W_QS_v2}, is not causally separable. In this case, we have three sites, $A$, $B$, and $F$, where $F$, having trivial output, cannot signal to any other site. Since $W_\textup{QS}^{+,\phi}$ is pure, it cannot be written as a nontrivial mixture of any other processes. Thus, to be causally separable, it would have to be compatible with one site being first. However, this is not the case, because all sites can receive signals from at least one other site: $A$ from $B$, $B$ from $A$, and $F$ from both $A$ and $B$. The non-separability of the open-past switch follows from the fact that it reduces to the fixed-past version by preparing the appropriate state at site $P$, while a causally separable process would always produce a causally separable one when fixing a CPTP map at one site.

As discussed above, tracing out the future control produces a causally separable process: the classical switch if we start from the open-past version and, starting from the fixed-past switch, the equal mixture of causal orders
\begin{align}
   \tr_{F_{\textup{c}}} W_\textup{QS}^{+,\phi}
   = & \frac{1}{2}\Big(\ketbra{\phi}{\phi}^{A_I} \kketbbra{\id}{\id}^{A_OB_I} \kketbbra{\id}{\id}^{B_O F_{\textup{t}}} \notag \\
   & \ \ + \ketbra{\phi}{\phi}^{B_I} \kketbbra{\id}{\id}^{B_OA_I} \kketbbra{\id}{\id}^{A_O F_{\textup{t}}} \Big).
\end{align}
On the other hand, it is interesting---and relevant for experiments---that tracing out the final target $F_{\textup{t}}$ \textit{does not} produce a causally separable process. This can be proven using the causal witness technique \citep{Araujo_2015, Branciard_2016}, reviewed in Sec.~\ref{subsec:causal_witness}. A particular class of witnesses, used by \citet{Araujo_2015} to prove the causal non-separability of this process, can be associated with the operational task of discriminating commuting vs anticommuting operations, introduced by \citet{Chiribella2012} and reviewed in Sec.~\ref{channel}. Indeed, accomplishing this task only requires measuring the control system.

An interesting aspect of the quantum switch is that its causal nonseparability cannot be verified in a fully device-independent way, without introducing any additional parties and assumptions: we need to rely on the quantum description of the process and on the knowledge of the particular local operations used to measure the witnesses. In a \textit{device-independent} approach, reviewed in Sec.~\ref{subsubsec:causal_correls} below, one only considers the conditional probabilities $P(a,b,c|x,y,z)$, without any assumption about the process or instruments that generates them, where $(x,a)$ are settings and outcomes for site $A$, $(y,b)$ are for site $B$, and $(z,c)$ are for $F$. As shown in \citep{Araujo_2015, OreshkovGiarmatzi_2016}, the conditional probabilities generated by the switch can always be reproduced by a causally ordered process. The argument can be understood as follows: in the quantum switch, $F$ cannot signal to $A$ and $B$, so only probabilities of the form $P(a,b,c|x,y,z) = P(c|a,b,x,y,z)P(a,b|x,y)$ can be generated, where $P(a,b|x,y)$ is obtained from the ``first part'' of the quantum switch, with the future site traced out---which, as discussed above, is causally separable. Therefore, $P(a,b|x,y)$ can be reproduced by a strategy with a definite causal order between $A$ and $B$. Since $P(c|a,b,x,y,z)$ can be reproduced by a causally ordered strategy too (where $F$ is last and receives all the variables $a,b,x,y,z$), all conditional probabilities generated by the switch can also be reproduced by causally separable strategies. The ``trick'' is that, in order to send all variables to $F$, it is, in general, necessary to use larger input and output spaces than in the quantum description. This is why it becomes possible to distinguish the switch from causally ordered scenarios if we constrain the local dimensions (or if we make other device-dependent assumptions). The argument extends to some semi-device-independent scenarios \citep{Bavaresco_2019}, Sec.~\ref{semiDI}, where we trust the quantum description of $F$ but not of the other parties, because any state prepared at $F$, as a function of $a,b,x,y,z$, can be reproduced by a causally ordered strategy where all the variables are sent to $F$. It further extends to all quantum circuits with quantum control of causal order (QC-QCs, see Generalizations below), and in particular to all two-slot unitary processes, which, as discussed in Sec.~\ref{subsec:unitary_processes}, can always be interpreted as such processes, similar to the switch.

Despite the impossibility of certifying indefinite causal order of the quantum switch in the standard device-independent scenario, such certification becomes possible by introducing additional parties and under extra assumptions. \citet{VanDerLugt_2022} showed that device-independent tests of indefinite causal order become possible if the switch is supplemented by an entangled state, and if the assumption of causal order is augmented by additional no-signaling constraints, see  Sec.~\ref{sec:vanderlugt} below for more details. Also, \citet{dourdent2023networkdeviceindependent} introduced the concept of network-device-independent certification of causal nonseparability, which can be used to certify the noncausality of the quantum switch, see Sec.~\ref{semiDI_QI}. 

\medskip

\subsection{Generalizations}
\label{subsec:Gen}

The basic control of causal order introduced above can be generalized in several ways. First, we can allow the target to undergo arbitrary unitary transformations when traveling between sites, with different unitaries for each order. The open-past switch, Eq.~\eqref{unitaryswitchvector} can then be generalized to
\begin{multline} \label{genswitch}
    \ket{w_{\textup{gen}}} \coloneqq \ket{0}^{P_\textup{c}} \kket{U_1}^{P_\textup{t}A_I} \kket{U_2}^{A_OB_I} \kket{U_3}^{B_OF_\textup{t}} \ket{0}^{F_\textup{c}}
    \\
+ \ket{1}^{P_\textup{c}} \kket{V_1}^{P_\textup{t}B_I} \kket{V_2}^{B_OA_I} \kket{V_3}^{A_OF_\textup{t}} \ket{1}^{F_\textup{c}} 
\end{multline}
for some arbitrary unitaries $U_1, U_2, U_3,V_1,V_2,V_3$. 

For a pair of Kraus operators $K_A, K_B$ (and, in particular, for unitaries) applied at sites $A, B$,  the generalized switch supermap returns the total Kraus operator (in particular, unitary)
\begin{multline} \label{genswitchsupermap}
   S_{\textup{gen}}(K_A,K_B) \\
   \coloneqq \ketbra{0}{0} \otimes U_3 K_B U_2 K_A U_1  + \ketbra{1}{1} \otimes V_3 K_A V_2 K_B V_1.
\end{multline}
If the unitaries connecting the sites are known, it is possible to ``absorb'' some of them in the choice of local operations \citep{scie3250}. For example, by redefining  $K_A = V_3^{\dagger} \Tilde{K_A} U_3 V_2^{\dagger}$ and $K_B = U_3^{\dagger} \Tilde{K_B} V_3 U_2^{\dagger}$, we can effectively transform the supermap \eqref{genswitchsupermap} into 
\begin{equation} 
  \!\!\!  \Tilde{S}_{\textup{gen}}(\Tilde{K_A},\Tilde{K_B}) = \ketbra{0}{0} \otimes K_B K_A \Tilde{U}  + \ketbra{1}{1} \otimes  K_A  K_B \Tilde{V},
\end{equation}
with $\Tilde{U}\coloneqq U_3V_2{^\dagger}$, $\Tilde{V}\coloneqq V_3U_2{^\dagger}$. Thus, for any task where the parties know in advance the intermediate unitaries, the generalized switch is equivalent to an ordinary switch, but where the target state undergoes different initial unitaries depending on the control's state  (alternatively, we could redefine the local operations to eliminate the initial unitaries on the target, at the cost of introducing different final unitaries). For the open-past, unitary switch, where we allow for arbitrary target-state preparation $\ket{\phi}$ at $P_{\textup{t}}$ (and arbitrary measurement at $F_{\textup{t}}$), no further simplification is possible, because the states $\Tilde{U} \ket{\phi}$, $\Tilde{V} \ket{\phi}$ are different unless $\ket{\phi} = \Tilde{V}^{\dagger}\Tilde{U} \ket{\phi} =  U_2 V_3^{\dagger} U_3V_2^{\dagger} \ket{\phi}$. However, if we are interested in a ``fixed-past'' switch, we can always choose the initial target state $\ket{\Tilde{\phi}}$ to be an eigenstate of $\Tilde{V}^{\dagger}\Tilde{U}$, so that $\Tilde{U} \ket{\phi} = e^{i \chi}\Tilde{V} \ket{\phi}$ for some $\chi\in [0,2\pi]$. The relative phase can be further absorbed in the initial control state, $\ket{\chi} = (e^{-i\chi} \ket{0} + \ket{1})/\sqrt{2}$, so any generalized switch can be mapped to a ``canonical'' fixed-past switch, Eq.~\eqref{eq:W_QS_v2}, through an appropriate choice of local operations and initial states.

Another generalization is the $N$-switch, which controls the order of $N$ sites, rather than just $2$ \citep{COLNAGHI20122940, Araujo2014}. In order to have control over arbitrary permutations, the control system needs to have $N!$ orthogonal states, although one can consider simplified versions where the control ranges over a subset of permutations \citep{Taddei2020}. One can further consider an $N$-switch where, for each permutation, the various sites are connected by different unitaries. In general, a redefinition of the $N$ local operations will not be sufficient to eliminate all the unitaries in the $N!$ components of the $N$-switch, although this can become possible under specific restrictions on the unitaries and/or permutations.

Generalizing further the idea of coherent control of causal order, \citet{Wechs2021} introduced \emph{quantum circuits with quantum control of causal order} (QC-QCs), which include the possibility of dynamical control of order, namely, processes where the order of all parties is not encoded in the control from the start, but gets established on the fly, during the implementation of the process.
This class of processes maintains some of the properties of the quantum switch, including the impossibility of violating causal inequalities and the fact that they can be seen as coherent extensions of circuits with definite---although possibly dynamically-controlled---causal order (referred to as \emph{quantum circuits with classical control of causal order, QC-CCs}). However, it also includes examples that depart in significant ways from the switch: a tripartite QC-QC process was found (called the ``Grenoble process'' in later works \citep{Vanrietvelde2022}) that is not causally separable, despite not having a global future (whereas the switch becomes causally separable if the final sites are traced out). QC-QCs currently represent the most general class of processes with a direct physical interpretation in terms of laboratory procedures that do not involve time traveling or other exotic physics, and respect the closed-lab assumption \citep{salzger2023,salzger2025,Salzger2026}. 
The (potential) dynamical properties of the control were further examined by \citet{mothe2025}, based on which different subclasses of QC-QCs and QC-CCs were defined and characterized.

\subsection{Quantum-circuit representations of the quantum switch}
\label{circuitswitch}

As discussed above, the quantum switch cannot be represented by any standard, causally ordered quantum circuit in which each operation corresponds to a single gate \citep{Chiribella2013}. 
However, by allowing multiple-gate encoding of each operation, representations of the switch within the ordinary quantum circuit formalism become possible. This implies that it is also possible to \textit{simulate} the switch using causally ordered protocols, where each of the operations is performed more than once. Whether specific experiments should be understood as implementations or simulations of the switch, and how to count the physical number of operations or events, is currently a matter of debate. In this section, we limit ourselves to review some known ways to design quantum circuits that reproduce the action of the switch, leaving a more detailed overview of the aforementioned debate to Sec.~\ref{sec:debates}.

\subsubsection{Quantum circuits with multiple black-box calls} \label{subsubsec:SwitchMultipleCalls}

\begin{figure}[hbt] 
	\begin{center}
		\includegraphics[width=.95\columnwidth]{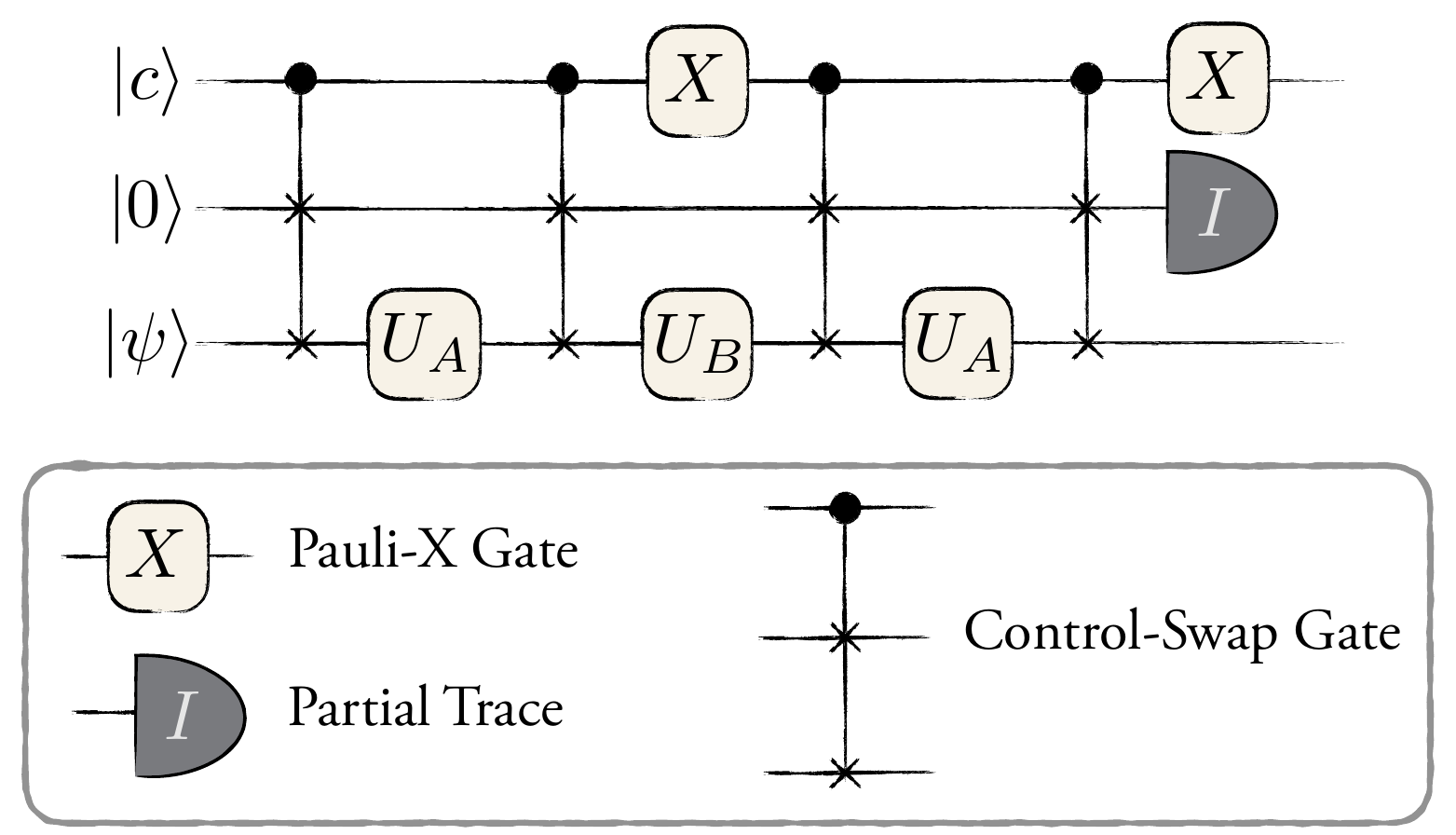} 
	\end{center}
\caption{
Quantum switch through multiple black-box calls to the unitaries.  Conditioned on the state of the control, this circuit swaps the target system with an auxiliary system of the same dimension, either before or after the action of $U_B$. In this way, the first call of $U_A$ acts on the target, resp., on the auxiliary system, if the control is in state $\ket{0}$, resp., $\ket{1}$, while the opposite happens for the second call. As the auxiliary system always undergoes a single $U_A$ transformation, tracing it out leaves control and target in the same output state as the quantum switch.}
\label{fig:Qcirc_impl}
\end{figure}

Recall that, in the unitary case, the quantum switch transforms a pair of unitaries $(U_A,U_B)$ into $S(U_A,U_B)=\ketbra{0}{0} \otimes U_B U_A + \ketbra{1}{1} \otimes U_AU_B$. As first presented in \citep{Chiribella2013}, for any unitary operations $U_A$ and $U_B$ the quantum circuit given in Fig.~\ref{fig:Qcirc_impl} behaves exactly as the quantum switch, where $\ket{c}$ is the control qubit, $\ket{\psi}$ is the target state and the control swap gate swaps the systems whenever the control state is $\ket{c}=\ket{1}$.

Therefore, if one extra call to $U_A$ is available, a causally ordered circuit can reproduce the action of the quantum switch on unitary operators. In a similar spirit, Ref. \citep{Araujo2014} (Appendix C) shows that, for unitary operations $(U_1,U_2,\ldots U_N)$, the $N$-quantum switch can be reproduced by a causally ordered circuit with $N^2$ calls of the input unitaries. 

If, however, instead of the unitary $U_A$, one inserts a non-unitary channel $C_A$ (such as the completely depolarizing channel) as a black box into the same circuit (Fig.~\ref{fig:Qcirc_impl}), 
the resulting operation of the circuit differs from what the quantum switch would output. A way to understand this is that, if we dilate each channel to a unitary interacting with an environment, every use of the channel requires, by definition, a distinct and independent environment, which can be traced out right after the channel is used. This allows the environment to gain information on whether the operation was performed the first or the second time, thus affecting the coherence of the control and target systems.

The problem of designing quantum circuits to reproduce the action of the quantum switch on non-unitary channels $\mathcal{C}_A$ and $\mathcal{C}_B$ is analyzed in Ref. \citep{bavaresco2024simulated,kristjansson2024exponential}. There, it is shown that,  for general channels, there is no standard ordered circuit which reproduces the action of the quantum switch even if we are allowed to have one extra call to each input channel. That is, for any $i,j,k,l\in\{A,B\}$, there is no fixed ordered process $\mathcal{F}$ such that $\mathcal{F}(\mathcal{C}_i,\mathcal{C}_j,\mathcal{C}_k,\mathcal{C}_l) = \mathcal{S}(\mathcal{C}_A,\mathcal{C}_B)$, for all pair of quantum channels $\mathcal{C}_A$ and $\mathcal{C}_B$. Furthermore, for an $n$-qubit target system, if one of the two channels is called once, then the switch cannot be reproduced if the other channel is called $k_B < d =2^n$ times. It is currently unknown whether the quantum switch can be reproduced with a finite number of calls to the channels.

Note that it is always possible to reproduce the action of the switch on an arbitrary CP map if one is given full access to its unitary dilation---i.e., if one represents the map through a unitary interaction with an environment that is eventually traced out (or, for a non-deterministic map, measured), see Thm.~1 of \citep{bavaresco2024simulated}. In this case, one can use again a circuit as in Fig.~\ref{fig:Qcirc_impl}, simply replacing the original target system with system and environment (which implies that, in this case, the auxiliary system has to be isomorphic to the joint target-environment system). In a physical implementation, where each gate in the circuit corresponds to a separate operation, this would require coherently swapping the environment with an auxiliary system of the same dimension, and using the same environment for all calls to each operation.

\subsubsection{Quantum circuits with quantum-controlled gates}\label{switchcontrolledU}

\begin{figure}[bt] 
	\begin{center}
		\includegraphics[width=.7\columnwidth]{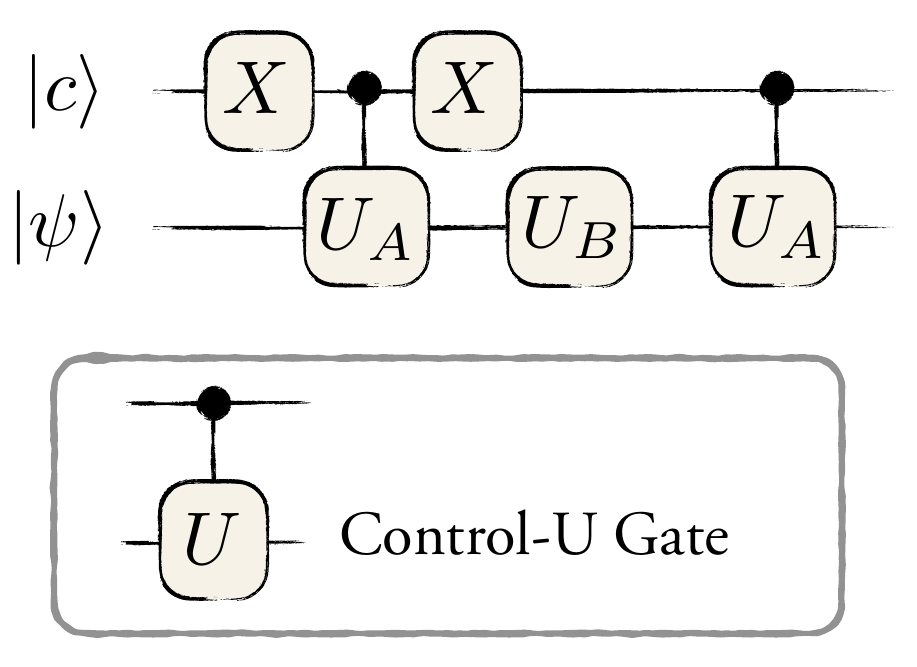} 
	\end{center}
\caption{Quantum switch through multiple calls to quantum-controlled unitaries. When the control is in state $\ket{0}$, the first gate acts as identity, and the last as $U_A$. For the control in $\ket{1}$, the first gate is $U_A$ and the last is identity. In this way, the circuit reproduces the action of the switch using two calls to a ``quantum-controlled'' version of one of the unitaries.}
\label{fig:Qcirc_impl_QCont}
\end{figure}

Another strategy to design a circuit that reproduces the same higher-order transformation as the quantum switch is to extend the local operations to quantum-controlled ones. For example, if we denote by $\textup{c-}U \coloneqq \proj{0}\otimes\id + \proj{1}\otimes U$  the controlled version of a unitary $U$, then the sequence of operations $\textup{c-}U_A\left(\sigma_x\otimes U_B \right) \textup{c-}U_A\left(\sigma_x\otimes \id\right)$ produces the same output as the quantum switch, see Fig~\ref{fig:Qcirc_impl_QCont}.
Notably, Fig.~\ref{fig:Qcirc_impl_QCont} is the representation of the quantum switch in the so-called ``causal reference frame'' \citep{Guerin2018a}, in which one of the operations is localized in time.

An advantage of this representation is that it does not require introducing an additional auxiliary system, as is the case for the black-box protocol in Fig.~\ref{fig:Qcirc_impl}.  
However, an implementation of the controlled unitary $\textup{c-}U$  does not count as a  ``black box'' use of $U$ within the standard quantum circuit formalism, as, by definition, this requires that $U$ acts on a subsystem as $U\otimes \id$, while $\textup{c-}U$ cannot be implemented in this way\footnote{An interferometer where a device applies a unitary $U$ in one arm is a natural implementation of a controlled unitary, which does not require knowledge of the particular unitary. However, such an implementation still requires knowing on which arm the unitary acts, which is a way to understand why such a unitary is not an entirely ``black'' box.} \citep{Araujo2014b}. 
Note also that this implementation of the switch still requires multiple calls 
to the quantum-controlled operations. A way to reproduce the switch within a circuit that uses only two gates to encode the two operations is by implementing  $\proj{0}\otimes U_A + \proj{1}\otimes U_B$ as first gate and $\proj{0}\otimes U_B + \proj{1}\otimes U_A$ as second. This departs even further from the notion of ``black box call'' in standard, causally ordered quantum circuits, and can be seen as an alternative representation of quantum control of the order of operations. Indeed, the most general ``quantum controlled quantum circuits'', introduced in \citep{Wechs2021} and reviewed in Sec.~\ref{subsec:Gen} above, are defined as a direct generalization of this construction.

Similarly to the black-box circuit in Fig.~\ref{fig:Qcirc_impl}, simply replacing the controlled unitaries with (suitably defined) quantum-controlled channels does not reproduce the action of the switch on those channels, because such a construction implies that different environment systems interact with the target at different times, potentially revealing the time at which the operation is preformed. Again as before, it is always possible to reduce the case of general operations to that of unitaries by adding to the target system a single environment common to all calls to the operation (where the environment is traced out or measured at the end of the circuit). A difference with the circuit in Fig.~\ref{fig:Qcirc_impl} is that, in the quantum-control case, there is no need to add and coherently control an extra auxliary system with the dimension of the environment.

Most generally, we expect of course that any experiment that can be implemented in the laboratory using current technology---including all reported implementations of the quantum switch---has some causally ordered representation (for example, the quantum state evolves in a well defined temporal order  in the Schrödinger equation or its relativistic generalizations), which may or may not align with one of those presented in this section. The interesting question is which representation gives a more accurate account of the number and type of operations implemented in the protocol, which may depend on the relevant foundational or applied context under consideration. We refer to Sec.~\ref{interpretationsection} below for a further discussion.

\section{Device-, semi-device- and theory-independent certifications of indefinite causal order}
\label{sec:causal_ineqs}
As discussed earlier, process matrices may not respect a definite causal order. One way to certify this property in a physical experiment is by implementing a causal witness, as detailed in Sec.~\ref{subsec:causal_witness}. This scenario is sometimes referred to as \textit{device-dependent}, since it is assumed that we have a perfect characterization of the instrument devices to certify indefinite causality with a causal witness violation. In this section, we discuss how to relax this instrument characterization assumption in a \textit{device-independent} scenario. Several terms used in this section, such as device-dependent and device-independent, also appear in the context of entanglement certification via entanglement witnesses and Bell nonlocality, and we will indeed discuss analogies and disanalogies with  Bell nonlocality in Sec.~\ref{subsubsec:analogy_noncausal_nonlocal}.

\subsection{Device-independent certifications} 
\label{subsec:DI_certifs}
\subsubsection{Causal and noncausal correlations}
\label{subsubsec:causal_correls}

Let us start by considering a scenario where two parties, Alice and Bob, share a process matrix $W\in\mathcal{L}\left(\mathcal{H}^{A_I}\otimes\mathcal{H}^{A_O}\otimes\mathcal{H}^{B_I}\otimes\mathcal{H}^{B_O}\right)$. In this scenario, Alice has access to a set of instruments given by $\{M_{a|x}^A\}$, $M_{a|x}^A\in\mathcal{L}\left(\mathcal{H}^{A_I}\otimes\mathcal{H}^{A_O}\right)$, where $x$ labels the settings and $a$ labels the outcomes. Analogously, Bob has access to a set of instruments given by $\{N_{b|y}^B\}$, $N_{b|y}^B\in\mathcal{L}\left(\mathcal{H}^{B_I}\otimes\mathcal{H}^{B_O}\right)$, where $y$ labels the settings and $b$ labels the outcomes. From the generalized Born rule~\eqref{processborn},  the probability that Alice and Bob respectively obtain the outcomes $a$ and $b$, when using the instruments labeled 
$x$ and $y$, is given by 
\begin{align}
P(a,b|x,y)=\tr\big[(M_{a|x}^A\otimes N_{b|y}^B)^T\,W\big].
\end{align}
A set of probability distributions given by $P(a,b|x,y)$ is often referred to as \textit{correlations}, a terminology that we shall use in this section.

If $W_{A\prec B}$ is an ordered process from $A$ to $B$, one may verify that probabilities $P_{A\prec B}(a,b|x,y)=\tr[(M_{a|x}^A\otimes N_{b|y}^B)^T\, W_{A\prec B}]$ respect the no-signaling constraint defined in Sec.\ref{signaling section}, that is, 
\begin{align} \label{eq:causal_AB}
   \sum_b P_{A\prec B}(a,b|x,y) = \sum_b P_{A\prec B}(a,b|x,y') \quad \forall \, a,x,y,y', 
\end{align}
in other words, Alice’s marginals do not depend on Bob’s choice of instrument.
Analogously,  if 
$W_{B\prec A}$ is an ordered process from $B$ to $A$, the probabilities $P_{B\prec A}(a,b|x,y)=\tr[(M_{a|x}^A\otimes N_{b|y}^B)^T\, W_{B\prec A}]$ respect the no-signaling constraint
\begin{align} \label{eq:causal_BA}
    \sum_a P_{B\prec A}(a,b|x,y) = \sum_a P_{B\prec A}(a,b|x',y) \quad \forall \, a,x,x',y.
\end{align}
It follows from linearity that the correlations $P(a,b|x,y)$ arising from a causally separable process $W=qW_{A\prec B} +(1-q)W_{B\prec A}$ necessarily respect the causal constraint
\begin{align} \label{eq:DIcausal_constraints}
	P(a,b|x,y) = q \, P_{A\prec B}(a,b|x,y) + (1-q) \, P_{B\prec A}(a,b|x,y).
\end{align}
Therefore, if we assume that Alice and Bob perform independent instruments, i.e., their joint instrument is given by $M_{a|x}^A\otimes N_{b|y}^B$, one may certify indefinite causality simply by analysing the probabilities $P(a,b|x,y)$. More specifically, if the probabilities do not respect the causal constraint of Eq.~\eqref{eq:DIcausal_constraints}, the process $W$ is necessarily causally nonseparable. Notice that, apart from independence from Alice and Bob, this certification does not assume any additional structure on the instruments $\{M_{a|x}^A\}$ and $\{N_{b|y}^B\}$, not even their dimensions. Since this certification holds independently of Alice and Bob device specificities, when a correlation $\tr[(M_{a|x}^A\otimes N_{b|y}^B)^T\, W]=P(a,b|x,y)$ does not respect the causal constraint of Eq.~\eqref{eq:DIcausal_constraints}, we have certified the causal non-separability of the process $W$ in a device-independent way.

Let us emphasize that the concept of no-signaling and causal probabilities is independent  of the formalism of process matrix or quantum theory. One may consider that the probabilities $P(a,b|x,y)$ arise from any experiment where two parties perform probabilistic operations on their systems. This provides a definition of (non)causality which goes beyond quantum mechanics or any particular physical theory: any bipartite correlation that can be decomposed in the form of Eq.~\eqref{eq:DIcausal_constraints} is said to be \emph{causal}. On the contrary, a set of probabilities $P(a,b|x,y)$ is \emph{noncausal} if it does not respect the constraint of Eq.~\eqref{eq:DIcausal_constraints}, regardless of how the correlations $P(a,b|x,y)$ were obtained \citep{OCB_2012,Branciard2016}.

The above definition of causal correlations was extended to the multipartite case by \citet{OreshkovGiarmatzi_2016}, where it was clarified what it means, for a general process, to be compatible with a well-defined---not necessarily fixed, but possibly dynamical---causal structure. It was shown that multipartite causal correlations could be characterized recursively in a similar manner to causally separable process matrices (see Definition~\ref{def:Wcsep_Npartite}) as convex combinations of correlations compatible with a given party acting first, and such that conditioned on that party's inputs and outputs, the remaining parties share some causal correlations \citep{OreshkovGiarmatzi_2016,abbott2016}. More formally, $N$-partite causal correlations are those that can be decomposed as 
\begin{align} \label{multicausalcorrelations}
    & P(\vec a|\vec x) = \sum_k q_k \, P_k(a_k|x_k) \, P_{\N\backslash k,x_k,a_k}^\text{causal}(\vec a_{\N\backslash k}|\vec x_{\N\backslash k})
\end{align}
with $q_k\ge 0$, $\sum_k q_k = 1$, where (for each $k, x_k, a_k$) $P_k(a_k|x_k)$ is a single-party distribution for party $k\in{\cal N}$ and $P_{\N\backslash k,x_k,a_k}^\text{causal}$ is an $(N-1)$-partite causal correlation for the parties in ${\cal N}\backslash\{k\}$ (with, as in Sec.~\ref{multipartite}, $\vec x_{\N\backslash k}$ and $\vec a_{\N\backslash k}$ denoting the list of inputs and outputs of all parties in $\N\backslash\{k\}$), conditioned on the $k^\text{th}$ party's input and output (and where for $N=1$, any probability distribution is considered to be causal).

We also note that some notion of \emph{genuinely multipartite} noncausality was defined and investigated in \citep{Abbott2017genuinely}, which more specifically captures the number of parties that are nontrivially involved in the noncausal character of some correlations.
Besides, the dynamical aspects were further analyzed in \citet{mothe2025}: while the intuition suggests that dynamicality comes from the fact that earlier parties can influence the order between future parties, it was found that there also exist correlations with dynamical, but still ``non-influenceable'' causal order---i.e., whose causal order cannot be fixed in advance (even probabilistically), but can also not be influenced by the choice of action of the parties.

\subsubsection{Causal inequalities} \label{subsec:causal_inequalities} 

A standard way to certify that a correlation is not causal is by violating a \textit{causal inequality}, which denotes a certain bound that must be satisfied by all causal correlations.

For instance, in a bipartite scenario with binary inputs ($x,y=0,1$) and outputs ($a,b=0,1$), the two parties cannot both guess the other party's input if they act each only once in a causally ordered manner. 
Consider indeed the so-called ``Guess Your Neighbour's Input'' (GYNI) cooperative game,
\footnote{Curiously, a multipartite version of the GYNI inequality appeared independently in the context of multipartite Bell nonlocality without quantum  violations \citep{almeida10_GYNI}.}
where the goal is that Alice's output equals Bob's input ($a=y$) and that Bob's output equals Alice's input ($b=x$). Assuming that the inputs are uniformly distributed ($P(x,y)=\frac14$), then the probability of success for this game is bounded, for causal correlations, by \citep{Branciard2016} 
\begin{align}
    p_\text{GYNI}\coloneqq P(a=y,b=x)\leq \frac{1}{2} \label{ineq:GYNI}
\end{align}
(with the short-hand notation $P(a{=}y,b{=}x) = \frac14 \sum_{abxy} \delta_{a,y}\,\delta_{b,x}\,P(ab|xy)$, which indeed implicitly assumes uniformly random inputs).
This bound on $p_\text{GYNI}$ can easily be understood as follows: if Alice is in the causal past of Bob, then she has no choice but make a random guess for Bob's input $y$, which will be correct half of the times. Bob, on the other hand, may always set $b=x$, since Alice can just send him her input $x$. Hence they can at most reach $p_\text{GYNI}=\frac{1}{2}$. The same holds if Bob is in the causal past of Alice, and by linearity all causal correlations are restricted to the same bound, as above.\footnote{In fact, for the GYNI game the causal bound of $\frac{1}{2}$ can also be achieved through a no-signaling strategy, e.g.\ if Alice and Bob simply output their own input, $a=x$ and $b=y$.} 

Eq.~\eqref{ineq:GYNI} is thus a simple example of a causal inequality. More general (linear) causal inequalities are defined by some set of real numbers $\{\gamma_{ab|xy}\}$ and a causal bound $\beta\in\mathbb{R}$, such that all causal distributions $P_{\text{causal}}(a,b|x,y)$ respect
\begin{align}
	\sum_{abxy} \gamma_{ab|xy} \, P_\text{causal}(a,b|x,y)\leq \beta .
\end{align}
Hence, if one verifies that $\sum_{abxy} \gamma_{ab|xy} P(a,b|x,y)> \beta $, this ensures that $P(a,b|x,y)$ is not causal.

The linear and convex structure of causal correlations allows for a convenient geometric approach to causal inequalities, that is directly inspired by that developed for Bell inequalities \citep{brunner_review}. For any scenario with a fixed number of inputs and outputs of each party, one may list all probabilities $P(a,b|x,y)$ in a vector, and see it as a point in some large-dimensional space. It is then easy to see that the set of such points constructed from causally ordered correlations $P_{A\prec B}(a,b|x,y)$ forms a convex polytope, and similarly for the points constructed from $P_{B\prec A}(a,b|x,y)$. By definition, the set of points constructed from causal correlations (or simply ``the set of causal correlations'') is then the convex hull of the two former polytopes and it is itself a convex polytope: the \emph{causal polytope} \citep{Branciard2016,OreshkovGiarmatzi_2016}.
The facets of this polytope define (tight) causal inequalities. This provides mathematical tools to decide whether some correlation $P(a,b|x,y)$ is causal or not: $P(a,b|x,y)$  is causal if and only if it respects all (finite) facet inequalities of its corresponding scenario.

The vertices of the causal polytope are given by the deterministic one-way signaling correlations $P(a,b|x,y)$, and listing such vertices can be done in a systematic manner \citep{Branciard2016}. Importantly, given the vertices of any polytope, one can in principle find and list all its facets via standard methods, such as the Fourier-Motzkin elimination. This means that, for any fixed number of settings and outcomes, one may obtain all causal inequalities. A caveat here is that the complexity of all known methods scales exponentially with the number of settings \citep{Branciard2016}, hence in practice, this general method is restricted to scenarios with a few settings and outcomes only.

The causal polytope for the bipartite scenario with binary inputs and outputs
\footnote{An even smaller scenario (in the sense of minimizing the number of inputs and outputs, and hence the dimension of the correlation space) is the bipartite ``lazy'' scenario where each party has a binary input, and a nontrivial binary output only for one of their input. In that case, the only nontrivial causal inequality is of the LGYNI type. \label{ftn:lazy}}
was completely characterized by \citet{Branciard2016}. Up to relabeling of settings and outcomes, there are only two types of facet inequalities in this scenario. One is the GYNI inequality of Eq.~\eqref{ineq:GYNI} above, while the other one, known as the ``Lazy Guess Your Neighbour's Input'' (LGYNI) inequality, reads
\begin{align}
p_\text{LGYNI}\coloneqq P(x(a\oplus y)=0,y(b\oplus x)=0)\leq \frac{3}{4} \label{ineq:LGYNI}
\end{align}
(where $\oplus$ denotes the addition modulo 2 and where, as in Eq.~\eqref{ineq:GYNI}, the notation above still implicitly assumes uniform inputs).
This can also be interpreted as the success probability of a game, where Alice and Bob's task is now to guess each other’s input only when their respective input is 1 (for an input 0, their output can be arbitrary). We note that a single bit of communication is sufficient to reproduce all causal correlations for binary inputs, so the resulting polytope coincides with that arising from Bell-local strategies augmented by one bit of communication from Alice to Bob or from Bob to Alice (or convex combinations of such), as considered (and fully characterized, for binary outputs) in \citep{bacon03}.

As just mentioned, because of the complexity of the problem, only small enough causal polytopes can be fully characterized.
\footnote{To the best of our knowledge, in the bipartite case only the causal polytopes for binary inputs and outputs \citep{Branciard2016} and in the tripartite case for the ``lazy'' scenario (cf. footnote~\ref{ftn:lazy}) \citep{abbott2016} could be fully characterized.}
This, however, does not prevent one from only trying to look for a limited number of facets in a given scenario or from considering causal inequalities that are not facets of the corresponding causal polytope. In fact, the first causal inequality that appeared in the literature, proposed by \citet{OCB_2012}, was not obtained through the causal polytope machinery. It was constructed for a scenario with a binary input ($x=0,1$) for Alice, a quaternary input (two input bits $y,y'=0,1$) for Bob, and binary outputs ($a,b=0,1$) for both, and reads:
\begin{align}
    p_\text{OCB} \coloneqq  \frac12 \big[P(a=y|y'=0) + P(b=x|y'=1)\big] \leq \frac34 \label{ineq:OCB}
\end{align}
(where the notation once again implicitly assumes that all inputs are uniform, so that $P(a{=}y|y'{=}0)=\frac14\sum_{abxy}\delta_{a,y}\,P(a,b|x,y,y'{=}0)$ and similarly for $P(b{=}x|y'{=}1)$).
This inequality can again be understood as a bound on the success probability of a cooperative game, where depending on Bob's second input $y'$, either Alice has to guess Bob's input (i.e., her output must be equal to Bob's input, so that $a=y$), or vice versa (so that $b=x$). While it is known that this inequality is not a facet of the corresponding polytope, to the best of our knowledge, no tighter causal inequality is known for that scenario.

We note also that beyond facet or other linear causal inequalities such as Eq.~\eqref{ineq:OCB}, \emph{entropic} causal inequalities have also been derived in \citep{Miklin2017}, following similar approaches to the ones developed for Bell inequalities \citep{BraunsteinCaves,ChavesFritz}. These provide the advantage of holding for arbitrary finite alphabets for the input and output (or just the output) variables, and of directly bounding information-theoretic quantities, which could be relevant for potential applications. Interestingly, however, no violation of entropic causal inequalities has been found within the process matrix formalism.

The multipartite scenario was analyzed in \citep{abbott2016}, were it was proven that, similar to the bipartite case, causal correlations for any given number of parties, inputs, and outputs form a polytope, which has deterministic correlations as vertices (and causal inequalities as facets). Note that this is not immediately obvious from the definition of multipartite causal correlations, Eq.~\eqref{multicausalcorrelations}, because the causal order between sets of parties can depend probabilistically on the settings of other parties. Nevertheless, \citet{abbott2016} proved that the most general, multipartite correlation can be written as 
\begin{equation}
 P(\vec{a}|\vec{x}) = \sum_j q_j P^{\mathrm{det}}_j(\vec{a}|\vec{x}),
\end{equation}
where $P^{\mathrm{det}}_j(\vec{a}|\vec{x})$ are deterministic correlations were the order is fixed for every choice of settings and the probabilistic weights $q_j\in[0,1]$ do not depend on the settings. With this result, \citet{abbott2016} characterized the ``lazy'' tripartite causal polytope, where each party has a binary setting and a binary outcome for one of the setting values (and constant for the other), listing all vertices and all facet inequalities. Some facet inequalities were generalized to more complex scenarios, although a full characterization of the corresponding polytopes exceeded computational capabilities. Other multipartite causal inequalities were also studied in the context of classical process matrices, as discussed in Sec.~\ref{violationclassicalprocesses} below. Besides, \citet{Abbott2017genuinely} extended the analysis to genuinely multipartite causal correlations, where the lack of causal order cannot be reduced to a subset of parties. They showed that the relevant correlation sets are still polytopes with deterministic vertices; they again fully characterized the lazy tripartite scenario and found example inequalities for more general scenarios. 

Refs. \citep{Tselentis2025mobius,Baumann2025no_quantum_advantage,Baumeler2025graphical,mothe2025} considered inequalities respected by convex combinations of fixed-order correlations, as well as (in the case of \citep{mothe2025}) by correlations with ``non-influenceable order'' (cf end of Sec.~\ref{subsubsec:causal_correls}). Such inequalities can still be violated by causal correlations, as defined in Eq.~\eqref{multicausalcorrelations}, because of the possibility of dynamical causal order.

\subsubsection{Violation of causal inequalities in the process matrix formalism} \label{sec:violations}

As discussed above, it is clear that within the process matrix formalism, causally separable processes can only generate causal correlations. A violation of a causal inequality can only be obtained from a causally nonseparable process. Since no assumption is made on the parties' instruments, and therefore on the actual experimental devices used to obtain the correlations under consideration, this provides a device-independent certification of causal nonseparability, as an alternative to the (device-dependent) use of causal witnesses discussed in Sec.~\ref{subsec:causal_witness}.

This is, in fact, precisely how the causal nonseparability of $W_\mathrm{OCB}$ from Eq.~\eqref{eq:W_OCB} was originally proven in Ref. \citep{OCB_2012}. It can indeed be verified that with the instruments
\begin{align}
M_{a|x}^A & = \Big(\frac{\id+(-1)^a\sigma_z}{2}\Big)^{A_I}\otimes\Big(\frac{\id+(-1)^x\sigma_z}{2}\Big)^{A_O}, \notag \\
N_{b|y, y'=0}^B & = \Big(\frac{\id+(-1)^b\sigma_x}{2}\Big)^{B_I}\otimes\Big(\frac{\id+(-1)^{y+b}\sigma_z}{2}\Big)^{B_O}, \notag \\
N_{b|y, y'=1}^B & = \Big(\frac{\id+(-1)^b\sigma_z}{2}\Big)^{B_I}\otimes\rho^{B_O}
\end{align}
(for any normalized state $\rho^{B_O}$), one obtains a value of $p_\text{OCB} = \frac{1+1/\sqrt{2}}{2} > \frac34$, in violation of the causal inequality of Eq.~\eqref{ineq:OCB}. 

As it turns out, this violation was proven to be the maximal allowed within the process matrix formalism for arbitrary finite-dimensional systems in Ref. \citep{Liu2024Tsirelson}. This extends the validity of the results in \citep{Brukner2015}, which considered the restricted case of local instruments based on measure-and-reprepare operations with traceless binary observables.
One can note that both this bound for process matrices and the causal bound of $\frac{3}{4}$ on $p_\text{OCB}$ coincide with the quantum \citep{Tsirelson} and the local bounds of the Clauser-Horne-Shimony-Holt version \citep{chsh} of Bell's inequality (when expressed as the success probability of a game where the goal is that $a \oplus b=x y$)---although this connection seems to be specific to this particular inequality, and breaks for other (simpler) inequalities like GYNI.

In the latter cases, indeed, some values of $p_\text{GYNI}\simeq 0.5694 > \frac12$ and $p_\text{LGYNI}\simeq 0.8194 > \frac34$ were obtained numerically with qubit systems, and even the larger violation $p_\text{GYNI}\simeq 0.6218 > \frac12$ was found, with 5-dimensional systems, for the GYNI inequality, while no increase was found for the LGYNI violation when going to larger systems \citep{Branciard2016}. On the other hand, \citet{Bavaresco_2019} obtained an upper-bound on the possible value of $p_\text{GYNI}$ for process matrices that depends on the product $d^I d^O$ of all local input and output 
dimensions, namely, $p_\text{GYNI} \leq 1-\frac{1}{d^I d^O+1}$. Therefore, the GYNI game cannot be won perfectly with finite-dimensional process matrices, and it was proven in Ref. \citep{Kunjwal2023_Nonclassicality} that the result holds even for infinite-dimensional systems.

Ref. \citep{Liu2024Tsirelson} introduced a methodology to derive nontrivial upper bounds for causal inequalities within the process matrix formalism, in any finite dimension. This method enables proving the bound $p_\text{LGYNI} \leq 0.8194$, which holds regardless of the dimension of the quantum system and is attainable by process matrices where local systems are two-dimensional. Also, Ref. \citep{Liu2024Tsirelson} shows that the bound $p_\text{GYNI} \leq 0.7592$ is respected by process matrices of any dimension, although this bound is possibly not tight and may be compared with the best-known process matrix violation found for $d=5$: 0.6218. \citep{Branciard2016}. 

Beyond the bipartite scenario, other process matrices violating multipartite causal inequalities were found, e.g., by \citet{baumeler2014,Baumeler2014PRA,abbott2016,Guerin2018a}, while violations of genuine multipartite causal inequalities were found in \citep{Abbott2017genuinely}.

\subsubsection{Causally nonseparable processes which do not violate causal inequalities}
\label{sec:no_violations}

As previously mentioned, it is immediate to check that causally separable process matrices cannot be used to violate causal inequalities. Causal non-separability is thus a necessary requirement for obtaining noncausal correlations within the process matrix formalism. In Sec.~\ref{sec:violations}, we saw that causally nonseparable matrices might lead to noncausal correlations and violate causal inequalities. However, there exist causally nonseparable process matrices in which, for any set of instruments, the resulting correlations are necessarily causal and do not violate causal inequalities. As discussed in Sec.~\ref{subsubsec:analogy_noncausal_nonlocal}, an analogy might be made with quantum states that are entangled but do not violate any Bell inequality---e.g. Werner states, for a certain range of parameters \citep{Werner89}.

The existence of causally nonseparable process matrices which do not violate causal inequalities was first noticed in Refs. \citep{Araujo_2015} and \citep{OreshkovGiarmatzi_2016} where it was shown that independently of the instruments performed by the parties, the quantum switch can only lead to causal correlations, hence not violating any causal inequality, see Sec.~\ref{switchnonseparability} above. This result was later extended by \citet{Wechs2021}, proving that any process matrix which can be written as quantum circuits with quantum control of the causal order, regardless the number of parties, cannot violate causal inequalities. A similar result (although not framed in the process matrix formalism) was independently obtained by Ref. \citep{Purves2021}. Under the assumption that quantum theory only allows for the use of quantum control of different parties' operations to generate superpositions of causal order, the result there was interpreted to imply that quantum theory cannot violate a causal inequality.

A general method to ensure that a process matrix cannot violate causal inequalities is presented in Ref. \citep{feix2016_model}, which shows that if a bipartite process matrix $W_{AB}$ has a positive partial transpose, i.e., $W_{AB}^{T_B}\geq0$, the correlations $P(ab|xy)=\tr[(M_{a|x}\otimes N_{b|y})^T W_{AB}]$ are causal, regardless of the instruments $\{M_{a|x}\}$ and $\{{N}_{b|y}\}$. Using this method, several examples of causally non-separable process matrices which do not violate any causal inequality were exhibited.

In Ref. \citep{OreshkovGiarmatzi_2016}, the authors consider a causal scenario where, in addition to a process matrix $W$, the parties may share auxiliary (possibly entangled) states $\rho$, a natural free resource from a causally ordered perspective. In this scenario, the parties may perform a joint instrument between their system described by $W$ and their corresponding system to $\rho$. The process is then described by an effective process matrix given by $W'\coloneqq W\otimes\rho$, where $\rho$ is shared by all parties. In the bipartite case, this corresponds to the correlations $P(ab|xy)=\tr\big[({M}_{a|x}^{A_IA_OA_{I'}}\otimes {N}_{b|y}^{B_IB_OB_{I'}}){^T} (W^{A_IA_OB_IB_O}\otimes \rho^{A_{I'}B_{I'}}) \big]$. Process matrices which do not violate any causal inequality even after attaching auxiliary states are said to be \emph{extensibly causal}. Ref. \citep{OreshkovGiarmatzi_2016} also shows that the quantum switch is extensibly causal, and Ref. \citep{Feix2017} shows that there exist causal processes which are not extensibly causal ---i.e.\ that can only generate noncausal correlations if one attaches an (entangled) state to them. 

\subsubsection{Violation of causal inequalities with classical processes}
\label{violationclassicalprocesses}

Since all two-party classical processes have been shown to be causally separable, it follows that they cannot exhibit any violation of causal inequalities. This implies that causal inequality violations for classial processes are exclusive to scenarios involving more than two parties. Next, we present two causal games that can be won using classical processes with indefinite causal order but cannot be won by any causal strategy.

The ``all-to-one signaling game'' \citep{baumeler2014} involves three parties, \(A\), \(B\), and \(C\), each with binary inputs and binary outputs, who attempt to determine the parity (sum modulo 2) of the inputs from other parties, based on a random selection \(m \in \{1, 2, 3\}\) that determines which party is chosen to receive the parity result. The success probability is defined as:
\[
\begin{aligned}
p_{\text{succ}} &= \frac{1}{3} \big[ P(a = y \oplus z \mid m = 1) + P(b = x \oplus z \mid m = 2)  \nonumber \\
 & + P(c = x \oplus y \mid m = 3) \big],
\end{aligned}
\]
where \(\oplus\) denotes addition modulo 2.

Under the assumption that operations are performed in a fixed causal order or a convex mixture therefrom, it is not possible to achieve a success probability of one. To determine the upper bound on \(p_{\text{succ}}\), note that only the last party in the causal sequence can receive the parity of the others, while the other two can only guess the parity with probability \(1/2\). This gives the upper bound:
\(
p_{\text{succ}} \leq \frac{1}{3} \left( 1 + \frac{1}{2} + \frac{1}{2} \right) = \frac{2}{3}.
\)

The causal strategies can be extended to those that exploit dynamic causal ordering, where the causal relationships between the parties can change dynamically based on the value of \(m\). In this case, since at least one party is not in the causal future of the others, this party can only guess the parity with a probability of \(1/2\). An example of such an adaptive strategy with $A$ being the ``first party'' is as follows:
\begin{itemize}
    \item If \(m = 2\), \(A\) ensures \(C \prec B\), allowing \(P(b = x \oplus z \mid m = 2)\) to reach 1.
    \item If \(m = 3\), \(A\) ensures \(B \prec C\), allowing \(P(c = x \oplus y \mid m = 3)\) to reach 1.
    \item If \(m = 1\), the success probability \(P(a = y \oplus z \mid m = 1)\) is limited to \(1/2\).
\end{itemize}
This strategy leads to an upper bound for \(p_{\text{succ}}\), given by \(p_{\text{succ}} \leq \frac{1}{3} \left( 1 + 1 + \frac{1}{2} \right) = \frac{5}{6}.
\)

By using the classical process with indefinite causal order~\eqref{loop1}, illustrated in~Fig~\ref{fig:W_3}a, the parties can win the game with certainty. For example, if \(m = 1\), party \(B\) sends out their input (\(o_B = y\)), party \(C\) sends out the parity of their input and the input received from the process (\(o_C = z  \oplus i_C\)), and party \(A\) uses the random variable received from the process as their guess (\(x = i_A\)). 
For the cases \(m = 2\) and \(m = 3\), the same strategy is used, with the parties' roles permuted accordingly.

The next causal game involves three parties (\(A\), \(B\), and \(C\)), each with binary inputs and outputs. The goal of the game is defined based on the majority function, \(\text{maj}(x, y, z)\), which determines whether the majority of the inputs is \(0\) or \(1\). If the majority is \(0\), the parties play a ``guess-your-neighbor's-input'' game: \(A\) guesses \(B\)'s input, \(B\) guesses \(C\)'s input, and \(C\) guesses \(A\)'s input. Conversely, if the majority is \(1\), the game is played in reverse order, but with the output bits flipped: \(A\) guesses flipped \(B\)'s input, \(B\) guesses flipped \(C\)'s input, and \(C\) guesses flipped \(A\)'s input. Hence the probability of success in the game is given by

\begin{align} 
p_{\text{succ}} &= \frac{1}{2} \big( 
P(a = z, b = x, c = y \mid \text{maj}(x, y, z) = 0) \nonumber\\
&+ 
P(a = \bar{y}, b = \bar{z}, c = \bar{x} \mid \text{maj}(x, y, z) = 1)
\big),
\end{align}
with (as introduced earlier) $\bar{x} = x \oplus 1$.
The game corresponds to a deterministic extremal point of the polytope with three parties and binary inputs/outputs.

The success probability of winning this game under a predefined causal order is upper bounded by \(3/4\). This limitation arises because at least one party must make a guess without having received any information from the others. For example, if \(A\) causally precedes \(B\) and \(C\), then \(A\) can only make a random guess, which causes the parties to lose in \(2\) out of \(8\) cases. Similar bounds apply when any other party is causally preceding the others.
However, by utilizing the ``Lugano process'' \citep{Baumeler2014PRA} [Eq.~\eqref{loop2} and Fig.~\ref{fig:W_3}b)], the game can be won perfectly. In this optimal strategy, the parties send out their inputs and use the bits obtained from the process as their guesses, winning the game with certainty. As discussed in Sec.~\ref{classicalprocesses},  
\citet{Araujo2017} showed that the Lugano process allows for a valid unitary extension, thus proving the existence of unitary processes that can violate causal inequalities.

Several works have addressed classical processes in multi-party causal games. \citet{Baumeler2022} introduced a causal game played among an arbitrary number $N$ of parties, and derived corresponding bi-causal inequalities. These inequalities are upper bounded by the winning probability for any bi-causal strategy, where the $N$ parties can be partitioned into two subsets that are causally ordered (i.e., their violation implies genuinely $N$-partite non-causality, as defined in \citep{Abbott2017genuinely}).
The maximum probability of winning the corresponding causal game tends toward \(1/2\), while it can be won deterministically using a classical deterministic process matrix. The games resemble the quantum non-locality games of~\citet{Ardehali1992} and~\citet{Svetlichny1987}, further indicating a possible link between the two problems. 

Further examples of causal inequalities violated by classical processes have been analyzed in \citep{abbott2016,Abbott2017genuinely} for genuinely multipartite causal inequalities. An interesting result is that several (although not all) facet inequalities (i.e., those defining the boundary of the causal polytope) can be violated by classical processes both in the ordinary and in the genuinely multipartite case. \citet{Tselentis2023} and \citet{Baumeler2025} further use graph-theoretic tools to analyze violation of causal inequalities with classical processes.

\subsubsection{Analogies and disanalogies with Bell nonlocality}
\label{subsubsec:analogy_noncausal_nonlocal}

Causal inequalities share some similarities with Bell inequalities \citep{bell64,brunner_review}. For instance, they both allow us to certify their associated property in a device-independent way. While Bell inequalities allow the device-independent certification of entanglement of quantum states, causal inequalities allow the device-independent certification of causal nonseparability. Additionally, for any  scenario with a finite and fixed number of settings and outcomes, the set of probabilities $P(a,b|x,y)$ respecting the causal constraint form a polytope where the facets correspond to all tight causal correlations of this scenario. This has a close analogy with the facets of the Bell local polytope being tight Bell inequalities.

Causally nonseparable process matrices which do not violate causal inequalities have a similar flavour to Werner states \citep{Werner89}, which, despite being entangled, cannot violate Bell inequalities. For such processes or states, the desired property (noncausality or nonlocality) manifests itself through device-dependent tests (that is, by violating a causal witness or an entanglement witness), but it cannot be certified from a device-independent perspective, since it cannot violate a causal inequality or a Bell inequality.

The set of causal correlations also has a one-to-one correspondence with a Bell scenario with (one-way) communication \citep{Branciard2016,bacon03}. The set of one-way signaling correlations from Alice to Bob is precisely a Bell local scenario where Alice can freely communicate to Bob, but Bob cannot communicate to Alice. Therefore, the set of causal correlations is given by the convex hull of the Bell correlations in which Alice can send messages to Bob and the Bell correlations in which Bob can send messages to Alice. Interestingly, because of this analogy, the simplest causal inequalities discussed in the previous sections (GYNI and LGYNI) have appeared in the literature in the context of Bell inequalities with (restricted) auxiliary communication.%
\footnote{\label{ftntBellcommunication} Note, however, that, in Bell scenarios, the amount of communication is typically restricted, while for causal inequalities this is not the case.}

The analogy between Bell inequalities and causal inequalities also has some limitations. For instance, while classical (no-signaling) systems cannot violate Bell inequalities, causal inequalities can be violated classically (recall Sec.~\ref{violationclassicalprocesses} above), so their violations do not witness non-classicality. 
Additionally, on one hand, the set of Bell local correlations (typically said to be classical), is precisely the convex hull of deterministic no-signaling correlations. On the other hand, the convex hull of deterministic classical correlations forms the set of so-called \textit{nomic} correlations \citep{Kunjwal2023_Nonclassicality,kunjwal2024Generalizing}, which is only a strict subset of correlations generated by arbitrary classical processes (see Sec.~\ref{classicalprocesses} and Fig.~\ref{fig:polytopes}). For this reason,   Refs. \citep{Kunjwal2023_Nonclassicality,kunjwal2024Generalizing} argue that correlations obtained by classical process that are \textit{antinomic} (i.e.\ not nomic) may also be perceived as non-classical.

Another difference is that while all entangled rank-1 quantum states (pure entangled states) violate a Bell inequality \citep{gisin91,popescu1992generic}, there are rank-1 causally nonseparable process matrices which do not violate any causal inequality, with the quantum switch being a notorious example~\citep{Araujo2017} (see Sec.~\ref{subsec:unitary_processes}). Moreover, while bipartite scenarios with binary inputs, where one input has a single output, can never produce Bell nonlocal correlations \citep{brunner_review}, this scenario aligns with the ``Lazy'' GYNI causal inequality \citep{Branciard_2016}, which process matrices are known to violate.

\subsubsection{Device-independent tests with additional locality assumptions}
\label{sec:vanderlugt}

In Refs. \citep{gogioso23,VanDerLugt_2022}, the authors propose a framework that goes beyond the standard causal inequalities and allows for a device independent certification of causal nonseparability of a larger class of process including the quantum switch, a phenomenon that cannot be observed in the standard causal inequality framework described in Sec.~\ref{sec:causal_ineqs} \citep{Araujo_2015,OreshkovGiarmatzi_2016}. The key idea  is to consider a multipartite scenario with a no-signaling bipartition, where each bipartition is subjected to the standard causal constraints presented in Eq.~\eqref{eq:DIcausal_constraints} and discussed in Sec.~\ref{sec:causal_ineqs}. The strategy is similar to the one previously adopted in \citep{zych2019, Rubino2022experimental}, and reviewed in Sec.~\ref{theoryindependenttests} below, although such results did not provide fully device-independent tests.

\begin{figure}[hbt] 
	\begin{center}
		\includegraphics[width=.9\columnwidth]{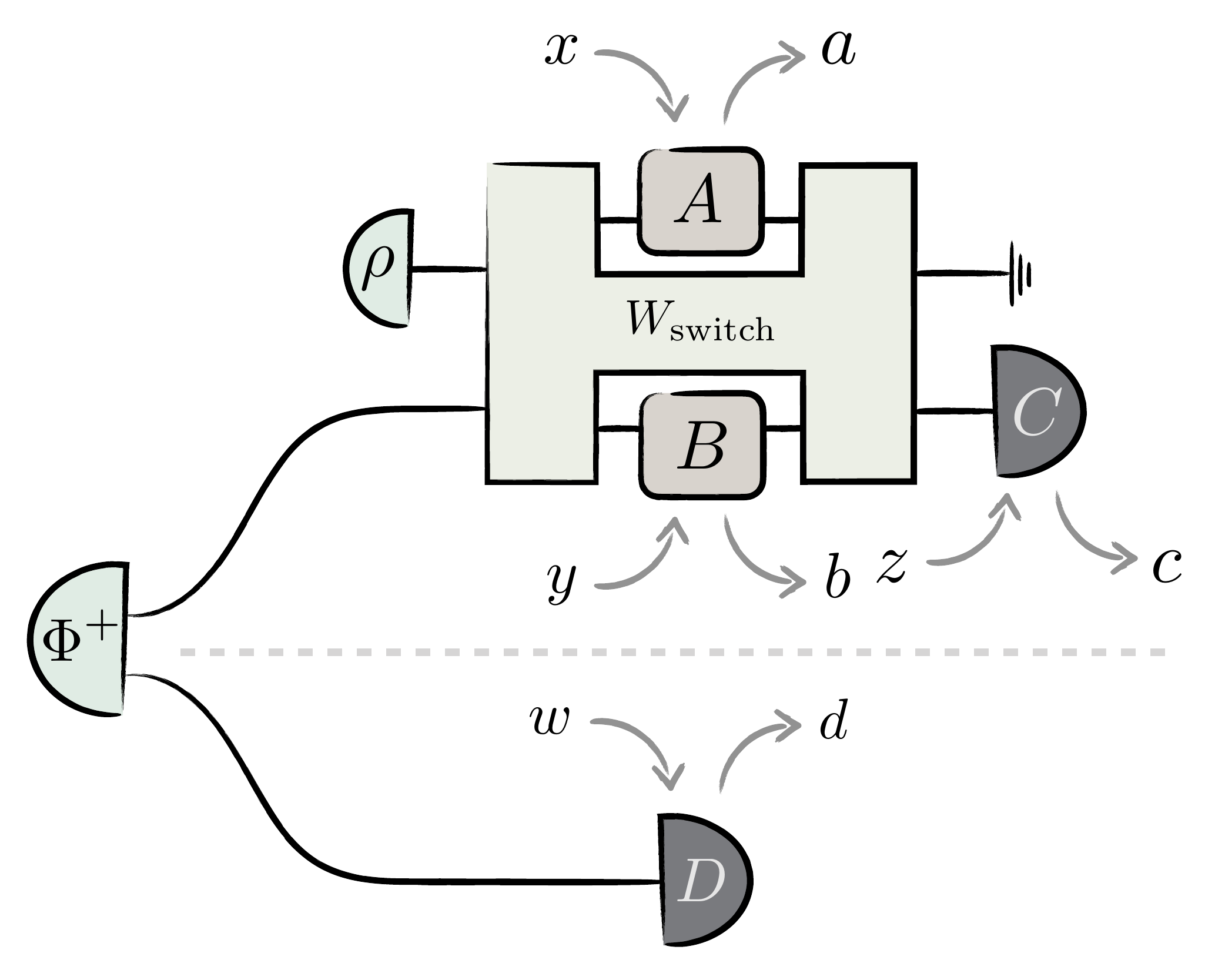} 
	\end{center}
\caption{Pictorial representation of the scenario considered in \citep{VanDerLugt_2022}, where the parties $ABCD$ are no-signaling in the bipartition $ABC/D$. For some particular choices of states, measurements and instruments, the quantum switch leads to probabilities which do not respect the $\mathcal{DRF}$ assumption, a combination of no-signaling and causal order as defined in Eq.~\eqref{eq:DRF}.
}
\label{fig:VanDerLugt}
\end{figure}

We now detail the scenario presented in Ref. \citep{VanDerLugt_2022}, which considers four parties, $A,B,C,D$ with probabilities $P(a,b,c,d|x,y,z,w)$, where $x,y,z,w$ and $a,b,c,d$ are the respective settings and outcomes for $A,B,C,D$ (see Fig.~\ref{fig:VanDerLugt}). Party $D$ is assumed to be space-like separated from $A, B, C$, which justifies augmenting the constraints due to causal order by adding a no-signaling condition, leading to a combined assumption referred to as $\mathcal{DRF}$, for ``$\mathcal{D}$efinite Causal Order, $\mathcal{R}$elativistic Causality, and $\mathcal{F}$ree Local Interventions''. We then denote by $ABC$ a single party which groups $A$, $B$, and $C$ all together, and the set of no-signaling probabilities across the bipartition\footnote{Note that the no-signaling condition itself is strictly weaker than the standard Bell locality assumption and includes non-quantum correlations such as PR-boxes \citep{Popescu1994}.} $ABC\vert D$ is denoted $\mathcal{NS}$ and it is formed by conditional probabilities which are compatible with both $ABC\prec D$ and $D\prec ABC$, see Eq.~\eqref{eq:nosig_B_to_A}.
The set of conditional probabilities respecting the causal order $A\prec B \prec C$ is denoted as $\mathcal{C}_{A\prec B\prec C}$ and the set of conditional probabilities which respect the causal order $B\prec A \prec C$ is denoted as $\mathcal{C}_{B\prec A\prec C}$. The $\mathcal{DRF}$ set  is then defined as the convex hull 
\begin{align} \label{eq:DRF}
\mathcal{DRF}\coloneqq\text{conv}\Big(\mathcal{NS}\cap(\mathcal{C}_{A\prec B\prec C}\cup \mathcal{C}_{B\prec A\prec C}) \Big).
\end{align}
``Free local interventions'' refers to the standard assumption in device-independent scenarios that allows one to single out the setting variables and define (no-)signaling and related conditions.

The authors present an inequality which is respected for all conditional probabilities of $\mathcal{DRF}$, but  for some particular choices of states, measurements and instruments, the quantum switch leads to probabilities that violate this inequality. Specifically, if $p \in \mathcal{DRF}$, they establish the bound
\begin{align}\label{eq:DRF_Ineq}
    &P(d=0,b=x\, \vert \, w=0) + P(d=1, a=y \, \vert \, w=0) \notag\\
    &\quad + P(c \oplus d = zw \, \vert \, x = y = 0) \leq \frac{7}{4}.
\end{align}
One can see that the first two terms are bounded by 1, which can be achieved by both the quantum switch and fully classical strategies. The third term, on the other hand, can be identified with a CHSH expression \citep{chsh}. Due to monogamy of Bell locality \citep{Barrett2006, Collins2002}, only a nonclassical process allows for the maximal violation of the three terms altogether.
Indeed, with the quantum switch and a strategy that violates CHSH, we can have a score of $\frac{1}{2}+\frac{1}{2}+\frac{2+\sqrt{2}}{4}\approx1.8536 > \frac{7}{4}$.
In this way, if one presupposes assumptions $\mathcal{R}$ and $\mathcal{F}$ then the indefinite causal property (the violation of $\mathcal{D}$) of the quantum switch may be certified in a device-independent manner. 

A different device-independent approach was proposed by \citet{dourdent2023networkdeviceindependent}, building on tests of causal nonseparablility that use quantum states rather than classical random variables as inputs (see Sec.~\ref{semiDI_QI} below). Here one uses different independent quantum sources to probe a given process, and one needs to trust the independence and the network structure of these sources---hence the denomination ``network-device-independent'' for this approach.

\subsubsection{No backwards in time signaling}
\label{nobackintime}
In the scenarios considered above, the only constraints on correlations come from assumptions about causal relations between parties.  \citet{Guryanova2019exploringlimitsofno} consider a modified scenario, which imposes additional constraints among each party's settings and outcomes. In this scenario, each party, say $A$, produces the local outcome $a$ \textit{before} having access to the setting $x$,  according to a local time ordering that is assumed to exist within each site. This scenario motivates the \textit{No backwards in time signaling} (NBTS) constraint, first formulated in \citep{Brukner14}: each party should not be able to signal to their own past, meaning that each party's outcome should be independent of their setting. For the bipartite scenario, the NBTS constraints are
\begin{subequations}
 \begin{align}
    P(a|x,y)&=P(a|y), \\
    P(b|x,y)&=P(b|x). 
\end{align}   
\end{subequations}

\citet{Guryanova2019exploringlimitsofno} further consider the correlations that can be produced by parties manipulating a classical or quantum system in a definite causal order, still respecting the NBTS paradigm with outcomes before settings\footnote{In the quantum case, the most general local operation in this class has the form $\M_{a|x} = \mathcal{C}_x\circ \mathcal{N}_a$, where $\mathcal{C}_x$ is CPTP for every $x$ and $\sum_a\mathcal{N}_a$ is CPTP.}. They find that the two classes coincide, namely, the most general causally ordered correlations produced in this scenario can be reproduced classically. More prominently, they prove that, in the bipartite case, even the most general process matrix produces causal correlations. This generalizes a previous result in \citep{BaumannBrukner} stating that, if only measurements in a fixed basis are considered, any bipartite process can be simulated by a causally separable one. Note that bipartite noncausal correlations that respect NBTS \textit{can} be generated by process matrices, for example, those violating the LGYNI inequalities discussed in Sec.~\ref{sec:violations}. However, these require operations where Alice's measurement choice depends on $x$, and Bob's on $y$, outside the NBTS paradigm. The situation changes for the tripartite case: as shown in \citep{Purves2019}, the ``Lugano process'', discussed in Sec.~\ref{classicalprocesses}, can generate noncausal correlations using measure-and-reprepare operations in a fixed basis.

\subsection{Semi-device-independent tests of causal nonseparability} \label{semiDI}

\subsubsection{Semi-device-independent certification via trusted quantum instruments}

As discussed in Sec.~\ref{sec:violations}, there are process matrices $W$ which, for a suitable choice of quantum instruments, $\{M^A_{a|x}\}_a$, $\{N^B_{b|y}\}_b$ lead to noncausal correlations, that is, the probabilities $P(a,b|x,y)=\tr[(M^A_{a|x}\otimes N^B_{b|y})^T\, W]$ violate a causal inequality. Within the process matrix formalism, if a set of probabilities $\{P(a,b|x,y)\}_{a,b,x,y}$ violates a causal inequality, the process matrix $W$ used in the experiment is necessarily causally nonseparable. Moreover, we can certify that $W$ is causally nonseparable in a \textit{device-independent} way by simply analysing the probabilities $P(a,b|x,y)$, without characterising or specifying the instruments involved in the experiment. This should be contrasted with the \textit{device-dependent} method of causal witness presented in subsection\,\ref{subsec:causal_witness}, where the analysis is made assuming that the involved instruments are known and fully characterized.

From this perspective, \citet{Bavaresco_2019}) consider an intermediary scenario and analyze causal correlations in a \textit{semi-device-independent} manner, where the instruments performed by some involved parties are completely characterized, and the instruments performed by other parties are uncharacterized\footnote{In the context of Bell nonlocality, scenarios where measurements performed by some parties are characterized and uncharacterized in others correspond to tests of EPR-steering \citep{Wiseman2007}.}. Correlations that are ``semi-device-independently noncausal'' are then ``stronger'' than correlations that simply violate a causal witness and ``weaker'' than correlations that violate a causal inequality. The authors have also shown that, although the quantum switch cannot lead to noncausal correlations when the instruments performed by Alice and Bob are uncharacterized, the same quantum switch leads to noncausal semi-device-independent correlations when the instruments of Alice (or Bob) are characterized. This shows that the quantum switch can lead to noncausal correlations which are stronger than a simple violation of causal witness.

As discussed in Sec.~\ref{sec:causal_ineqs} above, measurement of a causal witness yields device-dependent certification of indefinite causality, while violation of causal inequalities provides device-independent certification. In between these two extremes, Ref. \citep{Cao2022semiDI} introduces the concept of a semi-device-independent causal inequality, which in the tripartite scenario reads as
\begin{equation}
\label{eq:semiDI}
    S\coloneqq \sum_{a,b,c,x,y,z}\alpha_{x,y,z}^{a,b,c} \, p(a,b,c|x,y,z) \geq 0, 
\end{equation}
and the probabilities can be written as $p(a,b,c|x,y,z)=\tr[(M^A_{a|x}\otimes M^B_{b|y} \otimes M^C_{c|y})^T\,W_\mathrm{c\mhyphen sep}] $, where $W_\mathrm{c\mhyphen sep}$ is a causally separable process matrix, and some of the instruments $M^A_{a|x}$, $M^B_{b|y}$, $M^C_{c|z}$ are known and characterized, whereas other instruments are arbitrary.

The use of quantum switch for semi-device-independent certification of indefinite causality was verified experimentally by~\citet{Cao2022semiDI} (reviewed in Sec.~\ref{subsec:SemiDI_exp}), where the instruments of Alice are characterized but no assumption is made on the instruments of Bob, nor on the measurement on the control output state (and no final measurement on the target system is needed).

\subsubsection{Semi-device-independent certification via trusted quantum inputs}
\label{semiDI_QI}

\citet{Dourdent2022} proposed another semi-device-independent approach to causal nonseparability certification based on the use of trusted quantum inputs. The idea is directly inspired by the analogue proposal for testing entanglement. Indeed, \citet{Buscemi2012} showed that all entangled states could be certified in some variant of a Bell test, where classical inputs are replaced by quantum input states. This requires a trusted preparation of these input states, but the measurement devices do not need to be characterized: this scenario was also described as being ``measurement-device-independent'' \citep{branciard13}.

Adapting this idea to tests of causal nonseparability, it was shown that certain processes that do not violate causal inequalities ---most prominently, the quantum switch--- can be certified when using trusted quantum inputs (in what could be called an ``instrument-device-independent'' manner). This was done by introducing so-called ``causally nonseparable distributed POVMs'', thereby extending the notion of causal (non)separability to other types of objects, beyond process matrices.
However, it is worth noting here that Buscemi's result does not extend directly to the new scenarios under consideration: some causally nonseparable process matrices can still not be certified using trusted quantum inputs ---unless one makes some further assumption on the structure of the instruments, in which case one can recover the results, in the bipartite case, that all causally nonseparable process matrices can then be certified.

Exploiting further the analogies with entanglement and following \citep{bowles18}, it was then showed that the quantum inputs can be ``self-tested'' \citep{Supic2020selftestingof}, by replacing them with independent entangled states shared with some additional parties, so as to remove the necessity to trust their preparation \citep{dourdent2023networkdeviceindependent}. 
This then leads to a fully device-independent (but not theory-independent, see next subsection) certification of causal nonseparability that can be used, e.g., for the quantum switch. This approach was called ``network-device-independent'' to emphasize the fact that one still needs to trust the overall connectivity of the quantum resources used in the whole setup (specifically, that independent states are shared with separate parties for the self-testing part).

\subsection{Theory-independent tests of causal nonseparability}
\label{theoryindependenttests}

As discussed, indefinite causality can be certified either in a device-dependent or device-independent way. The latter is arguably more significant because an experimental violation of the causal inequality would show that indefinite causality is not a feature of a particular theory, but a fundamental property of nature, much as violation of a Bell inequality shows that nature is not locally causal. However, none of the experimentally realisable processes known to date can violate causal inequalities \citep{Branciard2016,OreshkovGiarmatzi_2016,Wechs2021, Purves2021} (these include all variants of the quantum switch and of quantum control of causal order discussed in Sec.~\ref{switchsection}). Motivated by the interplay between quantum theory and gravity (see Sec.~\ref{sec:Gravity}), \citet{zych2019} proposed an alternative certification of indefinite causality that aims to remove all theory-dependent assumptions, while maintaining certain assumptions about the characterization of the devices. In the proposed test, sites are grouped in two sets that are assumed to be causally independent (this can be enforced, for example, by embedding the sets in space-like separated regions). If, initially, the two sets can share at most a classically correlated state, and all sites are causally ordered, they can only produce correlations that satisfy a ``Bell inequality for temporal order''. On the other hand, the inequalities can be violated using quantum-control of causal order, which proves indefinite causal order as long as all other assumptions are satisfied. An experimental demonstration of the protocol (with a slightly modified formulation of the argument) was reported by  \citet{Rubino2022experimental}, see Sec.~\ref{nogotheorems}.

The Bell inequality for temporal order was shown to be  satisfied by a class of (quantum-like) generalized probabilistic theories (GPTs), in which the states and the laboratory operations are local and the operations are performed in a definite order. The class of theories considered are the so-called ``ball theories'', in which a two-level system is represented by a Bloch ball in an arbitrary dimension (the case of dimension $d=3$ is quantum theory) \citep{Dakic2010, Masanes2011}. However, \citet{debski_indefinite_2022} argued that the result relies on an additional, unstated assumption---that the system only evolves through the local operations, while remaining unchanged outside the sites---and that it is unclear how to formulate such an assumption in a theory-independent way.  Overcoming this limitation, \citet{VanDerLugt_2022} proposed a protocol based on a combination of locality and causality assumptions similar to \citep{zych2019, Rubino2022experimental}, but leading to a test that is both theory and device independent, see Sec.~\ref{sec:vanderlugt} above.
This proposal was then generalized to a scenario with more parties and which considers not only bipartitions, but tripartitions, allowing for a certification of device-independent indefinite causality in a ``possibilistic'' (rather than probabilistic)  scenario \citep{vanderLugt2023Possibilistic}.

\section{Applications of indefinite causal order}
\label{sec:applications}

A main line of research in the last few years concerns the study of quantum causal structure to model physical resources for information processing and other tasks. Results in this area include the formulation of tasks for which processes with indefinite causality outperform causally ordered strategies, or modeling communication scenarios in the absence of a fixed causal order. As the literature on this topic develops at an increasing pace, we will only give a broad overview of the main results.

\subsection{Channel discrimination tasks}
\label{channel}

Discriminating between different quantum operations is one of the first tasks in which indefinite causality was employed as a resource. \citet{Chiribella2012} pointed out an important property of the quantum switch: when the control state is $\ket{+}$ and the target state is an arbitrary pure state $\ket{\psi}$, if the switch process is applied on any of the pair commuting unitary operations, \textit{i.e.,} $U_A U_B=U_B U_A$, the output control state is still $\ket{+}$. Conversely, for the same input states, if we apply the switch on a pair of anti-commuting operations \textit{i.e.,} $U_A U_B=-U_B U_A$, the output control state is $\ket{-}$. By exploiting this property, \citet{Chiribella2012} constructed a pair of bipartite no-signaling channels that can be perfectly discriminated with a single use of the quantum switch, but cannot be perfectly discriminated with a single use of causally ordered processes. The switch property of identifying commuting and anticommuting unitary operators was also investigated by Ref.\,\citep{Araujo_2015} over a finite set of pairs of commuting unitary operators and a finite set of anti-commuting unitary operators. Specifically, while the quantum switch can always distinguish between commuting and anticommuting sets, \citet{Araujo_2015} employed SDP methods to show that a causally ordered process cannot distinguish these two sets with probability one. The problem of certifying commutation or anticommutation of a pair of unitaries was considered earlier in \citep{Andersson2005}, proposing a protocol where an interferometer ``recycles'' a photon in order to obtain different combinations of unitaries on the target system, similar in spirit to the switch, although without achieving deterministic success in the discrimination task.

The capabilities of indefinite causality were also investigated in standard quantum channel discrimination tasks, where multiple uses of a given channel are allowed. \citet{Bavaresco2020} used a general SDP-based computational approach to show that strategies with indefinite causality strictly outperform causally separable ones for discriminating between an amplitude damping channel and a bit-flip channel when two uses of the given channel are allowed. The authors also considered the problem of channel discrimination for randomly sampled channels and verified that strategies making use of indefinite causality strictly outperform causally separable ones for almost every choice of random channels, suggesting that the advantage of indefinite causality is typical. 
An advantage of indefinite causality is also observed when discriminating multiple uses of same unitary operations, but only in a scenario where three uses of the unitary operation are allowed \citep{Bavaresco2020b} (the usefulness of indefinite causality for discriminating unitary operations with two uses remains an open problem). It was also noted that, when discriminating multiple uses of the same unitary operations, the physically relevant class of quantum circuits with quantum control of order (QC-QCs, see Sec.~\ref{subsec:Gen}) cannot outperform causally ordered processes \citep{abbott2024query}, while QC-QCs do outperform causally ordered strategies in the non-unitary channel discrimination task \citep{Bavaresco2020b}.

A related task where indefinite causality was also shown to be a useful resource is the $K$-unitary equivalence determination problem \citep{Shimbo2018,Wechs2021,Soeda2021}. In this task, one is given $K$ different unitary operations which are uniformly sampled in the Haar measure and a further target operation which implements one of the $K$ reference unitary with probability $1/K$. The aim is to guess which of the $K$ reference operations is implemented by the target one while using each of the $K+1$ operations exactly once. \citet{Wechs2021} shows that, when $K=3$, indefinite causality provides a strictly better performance than causally separable ones. Moreover, some processes used leading to this advantage may be implemented with coherent control of causal order.

\subsection{Quantum computation}
\label{sec:QComp}

Quantum computation promises advantages over the best classical algorithms for certain tasks. Early investigations into indefinite causality \citep{Hardy2009,Chiribella2013} suggested that indefinite causal structures might, in principle, offer computational benefits, though without establishing a general, concrete route to speedup. In particular, \citet{Hardy2009} introduced the idea of a ``quantum gravity computer'' to frame his considerations, while \citet{Chiribella2013} presented the quantum switch as a specific computational resource, demonstrating limited oracle-query separations relative to standard, causally ordered quantum circuits.

These oracle separations are especially relevant because oracle problems provide a natural framework for quantifying and comparing computational resources. For instance, within the quantum circuit model—where the order of gates is fixed and independent of the input state—the problems introduced by \citet{Deutsch92rapidsolution}, \citet{Simon1997}, and \citet{Bernstein1997} demonstrate that a quantum computer can solve certain tasks with exponentially fewer queries than any classical computer. Indefinite causal structures enable quantum computation to be extended to more general scenarios in which the order of gates is controlled by an additional quantum state. This new resource has been shown to reduce the number of oracles in certain oracle problems beyond what any classical or quantum causal algorithm can achieve.

The simplest example of the problem where indefinite causal structures provide a constant advantage is deciding whether a pair of unitaries $U_A$ and $U_B$ commutes or anti-commutes, i.e., whether $[U_A, U_B]=0$ or $\{U_A, U_B\} =0$, under the promise that one of the two is realized \citep{Chiribella2012}.
As described in Sec.~\ref{channel}, an algorithm that exploits the quantum switch solves the task with a single call to each unitary. Solving the same task with a causally ordered quantum circuit, however, requires to call at least one unitary twice \citep{Chiribella2012}. This protocol has been experimentally demonstrated by~\citet{procopio2015}.

Building on the notion of the quantum switch, \citet{COLNAGHI20122940} and \citet{Facchini2015} introduced the $n$-switch as a resource: a generalization of the (classical or quantum) 2-switch to an arbitrary number of gates. Depending on the state of the control system, the $n$-switch applies a permutation of the gates to the target system, either incoherently (classical control) or coherently (quantum control). They formulated an associated ordering task—now standardly called the \emph{Unitary Permutation Problem} (UPP)—which asks: given black-box unitaries $U_1,\ldots,U_n$ and a permutation $\pi$ of $\{1,\ldots,n\}$, implement the ordered composition $U_{\pi(n)} \cdots U_{\pi(2)} U_{\pi(1)}$ using as few oracle calls as possible. They showed that the \textit{n}-switch reduces the query complexity of this task compared to fixed-order circuits (e.g., from $O(n^2)$ oracle calls in standard circuits to $n\log_2 n+O(n)$ calls in the switch-based model \citep{Facchini2015}). However, it is important to note that this separation can already be achieved with \emph{classical} (incoherent) control of the order, so these improvements do not arise from indefinite causal order.

In order to study the computational power of the indefinite causal structures in the asymptotic sense as the number of unitaries increases, a promise problem which generalizes the commutation/anticommutation task of Ref. \citep{Chiribella2012}, called {\it Fourier promise problem}
(FPP), was introduced in Ref. \citep{Araujo2014}. The problem concerns a set of $d$-dimensional unitary gates $\{U_0,...,U_{n-1}\}$. A  permutation $\pi_x$ of the $n$ unitaries is denoted as $\Pi_x = U_{\pi_x (n-1)}...U_{\pi_x (0)}$ and labeled by a
number $x \in \{0, 1, ..., n!-1\}$. It is promised that for some value $y \in \{0, 1, ..., n!-1\}$, the permutations satisfy the following relation:
$\forall x: \Pi_x = \omega^{xy} \Pi_0$, where $\omega\coloneqq e^{i\frac{2\pi}{n!}}$. The problem is to find the
value $y$ for which the above promise is satisfied. 

In Ref.~\citep{Araujo2014}, it was shown that there exist unitaries that satisfy the promise. Furthermore, it was shown that the task can be solved with the quantum $n$-switch with a single call to each gate (i.e., altogether $n$ queries). 
The best-known fixed order quantum circuit that simulates the $n$-switch requires $O(n^2)$ queries \citep{Araujo2014,Facchini2015}, and, under extra assumptions, any fixed order circuit would indeed require $\Omega(n^2)$ queries \citep{Facchini2015}. Interestingly,  a more efficient causal algorithm was found that solves the FPP with the unitaries from Ref. \citep{Araujo2014} with $O(n\log n)$ queries and a further causal algorithm that solves every FPP with $O(n\sqrt{n})$ queries \citep{Renner2021}. 

The physical requirements for achieving an advantage in the UPP are demanding, since for tasks with $n$ unitaries the dimensions of both the control and target systems must be at least $n!$. To address this, a new family of tasks with lower-dimensional control and target systems---called the \emph{Hadamard promise problems} (HPP)---was recently proposed \citep{Taddei2020,Renner2022}. In this setting, one considers $d$-dimensional black-box unitaries $\{U_i\}_{i=0}^{n-1}$ and a restricted set of permutations $\{\Pi_x\}_{x=0}^{n_x-1}$ with $n_x \leq n!$. The promise is that there exists a value $y \in \{0,\ldots,n_x-1\}$ such that, for all $x$, $\Pi_x = s(x,y)\,\Pi_0$, where $s(x,y) \in \{\pm 1\}$ are the entries of an $n_x \times n_x$ Hadamard matrix. The task is then to determine the hidden value of $y$.  
A specific instance of this problem with four gates was introduced and experimentally investigated by~\citet{Taddei2020}. \citet{Renner2022} generalized the construction to arbitrary $n$ and provided an $n$-switch--based solution requiring only qubit targets, thereby streamlining experimental implementations. Their work establishes a provable asymptotic gap in query complexity between quantum-controlled ordering of gates and causally ordered (fixed-order) quantum algorithms: while the problem can be solved with the quantum $n$-switch using a single call to each gate, any causal algorithm requires at least $2n-2$ queries. Moreover, they showed that the best known fixed-order algorithm uses $O(n\log n)$ queries and conjectured that no better causal solution exists. As a further generalization, \citet{EscandonMonardes2023} introduced \emph{Complex Hadamard promise problems} (CHPP). These are defined just as the HPPs described above, except the entries $s(x,y)$ of the matrix defining the promise can be complex, with $|s(x,y)|=1$. FPPs and HPPs are then particular cases of CHPPs. Also in this case, the problem can be solved with one query per gate by a switch controlling a limited number $n_x$ of permutations, and where the target's dimension depends on the matrix $s$ (most generally, a continuous-variable target is required, but a subclass of CHPPs was identified compatible with finite target dimension). The gap in query complexity with the best-known causally ordered solution (circuit simulation of the switch) was calculated for up to $n=4$ unitaries.

\citet{abbott2024query} analyze the quantum query complexity of Boolean functions, a standard computational task \citep{Ambainis2017} where the goal is to compute a Boolean function using a quantum circuit with access to an oracle, as in Grover's algorithm \citep{Grover1996}. There, it is shown that, while the quantum switch and QC-QC processes cannot outperform circuits with fixed order in this task, indefinite causal order can reduce the minimum error probability for certain 3-bit Boolean functions in the two-query scenario. Additionally, Ref. \citep{abbott2025classical_quantum} proves a polynomial separation between standard classical circuits and classical processes without definite causal order, identifying some Boolean functions whose quantum query complexity is reduced by causally indefinite computations.

In Ref. \citep{Baumeler2018}, a model of computation on indefinite causal structure with classical (as opposed to quantum) processes was formulated. In contrast to the process matrix formalism, where parties can apply arbitrary local operations, the gates are fixed in that model. The class of decision problems efficiently solvable by the classical noncausal circuit model was shown to be equal to UP $\cap$ coUP.
The class UP $\cap$ coUP contains all decision problems where, for each possible answer (‘yes’ or ‘no’), a unique witness exists (an example of such a problem is integer factorization). The class is believed to be strictly contained within NP, hence there are strong evidences that classical process matrices cannot efficiently solve NP-hard problems. This should be contrasted with the computational power of CTCs, both of  Deutschian CTCs and postselection CTCs type, which allow for the efficient solution of problems in PSPACE and PP, respectively.

\citet{Araujo2017} introduced a computational model for general quantum processes, defined as the subset of PostBQP (that is, the class of quantum computations with unrestricted postselection) restricted to circuits that reproduce valid process matrices. They further showed that there exist process matrices that can violate causal inequalities but
do not lead to any advantage in computational complexity.

\subsection{Transforming quantum operations}

As discussed in Sec.~\ref{subsec:open_past_future} and illustrated by the quantum switch, process matrices may be used as devices to transform quantum operations. A relevant instance of this is the following problem: given $k$ calls of an arbitrary $d$-dimensional unitary operation $U$, can we construct a quantum circuit---or a quantum process---that transforms these $k$ uses of $U$ into another unitary operation $f(U)$, where $f$ is an arbitrary function mapping unitaries into unitaries? \citet{Quintino2019} considered the case where $f(U)=U^{-1}$ to show that the probability of transforming $k$ calls of a $d$-dimensional unitary into its inverse with general process matrices is strictly greater than the success probability obtained by any ordered quantum circuits or causally separable process. In similar vein, \citet{Quintino2019b} showed that indefinite causality enhances the success probability for transforming $k$ calls of $U$ into its transpose $U^T$. The higher performance of process with indefinite causality over causally separable ones was also observed for deterministic transformations. \citet{Quintino2021} showed that, when transforming multiple calls of a unitary operation to its inverse or its transposition, process with indefinite causality offer strictly better values for the fidelity of the transformation and robustness to noise. Remarkably, when the input state for which one aims to implement the transpose/inverse will be implemented, the advantage of indefinite causal order seems to vanish \citep{brzic2025HOQO_KS}.
	
Interestingly, there are also cases where indefinite causality does not offer any advantage over causally ordered circuits. In Ref. \citep{Bisio2014} (see also \citep{Quintino2021}), the authors have shown that when transforming unitary operations $U$ into $f(U)$ and the function $f$ is a homomorphism, \textit{i.e.,} it respects $f(UV)=f(U)f(V)$ for every unitary $U$ and $V$, the optimal transformation is always attainable by parallel quantum circuits. This shows that processes with indefinite causality cannot outperform causally ordered ones for transformations such as unitary complex conjugation $f(U)=U^*$ \citep{Miyazaki2017,Ebler2022conj} and approximate unitary cloning $f(U^{\otimes k})\approx U^{\otimes (k+m)}$ \citep{Chiribella2008Cloning,Chiribella2015Replication}.

Indefinite causal order has also been proven to outperform causally ordered protocols in the tasks of isometry inversion and universal error detection, while no advantage is found when considering isometry adjointation \citep{Yoshida2025adjointation}. \citet{apadula2022nosignalling} developed a higher-order quantum theory framework for computation with indefinite causal structure and gave a full characterization of which compositions of higher-order maps are admissible. Their key result is that the permissible computational compositions are exactly those compatible with the no-signaling relations among the subsystems of the  higher order quantum maps.

\subsection{Quantum communication complexity}
\label{sec:QCommCompl}

Quantum resources allow certain communication and computation problems to be solved more efficiently than classically possible. In communication complexity problems, a set of separated parties must accomplish a common task while reducing the amount of communication between them \citep{yao1979,kushilevitz_nisan_1996}. The task can always be tailored in terms of computing a function whose inputs are nonlocally distributed among the parties. 

In general, one can distinguish two types of communication complexity problems related to the following two questions: (i) What is the minimum amount of communication required for all parties to determine the value of the function with certainty? (ii) What is the highest possible probability that the parties will arrive at the correct value of the function when only a limited amount of communication is allowed? The minimal amount of communication in problems of type (i) is called the {\em communication complexity} of the problem. Solutions to this abstract problem are relevant for minimizing time or energy consumption in distributed computation and computer networks, and in general whenever classical communication between distant nodes is costly. It is known that shared entanglement and communication through quantum channels can significantly, even exponentially, reduce the communication complexity as compared to protocols exploiting shared classical randomness and classical communication \citep{Buhrman2010}.

Recently, it was shown that for both types of problems the quantum switch used as a quantum control of the direction of communication is a valuable resource for communication complexity, above and beyond quantum (causal) channels and entanglement. In particular, tripartite communication tasks were found for which such a quantum control allows for an increase of the success probability in a problem of type (i) \citep{feix2015}, and an exponential saving in communication for a problem of type (ii), as compared to any one-way quantum (or classical) communication \citep{Guerin2016}. As detailed in Sec.~\ref{Sec:Exp_QComCompl}, the latter advantage has been experimentally demonstrated in a photonic realization of the quantum switch with $d$-dimensional target quantum systems with up to $d=2^{13}$ dimensions \citep{wei2019experimental}.

\begin{figure*}
	\begin{center}
		\includegraphics[width=\textwidth]{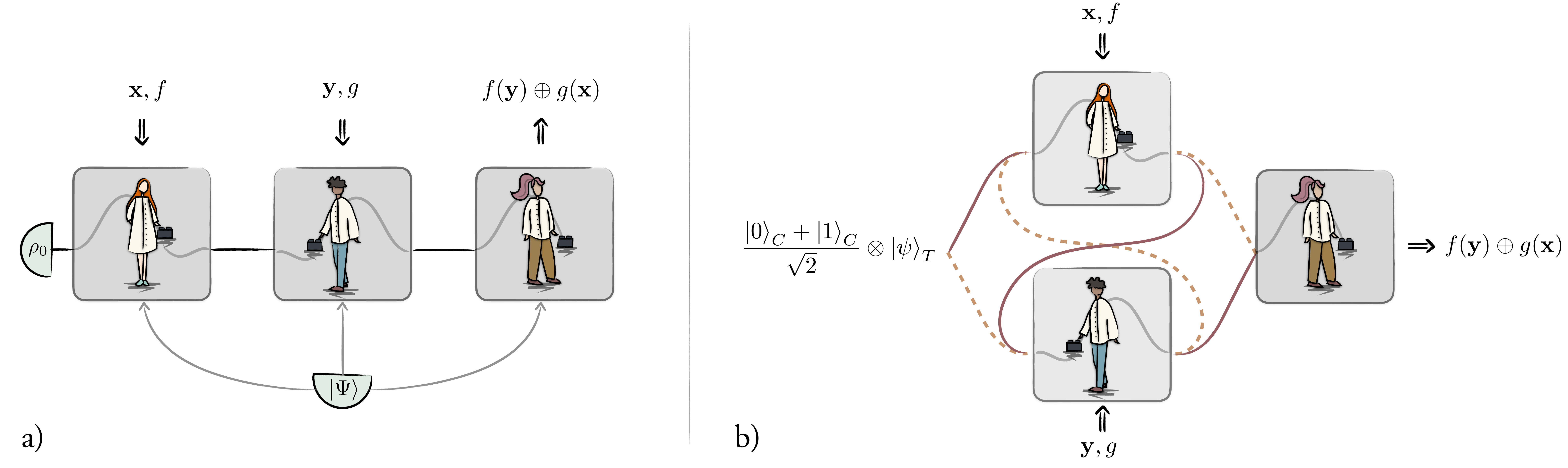} 
	\end{center}
\caption{Schematic of a communication complexity setup in a causally ordered
scenario and using the quantum
switch. a) Based on her input $(\textbf{x}, \textit{f})$, Alice sends a quantum state $\rho_{(\textbf{x},\textit{f})}$ to Bob. Bob processes it with a CP map dependent on his input $(\textbf{y}, \textit{g})$ and sends it to Charlie. Their shared unlimited entanglement is symbolized by $\ket{\Psi}$. The goal is to minimize the number of qubits in $\rho_{(\textbf{x},\textit{f})}$, setting a lower bound for communication complexity. b) A qubit in state $(\ket{0}_C+\ket{1}_C)/\sqrt{2}$ coherently controls
the path taken by a system of $N$ qubits in the initial state $\ket{\psi}_T$ which, depending on the state of the control qubit, are directed first to Alice's lab and then to Bob's, or vice versa. Alice and Bob receive the classical inputs $(\textbf{x}, \textit{f})$ and $(\textbf{y}, \textit{g})$, respectively, and Charlie (using the
control qubit) computes a binary function of their inputs $f(\textbf{y})\oplus g(\textbf{x})$.}
\label{fig:allard_prl}
\end{figure*}

The communication task of Ref. \citep{Guerin2016} is illustrated in Fig.~\ref{fig:allard_prl}. Alice receives input $({\bf x},f)$ and Bob $({\bf y},g)$, where ${\bf x}=\{x_1,...,x_n\},{\bf y}=\{y_1,...,y_n\} \in \{0,1\}^n$ are $n$ bit strings and $f,g: \{0,1\}^n \rightarrow \{0,1\}$ are Boolean functions from $n$ bit strings to a bit, such that $f({\bf 0})=g({\bf 0})=0$ where ${\bf 0}$ is the all-zero string. Another party, Charlie, must compute the exchange evaluation function $EE_n ({\bf x},f, {\bf y},g)= f({\bf y}) \oplus g({\bf x})$, where the symbol $\oplus$ denotes addition modulo 2. The important constraint in the task is that the communicating system can enter each party's laboratory only {\em once}, as also assumed in the process matrix formalism. If this was not the case, the task could be trivially accomplished using two-way classical communication: Alice and Bob would exchange their vectors ${\bf x}$ and ${\bf y}$, compute $f({\bf y})$ and $g({\bf x})$ locally, and then send both results to Charlie, using a total of $2n + 2$ bits.
In contrast, as also noticed in Ref. \citep{Guerin2016}, in the case where multiple rounds of communication are allowed, the advantage enabled by the quantum switch disappears.

Since the task requires (almost) no promise on inputs, it can be shown that the number of qubits required for deterministic success in the causally ordered communication is one-half of the logarithm of the dimension of the set of inputs plus one additional bit\footnote{For each party the dimension of the set of the input functions is $2^{2^n}-1$ (the $-1$ is due to the promise that $f({\bf 0})=g({\bf 0})=0$), the dimension of the set of the input vectors is $2^n$, and $\frac{1}{2}$ comes from using the quantum dense coding.}: $\frac{1}{2} 2^{2^n+n} +1$. By simply having Alice send $2^n+n-1$ qubits to Bob, followed by his evaluation of $f({\bf y}) \otimes g({\bf x})$ and communication of the result to Charlie they can complete the task. 

In contrast, it can be shown that the parties with the quantum switch can succeed at this task deterministically with only $n$ qubits of communication. This is achieved by encoding the inputs $\textbf{x}, f, \textbf{y}, g$ into local $n$-qubit unitaries that either commute or anticommute, and using the property of the switch in solving the promise problem (Sec.~\ref{channel}). To construct these unitaries, we define the following: the product of Pauli operators $X({\bf x}) = X^{x_1}_1 \otimes \cdots \otimes X^{x_n}_n$, where $X_i$ is the Pauli-$X$ operator and $X^0_i = \id$ is the identity, both acting on the $i$-th qubit; and a diagonal matrix $D(f) = \sum_{\bf z} (-1)^{f({\bf z})} |{\bf z}\rangle \langle {\bf z}|$, with $|{\bf z}\rangle$ the computational basis state satisfying $Z_i |{\bf z}\rangle = (-1)^{z_i} |{\bf z}\rangle$ for Pauli-$Z_i$, and where the sum runs over all strings ${\bf z}=\{z_1,...,z_n\}$ of $n$ bits. Alice will then encode her input in unitaries $X({\bf x}) D(f)$, and similarly Bob in $X({\bf y}) D(g)$. It can easily be seen that Alice's and Bob's unitaries either commute or anticommute with each other when acting on the $n$-qubit state $|{\bf 0}\rangle$, i.e., 
\begin{align}
    [X({\bf x}) D(f),X({\bf y}) D(g)]|{\bf 0}\rangle &=0 \mbox{ if } (-1)^{f({\bf y})}= (-1)^{g({\bf x})}, \notag \\
     \{X({\bf x}) D(f),X({\bf y}) D(g)\}|{\bf 0}\rangle &=0 \mbox{ if } (-1)^{f({\bf y})}= (-1)^{g({\bf x})\oplus 1}.
\end{align}
It then follows that, by discriminating between these two cases, Charlie can estimate the exchange evaluation function with certainty.

\subsection{Other applications of indefinite causal order}

\subsubsection{Noise reduction in communication}
\label{noisereduction}

Consider a communication between two parties through noisy channels. \citet{gisin2005} showed that placing the trajectory of the information carriers in quantum superposition can reduce the noise affecting the transmission. In particular, they proved that, if the trajectories were separated enough to assume the noise affecting each channel was independent, this allowed some of the noise to be filtered out by using quantum interference and post-selection. More recently, it was proposed that similar advantages could be obtained by placing the noisy channels in a quantum switch. This was shown to bring advantages both in the case of classical \citep{Ebler2018} and quantum communication \citep{salek2018quantum}, and even to allow total noise cancellation for some specific types of noise \citep{chiribella2018indefinite}. These findings triggered a number of subsequent proposals \citep{Procopio2019, Gupta2019, procopio2020sending, caleffi2020, Sazim2020, Wilson2020}, and some experiments \citep{ goswami2018communicating,guo2020experimental}. More recently, these studies have also been extended to the case of  random-receiver quantum communication \citep{Bhattacharya2021}, and quantum control of cyclic orders (for more than two noisy channels) \citep{Chiribella2021PRL}.

Following the growing interest sparked by the topic, the role played by indefinite causality in achieving the noise reduction has been further investigated \citep{abbott2018,chiribella2019quantum,guerin2019_communication,kristjansson2019,Grinbaum2020, Rubino2020}.  In this context, theoretical studies have indicated that indefinite causality is not required to reduce noise in classical and quantum communication \citep{abbott2018,guerin2019_communication,Grinbaum2020, Rubino2020}. In particular, a comparable or even higher noise reduction can be obtained by placing the noisy channels in parallel \citep{abbott2018}\footnote{\citet{pang2023} further explored the capacity enhancement achieved by combining two independent noisy qubit channels, noting that an expanded description of the channels is required to account for the additional control state (as already noted in \citep{kristjansson2019, Rubino2020}). This expansion process may not be unique and, in particular, in the case of an experimental realization of the channels, the physical implementation determines the expanded description that fully characterizes the channel behavior under superposition. This led to the conclusion that the capacity enhancement is due to the physical implementation of the channels rather than the superposition itself.}, or by placing the channels in series with quantum-controlled operations \citep{guerin2019_communication}. \citet{guerin2019_communication} showed that the Shor quantum error-correction code can be used to find an arrangement of channels in series with quantum-controlled gates, allowing any arbitrary noise to be completely eliminated. Finally, \citet{Rubino2020} highlighted that the common resource for all the considered schemes (and thus the only one necessary to achieve noise reduction in the noisy communication at stake) is the establishment of a coupling between the trajectory of the information carrier and the degree of freedom on which the noise acts. Only more recently a genuine advantage of indefinite causal order was shown in this type of scenario (which can be obtained with QC-QCs other than the quantum switch), when imposing further constraints on the allowed processes so as to exclude possible side-channels \citep{zhao2025}.

\subsubsection{Thermodynamic tasks}
\label{subsubsec:thermo_tasks}

Inspired by the results presented in the previous section \citep{Ebler2018,chiribella2018indefinite,salek2018quantum}, \citet{Felce_2020} transferred them into the quantum thermodynamics domain, with the aim of understanding whether similar advantages could be observed in the execution of thermodynamic tasks.

The authors first studied the case of two thermalizing channels, which are channels that map an input state $\rho$ to the thermal state $\rho_T$ (i.e., a state representing a system at thermal equilibrium at temperature $T$). By placing two of these channels in a quantum switch, the authors found that the effective temperature of the output target state does not correspond to that of the thermal state $\rho_T$ predicted by a sheer application of the thermal channels. Rather, the temperature of the final state is higher or lower depending on whether the control qubit is projected onto the $\ket{+}$ or $\ket{-}$ state.

The authors then used this result to establish a refrigeration cycle by exploiting indefinite causality. In the first stage (1), they placed two cold reservoirs in an indefinite causal order via a quantum switch. As in the previous case, the target qubit will not thermalize at the temperature of the reservoirs, but at a higher or lower temperature depending on whether the control qubit is measured in $\ket{+}$ or $\ket{-}$. If $\ket{+}$ is measured, the target system has dumped heat into the reservoir, which is the opposite of what is desired. It is therefore necessary to post-select in the case where $\ket{-}$ is measured, which is the case where the target system has extracted heat from the reservoir.

In the second stage (2), the system is classically thermalized with a hot reservoir. Finally, in the third stage (3), the system is classically thermalized with a cold tank and all cold reservoirs are classically thermalized with each other, so that they become identical.

In stage (1), heat is transferred from the cold reservoir to the working system. In stage (2), heat is transferred from the working system to the hot reservoir. In stage (3), heat is transferred from the working system to the cold reservoir, but this heat transfer is smaller in magnitude than in step (1), so overall heat is transferred out of the cold reservoir.
Although the success of this cooling cycle is based on post-selection, the authors take this into account when considering the average thermodynamic performance of the cycle.

Similar setups were then considered in \citep{Guha2020} and \citep{Simonov2022}, where it was shown that inserting some given channels into the quantum switch results in a higher final free energy and, respectively, ergotropy, as compared to a direct composition of the same channels.

However, similarly to the case of noise reduction discussed in sec.~\ref{noisereduction}, \citet{Capela_2022} showed that causally ordered processes can have an even higher performance by allowing the system to interact with some external environment (which is fair, considering that the switch-based protocols make critical use of the control system). At the simplest level, one can consider a process where, after the action of the thermal channels, the target system is simply replaced with a state from an environment with the desired properties (e.g., arbitrary free energy or ergotropy). A similar advantage can also be obtained with less trivial processes, where system and environment interact through some prescribed Hamiltonian.

Although it is currently unclear whether indefinite causality can offer a genuine advantage in thermodynamic tasks, the interplay between the two remains an active area of research
\citep{Mukhopadhyay2018, Dieguez2022, Liu_2022, Goldberg2023, Verma2024, Francica_2022, chen2021indefinite, Fellous_2022}.

\subsubsection{Quantum metrology}
\label{subsubsec:quantum_metro}

Indefinite causality has been proposed as a resource to improve the performance in quantum metrology tasks.
\citet{zhao2019} investigated a scenario involving $2N$ black boxes, each acting on a harmonic oscillator. Of these $2N$ boxes, $N$ will produce position displacements  $D_{x_j} = e^{-i x_j P}$ ($j; k = 1,\dots, N$), while the remaining $N$ boxes will perform momentum displacements $D_{p_k} = e^{i p_k X}$ (where $X$ and $P$ are the conjugate variables $X\coloneqq  (a^\dagger + a)/\sqrt{2}$ and $P\coloneqq  i(a^\dagger - a)/\sqrt{2}$, and $a$ and $a^\dagger$ are the creation and annihilation operators.). The values of $\{x_j\}$ and $\{p_k\}$ are unknown, and they independently span within the intervals $[x_{\text{min}}, x_{\text{max}}]$ and $[p_{\text{min}}, p_{\text{max}}]$, respectively. The goal is to measure the product $A\coloneqq  \overline{x} \cdot \overline{p}$ between the average displacements $\overline{x}\coloneqq  \sum_{j=1}^N x_j/N$ and $\overline{p}\coloneqq  \sum_{j=1}^N p_j/N$. When these black boxes are used sequentially, the authors prove that the root-mean-square error (RMSE) in predicting $A$ cannot decrease faster than $f(E)/N$, where $f(E)$ is a function of the average energy $E\coloneqq  \langle \psi| (X^2 + P^2) \ket{\psi}/2$  of the input state $\ket{\psi}$ used to probe the black boxes. This is consistent with the Heisenberg limit of quantum metrology \citep{Giovannetti_2006} for estimating $X$ and $P$ average displacements. In contrast, they show that a setup employing the quantum switch can achieve an error reduction scaling as $1/N^2$, regardless of the input state energy.

\citet{Liu2023a} presented an SDP approach to evaluate the quantum Fisher information of a finite set of finite-dimensional quantum channels with and without definite causal order. This approach is then used to show that there exist situations where strategies with indefinite causal order provide a strictly higher Fisher information than any definite causal order ones. This was further developed by~\citet{mothe2023reassesing}, who compared additional classes of quantum processes with or without definite causal order. In particular, this allowed them to reassess previous works which claimed to find some advantage for the quantum switch with respect to certain causally ordered quantum circuits in some metrological task consisting of estimating a parameter of a given type of channels \citep{Mukhopadhyay2018,Frey_2019,Frey21,Chapeau-Blondeau2021,Chapeau-Blondeau21coherentcontrol,ChapeauBlondeau2022,Chapeau_Blondeau_23, Goldberg2023b}. Some of these advantages were shown to vanish when a fairer comparison is made with all possible causally ordered strategies; although certain other scenarios indeed show an advantage of indefinite causal order, particularly through circuits with QC-QC.

Finally,~\citet{Kurdzialek2023} consider the attainable precision in asymptotic metrology scenarios through quantum Fisher information. There, it is proven that for finite-dimensional quantum systems, strategies which make use of the quantum $n$-switch lead to the same asymptotic performance as parallel strategies. Hence, differently to the infinite dimension advantage presented in Ref. \citep{Zhao_2022}, for finite-dimensional systems, the quantum $n$-switch and adaptive strategies cannot outperform parallel strategies on asymptotic metrological precision tasks. It remains to explore the finite-sampling regime, for which a framework including indefinite causal order has recently been put forward \citep{Meyer2025}.

\subsubsection{Other tasks}
\label{subsec:othertasks}

\paragraph{Quantum batteries}

Indefinite causal order has also been investigated as a resource in the context of quantum batteries. In this setting, the goal is to enhance the charging and storage capabilities of quantum devices beyond what is possible with classical approaches. \citet{chen2021indefinite} considered charging protocols in which two chargers act on a qubit battery within a quantum switch. They found that, while a single static charger cannot fully charge the qubit battery, placing two such chargers in a superposition of orders could achieve full charging, though only probabilistically via post-selection of measurement outcomes.

\paragraph{Measure the incompatibility of projective measurements}
When considering projective measurements, the concept of measurement compatibility is equivalent to commutativity \citep{heinosaari2016Invitation}. In Ref. \citep{gao2023measuring}, the authors introduce a measure to quantify the noncommutativity of pairs of projective measurements named ``\textit{Mutual Eigenspace Disturbance}''. This quantifier constitutes a metric on the space of von Neumann measurements that can be directly estimated via the quantum switch.

\section{Experimental approaches to indefinite causal order}
\label{sec:Experim}

The quantum switch, and variants thereof, is, to the best of our knowledge, the only causally nonseparable process that has been the object of experimental research to date. These experiments predominantly employ single photons and vary in aspects such as degrees of freedom (DOF), dimension of control and target systems, and number of operations. The only two exceptions to this trend are the experiments reported in \citep{Nie_2020}, which used an ensemble of nuclear magnetic resonance (NMR) spins, and \citep{Felce_2021}, which used the IBM quantum computing platform. Whether the experiments performed so far---and which of them---should be interpreted as `genuine' realizations of indefinite causal order is a debated subject, as discussed in Sec.~\ref{interpretationsection}.

In the following, we review the main experiments that have been carried out to date in the field of indefinite causality and categorize them according to the experimental system used and the specific DOF involved. Note that the order in which the experiments are presented does not necessarily correspond to the chronological order in which they were performed. \citet{Goswami_review} and \citet{Rozema2024} provide a dedicated review of the experimental realizations of the quantum switch.

\subsection{Quantum optics} 

This section focuses on quantum-optical implementations of the quantum switch. Table~\ref{tab:exp} provides a schematic summary, detailing the DOF, the dimensions of the control and target systems, and the number of operations in various experimental setups. The experiments discussed below are organized according to their experimental platforms and the DOFs used, and, where multiple experiments employ the same platform, they are further grouped by the specific tasks they demonstrate.

\providecommand{\exprefcell}[1]{%
  \begin{minipage}[c]{1.7cm}
  \vspace{2pt}
  \raggedright
  #1
  \vspace{2pt}
  \end{minipage}%
}

\begin{table}[tbh]
    \centering
    \renewcommand{\arraystretch}{1.25}
    \setlength{\tabcolsep}{6pt}
    \begin{tabular}{|c||c|c||c|c||c|}
     \hline
         \makebox[1.7cm][c]{}
         & \multicolumn{2}{c||}{Target} 
         & \multicolumn{2}{c||}{Control} 
         & \makebox[1.15cm][c]{Type of} \\
         \cline{2-5}
         & \makebox[0.90cm][c]{DOF} 
         & \makebox[0.55cm][c]{Dim.} 
         & \makebox[0.90cm][c]{DOF} 
         & \makebox[0.55cm][c]{Dim.} 
         & \makebox[1.15cm][c]{Interfer.} \\
         \hline\hline

         \exprefcell{
         \citet{procopio2015},
         \citet{rubino2017, Rubino2020, Rubino2022experimental},
         \citet{guo2020experimental},
         \citet{Cao2021}, \citet{Cao2022semiDI},
         \citet{Zhu2023, Qu_2025}
         }
         & \makebox[0.90cm][c]{Pol.} 
         & \makebox[0.55cm][c]{2} 
         & \makebox[0.90cm][c]{Path} 
         & \makebox[0.55cm][c]{2} 
         & \makebox[1.15cm][c]{MZ}\\
        \hline

        \exprefcell{
        \citet{stroemberg2022},
        \citet{Liu_2023}
        }
        & \makebox[0.90cm][c]{Pol.} 
        & \makebox[0.55cm][c]{2} 
        & \makebox[0.90cm][c]{Path} 
        & \makebox[0.55cm][c]{2} 
        & \makebox[1.15cm][c]{Sagnac}\\
         \hline

        \exprefcell{
        \citet{goswami2018, goswami2018communicating}
        }
        & \makebox[0.90cm][c]{TSM} 
        & \makebox[0.55cm][c]{2} 
        & \makebox[0.90cm][c]{Pol.} 
        & \makebox[0.55cm][c]{2} 
        & \makebox[1.15cm][c]{MZ}\\
        \hline

        \exprefcell{
        \citet{Antesberger2024},
        \citet{Richter_2025}
        }
        & \makebox[0.90cm][c]{Pol.} 
        & \makebox[0.55cm][c]{2} 
        & \makebox[0.90cm][c]{T.-B.} 
        & \makebox[0.55cm][c]{2} 
        & \makebox[1.15cm][c]{CPI}\\
        \hline

        \exprefcell{
        \citet{Guo_2025}
        }
        & \makebox[0.90cm][c]{T.-B.} 
        & \makebox[0.55cm][c]{2} 
        & \makebox[0.90cm][c]{Pol.} 
        & \makebox[0.55cm][c]{2} 
        & \makebox[1.15cm][c]{Sagnac}\\
        \hline

        \exprefcell{
        \citet{wei2019experimental}
        }
        & \makebox[0.90cm][c]{T.-B.} 
        & \makebox[0.55cm][c]{12} 
        & \makebox[0.90cm][c]{Path} 
        & \makebox[0.55cm][c]{2} 
        & \makebox[1.15cm][c]{Sagnac}\\
        \hline

        \exprefcell{
        \citet{Taddei2020}
        }
        & \makebox[0.90cm][c]{Pol.} 
        & \makebox[0.55cm][c]{2} 
        & \makebox[0.90cm][c]{Path} 
        & \makebox[0.55cm][c]{4} 
        & \makebox[1.15cm][c]{CPI}\\
        \hline

        \exprefcell{
        \citet{yin2023experimental}
        }
        & \makebox[0.90cm][c]{TSM} 
        & \makebox[0.55cm][c]{$\infty$} 
        & \makebox[0.90cm][c]{Pol.} 
        & \makebox[0.55cm][c]{2} 
        & \makebox[1.15cm][c]{MZ}\\
        \hline
    \end{tabular}
    \caption{Schematic overview of the eight different types of quantum-optical implementations of a quantum switch experimentally demonstrated thus far. CPI: common-path interferometer; Dim.: Dimension; MZ: Mach-Zehnder; Pol.: Polarization; T.-B.: Time-Bins; TSM: Transverse Spatial Modes. In all of these experiments, the dimension of the control system is equal to the number of operations in the quantum switch.}
    \label{tab:exp}
\end{table}
 
\subsubsection{Path and polarization of single photons in a Mach-Zehnder interferometer} 
\label{subsubsec:ExpPathAndPol}

In \citep{procopio2015, rubino2017}, quantum-optical switches were realized as follows: Single photons were generated via a heralded source and then injected into a two-loop Mach-Zehnder interferometer (MZI), which is a type of interferometer where a light beam is first split into two parts by a beam splitter (BS) and then recombined by a second BS.
In this optical implementation of a quantum switch, the control qubit was encoded in the path DOF, while the target qubit was encoded in the photons' polarization. The superposition of orders was achieved by routing the transmitted (reflected) branch of the interferometer through operation $A$ ($B$) first and then, after being looped back, through operation $B$ ($A$).

\begin{figure}
\centering
\includegraphics[width=.75\columnwidth]{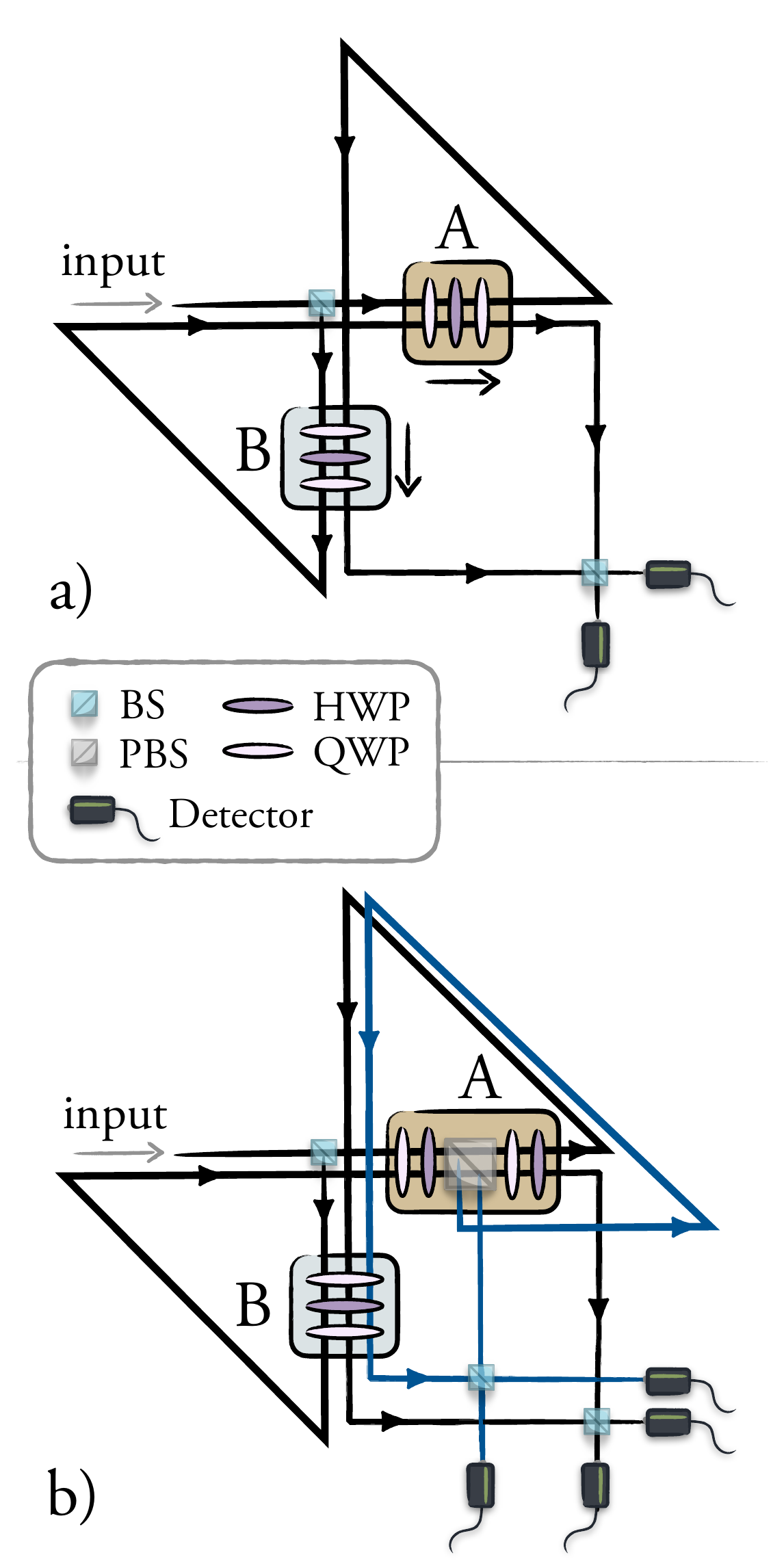}
\caption{Schematic of a two-party quantum switch with target system encoded in the polarization and control system in the path DOF. a) Quantum switch with unitary operations, from \citet{procopio2015}. b) Quantum switch with one measure-and-reprepare and one unitary operation, from \citet{rubino2017}. Single photons from a heralded source are injected into a two-loop MZI. Two unitary operations (Panel a) or one unitary operation and a measurement-and-reprepare apparatus (Panel b) are placed at locations corresponding to Alice and Bob's laboratories. In Panel b), the measurement-and-reprepare operation requires an additional optical branch for each polarization component selected by the PBS, so that the measurement outcome is not revealed and coherence between the two causal orders is preserved.}
\label{img:procopio_rubino1}
\end{figure}

Both operations $A$ and $B$ in \citep{procopio2015} were unitary operations implemented by a sequence of three waveplates: a quarter waveplate (QWP), a half waveplate (HWP) and a QWP (Fig.~\ref{img:procopio_rubino1}~\textbf{a)}). Conversely, in \citep{rubino2017},  operation $A$ was designed to function analogously to a `measure-and-reprepare' operation. This was achieved by a sequence of a QWP, a HWP and a polarising beam splitter (PBS) for measurement, followed by a QWP and a HWP for re-preparation (Fig.~\ref{img:procopio_rubino1}~\textbf{b)}). The significant departure from a standard ``measure-and-reprepare operation'', in this case, was the need to preserve coherence within the two branches of the quantum superposition by delaying the photon measurement until after both paths had been interfered in the second beam splitter.  This approach allowed the measurement result to be encoded in auxiliary systems (namely, additional paths of the single photons) entangled with the measured system.

In \citep{procopio2015}, this experimental setup was used to provide the first empirical evidence of indefinite causality, demonstrating a capability to solve a particular computational problem more efficiently than any ordered quantum circuit. The task, introduced by \citep{Chiribella2012} and reviewed in Sec.~\ref{sec:QComp}, was to determine whether two unitary gates, $U_A$ and $U_B$, commute or anti-commute---a result that, within the standard quantum circuit model, would typically require using at least one of the gates twice \citep{Chiribella2012}. Achieving this indirectly demonstrated that the experimental set-up used exhibits an indefinite causal structure.\footnote{Section~\ref{subsec:num-events} includes an in-depth discussion of the debate surrounding the estimation of gate uses in experimental implementations of the quantum switch.}
Specifically, the experiment achieved a success probability of 
$0.976 \pm 0.015$ in evaluating the commutation properties of $100$ pairs of commuting and anti-commuting gates, exceeding the maximum average success rate of $0.9288$ \citep{Araujo2014} achievable with a single use of each unitary gate in a fixed-order quantum circuit by 3 standard deviations (st.~dev.) \citep{procopio2015}.

In \citep{rubino2017}, an analogous setup was employed to demonstrate the causal nonseparability of a process by measuring a causal witness, as detailed in Sec.~\ref{subsec:causal_witness}. 
The quantity 
$-\tr[S^T \, W]$, for an appropriately constructed witness $S$, was measured, yielding a value of $0.202 \pm 0.029$. Since (according to Eq.~\eqref{eq:def_witness}) a negative value of $\tr[S^T \, W]$ is sufficient to certify causal nonseparability, this result effectively demonstrated that the experimental setup exhibited causal nonseparability (by about 7 st.\ dev.). Note that the definition of causal separability considered in \citep{rubino2017} did not allow for dynamical orders, see Subsec.~\ref{sectionmultiseparability}. As it turns out, however, the witness measured in that experiment is also a witness for causal nonseparability with the more general definition, when including dynamical orders.

The same interferometric architecture was later employed by~\citet{Zhu2023} to experimentally demonstrate the theoretical proposal in Ref. \citep{chen2021indefinite}, showing that charging quantum batteries under indefinite causal order can enhance both the stored energy and thermal efficiency. 

The following subsections review further experiments conducted on this optical platform. These experiments are categorized according to the specific task they were designed to demonstrate, ranging from certifying indefinite causality under minimal assumptions to exploring potential computational and thermodynamic advantages, and violating a semi-device-independent causal inequality.

\paragraph{Entangled quantum switch}
\label{nogotheorems}

Providing experimental evidence of indefinite causality via the measurement of a causal witness relies on several underlying assumptions. In particular, constructing a causal witness presupposes that the system and its operations are accurately described by quantum mechanics, and that the devices implementing the latter are carefully characterized. Accordingly, the results presented in \citep{rubino2017, goswami2018} are valid only insofar as these quantum assumptions hold. As discussed in Sec.~\ref{theoryindependenttests}, an alternative approach seeks to certify indefinite causality in a way that minimizes theory-dependent assumptions.

\begin{figure}
\centering
\includegraphics[width=.8\columnwidth]{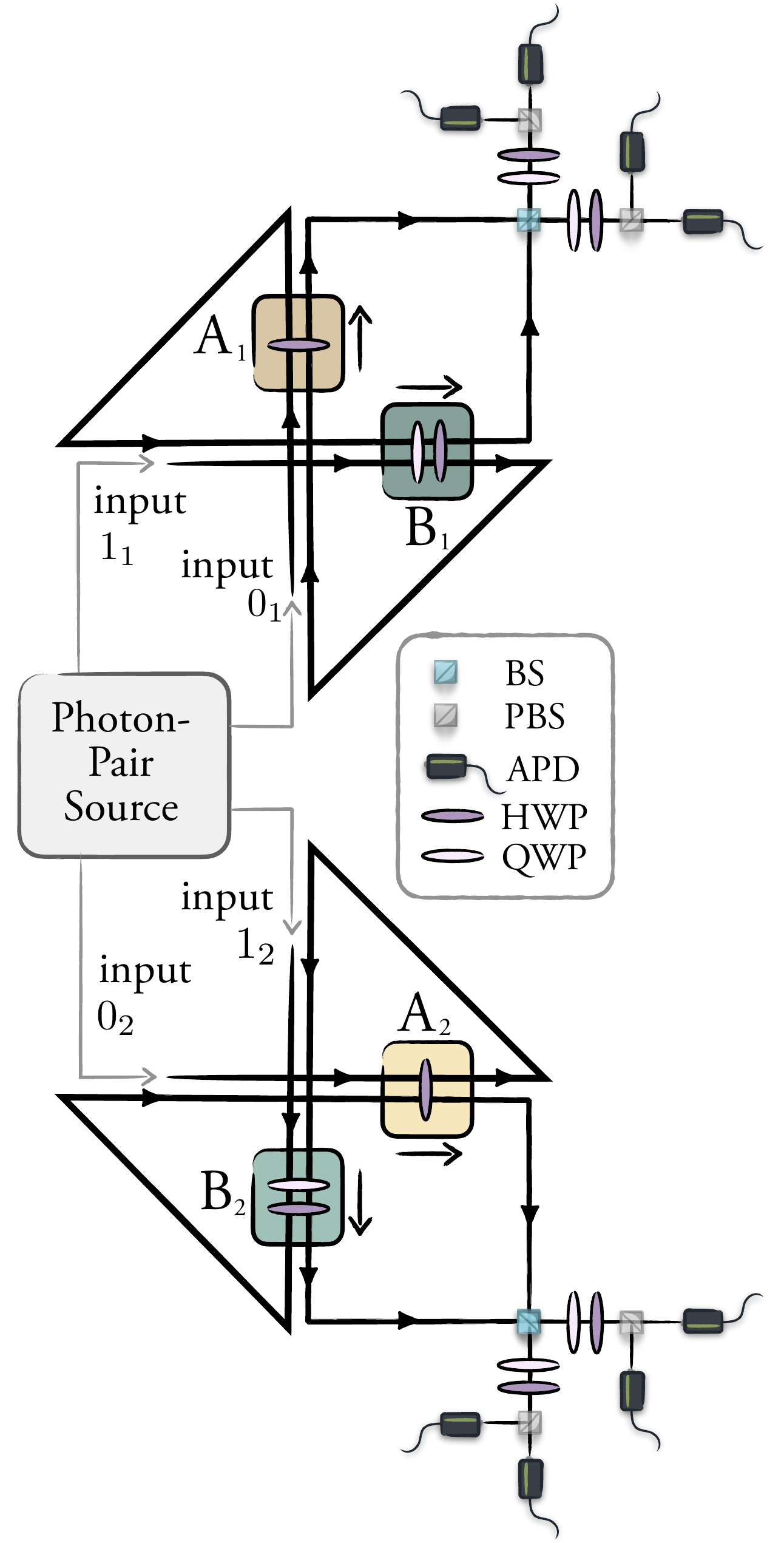}
\caption{Illustration of the entangled quantum switch configuration used in \citep{Rubino2022experimental}. Two quantum switches each implement an indefinite causal order between two operations ($A_i$ and $B_i$, $i=1,2$) applied to a target qubit (encoded in the polarization DOF of two separate photons, one in each switch). The order is controlled in each switch by a control qubit (encoded in the path DOF of the same two photons), with $\ket{0}_i$ corresponding to Alice-before-Bob and $\ket{1}_i$ to Bob-before-Alice. In the entangled configuration, the control qubits are prepared in an entangled state, such that the order of operations in the two switches is itself entangled. Measuring the control qubits in the $\bigl\{\ket{\pm}\bigr\}$ basis and performing measurements on the output target qubits can lead to a violation of a Bell inequality, even when the target qubits are initially separable and only local operations are performed within each switch.}
\label{img:rubino2}
\end{figure}

In this context, \citet{Rubino2022experimental} experimentally addressed whether one can devise a test whose validity extends beyond standard quantum mechanics. Building on the proposal of \citet{zych2019}, they adapted a Bell-like inequality for causal orders (originally conceived for gravitational masses in quantum superposition\footnote{\citet{zych2019} conceived the result to explore the implications of indefinite causal order within general relativity—introducing the so-called ``gravitational quantum switch.'' This topic, reviewed in Sec.~\ref{sec:Gravity}, was not part of the certification protocol implemented in \citet{Rubino2022experimental}.}) and implemented it in a quantum-optical experiment.

To test this inequality, the authors created an entangled quantum switch by entangling the causal order of two quantum switches, each with two pairs of agents and a target state passing through (shown in Fig.~\ref{img:rubino2}). More specifically, they used two copies of the setup shown in Fig.~\ref{img:procopio_rubino1}~a), injecting them with photons that were separable in the polarization DOF (which encoded the target system) and entangled in the path degree of freedom (which encoded the control). They then tested a Bell inequality on the output target system, obtaining a violation of 
$S_{\text{CHSH}} = 2.55 \pm 0.08$, where $S_{\text{CHSH}}$ is the CHSH parameter. 

This violation confirmed that the observed data could not be described by a class of generalized probabilistic theories (so-called ball-theories \citep{Dakic2010,Masanes2011}) under the assumptions that (a) the initial target states do not violate Bell inequalities, (b) the operations on the target states are local, and (c) these operations have a predefined order. Through additional tests, the authors verified that the initial target states were separable (hence, Bell local) and that the operations performed by Alice and Bob, when applied independently (i.e., outside the quantum switch), did not generate entanglement between the target systems. This left only one assumption unfulfilled: the requirement that Alice and Bob’s operations have a well-defined causal order, thereby proving the causal indefiniteness of the set-up without resorting to quantum mechanics, but considering the larger class of ball-theories.

\paragraph{Advantages in quantum computation}

Scaling up demonstrations of indefinite causality to quantum superpositions involving more than two gate orders significantly increases experimental complexity. For instance, extending the task performed in \citep{procopio2015}—whose generalization, known as the Fourier Promise Problem (FPP) \citep{Renner2021}, is reviewed in Sec.\ref{sec:QComp}—would require implementing a full quantum $N$-switch. However, in a quantum $N$-switch, the control system requires a Hilbert space of dimension $N!$, and the number of possible gate combinations grows rapidly, making the experiment increasingly complex and time-demanding.

To address this challenge, \citet{Taddei2020} proposed a simplified computational task known as the Hadamard promise problem, whose implementation does not require a full $N$-switch while still leveraging indefinite causal order. Their proposal is reviewed in Sec.~\ref{sec:QComp}, and the corresponding experiment is outlined below.

\begin{figure}
\centering
\includegraphics[width=\columnwidth]{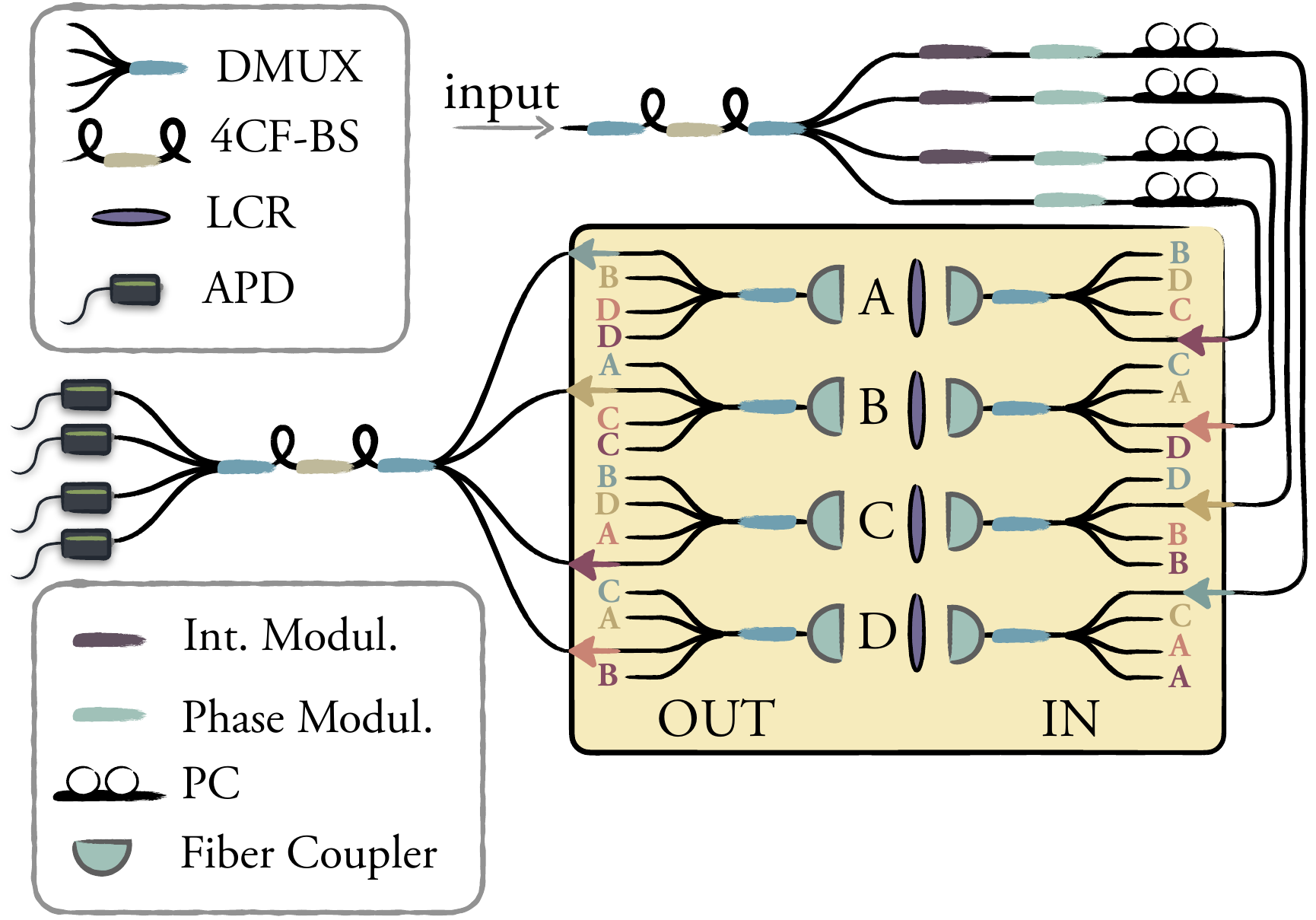}
\caption{Schematic of a 4-partite quantum switch with target system encoded in the polarization DOF \citep{Taddei2020}. In the figure, the fibers connecting the outputs in OUT with the inputs in IN is not depicted, but this connection is made clear via the letters and colours next to the input/output of each fiber. For example, a photon at the purple input undergoes the four unitary operations in the order $A \rightarrow D \rightarrow B \rightarrow C$, resulting in $U_C U_B U_D U_A$. Analogously, the other three orders are: $D \rightarrow C \rightarrow B \rightarrow A$ (blue), $C \rightarrow D \rightarrow A \rightarrow B$ (yellow), $B \rightarrow C \rightarrow A \rightarrow D$ (orange). DMUX: Multiplexer/Demultiplexer; 4CF-BS: Four-Core Fiber Beam Splitter; LCR: Liquid Crystal Retarder; APD: Avalanche Photo-Diode; Int./Phase Modul.: Intensity/Phase Modulator; PC: Polarization Controller.}
\label{img:taddei}
\end{figure}

In this implementation of the quantum 4-switch, the control system corresponds to the spatial mode of a single photon, while the target is its polarization. An attenuated pulsed laser injects light into a four-core fiber BS (4CF-BS), resulting in the Hadamard operation
\begin{equation}
    H_4 = \dfrac{1}{2} \left(\begin{smallmatrix}
1 & \phantom{-}1 & \phantom{-}1 & \phantom{-}1 \\
1 & \phantom{-}1 &        -1    &       -1     \\
1 &       -1     &        -1    & \phantom{-}1 \\
1 &       -1     & \phantom{-}1 &       -1
\end{smallmatrix} \right).
\end{equation}
The generated quantum state is coupled into a four-core fiber beam splitter (4CF-BS) via a single-mode to four-mode fiber multiplexer (DMUX), creating a coherent superposition of four spatial modes, and is then converted back to four single-mode fibers by a second DMUX. Each spatial mode is subsequently sent through a distinct permutation\footnote{Only four specific permutations are implemented, a subset of the full set of $4!$ possible permutations of four gates.} of the four unitary operations $U_i$, realized using controllable liquid crystal retarders (LCRs) acting on the photons’ polarization.
More specifically, each DMUX output is connected to a fiber launcher, where photons exit into free space, pass through the LCRs, and are then recoupled into another DMUX. After undergoing a sequence of all four unitaries (see Fig.~\ref{img:taddei}), a second Hadamard operation is applied to the control system using another set of DMUX and a 4CF-BS. Finally, the photons are detected.

Using this setup, the authors experimentally measured the probability of identifying the phase relations between four different permutations of four unitaries. Depending on the choice of gate sets a) $\{\mathbb{1},\sigma_x,\sigma_z\}$ or b) $\{\mathbb{1},\sigma_x, \sigma_z, (\sigma_x+\sigma_z)/\sqrt{2}\}$, they obtained an average success probability of a) $p_{\text{succ}} = 0.948 \pm 0.005$, and b) $p_{\text{succ}} = 0.959 \pm 0.008$.

However, this proof-of-principle experiment does not constitute a demonstration of indefinite causality, as no advantage over classical methods was shown \citep{Taddei2020}. In fact, if these probabilities had exceeded the classical limit, it would have provided evidence of a quantum advantage. Instead, a circuit with classical control of gate orders can deterministically identify the phase relations between the different permutations, outperforming the experimental results obtained with the quantum 4-switch.
A quantum advantage could have been observed by increasing the number of available gate choices and carefully selecting their prior probability distribution. This would have created scenarios where classical control of gate orders no longer guarantees unit success probability. However, the number of measurement settings required to probe such scenarios was prohibitively high for this experimental setup, and thus no such demonstration was performed.

\paragraph{Reducing noise in communication} 
\label{subsec:NoiseReduction}

\citet{Ebler2018}; \citet{salek2018quantum}; \citet{chiribella2018indefinite} have shown that quantum control over the order of noisy channels can enhance their capacity (see Sec.~\ref{noisereduction} for a detailed analysis of these studies). Experimental demonstrations of this effect have been reported in \citep{goswami2018communicating, guo2020experimental}. To achieve this, \citet{guo2020experimental} adopted a setup similar to the one discussed in Sec.~\ref{subsubsec:ExpPathAndPol}, while \citet{goswami2018communicating} employed the experimental approach detailed in Sec.~\ref{subsubsec:OAM}.
 
However, as discussed in Sec.~\ref{noisereduction}, this capacity enhancement is not unique to indefinite causality. Similar advantages have been observed in quantum superpositions of noisy channels in series (i.e., in a definite causal order) with quantum-controlled operations \citep{abbott2018, guerin2019_communication}, which were also experimentally demonstrated in \citep{Rubino2020}.

Given that these advantages can arise in scenarios devoid of indefinite causal order, we do not classify these experiments as providing evidence for indefinite causality.

\paragraph{Charging batteries and refrigerating}
\label{subsec:Refrig}

The interferometric architecture discussed in this section was employed by~\citet{Zhu2023} to experimentally demonstrate the theoretical proposal in Ref. \citep{chen2021indefinite}, showing that charging quantum batteries under indefinite causal order can enhance both the stored energy and thermal efficiency.

Furthermore, the effect introduced in \citep{Felce_2020} and reviewed in Sec.~\ref{subsubsec:thermo_tasks}, where a refrigeration cycle based on indefinite causality is introduced, has been experimentally explored in \citep{Cao2021, Nie_2020, Felce_2021}.

\begin{figure}
\centering
\includegraphics[width=0.8\columnwidth]{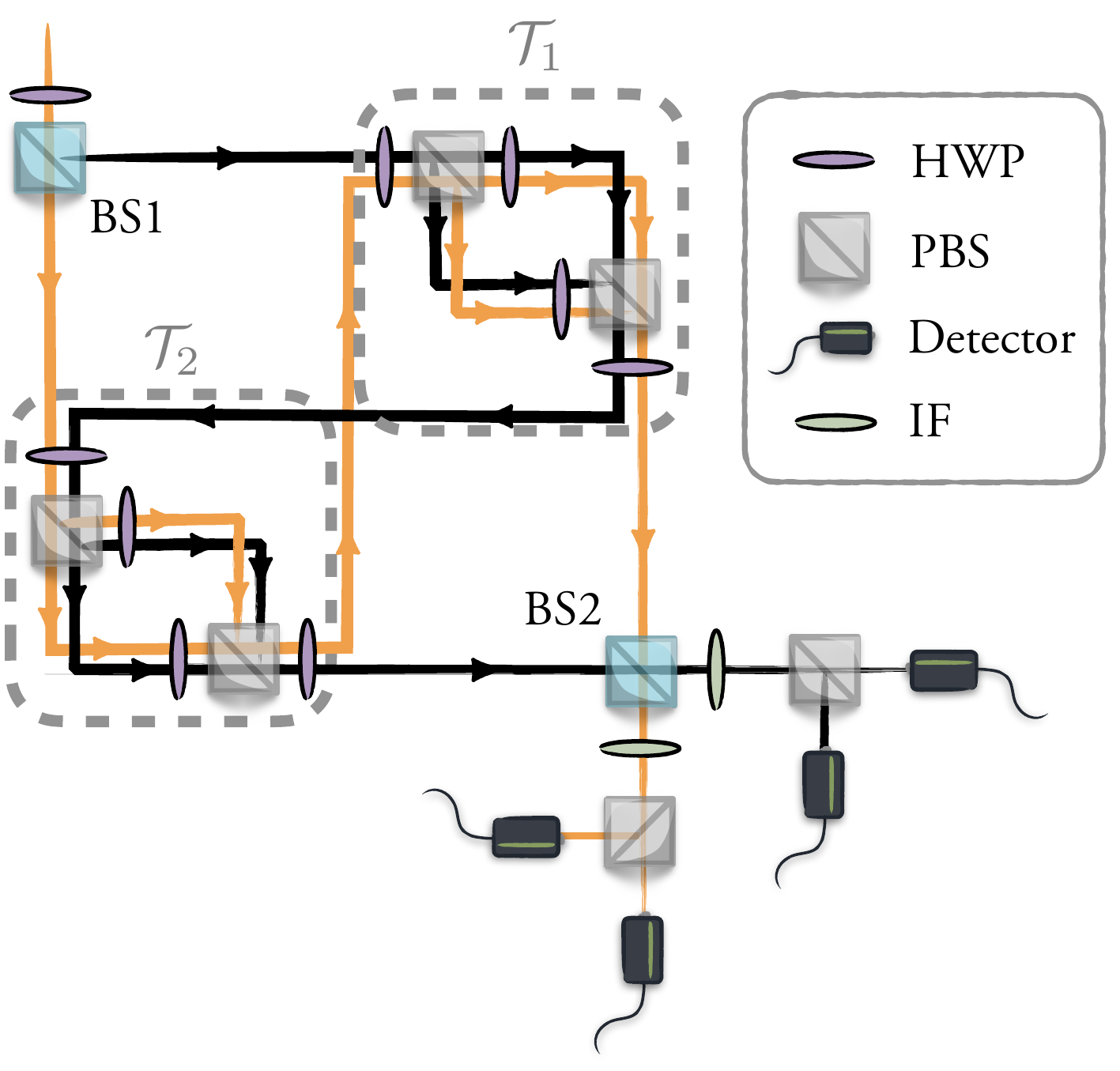}
\caption{Experimental setup from \citep{Cao2021} demonstrating a refrigeration cycle based on indefinite causality. A beam splitter (BS1) creates a superposition of two spatial modes, encoding the control qubit. In one mode, the polarization qubit undergoes the causal order $\mathcal{T}_2 \mathcal{T}_1$, while in the other, it experiences $\mathcal{T}_1 \mathcal{T}_2$. A second beam splitter (BS2) coherently recombines the spatial modes, projecting the control qubit onto the $\{\ket{\pm}\}$ basis. IF: Interference Filter.}
\label{img:cao}
\end{figure}

In \citep{Cao2021}, the interaction between a thermodynamic system and two identical thermal reservoirs was simulated using the quantum-optical setup shown in Fig.~\ref{img:cao}. In this experiment, photonic polarization encoded the energy levels of the target system, with horizontal ($H$) and vertical ($V)$ polarization states representing the ground and excited states, respectively. The system's temperature was simulated by preparing the photon in 
$H$ or $V$ with probabilities dictated by a Boltzmann-like distribution. The interaction with the thermal reservoirs was simulated by reconstructing a generalized amplitude damping channel via Kraus-operator post-processing of polarization measurement data, thereby capturing both energy absorption and dissipation processes. Energy measurements were then inferred from statistics in the Pauli-$Z$ basis.

An independent experimental investigation of the same effect was reported by \citet{Felce_2021} using IBM's cloud-based quantum computing platform. The experiment was carried out on the \texttt{ibmq\_5\_yorktown} processor, part of IBM's Canary family of superconducting quantum processors. This device consists of five superconducting qubits with high connectivity and relatively high gate fidelities. However, since \citet{Felce_2021} do not provide a detailed account of the internal operation of IBM’s Canary family of quantum processors, a more detailed analysis of their experimental implementation is beyond the scope of this review.

We will analyze the implementation in \citep{Nie_2020} in Sec.~\ref{subsec:NMR}. 

It should be noted that, as discussed in Sec.~\ref{subsubsec:thermo_tasks}, \citet{Capela_2022} pointed out that simple causally ordered processes can lead to enhancements similar to or beyond those observed in studies on indefinite causal order \citep{Cao2021, Nie_2020}. In particular, a scheme where arbitrary interactions between target and control systems are allowed can outperform the switch even when all operations occur in a definite causal order. For this reason, just as in the previous subsection, we do not classify these experiments as providing evidence for indefinite causality, as similar advantages could, in principle, be achieved without it.

\paragraph{Semi-device-independent certification and violation of a 
\texorpdfstring{$\mathcal{DRF}$}{DRF} inequality}
\label{subsec:SemiDI_exp}

\citet{Cao2022semiDI} reported an experimental realization of a photonic quantum switch (again, with path as control and polarization as target) designed to test a semi-device-independent certification of indefinite causality introduced in \citep{Bavaresco_2019} and reviewed in Sec.~\ref{semiDI}. 
The interferometric structure is that of an MZI: a first BS creates the path superposition, and a second BS recombines the paths, effectively projecting the control qubit onto the $\{\ket{\pm}\}$ basis. 

\begin{figure}
\centering
\includegraphics[width=\columnwidth]{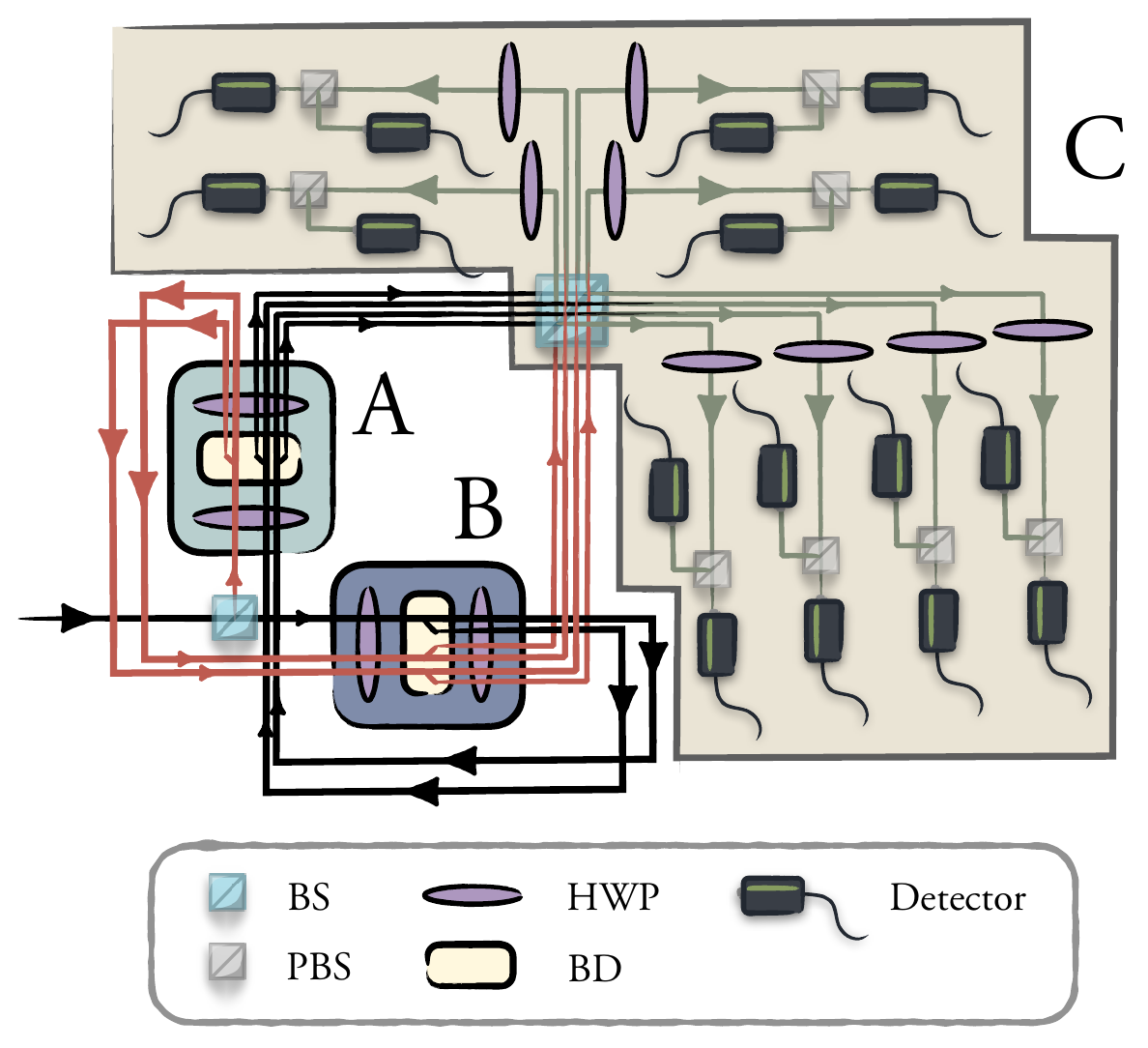}
\caption{Schematic of a quantum-optical switch implementing multi-outcome instruments for all three parties \citep{Cao2022semiDI}. A heralded single photon is prepared in a polarization state via a HWP and injected into a Mach–Zehnder interferometer, where the control qubit is encoded in the path DOF and the target in the polarization. The first BS creates a superposition of propagation paths, which are coherently recombined at a second BS to complete the quantum switch. $A$ and $B$ each implement a two-outcome measure-and-reprepare instrument by coupling polarization to additional spatial modes via HWPs and BDs, with orthogonal deflection directions to separate their outcomes. Finally, $C$ applies one of two four-outcome projective measurements on the joint control-target system.}
\label{img:cao2023}
\end{figure}

A key feature of this experiment is the integration of multiple-outcome instruments for all three parties $A, B$ and $C$. Previously, the only experimental implementation that generated local outcomes within the quantum switch was \citep{rubino2017} (reviewed above), where two interferometric loops were used to coherently recombine the outputs of a single party’s measurement, resulting in four joint outcomes. In the present experiment, by contrast, parties $A$ and $B$ each perform one of two measure-and-reprepare type of instruments with two outcomes, while $C$ applies one of two projective measurements with four outcomes on the joint control–target system. This leads to a total of 16 joint outcomes.

Incorporating multiple-outcome instruments gives rise to the same challenge encountered in \citep{rubino2017}: how to record the outcomes of Alice and Bob without disturbing the coherent superposition of causal orders. This is addressed using an analogous strategy, namely, coupling the polarization (target) degree of freedom to additional spatial modes. Specifically, in this experiment, each of $A$’s and $B$’s instruments is implemented using a sequence of a HWP, a beam displacer (BD), and a second HWP. Outcome information is encoded in the deflection direction of the BD (horizontal for $A$ and vertical for $B$). The resulting interferometric paths form a 2$\times$2 array, arranged so that which-path information is erased upon recombination, thereby preserving coherence across all branches (see Fig.~\ref{img:cao2023}).

With this setup, an average interference visibility exceeding $99.1\%$ was maintained for over 30 minutes. 
Based on the experimental data, the semi-device-independent causal inequality parameter was found to be $S_{\mathrm{exp}} = -0.0673 \pm 0.0003$ [see Eq.~\eqref{eq:semiDI}]. This result corresponds to a violation of the inequality by more than 224 st.~dev.

A variant of this setup was subsequently employed in \citep{Qu_2025} to obtain an experimental violation of the $\mathcal{DRF}$ inequality~\eqref{eq:DRF_Ineq} (see Sec.~\ref{sec:vanderlugt}).

\subsubsection{Path and polarization of single photons in a Sagnac interferometer} 

\begin{figure}
\centering
\includegraphics[width=\columnwidth]{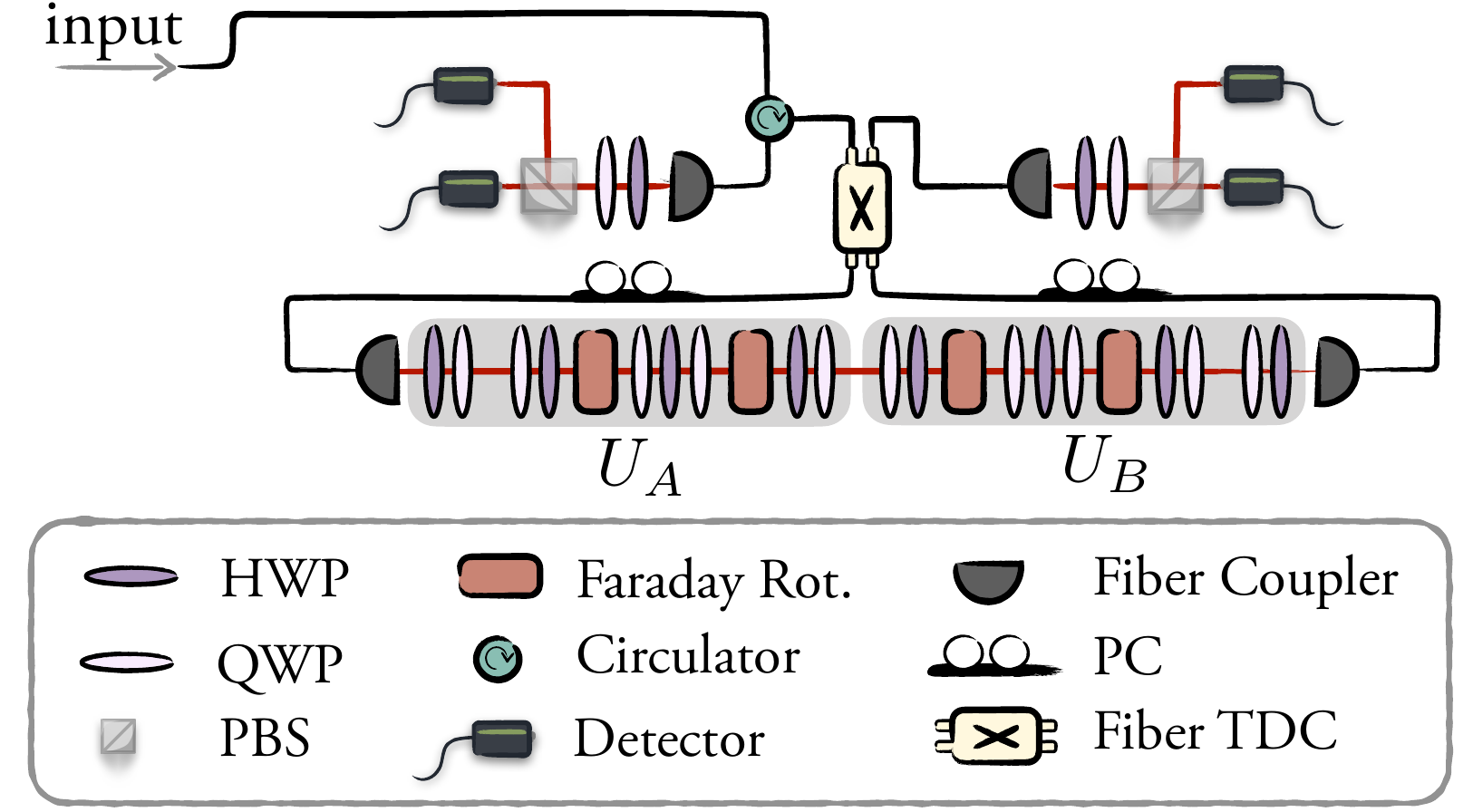}
\caption{Schematic of a quantum-optical switch in a Sagnac configuration \citep{stroemberg2022}. A tunable directional coupler (TDC), set to a balanced splitting ratio, sends photons into a free-space path in a superposition of two propagation directions. This path includes two reciprocal polarization gadgets, each composed of Faraday rotators, quarter-wave plates (QWPs), and half-wave plates (HWPs), which implement the operators $U_A$ and $U_B$. The photon’s exit port from the TDC depends on whether $U_A$ and $U_B$ commute or anticommute, and it is subsequently detected via a polarization-resolving measurement. Additionally, a fiber circulator routes photons exiting the Sagnac loop from the tunable directional coupler to the detection stage.}
\label{img:stroemberg}
\end{figure}

Most experiments on quantum switches carried out so far (with the exception of \citet{Felce_2021, Nie_2020}) have been conducted using single photons, with a common approach involving folded MZIs and polarization optics. One of the challenges in these setups is maintaining a constant phase in the interferometer, which is crucial for the accurate functioning of the quantum switch, especially when different unitary transformations are applied to the target qubit.

To mitigate phase instability, some earlier works explored the use of polarization as a control parameter in a common-path geometry \citep{goswami2018}. While this approach improved phase stability, it also introduced new challenges (see Sec.~\ref{subsubsec:OAM} for a detailed discussion).

A promising solution was proposed by \citet{stroemberg2022}, who implemented the quantum switch in a passively stable Sagnac interferometer. The Sagnac geometry inherently provides common-path phase stability, but it poses a challenge for gate implementation: standard polarization optics (such as the combination of two QWPs and one HWP used to realize arbitrary single-qubit unitaries) are non-reciprocal. Since the same gate is traversed in opposite directions within the Sagnac loop, using non-reciprocal elements would result in different operations depending on the direction of propagation, thereby preventing a truthful implementation of the switch.

To address this, \citet{stroemberg2022} developed a modified polarization gadget based on the Simon–Mukunda scheme \citep{Simon_1990}, incorporating magneto-optic Faraday rotators. Although Faraday rotators are intrinsically nonreciprocal (breaking Lorentz reciprocity due to the magneto-optic effect), they enable the construction of composite devices that are reciprocal in their overall transformation. Specifically, by arranging elements in a palindromic sequence, the resulting gadget ensures that the same single-qubit unitary is implemented regardless of the direction of propagation. This design enables fully reciprocal and universal polarization transformations, overcoming the limitations of conventional setups.

The use of this reciprocal gadget allowed the authors to preserve the polarization-encoded target qubit while ensuring spatial mode overlap, thereby achieving both high-fidelity target operations and phase stability. Their implementation—combining free-space and fiber optics—achieved interferometric visibility exceeding 0.9995 (see Fig.~\ref{img:stroemberg}).

To demonstrate the capabilities of their platform, the authors applied it to the commuting-vs-anticommuting quantum channel discrimination task \citep{Chiribella2012,Araujo_2015} (reviewed in Sec.~\ref{channel}), previously implemented by \citet{procopio2015}. In this task, the quantum switch enables perfect discrimination where causally ordered strategies cannot. Over a continuous five-hour measurement run, the experiment consistently surpassed the bounds for causally separable processes.

This work also builds on an earlier Sagnac-based implementation of the quantum switch by the same group, which focused on a different operational task, namely, universal time-reversal (`rewinding') of qubit dynamics \citep{Schiansky23}. Since that implementation only required a restricted set of polarization unitaries (linear combinations of $\sigma_y$ and $\sigma_z$), which can be realized with standard waveplates under counterpropagation, the extension to arbitrary single-qubit operations motivated the fully universal reciprocal-gadget approach developed in \citep{stroemberg2022}.

\subsubsection{Transverse spatial mode and polarization of single photons}

This subsubsection covers implementations where the target system is encoded in a photon's transverse spatial mode, with the control system encoded in its polarization. The experiments are grouped according to whether they employ discrete or continuous-variable encodings of the target system, and are further distinguished by the specific tasks they demonstrate.

\paragraph{Discrete variables}
\label{subsubsec:OAM}

\begin{figure}
\centering
\includegraphics[width=.75\columnwidth]{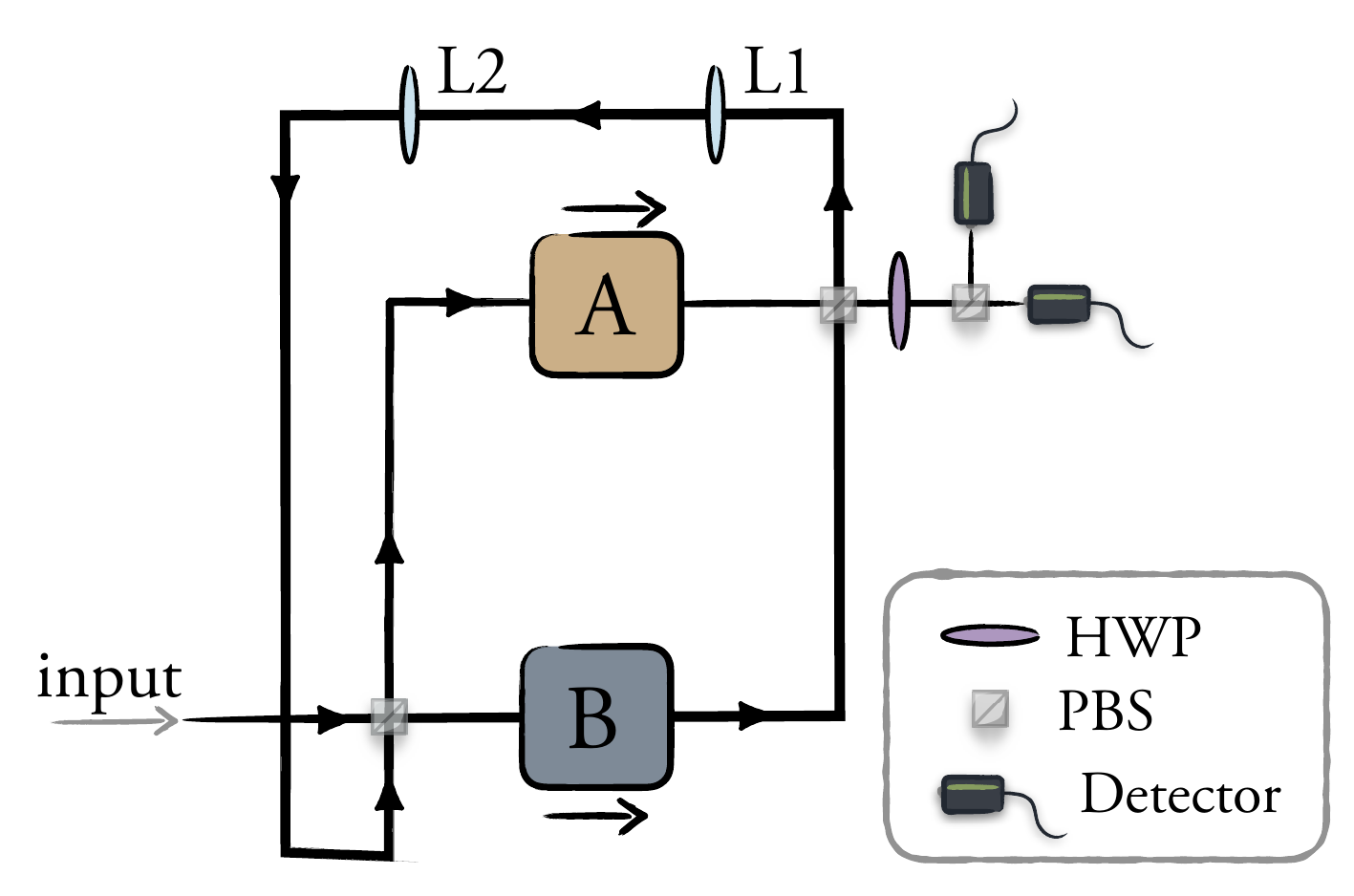}
\caption{Schematic of a two-party quantum switch with the target system encoded in the OAM degree of freedom \citep{goswami2018}. Photons from an attenuated laser source are injected into one of the two input ports of a common-path interferometer. Depending on their polarization, they are routed to operation $A$ or $B$, each implemented using special inverting prisms and pairs of cylindrical lenses. Two lenses, $L1$ and $L2$, act as a telescope for mode matching. After a second polarising beam splitter (PBS), the photons are sent towards the operation they have not yet undergone. Finally, the photons reach the final PBS, where they interfere and are then measured.
}
\label{img:goswami}
\end{figure}

In \citep{goswami2018}, an attenuated laser is injected into one of the two input ports of an MZI made up of PBSs. Depending on their polarization, the photons are directed towards one or the other unitary operation ($A$ and $B$ in Fig.~\ref{img:goswami}), performed through a set of special inverting prisms and cylindrical lenses acting on the photons' transverse spatial mode. Moreover, two lenses are used as a telescope for mode matching ($L1$ and $L2$ in Fig.~\ref{img:goswami}). The photons then reach a second polarising beam splitter that loops them back into the interferometer, directing them towards the operation to which they have not yet been subjected. Finally, the photons are interfered with each other and measured.

Using this setup, the authors measured the quantity $-\tr[S^T \, W]$ (Eq.~\eqref{eq:def_witness}), for an appropriate causal witness $S$ (recall Sec.~\ref{subsec:causal_witness}) obtaining $0.171 \pm 0.009$, which, being above zero by 18 st.~dev., provided a direct demonstration of the indefinite causality of the setup \citep{goswami2018}. It should be noted that, to improve phase stability, \citet{goswami2018} adopted a common-path geometry by encoding the control system in the polarization degree of freedom. This choice effectively mitigated phase instability, but it introduced practical limitations: implementing operations on the transverse spatial mode (now used as the target) became more cumbersome and suffered from reduced fidelity, as reflected in the experimental data (see Table~I in the Suppl.~Material of \citep{goswami2018}). The same setup was then also used to demonstrate channel noise reduction in quantum communication \citep{goswami2018communicating}, as already outlined in Sec.~\ref{subsec:NoiseReduction}.

The interferometric setup used in \citep{goswami2018, goswami2018communicating} bears some similarity to a theoretical scheme previously proposed in \citep{Andersson2005}, where an MZI was also used to probe properties of pairs of operations. However, the two setups are not equivalent. In particular, the setup in \citep{goswami2018} realizes a quantum switch, where the order of operations is coherently controlled, allowing for applications such as perfect channel discrimination, as proposed in \citep{Chiribella2012} and reviewed in Sec.~\ref{channel}, and as experimentally demonstrated in \citep{procopio2015}. In contrast, the scheme in \citep{Andersson2005} does not use a control degree of freedom in the same way as the quantum switch does, and the internal wiring is different. As a consequence, the setup in \citep{Andersson2005} would not enable (and, indeed, was not intended to allow for) perfect channel discrimination and can only witness either non-commutativity or non-anticommutativity, depending on the chosen setting, and only with bounded probability.

\paragraph{Continuous variables}

\begin{figure}
\centering
\includegraphics[width=\columnwidth]{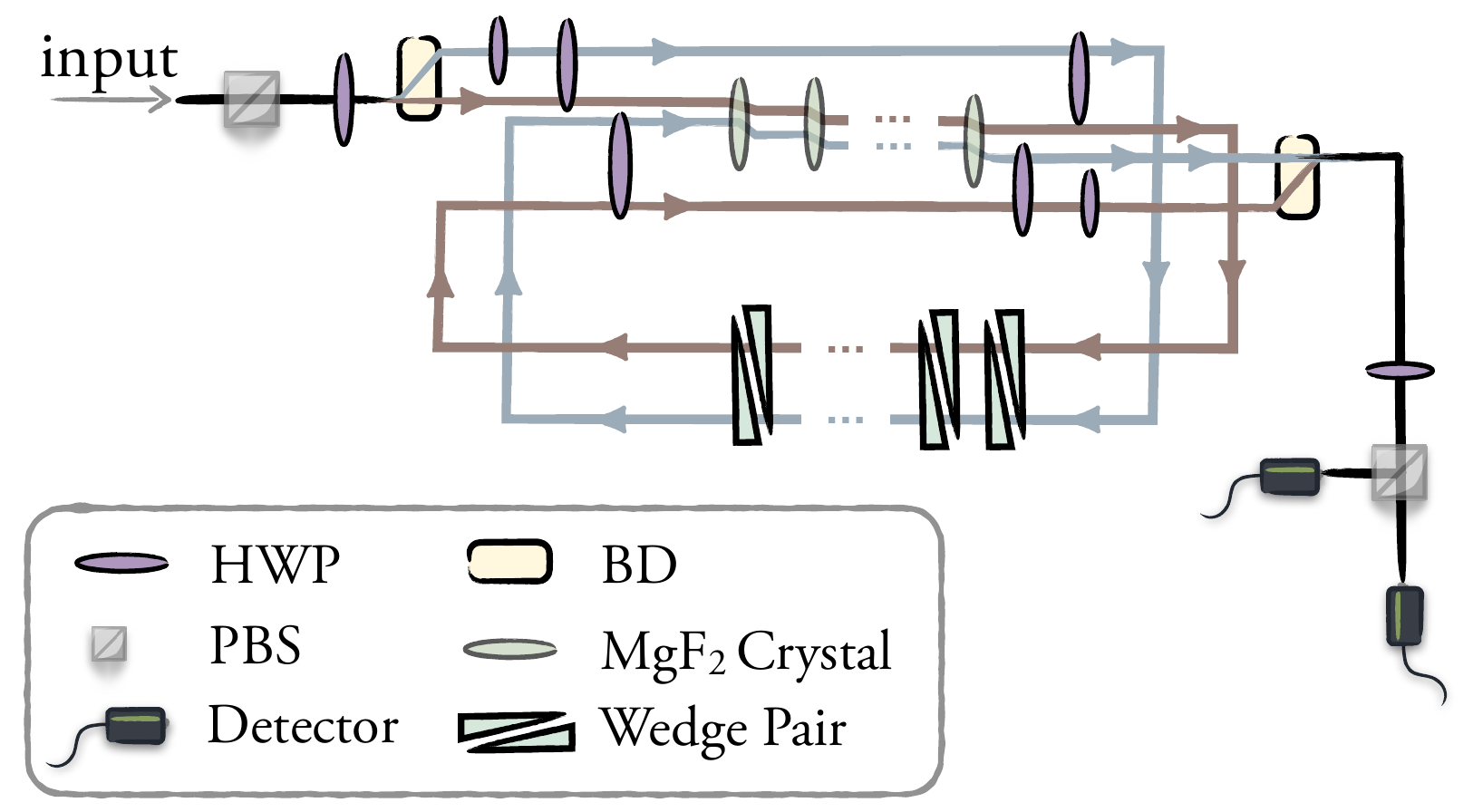}
\caption{Schematic of a two-party switch where the target qubit is encoded in the continuous variable DOF \citep{yin2023experimental}. The polarization state of single photons is first prepared in the $\ket{+}$ state through a combination of a PBS and a HWP. This polarization state is then converted into a coherent superposition between two distinct paths (blue and brown arms) within an MZI using a BD. Within the interferometer, the displacements in $X$ and $P$ are generated by $\text{MgF}_2$ crystal plates and by wedge pairs, respectively. Upon reaching the end of the interferometer, the two arms are merged back together using a BD, and they are then transformed back into a polarization qubit. This polarization qubit is projected onto $\ket{\pm}$ using a HWP and a PBS. Finally, the two output paths of the PBS are linked to two single-photon detectors.}
\label{img:yin}
\end{figure}

\citet{yin2023experimental} realized a photonic quantum switch where the target system is encoded on a continuous-variable quantum system. The experiment aimed to demonstrate a photonic implementation of a quantum metrology protocol beyond the Heisenberg limit by probing two sets of independent phase displacements in a superposition of their causal orders \citep{zhao2019} (outlined in Sec.~\ref{subsubsec:quantum_metro}). The setup (illustrated in Fig.~\ref{img:yin}) employed a single-photon quantum system, wherein the control system was encoded using the photon's polarization and the target system using the photon's transverse spatial modes. The experimental setup started with the preparation of single photons in the polarization state $\ket{+}$. An MZI, constructed using BDs and waveplates, transformed this polarization qubit into a coherent superposition of two spatial paths, defining a control qubit over the two causal orders.
This was employed to create a coherent superposition of the order of phase displacements in $X$ and $P$. Specifically, the $X$-displacement stage was implemented using a sequence of birefringent $\text{MgF}_2$ plates, while the $P$-displacement stage was realized using pairs of optical wedges; the two interferometer arms applied these two displacements in opposite orders.
These two orders defined two distinct phase-space trajectories, one involving $X$ displacements followed by $P$, and the other involving $P$ displacements followed by $X$. These trajectories formed a loop in phase space (which was assumed to be non-intersecting), and enclosed an area denoted as $\mathcal{A}$. After traversing the interferometer, the paths were recombined and converted back into a polarization qubit, which was measured in the $\bigl\{\ket{\pm}\bigr\}$ basis.

Each displacement in the protocol is implemented by a physical optical element and can, in principle, vary independently. Although the experiment used fixed, well-characterized displacements, the theoretical framework accommodates arbitrary displacement parameters---including non-canonically conjugate generators---allowing robustness to fluctuations even under imperfect or variable conditions.

Using their experimental setup, the authors evaluated the normalized area $A = \mathcal{A}/N^2$, where $N$ is defined as half the total number of displacement operators applied in the experiment, leading to $2N$ displacement operators in total ($N$ for position and $N$ for momentum displacements) as detailed in Subsec.~\ref{subsubsec:quantum_metro}. This evaluation was conducted by measuring the photon's polarization in the $\{\ket{\pm}\}$ basis, with the probabilities of outcomes expressed as  $P_{\pm} = [1\pm \mathrm{cos}(N^2 A+ \phi_0)]/2$.
They collected data for $P_-$ over different values of $N$ and performed statistical analyzes on the results. Their results showed that $\delta A$ decreased according to the super-Heisenberg scaling $\delta A \propto N^{-2}$, which is consistent with the predictions made in \citep{zhao2019}.

\subsubsection{Path and arrival time of single photons}
\label{Sec:Exp_QComCompl}

\citet{wei2019experimental} provided the first demonstration of a quantum switch involving a qudit as target system. In this work, the authors experimentally implemented an adaptation of the theoretical proposal of \citep{Guerin2016} called the exchange evaluation game as presented in Sec.~\ref{sec:QCommCompl}. In this adaptation, instead of using multiple target qubits (which requires the manipulation of $>10$ entangled qubits to enter the regime in which the causally indefinite advantage begins), the advantage is obtained by using only one target qudit, thereby reducing the experimental demand of the implementation. 

As illustrated in Sec.~\ref{sec:QCommCompl}, the theoretical protocol exploits the quantum control of the direction of communication between Alice and Bob to solve the exchange evaluation game with unit probability of success, using an amount of communication that scales exponentially better with the size $n$ of the input bit string than any causally ordered protocol. The details of the game are illustrated in Sect.~\ref{sec:QCommCompl} and Fig.~\ref{fig:allard_prl}, and are briefly recalled below. Alice and Bob receive, respectively, the inputs $({\bf x},f)$ and $({\bf y},g)$, where $x,y \in \{0,1\}^n$ are bit strings and $f,g: \{0,1\}^n \rightarrow \{0,1\}$ are Boolean functions such that $f({\bf 0})=g({\bf 0})=0$. Finally, Charlie, has to compute the exchange evaluation function $EE_n ({\bf x},f, {\bf y},g)= f({\bf y}) \oplus g({\bf x})$ subject to the one-way communication condition.

For the experimental realization of this protocol, a quantum switch was used in which the control system was a qubit in the path degree of freedom and the target system was a $d$-dimensional qudit with $d = 2^{n+1}$ (equivalent to $(n+1)$  qubits) encoded in the arrival time of the photon, and divided into a set of time bins of $2^{n+1}$ forming the basis ${\ket{\mathbf{z}}; \mathbf{z} \in \{0, \dots, 2^{n+1}-1 \}}$, with $n = 12$.

\begin{figure}
\centering
\includegraphics[width=0.85\columnwidth]{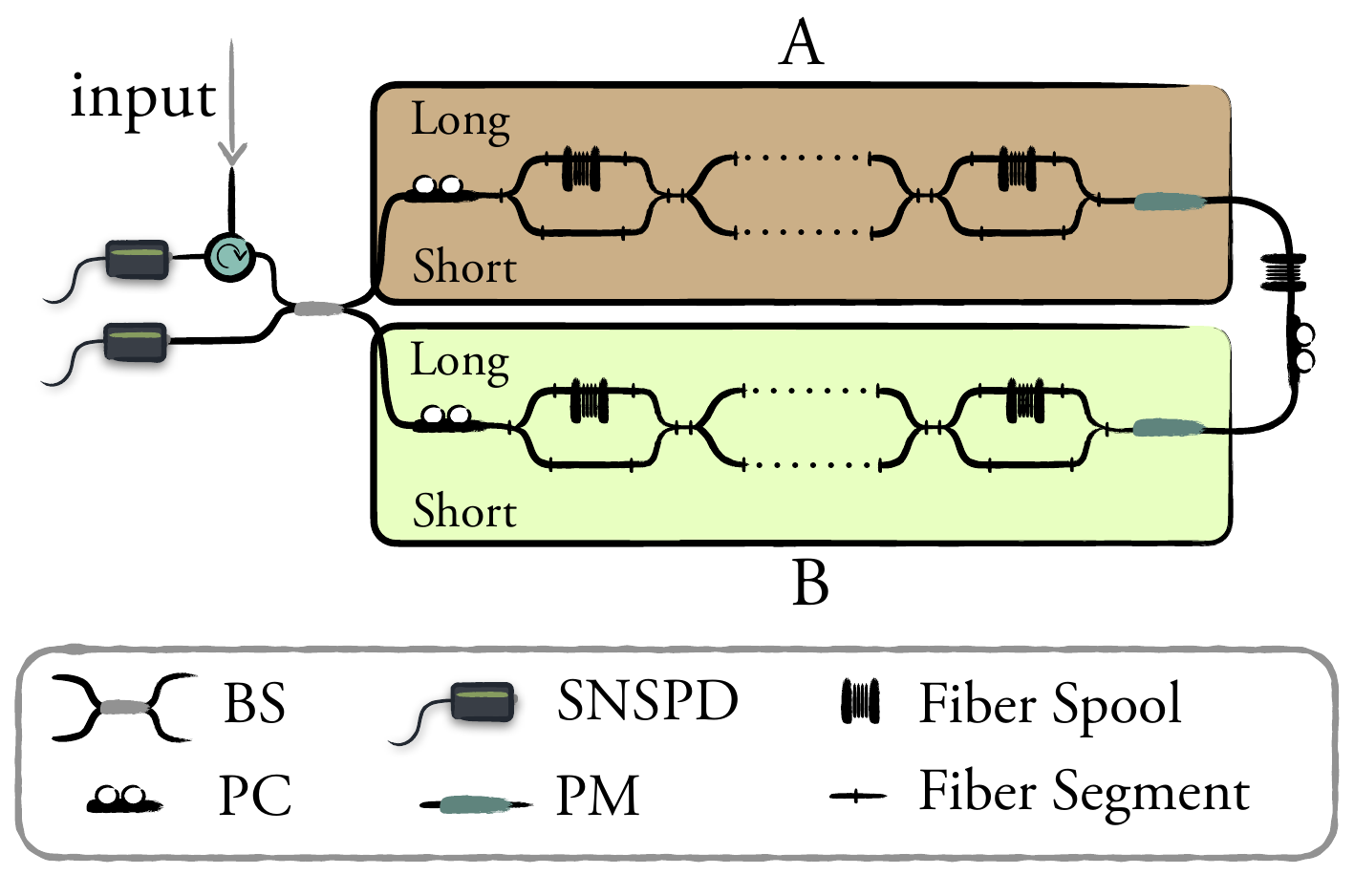}
\caption{Schematic of a two-party quantum switch with target system encoded as a time-bin qubit \citep{wei2019experimental}. Heralded single photons are sent to a circulator and then, after being initialized through a polarization controller (PC), enter the Sagnac loop through a beam splitter (BS). Within the loop, the operations at the Alice and Bob stations are realized using phase modulators (PMs) and programmable variable-delay lines. Each variable-delay line is composed of $n$ fiber segments, each offering a short or long option. The selection of short/long segments is set by the bits of $x$ (at Alice) and $y$ (at Bob), thereby realizing one of $2^n$ possible time-bin delays at each station. A 7-km fiber spool (FS) placed in the loop serves as a storage fiber for the pulse train between the two stations. The interference results are monitored by SNSPDs.}
\label{img:wei}
\end{figure}

The experiment was carried out with a fiber-optic Sagnac interferometer, illustrated in Fig.~\ref{img:wei}. A single photon, initialized in the state $\ket{0}$ (i.e., first-time interval), passed through a circulator and then entered the Sagnac loop in a two-way superposition via a BS. Alice and Bob each possessed a variable delay line and a time-dependent phase modulator to realize the unitary transformations $U_A = X(\mathbf{x}) D (f)$ and $U_B = X(\mathbf{y}) D (g)$, respectively.  The transformations $X(\mathbf{x})$ and $X(\mathbf{y})$ were implemented as delays of $\mathbf{x}$ ($\mathbf{y}$) time bins in Alice's (Bob's) lab.\footnote{As such, this implementation relied on a restricted family of target operations (rather than arbitrary single-qubit unitaries) tailored to the exchange-evaluation protocol of \citep{Guerin2016}.} $D(f)$ ($D(g)$) were executed in Alice's (Bob's) lab using a phase modulator that applied the $0$ phase if $f(\mathbf{z})=0$ ($g(\mathbf{z})=0$) or $\pi$ if $f(\mathbf{z})=1$ ($g(\mathbf{z})=1$) to each time bin.

Depending on the operations carried out by Alice and Bob, the photon exited from one or the other of the interferometer's outputs, and the outputs were monitored by two superconducting nanowire single photon detectors (SNSPDs). The experimental results obtained through this setup demonstrated an advantage in terms of communication complexity by requiring
the transmission of only $(34.8 \pm 0.3)\%$ of
the information of any classical protocol and $(69.6 \pm 0.6)\%$ of the information of any causally separable quantum protocol.

\subsubsection{Polarization and arrival time of single photons}

This subsubsection reviews implementations in which the target and control qubits are encoded in the polarization and arrival time of single photons. The experiments are organized according to whether the polarization encodes the target qubit or the control qubit.

\paragraph{Polarization as target}
\citet{Antesberger2024} reported the first experimental implementation of higher-order quantum process tomography, applied to a quantum switch realized in a fiber-based architecture. The control system was encoded in a time-bin qubit and the target in the polarization of the same photon. The routing through Alice's and Bob's operations in a superposition of orders was implemented using a sequence of ultrafast fiber optical switches (UFOSs), following the theoretical proposal of \citep{Rambo_2016}.

\begin{figure}
\centering
\includegraphics[width=\columnwidth]{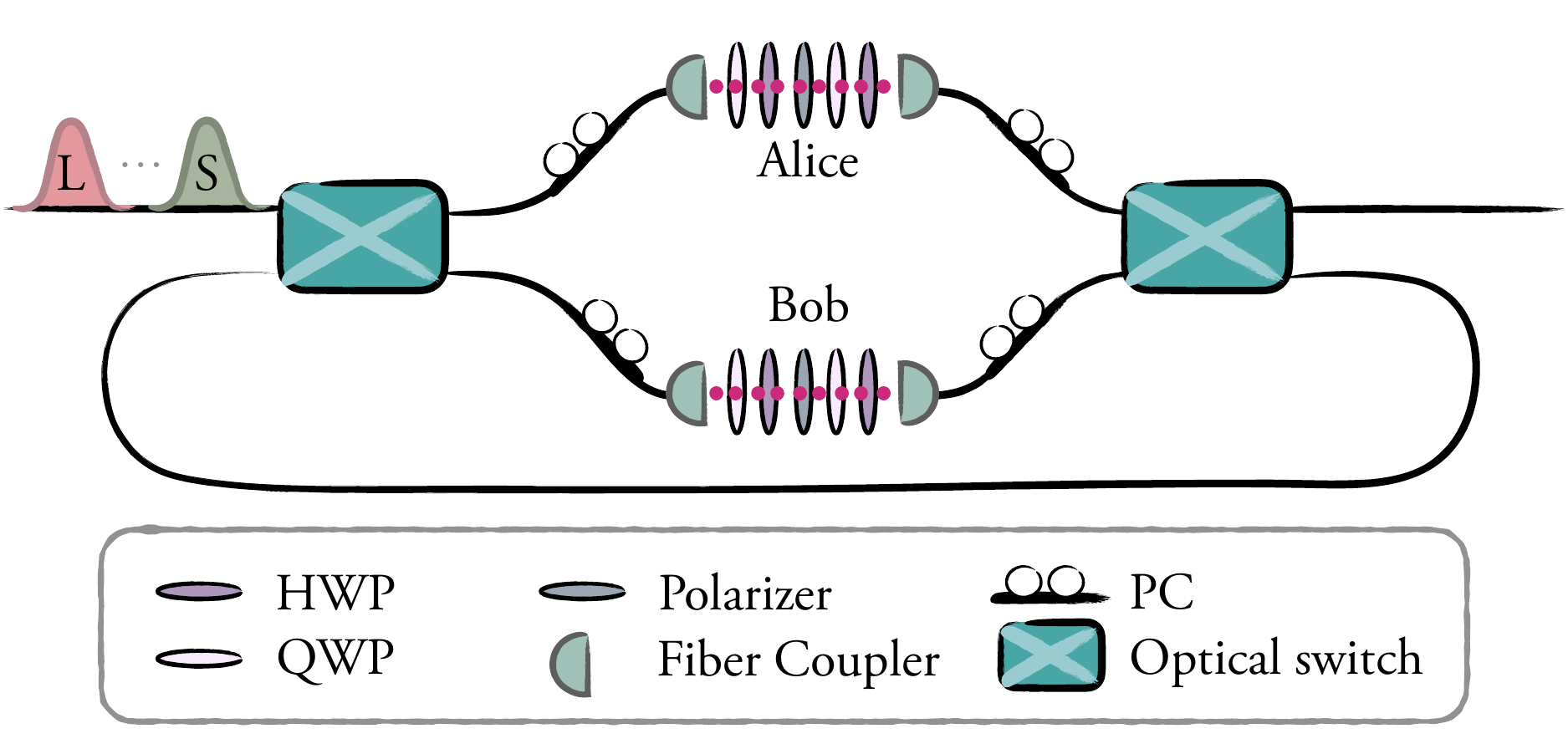}
\caption{Schematic of the fiber-based quantum switch from \citep{Antesberger2024}. The setup uses a sequence of UFOSs to coherently route two time-bin components of a single photon through Alice’s and Bob’s operations in different orders, depending on the control state. Alice's and Bob's operations are implemented using a sequence of QWP, HWP, polarizer, QWP, and HWP to reproduce the statistics of a measure-and-reprepare type of instrument. L: long time bin; S: short time bin; PC: polarization controllers.}
\label{img:antesberger}
\end{figure}

The time-bin control qubit was prepared using a Mach-Zehnder-like interferometer with unbalanced fiber arms and a UFOS for deterministic recombination (see Fig.~\ref{img:antesberger}). The resulting state, a coherent superposition of the short and long path modes, was then directed through the quantum switch. A voltage pulse sequence controlled the state of the fiber switches, such that the short and long time-bin components passed through the operations of Alice and Bob in opposite orders. Alice's and Bob's operations were realized in short free-space sections using standard polarization optics (namely, a sequence of waveplates and a polarizer for the measurement, followed by a sequence of waveplates for the re-preparation\footnote{It should be noted that this does not allow access to all possible outcomes of the experiment, as the orthogonal component selected out by the polarizer is always discarded. The experiment, therefore, had to be run twice for each setting to reconstruct the full statistics.}), acting on the polarization-encoded target qubit. After passing through the switch, the control qubit was measured by re-entering the same interferometric structure used for its preparation, providing passive phase stability.

While passive phase stability had been demonstrated in earlier work \citep{stroemberg2022}, the present scheme achieves it in a different way, namely, by reusing the preparation interferometer for measurement, thereby eliminating the need for active stabilization and enabling the acquisition of the large data sets required for higher-order tomography. The measurement scheme allowed for projections in two Pauli bases ($Z$ and $Y$), which, while not tomographically complete, sufficed to reconstruct the process matrix of the switch under the chosen input settings.

The tomography protocol developed in this work generalizes standard quantum process tomography to arbitrary higher-order processes, by preparing a tomographically complete set of input states and operations and measuring a tomographically complete set of observables on the output, within the available basis settings. The reconstructed process showed good agreement with the theoretical ideal of a quantum switch. As a further benchmark, the implementation was used to perform the anticommuting-commuting gate discrimination game \citep{Chiribella2012,Araujo_2015}, achieving a success probability of $0.974 \pm 0.018$.

The same setup was later used in \citep{Richter_2025} to demonstrate an experimental violation of the $\mathcal{DRF}$ inequality~\eqref{eq:DRF_Ineq} (reviewed in Sec.~\ref{sec:vanderlugt}).

\paragraph{Polarization as control} {\citet{Guo_2025} reported an experimental violation of the $\mathcal{DRF}$ inequality \citep{VanDerLugt_2022}, reviewed in Sec.~\ref{sec:vanderlugt}, using a quantum switch. The experiment involved four parties, $A$, $B$, $C$, and $D$, with $D$ space-like separated from the other three. Polarization-entangled photon pairs were generated, with one photon entering the quantum switch and measured by $C$, and the other sent directly to $D$ via a 3~km optical fiber to mimic space-like separation. In this implementation, the polarization DOF served as the control qubit and the time-bin DOF as the target qubit.

\begin{figure}
\centering
\includegraphics[width=\columnwidth]{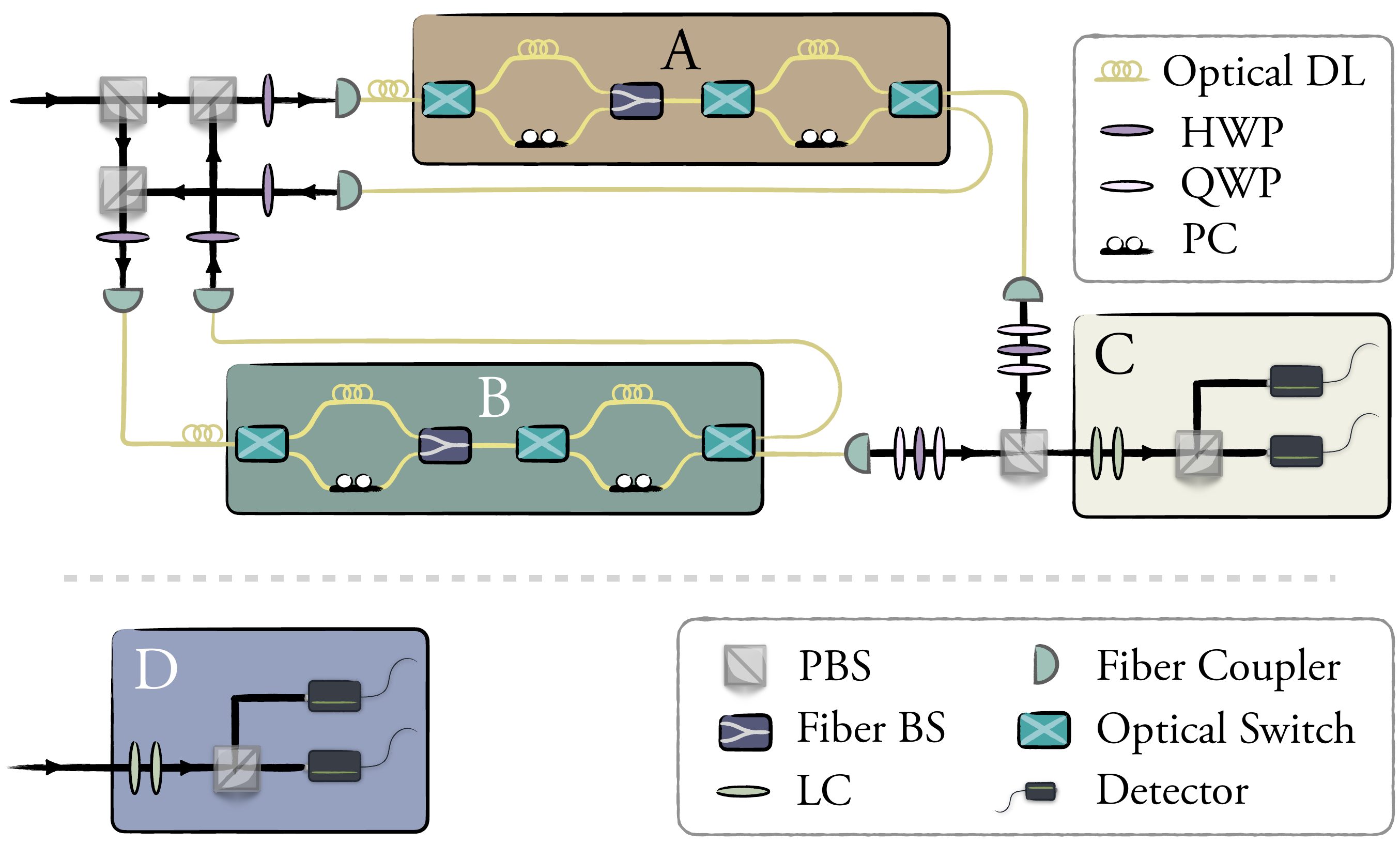}
\caption{Schematic of a two-party quantum switch with control system encoded as a time-bin qubit \citep{Guo_2025}. A pair of polarization-entangled single photons is produced. One photon from the pair is sent to the quantum switch, then measured by $C$; the other photon is sent directly to $D$. In the quantum switch, the photons' polarization and time-bin serve as the control and target qubits, respectively. At $A$'s and $B$'s stations, a measure-and-reprepare operation on the time-bin is implemented using optical switches, fibre-integrated beam splitters (BS) and polarization controllers (PC). An optical delay line (DL) and a liquid crystal retarder (LC) are used to adjust the path length and relative phases of the interferometer. The settings chosen by $A$ and $B$ are randomized using random number generators.}
\label{img:guo}
\end{figure}

Within the switch, $A$ and $B$ each performed a measure-and-reprepare operation on the time-bin qubit, implemented using optical switches, fibre-integrated BSs, and polarization controllers. Measurements were performed using asymmetric MZIs with a 200m path-length difference, matching the temporal separation of the time-bin states. To avoid revealing the order of operations (thus preserving the superposition of causal orders), the measurement outcomes of $A$ and $B$ were extracted following the approach of \citep{rubino2017}, delaying readout until after measurement of the control qubit. This required duplicating the measurement apparatus for each possible outcome, achieved here by introducing an auxiliary time-bin qubit to encode the result via additional 200~m fibre delays. Polarization preservation across the setup was ensured using sequences of wave plates and fibre controllers, while $C$ and $D$’s measurements on the polarization qubit were implemented using HWPs, PBSs, and SNSPDs.

In each run, the target qubit was initialized in the early time-bin state $\ket{e}$, and the control qubit in the Bell state $(\ket{HH} + \ket{VV})/\sqrt{2}$. The settings for $A$ and $B$'s measurements and repreparations, as well as $C$ and $D$'s measurements, were chosen independently using random number generators. The experiment achieved an observed value of $1.807\pm 0.010$ for the left-hand side of the $\mathcal{DRF}$ inequality~\eqref{eq:DRF_Ineq}, exceeding the bound of $7/4$ for all $\mathcal{DRF}$ correlations by 5.7 st.~dev.. No-signaling checks confirmed invariance of each party's outcome distributions within three standard deviations.

It should be noted, as discussed in \citep{VanDerLugt_2022}, that in photonic implementations such as the present one, the delayed readout of $A$ and $B$'s outcomes (necessary to preserve the causal superposition) means that these outcomes only acquire definite values in the intersection of the future lightcones of the points where $x$ and $y$ are chosen. From the perspective of classical relativity, this allows for mutual influence between $x$ and $a$ as well as between $y$ and $b$, and therefore a violation of the $\mathcal{DRF}$ inequality in this context does not unambiguously demonstrate an ``indefinite causal order'' in the intended device-independent sense. This limitation is common to other photonic realizations of such tests \citep{Richter_2025, Qu_2025}.}

\subsubsection{Integrated photonics}

\citet{Deng2025} used an integrated programmable photonic quantum chip that encodes both the control and target systems in distinct spatial modes of the waveguides. Depending on the state of the four-dimensional control system, the target system (also of dimension up to four) undergoes one of up to four possible unitaries, $\Pi_i$ ($i=1,\dots,4$). Each $\Pi_i$ is implemented via the universal encoding scheme of Ref.~\citep{Clements2016} and corresponds to a product of $M$ selected unitaries $U_1,\dots,U_M$ (with $M \le 3$ in the experiment), arranged in a different order for each $i$. As a result, the overall unitary transformation coincides with that of a multi-partite switch applied on the unitaries $U_j$. Based on this, various witnesses of causal non-separability are measured and a protocol for distillation of causal non-separability is implemented. However, since the experiment does not implement the individual unitaries $U_j$, but directly their products, it cannot be considered as a demonstration of indefinite causal order and cannot be used for any of the associated advantages. Despite this, the experiment demonstrates a high degree of quantum control, highlighting integrated photonics as a promising avenue for future experimental investigations of causal structures.

\subsection{Nuclear Magnetic Resonance}
\label{subsec:NMR}

\begin{figure}
\centering
\includegraphics[width=0.85\columnwidth]{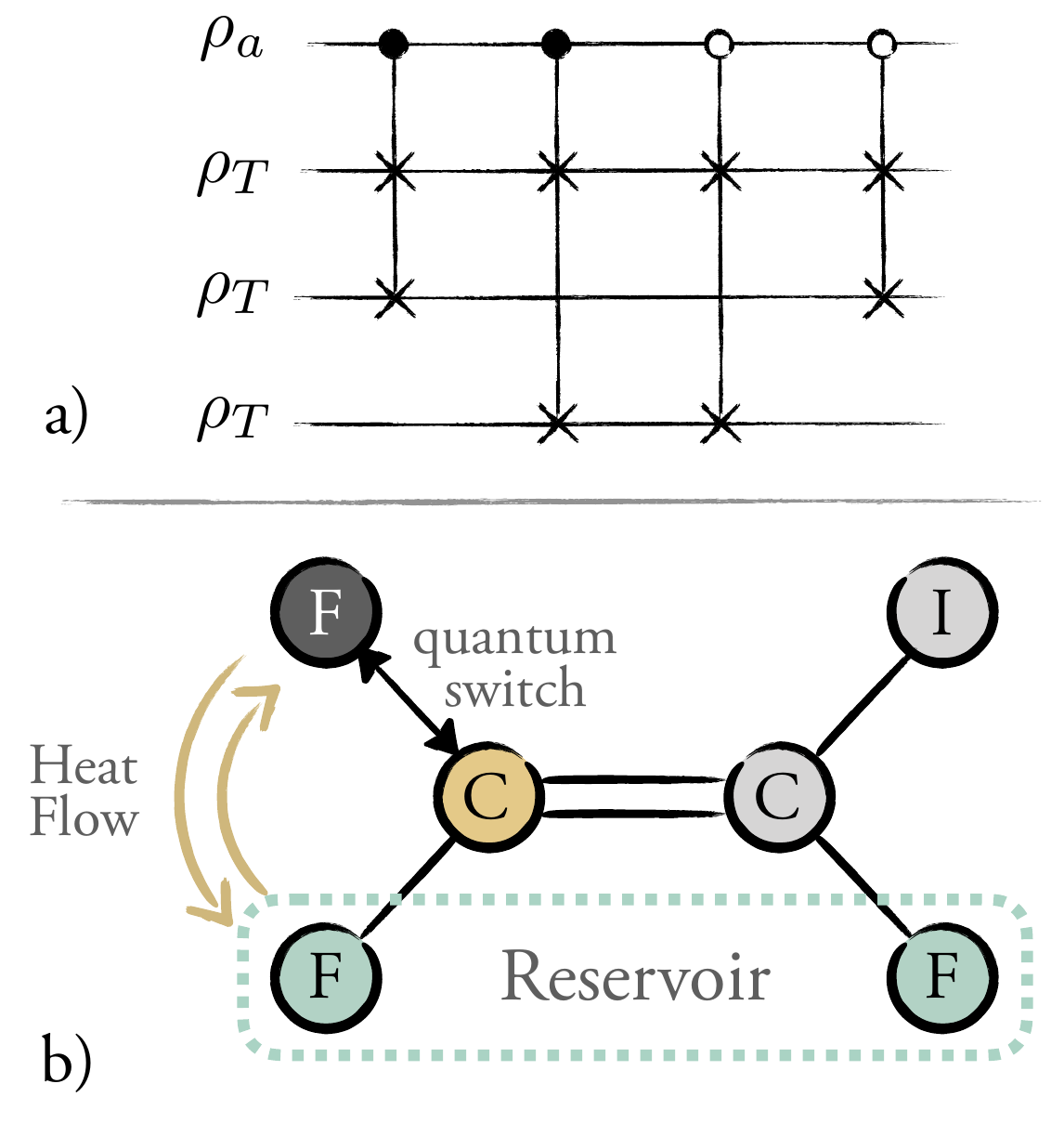}
\caption{a) Quantum circuit reproducing a quantum switch of thermalizing channels. The first qubit (encoded in the $\prescript{13}{}{\text{C}}$ nucleus) acted as the control system, the second qubit ($\prescript{19}{}{\text{F}}$) as the target system, and the last two qubits (also $\prescript{19}{}{\text{F}}$ nuclei) were used to simulate the effect of reservoirs. Each control SWAP gate was realized through the decomposition into three Toffoli gates. The second and third controlled-SWAP operations could, in principle, be replaced by a simple SWAP operation. However, we have chosen to retain them in this form to remain consistent with the schematic reported in \citep{Nie_2020}. (Note that this circuit also corresponds to the quantum-control representation of the switch shown in Fig.~\ref{fig:Qcirc_impl_QCont}.)
b) Implementation of the quantum switch on nuclear spins. In the $\text{C}_2\text{F}_3\text{I}$, the $\prescript{13}{}{\text{C}}$ nucleus was used as the control system, the upper $\prescript{19}{}{\text{F}}$ nucleus as the working substance, and two remaining $\prescript{19}{}{\text{F}}$ nuclei as reservoirs. The direction of heat flow was shown to depend on the measurement result of the control system.}
\label{img:Nie}
\end{figure}

The effect presented in \citep{Felce_2020} and reviewed in Sec.~\ref{subsubsec:thermo_tasks}, where a refrigeration cycle based on indefinite causality is introduced, has been experimentally investigated in \citep{Nie_2020, Cao2021, Felce_2021}. The implementations of~\citet{Cao2021, Felce_2021} were reviewed in Sec.~\ref{subsec:Refrig}, while the work of~\citet{Nie_2020} is reviewed below.

\citet{Nie_2020} used an ensemble of NMR spins as the working substance and showed that it can be heated or cooled despite being in contact with reservoirs at the same temperature. To achieve this, they implemented a quantum-controlled swap of the state of the reservoirs and the target system in different orders based on the state of the control system (Fig.~\ref{img:Nie}), using four nuclei in $\prescript{13}{}{\text{C}}$-iodotrifluoroethylene ($\text{C}_2\text{F}_3\text{I}$) dissolved in acetone-d6. The four nuclei represented the four qubits, with the $\prescript{13}{}{\text{C}}$ nucleus being the control system, the upper $\prescript{19}{}{\text{F}}$ nucleus being the target system, and the two remaining $\prescript{19}{}{\text{F}}$ nuclei being the two reservoirs (Fig.~\ref{img:Nie}). The control SWAP gates were implemented by decomposition into three Toffoli gates, as illustrated in \citep{Nie_2020}.

While this experiment demonstrates a thermal process conditioned on a quantum control system, the realization is tailored to a specific choice of operations (namely, thermalization channels) rather than a fully general implementation of a quantum switch. In particular, the specific design allows the process to behave like a quantum switch for these channels, but it does not reproduce the output of a quantum switch for arbitrary operations. This contrasts with interferometric realizations, where the ordering of arbitrary processes can be coherently superposed. Instead, the NMR implementation can be seen as a circuit-based simulation of a quantum switch, with four individually controlled operations applied in a sequence conditioned on the state of the control system (as highlighted in Fig.~\ref{img:Nie}a)).

To demonstrate the claimed advantage of placing the thermalization processes in an indefinite causal order, the authors compared this scenario with what they considered to be its classical counterpart, namely, setting the control system in the $\ket{0}$ state and thus observing only one order of thermalization. Under these conditions, they showed that a non-zero heat flow only occurs in the presence of an indefinite causal order, and that the direction of heat flow is determined by the measurement outcome of the control system. In particular, when the control system was measured in the $\ket{+}$ state, the reservoir transferred heat to the working system, and the opposite was observed when the control system was measured in the $\ket{-}$ state. This result is consistent with the theoretical predictions, where the direction of heat flow is determined by the projective measurement of the control system.

We note that similar thermodynamic advantages can also be achieved using causally ordered processes, provided that one allows for access to system–environment interactions, as discussed in Sec.~\ref{subsubsec:thermo_tasks} and shown by~\citet{Capela_2022}.

\section{Indefinite causality in relativity and quantum gravity}
\label{sec:Gravity}

The interplay between quantum theory and general relativity is a natural setting for the investigation of indefinite causality and provides one of the driving motivations for the field. In most approaches to quantum gravity, where spacetime geometry is treated quantum mechanically, some form of non-classical or fluctuating causal structure is implied, although a direct analysis of such features is not the primary focus of traditional quantum-gravity research. The suggestion that quantum gravity could make light cones non-classical was already discussed in the fifties \citep{Rickles2018}, while the idea that events lose a definite space-like or time-like relation has been emphasized, for example, in \citep{Hartle1993}. Several lines of research have directly addressed uncertainties in causal structure, such as fluctuations of light cones \citep{Ford1995, Yu1999, Jia2022, Cintia2026, Fujita2026}. However, these investigations typically do not treat causal relations operationally by modeling interventions, nor do they distinguish between statistical mixtures and genuinely indefinite causal structure as discussed in this review. Hence, an organic exploration of indefinite causal structure within quantum gravity remains a relatively underexplored topic.

In this section, we focus on recent works that treat causal structure operationally---in the spirit discussed throughout this review---within a context where classical or quantum relativistic spacetime plays a prominent role.

\subsection{Quantum control of spacetime metric}
\label{sec:Control_spacetime}

\subsubsection{Spacetime with indefinite order of time-like events}
\label{gravitationalswitch}

\begin{figure}
\centering
\includegraphics[width=.85\columnwidth]{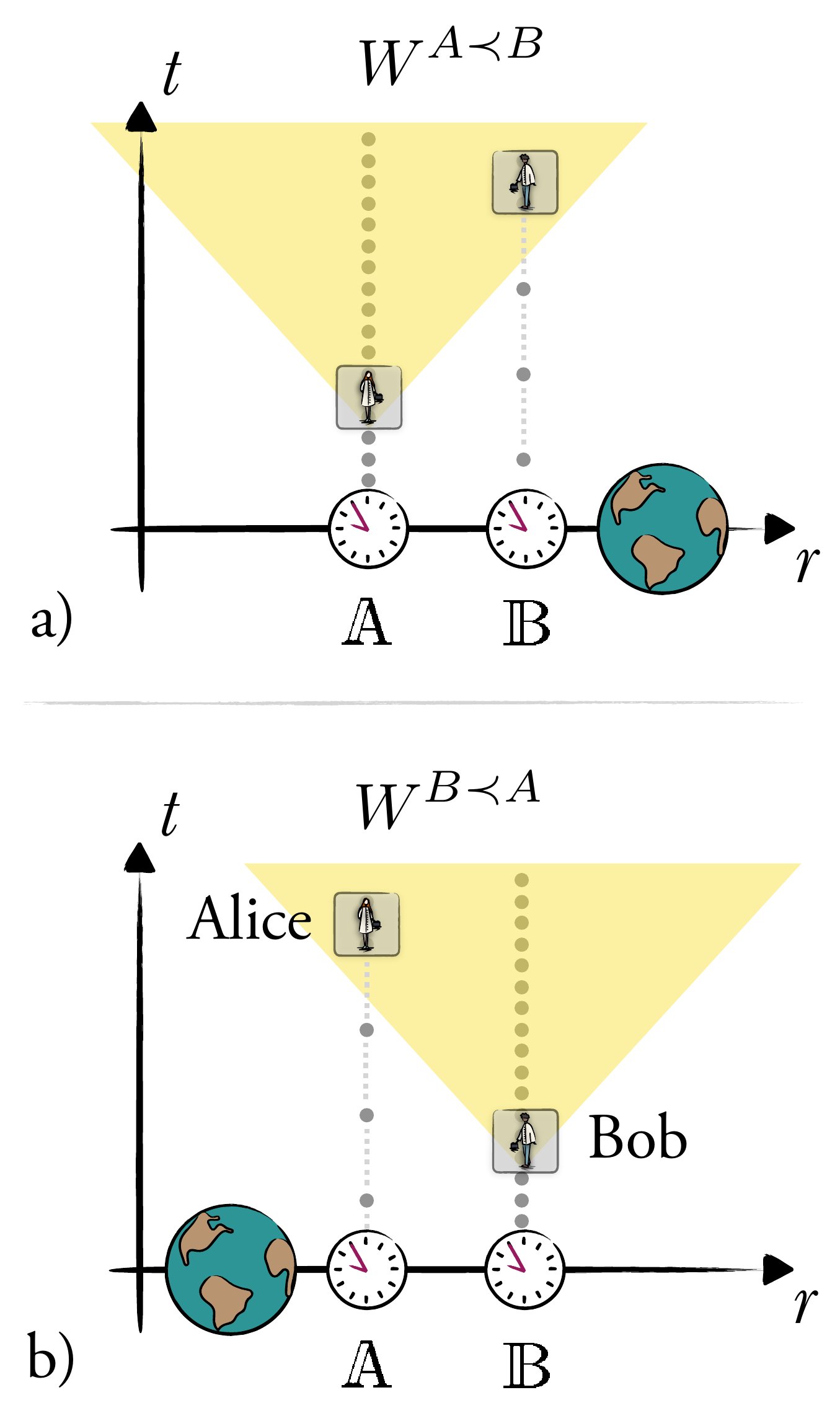}
\caption{Quantum control of causal order due to gravitational time dilation. In the scheme introduced in \citet{zych2019}, two quantum operations take place at fixed proper times along two fixed world lines, labeled $\mathbb{A}$, $\mathbb{B}$, but their causal order depends on the position $r$ of an external mass: $A\prec B$ if the mass is closer to $\mathbb{B}$, panel \textbf{a)}, or $B\prec A$ if it is closer to $\mathbb{A}$, panel \textbf{b)}. Quantum control of the mass naturally leads to quantum control of causal order. The pictures employ a common background time coordinate, defined independently of the mass configuration, which can be interpreted as the time of a distant observer.}
\label{zych}
\end{figure}

The first explicit connection between indefinite causal order and non-classical spacetime was introduced by \citep{zych2019}, in the form of a realization of the quantum switch using gravitational time dilation and quantum superposition of masses. The general idea, illustrated in Fig.~\ref{zych}, is the following: consider two spatially separated laboratories\footnote{Here, ``laboratory'' refers to a spatial location (at all times)---namely, a worldline---not (as sometimes used in the literature) to a site.} $\mathbb{A}$, $\mathbb{B}$ where, at a given time $\tau^*$ according to local clocks, operations are performed on some local degrees of freedom. In the absence of any external mass, the protocol takes place in Minkowski spacetime and the two operations are space-like separated, so that the time $\tau^*$ can be identified with a global Minkowski time coordinate. If we now introduce a massive system and bring it closer to laboratory $\mathbb{A}$, time dilation will make the local clock tick slower than the clock at $\mathbb{B}$. If sufficient time dilation is accumulated, the operation at $\mathbb{A}$ (which is still taking place at the local time $\tau^*$) will be in the future light cone of the operation at $\mathbb{B}$ (which is also always at local time $\tau^*$). On the other hand, bringing the mass closer to $\mathbb{B}$ can result in the operation at $\mathbb{B}$ happening in the future of the operation at $\mathbb{A}$. Given the ability to prepare arbitrary quantum states of the mass, it is in principle possible to produce a quantum control of causal order, akin to the quantum switch.

To ensure a scenario where the indefinite causal order is only due to the spacetime metric, \citet{zych2019} implement the above idea through a scheme where the relative causal order is only due to the time dilation induced by the mass and not (as in non-relativistic implementations) correlations between the control and the trajectory of the particle encoding the target. In such a scheme, the target is prepared in a fixed state $\ket{\psi}$ and sent along a fixed trajectory (independent of the mass configuration), while the operations $A$ (in laboratory $\mathbb{A}$) and $B$ (in laboratory $\mathbb{B}$) always take place in fixed locations relative to the respective local reference frames. 
Furthermore, spherical mass configurations can be arranged in such a way that they produce no force on the local laboratories---so that the state of the control cannot be revealed by local measurements.

As opposed to the instances of indefinite causal order discussed earlier, the above scenario cannot be analyzed with standard tools using a single, classical spacetime. However, the complete protocol requires recombining the mass configurations in order to perform an interference measurement, after which we can again use a single coordinate chart defined in a single spacetime.%
\footnote{Alternatively, as discussed in \citep{zych2019}, in this scenario one can define \textit{metric-independent coordinates} (e.g., interpreted as those used by a far-away observer, not affected by the mass), and use equal-time slices defined by such coordinates to define quantum states, as in Fig.~\ref{zych}. This description breaks general covariance but remains valid as long as the spacetime metric is sufficiently regular for each superposition ``branch''. Another choice of coordinates (i.e., causal reference frame) is possible according to which one of the two events in the gravitational switch is localized, while the other is delocalized and has a time coordinate that is entangled with the position of the mass \citep{Guerin2018a,ApadulaGrinbaumBrukner2026}. Finally, it is possible to choose coordinates in which both Alice's and Bob's events are localized, yet the causal order between them remains indefinite \citep{delahamette2022quantum,Kabel24}.}
Accordingly, assuming that control and target systems do not undergo any evolution apart from the unitaries $U_A$, $U_B$ that describe the local operations, the final state of the system at some time after the mass configurations are recombined (but before the measurement) is
\begin{multline} \label{graviswitchfinalstate}
  \ket{\psi}_{\textrm{out}}  = \alpha\ket{0} U_BU_A\ket{\psi} \ket{\tau^A_0}  \ket{\tau^B_0} \\
  +\beta \ket{1} U_AU_B\ket{\psi} \ket{\tau^A_1}  \ket{\tau^B_1},
\end{multline}
where the initial control state was $\alpha\ket{0} +\beta \ket{1}$ (with $\ket{0}$, $\ket{1}$ denoting the two mass configuration states) and $\ket{\tau^{A,B}_{0,1}}$ are the states to which the two local clocks have evolved by the end of the protocol, for each of the two respective configurations. \citet{zych2019} further show that, by appropriately rearranging the mass configuration, any differential time dilation between the two labs can be undone, so that $\ket{\tau^A_0}  \ket{\tau^B_0} = \ket{\tau^A_1}  \ket{\tau^B_1}$. In this way, the clock degrees of freedom factor out, and we are left with an output state formally equivalent to that produced by the quantum switch.

The gravitational implementation of the switch can be used, in principle, to realize a demonstration of indefinite causal order, for example by using the causal witnesses discussed in Sec.~\ref{subsec:causal_witness}. However, in order to obtain a stronger, theory-independent demonstration, \citet{zych2019} propose a modified protocol where four parties act in two ``entangled'' switches, namely a process where the order between two sites is quantum-correlated with the order of two other sites. The protocol produces entanglement between two target systems and, upon subsequent measurement, enables the violation of the ``Bell inequality for temporal order'' reviewed in Sec.~\ref{theoryindependenttests}, which would be impossible if the sites were in a definite order. 
\citet{Rubino2022experimental} realized the entangled switches in an optical experiment and verified violation of the  Bell inequality for the temporal order for certain class of theories, suitably adapted for the setup. As discussed in Sec.~\ref{theoryindependenttests}, the full theory independence of the protocol was questioned in \citep{debski_indefinite_2022}. The issue may be overcome by a gravitational implementation of the theory- and device-independent protocol introduced by~\citet{VanDerLugt_2022} and reviewed in Sec.~\ref{sec:vanderlugt} above.

Finally, \citet{Moller2024gravitational} devised a different implementation of the switch in quantum-controlled spacetime where, using the radius of a mass shell as a control system, spacetime remains fully classical outside a spherical region. One party (Bob) always remains in the classical region, while the other (Alice) travels through the superposed shells in free fall. Alice's trajectory begins and ends along a classical worldline in the classical region outside the shell, in such a way that the total elapsed proper time does not depend on the mass configuration (hence removing the requirement to swap the masses to erase which-order information encoded in the local clocks). Alice also passes through the classical region mid-journey, where she meets the target system either before or after Bob depending on the shell's radius, hence realizing a quantum control of causal order.

\subsubsection{Quantum control of space-like and time-like separation}
\label{spacetimelike}

Following \citet{zych2019}, \citet{Feix2017} use the quantum control of spacetime metric to quantum-control the situations of ``direct cause'' and ``common cause'', verifiable through witnesses that can be computed efficiently. More precisely, the position of the mass in a spatial superposition and the time dilation induced are set to control whether event $B$ is inside---in which case, there is a channel from $A$ to $B$---or outside the future light cone of event $A$---in which case, there is no channel but laboratories of $A$ and $B$ can share an entangled state. Interestingly, the time dilation required to move $B$ into or out of the future light cone of $A$ can in principle be made arbitrarily small.  

In a similar vein, in \citep{MacLean} a separation was made between processes with ``direct cause'' and processes with ``common cause''. The first set consists of channels without memory, the second of shared states. In addition, the authors experimentally demonstrated causally ordered processes that cannot be written as a convex mixture of processes with ``direct cause'' and ``common cause'' and call them ``physical mixtures'' of the two. In our terminology, these are channels with memory that cannot be written as a mixture of Markovian channels and shared states. 

\subsubsection{Indefinite causal order from time dilation on a classical background}\label{timedilationclassical}

\citet{aleks2017simulating}  propose a way to implement the quantum switch by exchanging a photon between two Rindler observers moving in an entangled  state of different proper accelerations. Exploiting the equivalence between any sufficiently well-localized stationary observer in Schwarzschild geometry and a uniformly accelerated observer in Minkowski space, they show how the quantum switch can also be  realized for two observers that are in a state of entangled proper distances from the Schwarzschild horizon (e.g., of a black hole) and fixed relative distance from each other.
Another special-relativistic implementation was considered in \citep{debski_indefinite_2022}, with an explicit description of a protocol where the local operations are implemented by particles interacting with optical cavities. 
\citet{moller2021} also consider a relativistic implementation of the quantum switch, where the clock determining Alice's operation on a target photon moves in a superposition of being raised to a given height above Earth at an earlier or later time. This results in a time dilation between the two superposed paths, so that Alice's operation---always at a fixed proper time---takes place before or after Bob's operation---which is at a given coordinate time since Bob's position is fixed. The authors estimate that such an experiment could be feasible with spatial superpositions of $\sim 1$ m and coherence times of $\sim 10$ s, which, although challenging, does not transcend foreseeable technological possibility.

\subsection{Indefinite causal structure from temporal reference frames} \label{causaluncertainty}

Indefinite causal order is considered within the ``timeless'' formalism in Ref. \citep{Castro-Ruiz2020}. In such a formalism, time does not enter as an external parameter but rather emerges through correlations between ``what a clock shows'' and the state of the system \citep{Page1983}. \citep{Castro-Ruiz2020} consider gravitationally interacting clocks and describe the evolution of a quantum system according to individual clocks. Each clock defines a ``temporal reference frame''
with respect to which a notion of time evolution can be defined. Since the clocks are quantum systems in a superposition of energy eigenstates the induced spacetime metric is in general indefinite \citep{CastroRuizE2303}. \citep{Castro-Ruiz2020,Giacomini2021spacetimequantum} study the temporal localization of events
with respect to different temporal reference frames and find that,
when quantum clocks interact with gravitating quantum systems,
the temporal localizability of an event becomes relative, depending on the temporal reference frame. This provides a 
concrete physical realization of the result, that the localisability of events is observer-dependent (or, equivalently, dependent on the ``causal reference frames'') \citep{Guerin2018a},
and that operations can be performed on time-delocalized
subsystems \citep{Oreshkov2019}. Furthermore, \citep{Castro-Ruiz2020} argue that the impossibility to find a temporal reference frame in which all events are local characterizes an indefinite causal order. This is closely related to the result within the process matrix formalism \citep{Guerin2018a}, where it was shown that a pure process matrix is causally nonseparable if and only if there is no causal reference frame in which all events are local.

Still within the timeless formalism, \citep{Baumann2022} show how with
varying clock ticking rates, arbitrary quantum-controlled causal orders can be implemented, and discuss additional constraints that might provide an explanation for why some processes that violate causal inequalities---such as the Lugano process discussed in Sec.~\ref{classicalprocesses}---might not be implementable in the formalism.

\subsection{Background independence and quantum coordinate transformations}
\label{quantumcoordinates}

One of the key issues underlying a quantum theory of gravity is general relativity's background independence, namely, the fact that metric tensors that differ by a diffeomorphism represent the same physical spacetime (where a diffeomorphism is a smooth coordinate transformation). The consequence is that spacetime points on a manifold have no physical meaning: events need to be defined physically, for example in terms of the intersection of worldlines of physical objects.

In the process matrix formalism, events are labeled, in the sense that there is an a-priori identification of what is site $A$, site $B$, etc. These labels can be understood operationally, e.g., as denoting the passage of a particle through a measurement device, and can be seen to play the same role as coordinates in a background-dependent theory. However, in a fundamental theory, such an identification should not be fixed a priori: as in general relativity, it should emerge from an background independent underlying structure. In this view, \citet{parker2021background} consider a version of the formalism that is invariant under site permutations (i.e., arbitrary site relabeling, which can be seen as a discrete coordinate transformation). Using methods from the theory of quantum reference frame \citep{bartlett:2007}, they show that ordinary, labeled, processes can be recovered by taking chosen degrees of freedom at each site (such as `rods and clocks') as physical reference frames.

It has further been suggested that, in quantum theory, coordinate transformations should not be limited to classical ones, but that \textit{quantum coordinate transformations} should be considered as well, which would include ``superpositions'' of coordinate transformations \citep{hardy2018_road, zych2018_rel}. 
A related approach is that of so-called quantum reference frames \citep{Giacomini2019, vanrietvelde_hoehn_giacomini_castro-ruiz_2020_change_of_perspective, carette_glowacki_loveridge_2025_operational_qrf,castro-ruiz_oreshkov_2025_relative_subsystems}, where a physical system (a particle or a field) acts as a reference frame, which can be in a general quantum state.  
More recently \citep{delahamette23,delahamette2022quantum,Kabel24}, the view has been adapted that the same formalism applies when the reference-frame fields are considered as mere ``coordinate fields'', namely abstract labels without direct operational significance. Accordingly, a background-independent theory should be invariant under such quantum coordinate transformation (also named ``quantum diffeomorphism''). Indeed, it has been proposed that they could be used to define a quantum extension of Einstein's equivalence principle: ``For any given point, it is possible to find a quantum coordinate system with respect to which we have definite causal structure in the vicinity of that point'' \citep{hardy2019_impl}. However, even though in the presence of quantum-controlled  spacetimes it is possible to define quantum coordinates such that the non-classical spacetime appears locally Minkowskian and hence classical \citep{Giacomini20,Giacomini21,CepollaroGiacomini2024quantumEEP}, the nature of causal relations cannot be changed: if one defines events as the intersection of (possibly quantum) worldlines, it is not possible to turn indefinite causal relations into definite ones through quantum-controlled coordinate transformations  \citep{delahamette2022quantum, Kabel24}. The reason is that the existence of a local quantum frame in which spacetime is flat is a local property, while the causal order between events is not. 

\citet{Fedida2025knotinvariants} introduced quantifiers to measure the degree of indefiniteness of causal order for an arbitrary finite number of events and spacetime configurations. By mapping causal-order structures to knots and using knot invariants, they showed that definiteness and maximal indefiniteness of causal order are not only diffeomorphism-invariant but also topologically invariant and extended this framework to include quantum-coherence–based quantifiers.

\vspace{7mm}

\section{Physical interpretation of indefinite causal structures}
\label{interpretationsection}

The physical interpretation of indefinite causal structure has been a subject of debate, with different views on whether and to what extent certain physical scenarios---in particular, table-top experiments---represent genuine instances of indefinite causal structures. In this section, we gather and briefly discuss some of the main arguments that have been put forward in this debate.

\subsection{Temporal localization of events}
\label{delocalisedsubsystems}

A core question in the analysis of causal structure is how to identify the relevant \textit{events} between which causal relations are established. In the context of general relativity, causal structure is typically associated with points in a manifold, which can be interpreted as locations of potential physical events---where and when something \textit{can} happen---defined relative to some classical, arbitrarily sharp reference frame.\footnote{A well-known ambiguity in terminology arises from the fact that points in spacetime are often called ``events'' in general-relativistic treatments, without necessarily having a direct reference to physical events. Issues in the physical interpretation of such mathematically defined points were already discussed by Einstein in the context of the ``hole argument'', where he argued for an identification of events with physical occurrences, such as the crossings of particles' worldlines \citep{Stachel2014}. The terminology in this review follows a similar spirit, identifying events with physical operations, unless stated otherwise.}
As already discussed in Sec.~\ref{locatingevents}, both foundational and applied considerations motivate relaxing this identification, allowing for physical events that are not localized in a classical spacetime but are instead defined operationally as the action of a quantum instrument, whose spacetime location may be uncertain or indefinite. The interpretation of physical scenarios as instances of indefinite causal order crucially hinges on how this identification is made, in particular on whether the action of some device over an extended region of spacetime---or over multiple regions---counts as a single event or as a concatenation of multiple ones.

Indeed, lack of definite causal order can occur trivially for events extended in time. This is common in classical system with feedback loops, which appear naturally in economics, biology, etc.~\citep{Bongers2021}. Such scenarios are typically not regarded as instances of indefinite causal order in the sense considered in this review, because they allow systems to freely exchange information while the relevant events take place.
Separating such scenarios from temporally delocalized events requires distinguishing the time interval involved in the description of an event from its actual duration. To illustrate this, consider an event defined by a particle going through a device. If there is classical uncertainty about the arrival time, then the time during which the particle \textit{can} go through the device is longer than the actual traversal time. In particular, a probabilistic mixture of two, well separated arrival times is clearly distinct from a situation where the particle goes through the device \textit{twice}.
A similar distinction applies for a particle in a superposition of arrival times: the particle arrives only once, but at an indefinite arrival time in the quantum-mechanical sense. This scenario is still distinct from the double-passage case. \citet{Ormrod2023} introduced the term \emph{implementation events} (or \emph{regions}) to denote all the spacetime points (resp., extended regions) involved in the execution of a quantum operation.

Identifying a general, operationally verifiable criterion to characterize a \textit{single physical event} without referring to its spacetime location---and to distinguish temporally delocalized events from multiple or genuinely extended ones---is one of the challenges for experimental implementations and interpretations of indefinite causal order. To this end, \citep{OCB_2012} introduced the notion of a \textit{closed laboratory}, which stipulates that the device should not exchange signals with the outside while implementing an operation. Although in principle applicable to device-independent scenarios, verifying such a condition likely requires a device-dependent approach, whose specific implementation is an active open problem.

Within the process matrix formalism, temporally delocalized events can be described using \textit{temporally delocalized subsystems} \citep{Oreshkov2019}, namely tensor factors of the Hilbert space that are not associated with definite times. In the example of two distinct time intervals, the space of all possible operations on the internal states of a particle---including the `double passage' scenario---lives on the Hilbert space $\H^{A^1}\otimes \H^{A^2}$, where $\H^{A^j}= \H^{A^j_I}\otimes \H^{A^j_O}$ represents the site for operations performed at time $j=1,2$. For an operation performed \textit{once}, but in a coherent superposition of being at time $1$ or $2$, it is possible to identify a different tensor product decomposition, such that the operation acts non-trivially only on one tensor factor isomorphic to the space of single-site operations. This example is discussed in further detail in Appendix~\ref{app:subsystems}. 

\subsection{Common arguments on the physical realizations of indefinite causal order} \label{sec:debates}

The majority of the literature on interpretations of indefinite causal order focuses on the quantum switch (or closely related processes with quantum control of the causal order). In the following discussion, unless otherwise stated, indefinite causal order will always refer to the bipartite switch with fixed initial state, sending a pair of operations $U_A$, $U_B$ to the state $\frac{1}{\sqrt{2}}\left(\ket{0}\otimes U_BU_A\ket{\phi}+ \ket{1}\otimes U_AU_B\ket{\phi}\right)$, for some input state $\ket{\phi}$ of the target system; see Sec.~\ref{subsec:W_QS}. 

\subsubsection{Implementation vs simulation} \label{simulations}

A recurring question is whether the table-top experiments performed so far can be interpreted as genuine implementations of indefinite causal order or should rather be understood as simulations. For example, \citet{MacLean} pointed out that optical implementations of the quantum switch are equivalent to ``unfolded'' interferometers, where in one arm unitaries $U_A$ and $U_B$ are performed sequentially and in the other $U_B$ and $U_A$, concluding that these are causally ordered simulations of a quantum switch. Similar conclusions were reached in \citep{Paunkovic2019} for the quantum switch, and for arbitrary quantum processes realizable in spacetime via no-go theorems derived in \citep{Vilasini2022, Vilasini2024}. More fundamentally, it is clear that all current experiments---and, arguably, all those feasible in the foreseeable future---have a ``causal description'' in terms of a quantum state evolving in time, compatible with established quantum physics on classical spacetime. Having said this, it is important to stress that such a ``causal description'' does not preclude the possibility of a causally indefinite order of well-defined operational events.

A key question is what should count as a ``genuine'' implementation of indefinite causal order, something that is not settled by the mere mathematical definition of the switch---or of a general process matrix. A recurring point of view is that the relevant events associated with the experiment should be localized in classical spacetime, leading to the requirement that all operations should
take place within a very short time.  This interpretation not only rules out any implementation of indefinite causal order within conventional physics, but it is also in tension with any thought experiment involving ``quantum spacetime'', such as those discussed in Sec.~\ref{gravitationalswitch}, because there is no classical spacetime over which operations can be localized. Following this strict view, the only viable route to realize indefinite causal order is through closed time-like curves, as first discussed by \citet{Chiribella2013}, or some other exotic classical spacetime. 
In contrast, much research on indefinite quantum causality is motivated by the possibility of novel causal structures enabled by quantum theory, either through non-classical spacetime or, as discussed in Sec.~\ref{delocalisedsubsystems} above, temporally delocalized events.

At the formal level, experiments that reproduce the action of the switch---i.e., that prepare the correct final states as a function of the input unitaries---typically admit a description in terms of temporally delocalized events with indefinite causal order.\footnote{Exceptions include experiments, such as Ref.~\cite{Deng2025}, which do not implement the individual input unitaries but rather their product directly.} However, this also applies to direct implementations of the circuit representation of the switch, as in Fig.~\ref{fig:Qcirc_impl}, in which each gate of the circuit is realized as a temporally localized and independently controllable operation. In the latter case, it is reasonable to count each gate in the circuit as a distinct operation and to interpret the experiment as a simulation of indefinite causal order through repeated, causally ordered operations. The key question is then how to count events (or operations) in an experiment, which we discuss next.

\subsubsection{Number of events}
\label{subsec:num-events}

As discussed above, a key question related to whether experiments implement or simulate an indefinite causal order is how to identify and count operations (or events). As discussed in Sec.~\ref{circuitswitch} above, the process-matrix representation of the quantum switch involves two operations in an indefinite causal order, while causally-ordered descriptions require at least one operation to be counted more than once. 

\citet{MacLean} and \citet{Paunkovic2019} identify the events to be counted are the spacetime points---or sufficiently localized regions---at which the operations take place, hence arguing for a four-event description of optical quantum-switch experiments. \citet{Paunkovic2019} further propose an operational protocol to count the number of spacetime events, discussed in Sec.~\ref{sec:classicalvsquantum} below. Although it is clear that, in quantum-optical implementations of the switch, each operation must extend beyond a single, localized spacetime region, the specific count as two events per operation is rather artificial, as it presumes that the setup involves very short pulses that allow one to distinguish a ``first'' from a ``second'' time at which a photon passes through the optical element implementing the operation. 
Such an assumption is never invoked in experiments, which are typically insensitive to the pulse duration. In fact, realizations as in \citep{goswami2018, goswami2018communicating} use photon coherence times that far exceed the travel time within the whole setup, so that it is not possible---even in principle---to identify two well-separated time intervals at which a photon's wave packet traverses the optical elements. Instead, the first and second passages extend over almost completely overlapping time intervals. Of course, it is always possible to break down the intervals in a sufficiently large number of individual events to recover a causally ordered description, but the specific count has no direct relevance for the experiment. Indeed, the experimental proposal for counting the number of events in Ref.~\cite{Paunkovic2019} does not count the number of operations directly. Rather, it tests whether the two operations delocalized over four rather than two spacetime points.
\citet{Vilasini2022, Vilasini2024} considered a broader notion of event, which can be localized or delocalized in spacetime, and observe that the number and localization of events associated with an experiment depends on a chosen ``level of grain'' of the description. They discuss how every experiment in spacetime can be given a ``fine-grained'' description in terms of---sufficiently many---causally ordered events, concluding that experiments in classical spacetime cannot implement a fundamentally indefinite causal structure. The analysis is further extended in Ref.~\citep{Vilasini2025}, where different levels of descriptions are associated to physical ``labs'' and reference frames, rather than abstract regions of spacetime. In these works, the fine-grained description aims to capture causal relations between all conceivable interventions. 
A very similar argument was proposed in \citep{Ormrod2023}, where it is argued that event counting should not be based on spacetime localization, but on ``which transformations are possible, rather than just those transformations that actually took place''. The view is that it is possible in principle (although not necessarily practically feasible) to let the operations in a quantum switch act differently depending on the causal order in which they occur. For example, in the optical implementations where position is the control degree of freedom \citep{procopio2015}, Fig.~\ref{img:procopio_rubino1}, two parallel, non-overlapping beams traverse the setup, one in the order $A\prec B$ and one in $B\prec A$. These beams impinge on distinct locations of the waveplates implementing the operations on the target (polarization). By modifying the waveplates, one could implement a different operation for the two spatially separated beams and, therefore, for the two states of the control, resulting in a final state $\frac{1}{\sqrt{2}}(\ket{0}\otimes U^{0}_B U^{0}_A\ket{\phi} + \ket{1}\otimes U^{1}_A U^{1}_B\ket{\phi})$, which depends on four, rather than two, unitaries.\footnote{Even in this case, one can still interpret the experiment as a quantum-controlled superposition of two events, with \textit{different} events in the two branches of the superposition: $U^{0}_A$, $U^{0}_B$ in one and $U^{1}_B$, $U^{1}_A$ in the other (which is different from applying all four operations as black-boxes in a quantum circuit, which would be a four-event case). Nonetheless, such a modified experiment would anyway differ from a quantum switch, pointing out that appropriate event-counting is not the sole requirement for the implementation of indefinite causal order.} Experiments that use polarization \citep{goswami2018} or other photon degrees of freedom as control system could also be modified in a similar way, resulting in different operations depending on the state of the control. 

An open question arising within this perspective is how much a given scenario should be extended to include \textit{all possible transformations}. By changing operation settings multiple times for each run of the experiment, or by acting on other modes not involved in the original setup, one could arbitrarily multiply the number of events (and drastically change the process connecting them). In the limit, this pushes the analysis of the causal structure of any setup to the underlying classical spacetime, effectively recovering the identification of events with spacetime points. Although suitable for the description of the physics in the experiment, this view looses the  identification of events with the relevant and accessible operations in a given setup, which is the core element in the operational approach to causal relations.
It remains true, however, that current experiments have not reached the necessary level of idealization required to remove such ambiguities, as they need specific adjustments to ensure one and the same operation is implemented on the target, regardless of the state of the control. \footnote{For example, in \citep{goswami2018}, the Dove prisms acting on the target degrees of freedom---the spatial modes of the beam---also affect the control system---polarization---requiring the insertion of additional correcting waveplates.} Hence, even in a perspective where only the ``naturally available'' interventions are considered, the counting of operations remains somewhat ambiguous in current implementations. A proposed solution to this problem is to physically separate the control from the target system, as proposed in \citep{CostaRQI24}. 

To address the problem of comparing the number of events in different possible realizations of the quantum switch, \citet{delahamette2022quantum,Kabel24,delaHamette2024} introduced a relativistic definition of causal order between operationally defined events, which yields a meaningful observable in both the general relativistic and quantum mechanical regimes. They provide an explicit map—a quantum-controlled diffeomorphism preserving indefinite causal order—from a switch realized at two space-time points to one realized at four. 
This construction directly supports the view that the optical quantum switch is just as much a realisation of indefinite causal order as its gravitational counterpart, as far as causal order is concerned. 

An approach to counting operations, independent of their spatiotemporal location, was proposed in Ref. \citep{Araujo2014}, arguing that, if one were to install a ``counter'' that counts how many times each operation is performed (e.g., counting the number of photons passing through an optical element, without disturbing the overall coherence), it would show a total of one use per operation in typical quantum-switch experiments. 
In this approach, the event count does not depend on the particular representation of the experiment (through a causally non-separable process or a quantum circuit with multiple calls to the operations). In a temporal description, the counter may get entangled with the other degrees of freedom at intermediate times, but eventually it factors out displaying exactly one use per operation.
Possible limitations of this approach are that it could depend on the physical implementation of the counters, and that it might be unclear how to apply the reasoning to actual experiments, where the counters are not implemented. 

It is sometimes argued that operations on the vacuum should also count as events, giving four events in the optical quantum switch. However, it is unclear operationally how such ``vacuum events'' are to be counted. If every vacuum mode in a lab counts, the number of events becomes unbounded, even without an injected photon. If they count only when the photon is present somewhere in the setup, then obtaining four events would require a local counter in one lab to register an event depending on the photon being located outside that lab. Thus, the operational meaning of counting vacuum events remains ambiguous.

Finally, although most discussions focus on unitary operations, the switch can be applied to more general operations as well, e.g., quantum channels, quantum instruments, or operations that also act on auxilliary systems. As discussed in Sec.~\ref{subsubsec:SwitchMultipleCalls} and proved in Refs. \citep{bavaresco2024simulated,kristjansson2024exponential}, common causally-ordered representations of the switch with one extra use of the operations do not extend directly to the non-unitary case, while the quantum switch can be applied to arbitrary non-unitary operations. In such cases, any causal description would require a number of events that scales at least exponentially with the number of target qubits. 

\subsubsection{Causal loops}
Building upon the framework of classical causal models, where causal structure is represented by a directed acyclic graph (DAG), \citep{MacLean} argue that the causal structure of the switch (classical or quantum) contains a loop, and ``it is unclear how to make sense of such a graph''.  Furthermore, it is argued that a causal cycle is a closed time-like curve, and therefore the quantum switch cannot be realized in an ordinary spacetime.

The assumption underlying the above conclusions is that the ``nodes'' in a causal structure, i.e., the relevant events, must correspond to ``fixed spatio-temporal regions''. This is stronger than the assumption of definite causal order, as it implies that the location of each instance of an event is fixed with respect to a fresh set of coordinates that is reset at each run of the experiment, independently of any other variable (thus ruling out also probabilistic and dynamical causal order). This diverges from the typical assumptions within classical causal models, which allow for a more general interpretation of events whose coordinates need not be specified in advance (e.g., if one asks what causes a train to arrive sooner or later, the arrival time cannot be fixed a priori). Even in typical quantum-optics experiments---such as the one in \citep{MacLean} that superposes a ``common cause'' and a ``direct cause''---the time of events is not fixed a priori, but depends on the probabilistic process of photon production. This all suggests that the identification of nodes in a causal model with fixed spacetime regions is an excessively rigid requirement, calling for generalizations. Works in this direction include \citep{Barrett2020}, which proposed an extension of quantum causal models to include loops, see Sec.~\ref{causalmodels}. Consistency conditions for so-called ``routed quantum circuits'' with loops were investigated further in \citep{Vanrietvelde2022}, Sec.~\ref{sec:routed}. 

\subsubsection{The role of the vacuum} \label{subsec:role_vacuum}
It is sometimes argued that the indefinite causal order in the quantum switch is a spurious consequence of not including the vacuum in the local operations, and that doing so naturally restores the universality of causally ordered quantum circuits \citep{Portmann2017}. Adding the vacuum to process matrices is also prominent in Refs. \citep{Paunkovic2019, Ormrod2023}, and is further emphasized by \citet{Faleiro2022} in connection with the ``two-way communication with a single particle'' introduced by \citet{DelSanto2018}. 
The argument stems from the fact that common quantum-optical operations act trivially on the vacuum, and their theoretical representation is often confined to the non-vacuum subspace, while causally ordered descriptions of optical implementations of the switch typically re-introduce the vacuum explicitly. 

A further reason to consider the role of the vacuum in quantum switch experiments is the analogy with other quantum-information protocols with similar interferometric implementations. For example, an interferometer where a unitary (acting, say, on polarization) is applied to one arm (labeled $\ket{1}$) and the identity on the other arm (labeled $\ket{0}$) implements a ``controlled unitary'', $\proj{0}\otimes \id + \proj{1}\otimes U$. This cannot be realized by a quantum circuit with a single call to the unitary (where the unitary acts on a subsystem, i.e., on a wire in the circuit), but it becomes possible if the unitary is extended to act trivially on an extra state, such as the vacuum \citep{Soeda2013controlisation, kitaev95, Araujo2014b, Friis2014, Thompson2018,dong2019control,Gavorova2020control}. This has motivated the development of extensions of the quantum circuit formalism, including superposition of trajectories \citep{Chiribella2019trejectories}, ``PBS diagrams'' \citep{clement_MFCS2020,branciard_MFCS2021}, or routed quantum circuits \citep{Vanrietvelde_2020}; see Sec.~\ref{sec:routed}.

The situation for indefinite causal order is markedly different to that of controlled unitaries, or other ``superpositions of circuits'': including the vacuum is neither necessary nor sufficient for a definite-order representation of the switch. In general, the Hilbert spaces defining the sites of a process can represent any quantum degree of freedom, including field modes with a vacuum state. In a quantum switch with a $d$-level non-vacuum target, the local operations can always be extended to a $(d+1)$-level system that includes the vacuum. However, it still remains impossible to represent the resulting process using a single call to each (extended) operation in a causally ordered quantum circuit\footnote{To see this, note that the switch allows signaling in both directions, from $A$ to $B$ and from $B$ to $A$, which is not possible through a single use of causally-ordered operations, even if extended to the vacuum.}---one still needs to call at least one of the operations more than once, see Sec.~\ref{switchcontrolledU}. 

On the other hand, as discussed in Sec.~\ref{subsubsec:SwitchMultipleCalls}, when considering unitary operations, the quantum switch \textit{can} be represented by a quantum circuit without extending the operations to the vacuum, at the cost of adding multiple copies of each operation and introducing an auxiliary system. Therefore, albeit relevant in the analysis of quantum-optics setups, the vacuum  does not appear to play a central role in the physical interpretation of indefinite causal order.

Furthermore, it is possible in principle to demonstrate indefinite causal order in quantum-switch experiments where the local operations act non-trivially on the vacuum, as shown in the supplementary material of \citep{zych2019} for a quantum-controlled spacetime and in \citep{debski_indefinite_2022} for a classical relativistic spacetime.

\subsubsection{Realizations in classical vs quantum spacetime}
\label{sec:classicalvsquantum}

Given that causal structure in a quantum spacetime is one of the primary motivations for the introduction of indefinite causal order, it is natural to ask what are the differences between laboratory vs gravitational---as in \citep{zych2019} and \citep{Moller2024gravitational}---realizations, and if quantum gravity can lead to fundamentally different causal structures, beyond what is possible with delocalized events in classical spacetime.

Clearly, the physics underlying a photonic quantum switch is not equivalent to the physics in a quantum-gravitational implementation. The former case is fully captured by quantum electrodynamics on Minkowski spacetime, while the latter involves non-classical spacetime, beyond currently tested regimes. However, distinguishing the two scenarios would require probing additional observables---for example, carrying reference-frame information---beyond the bare implementation of an indefinite-order protocol. As far as the \textit{causal relations} between physically-defined events are concerned, the two scenarios appear indistinguishable. This observation may be put in close correspondence with the fact that, on its own, the causal relation between a pair of events cannot distinguish Minkowski from any other causal (i.e., with no CTCs) spacetime: in all cases, an event $A$ can only be in the past, in the future, or space-like separated from another event $B$. Similarly, the quantum causal relation between a pair of events is not sufficient to distinguish a classical from a quantum spacetime. 

The view expressed above is supported by the analysis in \citep{delahamette2022quantum,Kabel24} following \citep{Guerin2018a}, see Sec.~\ref{quantumcoordinates}. The authors introduce an observable to measure definite vs. indefinite causal order by the sign of the proper times elapsed between two (or more) events in different ``branches'' of quantum controlled orders of events. They show that, while the localization of events depends on the choice of quantum coordinates, the observable associated with causal order \textit{is} independent (i.e., it is quantum-diffeomorphism invariant, thereby extending the general-relativistic requirement of coordinate-independent observables to the quantum-controlled domain) and is operationally accessible. It verifies indefinite causal order in both optical and gravitational realization of the quantum switch. Interestingly, the operational observable for causal order does not extend to more general scenarios, where multiple events are involved or the time interval between events is not fixed in advance \citep{Maguire2026}.

An important aspect, highlighted in \citep{Paunkovic2019, Ormrod2023} and mentioned in Sec.~\ref{subsec:num-events} above, is that which-order information is in principle available at the local operations in current laboratory implementations. This is because the control degree of freedom (such as position, polarization, or time bin) travels ``together'' with the target one. 
\citet{zych2018_rel} and \citet{Moller2024gravitational} show that, using superpositions of spherical shells and agents in free fall, any potential local which-order information can be removed in a quantum-spacetime implementation. However, this does not seem to be a fundamental difference between classical and quantum spacetime. For example, the relativistic, classical-background versions of the switch in \citep{aleks2017simulating, moller2021}, discussed in Sec.~\ref{timedilationclassical}, remove local which-order information too. Laboratory implementations with this property are also under investigation \citep{CostaRQI24}.

Finally, \citet{Paunkovic2019} propose a protocol to distinguish operationally whether an experiment involves four or two spacetime events, and that only a quantum-gravity implementation can yield a two-event count. When applied to quantum-optics experiments, the protocol relies on an observer using a classical-spacetime reference frame, who would be able to assign distinct, operationally-defined coordinates to the different ``passages'' of a photon through an optical element implementing a local operation. However, a similar counting would be possible in a quantum gravitational implementation too: an observer in a classical region of spacetime [e.g., asymptotically far from the superposed masses or, for the protocol in \citet{Moller2024gravitational}, at a finite distance outside the spherical shell] can use reference systems that give operational meaning to the common-coordinate description of the protocol, where events take place at different coordinate locations depending on the metric, as depicted in Fig.~\ref{zych}. According to such coordinates,  the gravitational switch involves four spacetime points just as the optical quantum switch.

Given that causal relations in switch-like experiments do not appear to be sufficient to distinguish a quantum from a classical spacetime, it remains an open question what are the minimal requirements for giving operational meaning to such a distinction, and whether a distinction based uniquely on causal structure, possibly in more complex scenarios, is at all possible.

\section{Adjacent results}
\label{sec:AdjRes}

\subsection{Quantum causal models}\label{causalmodels}
\begin{figure}
\centering
\includegraphics[width=.35\columnwidth]{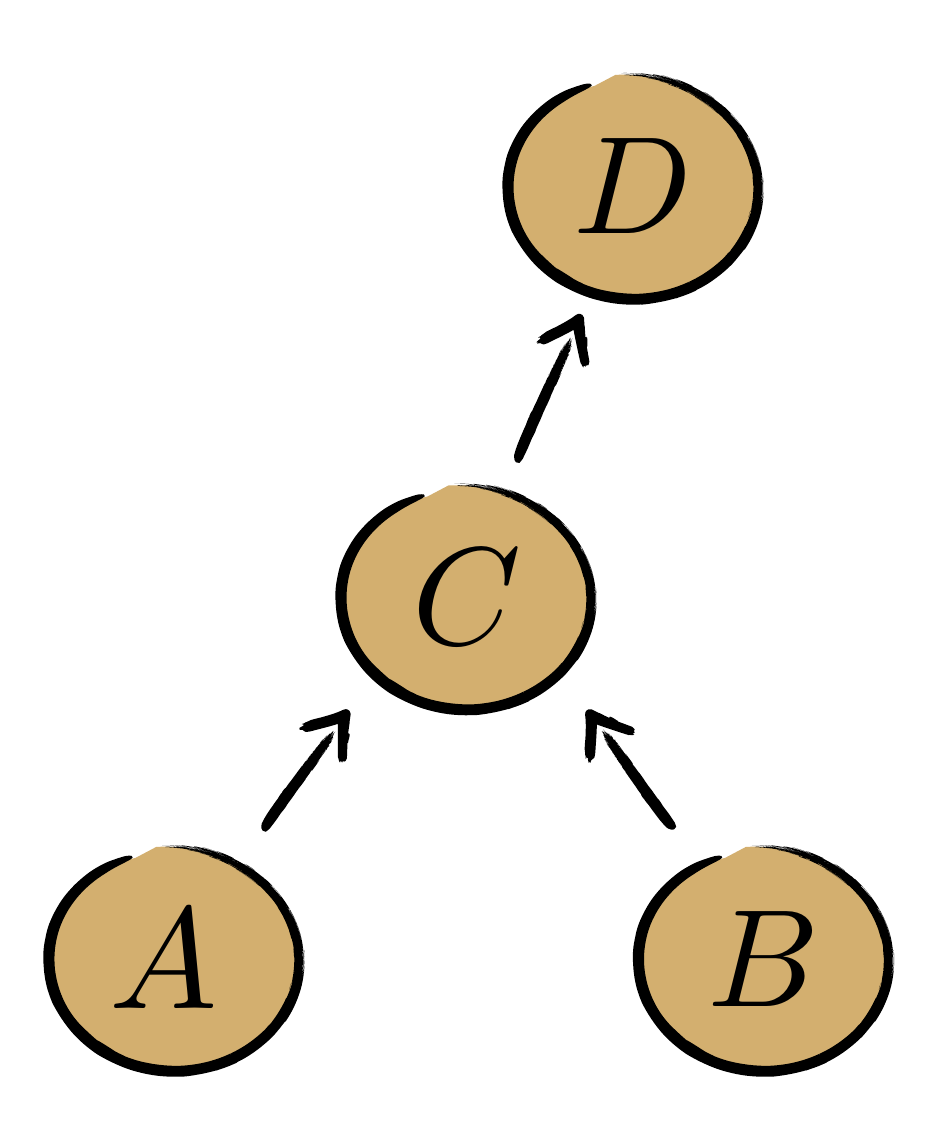}
\caption{A \textbf{Directed Acyclic Graph (DAG)} comprises nodes connected by directed edges, with no directed loops allowed. In a causal model, nodes represent observable variables and edges direct cause-effect relations.}
\label{DAG}
\end{figure}

In the most common modern approaches, causal models represent causal structure through a directed acyclic graph (DAG), Fig.~\ref{DAG}, where the nodes represent random variables and the directed edges denote a relation of cause-effect, which imposes constraints both on the correlations between variables and on the effect of an intervention \citep{spirtes2000causation, Pearl2009, Koller2009}. The constraints imposed by a DAG are captured by the \emph{causal Markov condition}, which states that a variable conditioned on its direct causes is independent of all other past variables. For instance, a DAG $B{\leftarrow}A{\rightarrow}C$ represents $A$ as a complete common cause of $B$ and $C$, leading to the conditional independence $P(B,C|A)=P(B|A)P(C|A)$. This condition is not compatible with quantum common causes in general: even a pure bipartite quantum state can give rise to correlated measurement outcomes. Indeed, common-cause conditional independence is closely related to Bell's local causality condition \citep{bell64}. More generally, quantum correlations violate the classical causal Markov condition unless one allows fine-tuned causal explanations, namely causal influences that are not manifest at the level of observed statistics \citep{wood2015, Cavalcanti2018}. This has motivated several approaches to extend causal modelling to quantum systems \citep{Laskey2007, leifer2013, cavalcanti2014modifications, Henson2015, pienaar2014graph}, eventually resulting in a more unified picture based on the process-matrix formalism or equivalent frameworks \citep{costa2016, Allen2016, Barrett2019}.

Process-matrix-based approaches provide a natural way to formulate such models. In the framework of~\citet{costa2016}, the nodes of a DAG are identified with local sites of a causally ordered process, and the quantum analogue of the causal Markov condition requires the process matrix to factorize into the tensor product of channels associated with the parents of each node. In this formulation, outputs of common-cause nodes decompose into subsystems corresponding to the outgoing arrows. This structure has also supported the development of causal-discovery algorithms for quantum experiments \citep{Giarmatzi2019, Bai2020} and recovers the operational notion of Markovianity developed independently in the study of open quantum systems \cite{Pollock2018}.

A related formulation was developed by~\citet{Allen2016} and~\citet{Barrett2019}, where quantum common causes are characterized through the existence of an underlying unitary interaction giving rise to the observed process, resulting in a quantum Markov condition similar to \citep{costa2016}. The main difference is that, in \citep{Allen2016, Barrett2019}, the outputs of sites do not factorize in subsystems according to the outgoing arrows and the process matrix decomposes into a product of commuting Choi operators of channels, rather than a tensor product. This coincides with the condition previously proposed in \citep{Cavalcanti2018}.

\citet{Barrett2020} extended the framework of quantum causal modeling to cyclic causal models, allowing for the description of processes with indefinite causal structure, including the quantum switch, when suitable global-past and global-future systems are included. More recent work has developed operational frameworks for cyclic and fine-tuned causal models and their compatibility with spacetime \citep{Vilasini2022_cyclic}, as well as a cyclic quantum causal-modelling framework with a graph-separation theorem for finite-dimensional systems \citep{Ferradini2025}. These developments extend causal-modelling tools beyond acyclic, faithful scenarios and may be relevant for causal discovery in settings with feedback, fine-tuning and non-classical causal influences.

\subsection{Channel capacities of general processes}\label{processchannelcapacity}

In light of the advantages offered by indefinite causal structures in communication tasks, it is natural to ask whether these advantages can be captured within a systematic communication theory for general quantum processes. The standard figures of merit in classical and quantum information theory are channel capacities, which quantify the asymptotic rate at which noiseless information can be transmitted using many copies of a communication resource. Different capacities correspond to different types of information (e.g., classical or quantum) and to different auxiliary resources (e.g., joint measurements, shared entanglement, etc.). In the standard setting, however, the sender is assumed to act before the receiver in each run of the protocol.\footnote{This is also the case for the proposed communication advantages from indefinite causal order, reviewed in as in Sec.~\ref{noisereduction}, where the quantum switch is used as an ordinary channel from its global past to its global future.}
For a general process, by contrast, the causal order between sender and receiver need not be fixed. This motivates extending the notion of channel capacity to processes with no fixed causal order, and asking whether such processes can outperform causally ordered communication resources. These questions were addressed in \citep{Jia2019} for quantum capacity and in \citep{Goswami2020} for classical capacity. The general setting involves two agents, Alice and Bob, who have access to $n$ independent copies of the same bipartite process $W$. For each use of the process, Alice and Bob apply local operations, possibly involving auxiliary systems used for encoding and decoding. Together with the $n$ copies of $W$, these operations induce an effective $n$-round channel on the auxiliary systems. The capacity of the process is then defined by applying the usual asymptotic notions of channel capacity to this induced channel, with an additional optimization over Alice's and Bob's local operations. A technical subtlety is that, although the underlying resource (i.e., the process) is stationary, the effective channels associated with different uses need not be identical, since the local operations may vary from one use to the next. Thus the problem maps to a non-stationary asymptotic setting \citep{Verdu_Han_94, Hayashi_non-stationary}. This issue is handled in \citep{Jia2019,Goswami2020} by showing that a fixed choice of local operations can reproduce the optimal asymptotic behavior for the purpose of one-way communication, but may require separate treatment in other capacity scenarios.

Only a few results are known for the capacities of general processes. \citet{Jia2019} proved that, for a causally separable process of the form $q W^{A\prec B} + (1-q) W^{B\prec A}$, with $0\leq q \leq 1$, the quantum capacity from Alice to Bob is zero unless $q>\frac{1}{2}$. Thus no quantum communication is possible in either direction when the causal order is completely unknown, $q=\frac{1}{2}$. In this sense, causal order itself is a resource for quantum communication, although it remains open whether an analogous statement holds for causally nonseparable processes.

\citet{Goswami2020} extended several bounds on the classical capacities of quantum channels to general processes, including causally nonseparable ones. In particular, they proved a process-matrix analogue of the Holevo bound \citep{Holevo1973, Schumacher1997, Holevo1998}: the one-way classical capacity from Alice to Bob is bounded by $\log d^{B_I}$, where $d^{B_I}$ is the dimension of Bob's input space. This bound implies a negative result for one-way communication: indefinite causal order alone does not increase channel capacity. However, it remains an open question whether any advantage is possible for two-way or genuinely multipartite communication.

\subsection{Transformations of process matrices} 
\label{subsec:TransfPM}

Similarly to quantum states and quantum operations, process matrices may also be subjected to transformations. One way to define general transformations between process matrices is to consider all completely positive linear maps which transform process matrices into process matrices. This approach was in Ref. \citep{CastroRuiz2018}, which provides linear projectors to characterize such transformations which are correct up to a missing term. Motivated by developing a resource theory for causal connections, Ref. \citep{Milz2022Resource}---expanding on previous related work \citep{Taddei2019}---analyzes transformations between general process, there named adapters, and presents a characterization for these transformations in terms of projectors. Ref. \citep{Milz2022Resource} also analyzes the concept of completely admissible adapters, which are transformations which in addition of mapping process matrices into process matrices, should also respect a completeness property of being able to map process matrices into process matrices even when applied to part of their systems. There, the authors show that different notions of completely admissibility may lead to different classes of allowed transformations. Transformations between process matrices and even more general ``sets of quantum objects'' are also analyzed in Refs. \citep{Hoffreumon2022_proj,milz2023_transformations}, papers which also characterize such transformations in terms of linear projectors, and analyze their properties.
General transformations between process matrices and general quantum objects have also been studied from the perspective of type systems \citep{perinotti16higher,bisio19higher}, category theory \citep{Kissinger2019,Simmons2024,Wilson2022Polycategorical}, and linear logic \citep{Simmons2022Higer,Hoffreumon2022_proj,Hefford2025BV}. Mathematical formalism which is motivated for developing a quantum analogue of functional programming.

\subsection{Indefinite causality in quantum foundations}
\label{foundations}

\subsubsection{Operational models and reconstruction of quantum theory} \label{reconstructions}

An important trend in the foundations of physics seeks to understand quantum mechanics within a larger set of operationally well-defined theories \citep{hardy01, Barrett2007, Chiribella2010}. Common goals in this field include identifying a small and physically motivated set of physical axioms that recovers quantum theory, as well as understanding what fundamental aspects would have to change to yield distinct theories \citep{hardy01, Chiribella2011, Dakic2010, Masanes_2011, Selby2021reconstructing}. See  \citep{Grinbaum2007-GRIROQ} for a critical review of historic and contemporary reconstruction programmes, arguing that the shift from interpreting to reconstructing quantum theory marks a paradigmatic change in how its foundations are understood.

Most approaches assume, more or less explicitly, a definite causal or temporal structure in the construction of general operational theories. An axiom of ``causality'' is often introduced to ensure that the (a-priori defined) future cannot influence the past \citep{Chiribella2011}. The \textit{causaloid} framework was introduced in \citep{Hardy2008} as a generalized probabilistic framework without a definite causal structure. A similar framework was introduced in \citep{OreshkovGiarmatzi_2016}, based on a distinction between ``processes'' and ``operations'' similar to the process matrix formalism. In Ref. \citep{Ibnouhsein_2015}, information-theoretic principles were proposed to constrain bipartite correlations in the GYNI game within generalized probabilistic theories without global causal ordering to match the bound derived in the process matrix formalism. In particular, the principles extend the usual inequality of data processing and correspondingly constrain the mutual information between parties. A reconstruction of quantum theory without definite causal structure (i.e., of the process matrix formalism) based on operational axioms was proposed in \citep{Jia2018a}.

Other approaches to generalized probabilistic theories without definite causal structure include the \textit{positive formalism} \citep{Oeckl2019} and the category-theory approach of \citep{Kissinger2019}. 

\subsubsection{(Non-)contextuality without causal assumptions}
\label{subsubsec:noncontextuality}

One of the open questions in the foundations of physics is whether quantum theory can be expressed in terms of some underlying elements of reality, such as classical hidden variables. No-go theorems impose strong limitations on such models as, under a few assumptions, hidden variables have to be contextual and nonlocal in order to reproduce quantum predictions \citep{bell64, kochen67}. One of the main assumptions appearing in most treatments is \textit{forward causality}, namely that causal influence, at the hidden variable level, is one-directional, from the past to the future. An approach that has been gaining increasing popularity in recent years is to consider more general models, featuring some form of retrocausality. The idea is that such models might retain a classical picture of reality (including locality and non-contextuality) where quantum statistics---and all associated quantum features---are the result of ignoring relevant future causes, as well as the particular state of the hidden variables; see \citep{Wharton2020} for an overview of the approach.  

\citet{Shrapnel2018causation} have shown that, even by allowing arbitrary, exotic, causal structure, realistic models that aim to reproduce quantum predictions are still subject to some form of contextuality. The key observation is that, as we have seen in Sec.~\ref{indefiniteaxiomatic}, the generalized Born rule, Eq.~\eqref{simpleBorn2}, is the most general \textit{non-contextual} probability assignment to quantum measurements, namely it only depends on the CP maps associated with the observed outcome, and not on other CP maps in the local instruments.
The implication is that any underlying model giving better predictions (i.e., with less uncertainty) for quantum experiments  has to be instrument \textit{contextual}: the hidden variables---or the laws governing them---have to know about events that did not occur (namely the CP maps completing the instrument) or on implementation details that are irrelevant for the observed statistics.

\subsection{Related frameworks}\label{subsec:related} 

Over the years, a broad and diverse range of work has been developed that is related to the process matrix formalism in some formal or conceptual ways. Some of this work was motivated by similar challenges, such as unifying spatial and temporal concepts in quantum theory, while other work converged on similar ideas from different directions. Here we provide a brief overview of some related formalisms for reference, without attempting to describe them in detail.

\subsubsection{Causally ordered processes }
\label{otherorderedprocesses}

\citet{Kretschmann2005} developed a framework to include memory effects in quantum channels, generalizing earlier work \citep{Knill2000}. In this context, a \textit{quantum channel}, similar to its counterpart in classical information theory \citep{csiszár_körner_2011}, is understood in an asymptotic communication setting, namely as a map from (arbitrarily many) copies of Alice's state space to (an equal number of) copies of Bob's state space, $\mathcal{W}^{(n)}:\mathcal{L}(\mathcal{H}^A)^{\otimes n}\rightarrow \mathcal{L}(\mathcal{H}^B)^{\otimes n}$, where the $j$-th copy refers to the use of the channel at a time $t_j$ to send information from Alice to Bob. In our language, the $j$-th input of the channel is the output of a local laboratory, which has the $j{-}1$-th output of the channel as an input.%
\footnote{Note the swap in terminology, where an ``input'' of the channel is an ``output'' of a local laboratory. Note also that, in \citep{Kretschmann2005}, each of Alice's copies can only send into the channel states that are prepared independently of Bob's measurements, while, in the broader context of causally ordered processes, we allow arbitrary operations from Bob's space at time $t_{j-1}$ to Alice's space at time $t_j$ (which we interpret as, respectively, input and output of a single laboratory).}
In defining a channel with memory, \citep{Kretschmann2005} impose a ``causality condition'' on $\mathcal{W}^{(n)}$ to ensure that the output of the channel up to a time $t$ does not depend on future operations. This makes $\mathcal{W}^{(n)}$ equivalent to the channel representation of a causally ordered process, as per Sec.~\ref{processaschannel}, where each site has Bob's space at a given time as input and Alice's space one time step later as output. 

In \citep{Gutoski2006}, a causally ordered process represents a  strategy in a multi-round quantum game: at each round of the game, each player receives a quantum system, performs some action on it, and sends the system back to the game---in other words, a player's action in a given round takes place at a ``site'' in our terminology. If the measurement outcomes obtained by other players are not available, then their collective strategies (together with any environmental effect) are described by a deterministic, causally ordered process. In addition to the local operations we discussed, here each player is allowed, at each round, to let the system interact and be jointly measured with a local memory system, which can correlate operations at different times. As it turns out, the most general operation of this type, called a ``measuring strategy'', is equivalent to a deterministic strategy (i.e., a causally ordered process) with a single measurement at the end. The most general $n$-turn measuring strategy with outcomes $a$ is represented in Choi form by a set of operators $M^{A^1\dots A^n}_a$, such that $\sum_a M^{A^1\dots A^n}_a$ is a (deterministic) causally ordered process. The probability to obtain an outcome $a$ is given by a direct extension of the  Born rule: $P(a) = \tr\left(M^T_a W\right)$. 

Causally ordered processes have been studied extensively within a framework for quantum networks \citep{Chiribella2008architecture, Chiribella2009, Bisio2011}. In this context, a causally ordered process matrix is called a comb, while measuring strategies are called testers. Single-site testers were also introduced in \citep{ziman2008} as compact representations of measurements of quantum channels, under the name of process positive-operator-valued measures.

Another field where the framework has proven extremely fruitful is the study of non-Markovian open quantum systems \citep{modioperational2012, Pollock2018, Milz2017}. The traditional approach describes open system in terms of a time-dependent channel acting on the state of the system. However, such description is inherently oblivious to multi-time correlations---which are crucial to describe memory effects---and and can become ill-defined when initial system-environment correlations cannot be ignored \citep{Pechukas1994, Stelmachovic2001, Shaji2005, Schmid2019}. On the other hand, causally ordered processes provide a direct generalization of classical stochastic processes and offer a simple characterization of Markovianity: a Markovian process has the product form $W^{ABC\dots} = \rho^{A_I}\otimes T_1^{A_OB_I}\otimes T_2^{B_OC_I}\dots$, where $\rho$ is the initial state and $T_1, T_2,\dots$ are (independent) channels mapping each time step to the next. This reproduces the original definition of Markovian Quantum Stochastic Process in \citep{Lindblad1979}.

Finally, \citet{Milz2018} show that, any process matrix, including ones with an indefinite causal order, can be probabilistic simulated by fixed-order, non-Markovian processes via postselection. In order to reproduce indefinite causal order, one needs tri-partite entanglement between the two local systems and the environment and a non-local unitary. Ref. \citep{PhysRevResearch.3.023028} further presents an example where every outcome of final pure measurements yields a conditional process with indefinite causal order.

\subsubsection{Higher-order transformations }

As discussed in Sec.~\ref{subsec:open_past_future}, quantum processes can be seen as higher-order transformations, mapping product operations to probabilities (in the original formulation) or to channels (in the formulation with ``open past and future''), perspective adopted in the review article \citep{taranto2025higher}. \citet{Perinotti2017,bisio19higher, Wilson2021} introduce an extension of the framework to analyze higher-order transformations between arbitrary types of objects (e.g., transformations from combs to combs), deriving the linear constraints that characterize each class of transformations,  results that were also investigated and generalized in \citep{milz2023_transformations,Hoffreumon2022_proj,Milz2022Resource,jencova2024higher}. Similar constraints were also derived in \citep{Jia2020}. The approach was further expanded to a category-theory framework, extending beyond quantum theory, in \citep{Kissinger2019}, enabling a systematic study of composition and compatibility of general processes with arbitrary causal constraints.

Higher-order transformations were also studied in \citep{Zyczkowski2008}, although with a different interpretation as an extension of the quantum-mechanical description of states.

\subsubsection{Beyond valid process matrices: two-state vectors and other nonlinear versions of the formalism} \label{postselectedprocesses}

Other related frameworks incorporate, and give physical interpretation to, ``non-valid'' process matrices; i.e., to process matrices that do not satisfy the linear constraints imposed by normalization and can thus contain arbitrary terms of ``non-valid'' types in their HS decomposition, see Sec.~\ref{para:HS_terms}.

A prominent example is the two-state vector formalism, which describes processes with pre-selection, encoded into ``forward-evolving'' states $\ket{\psi}$, and postselection, given by ``backwards evolving'' dual states $\bra{\phi}$ \citep{Watanabe1955, aharonov1964, Aharonov2008}. The framework extends to include arbitrary pre- and postselected operations over an arbitrary number of times \citep{Aharonov2009, Silva2014}. The resulting ``multi-time states'' are equivalent to conditional processes \citep{Chiribella2009,Oreshkov_2016, Silva_2017}, where, in process matrix terms ``forward evolving'' states are defined on input spaces, while ``backward evolving'' ones live on outputs. As discussed in Sec.~\ref{conditionalprocesses}, conditional process matrices can be proportional to arbitrary positive semidefinite operators, so one can formally view valid process matrices as a strict subset of multi-time states.

Another approach to ``states over time'' is that of \textit{entangled histories} \citep{Cotler2016}, whose connection with the two-state vector formalism has been discussed in \citep{Nowakowski2018}. \emph{Superdensity operators}, an extension of entangled histories, can be mapped directly to arbitrary conditional process matrices \citep{Cotler2017}.

It should be stressed that multi-time states and superdensity operators are still defined relative to a global background time, with ``temporal subsystems'' corresponding to temporally ordered events. Therefore, even though these formalisms can reproduce arbitrary process matrices, including conditional ones, these are obtained through postselection of causally ordered processes, and thus they do not represent indefinite causal order.

As discussed, for example, in \citep{Silva_2017}, indefinite causal order with definite temporal order would be possible in the presence of ``fundamental postselection'', i.e., in a world where certain future boundary conditions are guaranteed to occur with probability one---as, for example, in the final-state model of black holes \citep{Horowitz2004}. Fundamental postelection turns out to also be equivalent to the ``P-CTC'' model of CTCs \citep{Lloyd2011}, originally formulated in the language of path integrals \citep{Politzer1994}.
In such scenarios, a set of sites could be connected through a non-valid process, even without conditioning on future events, yielding the modified probability rule 
\begin{equation} \label{postselection}
    P(a, b\ldots) = \frac{\tr \left[\left(M^{A}_{a}\otimes N^{B}_{b} \cdots \right)^T \! W^{A B\ldots}\right]}{\sum_{ab} \tr \left[\left(M^{A}_{a}\otimes N^{B}_{b} \cdots \right)^T \! W^{A B\ldots}\right]}
\end{equation}
for a set of CP maps with Choi operators $M^A_a, N^B_b, \dots$
By definition, the denominator is equal to $1$ only for valid processes, while in the more general case probabilities would be nonlinear functions of the CP maps. This implies a violation of the assumption of ``local quantum theory'', discussed in Sec.~\eqref{indefiniteaxiomatic}, from which the linear Born rule is derived. 
Hence, local agents performing temporally-localized operations in causal regions of spactime, far away from ``anomalies'' such as black holes or CTCs, would be able to observe a departure from quantum physics through their local observations. This is also true for other nonlinear models for CTCs \citep{deutsch1991quantum}, in which probabilities do not have a conditional form as in Eq.~\eqref{postselection}. On the other hand, \citet{Araujo2017} have shown that process matrices correspond to a linear particular case of P-CTCs, and linear models of CTCs have been further investigated in \citep{Baumeler_2019, Tobar_2020}.

All the frameworks above, including the process matrix formalism, assume that it is possible to assign a time direction to local operations, formalized in the distinction between input and output spaces. Relaxing this distinction, one can associate operations with positive semidefinite operators that live on the tensor product of an arbitrary number of ``boundary spaces''. This approach was introduced by \citep{Oeckl2003318} in the ``general boundary formulation'' of quantum theory, where the boundaries were further associated with regions of spacetime (but without assuming any metric or causal structure). A similar idea was developed in \citep{Oreshkov_2016}, as a modification of the process matrix formalism to remove a pre-defined local time. Pictorially, an operation in this framework is a box with an arbitrary number of open wires---representing the boundary spaces---which can be combined to the wires of other operations through a version of the link product without any distinction between input and output wires. Probabilities are obtained when all wires are connected and, as for conditional processes, they are generally nonlinear functions of the local operations. An extension of the general boundary formalism beyond quantum theory was formulated in \citep{Oeckl2019}. 

\subsubsection{Other approaches to states over time} 

One of the common features of the frameworks described so far is that sites of potential events are associated with two Hilbert spaces: an input and an output space of local operations. Other approaches aim to define  ``states over (space)-time'' by associating a single Hilbert space to each site. However, such approaches generally face limitations. For example, the causal joint states introduced in \citep{leifer2013}---based on prior work on quantum conditional states \citep{Leifer2006, Leifer2008}---do not generalize well beyond two time steps. In fact, in order to extend the framework to an arbitrary number of sites, \citep{Allen2016} decompose each site into input and output spaces, effectively recovering causally ordered process matrices. Another example of single-Hilbert-space states over time are Pseudo-density matrices \citep{Fitzsimons2015, Pisarczyk2019}. These can be defined for an arbitrary number of time steps; however, they can only reproduce temporal correlations for projective measurements of dichotomic, traceless observables. Only by adding input and output spaces for each site can pseudo density matrices account for arbitrary operations \citep{Liu2023}. Finally, the quasi probability distributions over time introduced in \citep{Horsman2017} fail to reproduce the correct joint probability distributions in the classical limit.

Some of the limitations behind these approaches have been encapsulated in a general no-go theorem, proven in \citep{Horsman2017}, which states that states over time with an individual Hilbert space at each site cannot simultaneously fulfill a set of natural requirements. It remains interesting that the approaches above do have a range of applicability, which can facilitate the analysis of particular scenarios.

Yet another approach, initiated by \citep{Page1983}, is to encode the ordinary time evolution of a (pure) quantum state, $\ket{\psi(t)}$, into a ``timeless'' state that includes a ``clock'' degree of freedom, $\kket{\Psi} \coloneqq  \int dt \ket{\psi(t)}\otimes\ket{t}$ (where the ``double ket'' notation is independent to the one used in this review), so that the time-dependent state can be recovered by conditioning on observing the clock at a given time: $\ket{\psi(t)} = \left(\id\otimes\bra{t}\right)\kket{\Psi}$, see \citep{Giovannetti2015}. This formalism naturally reproduces predictions for measurements at a given time as conditional measurements on the timeless state. However, in order to describe multiple, temporally-separated measurements on the same system, it requires an explicit modeling of the system-apparatus interaction. This means that different choices of operations are mapped to different timeless states. Therefore, unlike the process matrix, a single $\kket{\Psi}$ does not encode predictions for arbitrary operations, so, in this sense, it does not generalize ordinary quantum states to the time domain.
Nonetheless, the timeless formalism can model a broad range of processes with indefinite causal order, where the choice of operations is encoded in free parameters of the system-apparatus interaction. In particular, \citep{Castro-Ruiz2020} have shown how to represent the quantum switch in the timeless formalism, and \citep{Baumann2022} extend the construction to more general quantum control of causal order, see Sec.~\ref{causaluncertainty} above.

\subsubsection{Indefinite time arrow and quantum time flip} \label{subsec:time_flip}

\citet{Rubino2021_TimeArrows} and \citet{Chiribella2022TimeFlip} introduced the notion of quantum transformations with indefinite time direction, or indefinite input-output direction. Examples include quantum superpositions of thermodynamic time \citep{Rubino2021_TimeArrows} and the closely related ``quantum time flip'' \citep{Chiribella2022TimeFlip}. Hereafter, we focus on the latter, as it lends itself more naturally to analysis within the framework presented in this review and does not require the introduction of additional concepts from quantum thermodynamics.

The quantum time flip enables coherent control between an arbitrary unitary operator $U$ and its transpose\footnote{Processes that reverse a composition of such unitary operators can be expressed by anti-homomorphisms, that is, a function $f$ satisfying $f(UV)=f(V)f(U), \forall \, U,V$. Up to unitary equivalence, there are only two non-trivial such functions $f:\text{SU}(d)\to\text{SU}(d)$, unitary inversion $f(U)=U^{-1}$ and unitary transposition $f(U)=U^T$ \citep{fulton-harris,Chiribella2022TimeFlip}. Due to positivity constraints, unitary inversion cannot be realized even probabilistically when $d>2$ \citep{Quintino2019b}. As a result, \citet{Chiribella2022TimeFlip} adopt unitary transposition as the operational notion of time reversal.} $U^T$, by acting on unitary operators via
\begin{align}
    U\mapsto \ketbra{0}{0}_c\otimes  U + \ketbra{1}{1}_c\otimes  U^T.
\end{align}
As in the case of the quantum switch, control is implemented via a qubit system. Also, similarly to the switch, the quantum time flip serves as a resource for certain channel discrimination tasks. In particular, the quantum time flip allows one to determine (with probability one) if a pair of unitaries $(U,V)$ respect the property $UV^T=U^TV$ or $UV^T=-U^TV$ with a single call of each operation, a task that cannot be achieved with unit success probability even using general process matrices.

From a broader perspective, \citet{Chiribella2022TimeFlip} propose an extension of the process matrix formalism to accommodate processes with indefinite input-output direction. This framework parallels that of standard process matrices, except that the requirement for validity across all quantum instruments or channels (as in Subsec.~\ref{subsec:open_past_future}) is relaxed to validity over unitary channels only, and by linearity, to unital channels (also referred to as bi-stochastic quantum channels). Since the linear space spanned by unitary channels is strictly smaller than that spanned by arbitrary quantum channels, the resulting set of admissible processes---also characterized by affine and positive semidefinite constraints \citep{Chiribella2022TimeFlip,milz2023_transformations}---is strictly larger than that defined by standard process matrices with equal input and output dimension at each site (different dimensions are allowed in ordinary process matrices, but not for indefinite input-output direction). In particular, this extended framework permits processes that map a single use of an arbitrary unitary to its transpose, something unattainable within the standard process matrix formalism \citep{Quintino2019b,Quintino2021}. Moreover, this extended framework, as well as its classical analogue, allows for a maximal violation of causal inequalities such as the GYNI and the LGYNI \citep{Liu2024Tsirelson} (see Subsec.~\ref{subsec:causal_inequalities}).

Because the framework in Ref. \citep{Chiribella2022TimeFlip} assumes the input operation is unitary (or a linear combination thereof), a deterministic implementation of transformations like the quantum time flip is only possible in systems where the input is guaranteed to be unitary. As such, it cannot be realized under fully black-box assumptions. However, real experimental scenarios typically align more closely with so-called grey-box assumptions, where some knowledge about the input operations is available. Under such conditions, optical implementations of the quantum time flip have been demonstrated \citep{Guo2024TimeFlip,Stromberg2024TimeFlip}.
Specifically, \citet{Guo2024TimeFlip} developed a general method for witnessing input-output indefiniteness, derived witnesses with maximum robustness to noise and used the method to demonstrate the incompatibility of their experimental setup with a definite input-output direction. \citet{Stromberg2024TimeFlip}, on the other hand, certified the power of the quantum time flip by succeeding at the channel discrimination task described above with a probability very close to one.

\subsubsection{Routed quantum circuits} \label{sec:routed}
When analyzing interferometer diagrams, which are common in quantum-optical setups involving photons in superpositions of modes or trajectories, the standard quantum circuit can appear inconsistent or inappropriate. As discussed in Sec.~\ref{subsec:role_vacuum}, while it is not possible to design a quantum circuit that transforms an arbitrary unitary operation---applied as a black box on one of the circuit's wires---into its controlled version, this can be achieved by performing the unitary in an arm of a Mach-Zehnder interferometer \citep{Soeda2013controlisation, kitaev95, Araujo2014b, Friis2014, Thompson2018,dong2019control,Gavorova2020control}. This apparent mismatch suggests that the standard quantum circuit framework may not be the most suitable for faithfully describing quantum-optical or interferometric setups.

In order to address this issue, \citet{Vanrietvelde_2020} introduced the concept of routed linear maps and routed circuits, extending an earlier framework for the analysis of causal and compositional structure in unitary channels \citep{Lorenz2021}. A routed linear map is a linear map combined with a partition of its involved linear spaces into subspaces, allowing one to define its possible routes.  This extends the standard quantum circuit formalism, incorporating the control of unitary operations and the superposition of trajectories \citep{kristjansson2019}. Routed circuits have also proven useful to analyze causal structures and to describe the quantum switch \citep{Ormrod2023}.
In addition, routed circuits offer a means of creating consistent processes---not only quantum circuits with quantum control of causal order \citep{Grothus2025}, but also certain processes that violate causal inequalities---based on certain types of directed graphs. These graphs can be used to to provide a compositional circuit-style framework to analyse causal relations in the presence of `loops', offering an interpretation as to why processes with indefinite causal order do not lead to inconsistencies \citep{Vanrietvelde2022}.

\subsubsection{Causal boxes} 
The causal boxes formalism introduced by \citep{Portmann2017} provides a model in which quantum information-processing units can be composed in a modular way to build larger networks. Unlike the circuit model or the process-matrix formalism, where each box represents a single use of an operation on a quantum system, a causal box represents a device that can be used multiple times and can internally process sequences or superpositions of messages. This is modeled by attaching timing information to the systems on which the box acts and by requiring that the output of the box at any given time depends only on inputs at earlier times. This ensures that the formalism is closed under arbitrary wiring, including loops, and is therefore well suited for a time-local description of laboratory implementations of information-processing protocols, such as the quantum switch. \citet{Vilasini2022} showed that the most general experiments on finite-dimensional quantum systems that can be conducted by agents in a classical background spacetime are describable by causal boxes. A certain subclass of these, in which each party acts on a quantum message once and only once---thus respecting the ``closed laboratory'' assumption---was shown to be equivalent to quantum circuits with quantum control of causal order \citep{salzger2023,salzger2025, Salzger2026}.

\subsubsection{Quantum causal graph dynamics and addressable quantum gates}

\citet{Arrighi_2024} introduced a framework for discrete quantum systems connected through a network, where the network connectivity is treated dynamically, allowing for unitary evolutions that generate superpositions of distinct network configurations.  A motivation for this line of research is connecting research in indefinite causal order with candidate quantum gravity theories, such as loop quantum gravity \citep{Rovelli_2011} and causal dynamical triangulations \citep{Loll_2004}, which describe Feynman-path-like evolutions of graphs that encode quantum geometry.

A related, more circuit-oriented construction is provided by ``addressable quantum gates'', where the wiring between gates is encoded in internal registers that can be coherently manipulated \citep{Arrighi_2023}. This allows one to represent processes such as the quantum switch within an extended circuit description, as well as other transformations that do not admit a straightforward representation in the standard fixed-wiring circuit model.

\subsubsection{Indefinite causality in supraquantum theories} 

\citet{bavaresco24boxworld} constructed a higher-order theory for \textit{``boxworld''}, an archetypical type of generalized probabilistic theory that allows for maximal nonlocality \citep{Barrett2007}. To derive nontrivial bounds on the set of correlations that can be obtained in such a theory, the authors imposed two physically motivated principles in addition to the boxworld basic requirements: ``no-signaling preservation'' and ``no signaling without system exchange'' (NSWSE). The boxworld framework, combined with these two principles, allows for violations of causal inequalities to a degree that is not allowed by quantum process matrices. The authors then conjecture that the set of correlations of this proposed higher-order boxworld theory is an outer approximation to the set of correlations produced by higher-order quantum theory.

\citet{wilson2026} further demonstrated how the category-theoretical approach~\citep{Wilson2022,Wilson2022Polycategorical,James2024Profunctorial,Hefford2025BV} can be used to define higher-order processes in theories beyond finite-dimensional quantum mechanics. Additionally, they showed how a weakened NSWSE principle, arising from its restriction to a narrower class of Boxworld instruments, can be derived using the categorical notion of completeness, i.e., the property that higher-order operations can act locally on a subsystem. In a similar spirit, \citet{Wilson2022} showed that, under monoidal category assumptions, locally-applicable transformations on quantum channels are in one-to-one correspondence with deterministic quantum supermaps.

\citet{wilson2026} further demonstrated how the category-theoretical approach \citep{Wilson2022,Wilson2022Polycategorical,James2024Profunctorial,Hefford2025BV} can be used to define higher-order processes in theories beyond finite-dimensional quantum mechanics. Additionally, they showed how a weakened NSWSE principle, arising from its restriction to a narrower class of Boxworld instruments, can be derived using the categorical notion of completeness, i.e., the property that higher-order operations can act locally on a subsystem. In similar spirit, \cite{Wilson2022} shows that, under monoidal category assumptions, locally-applicable transformations on quantum channels are in one-to-one correspondence with deterministic quantum supermaps.

In a similar spirit to the approach of \citet{bavaresco24boxworld}, \citet{sengupta2024maximal} constructed another explicit higher-order boxworld-type theory, and explored the possible violations of DRF inequalities (recall Sec.~\ref{sec:vanderlugt}), showing that some of these can be violated up to their algebraic maximum.

\section{Outlook}
\label{sec:Concl}
The field of indefinite quantum causality is a relatively new and dynamic area of research, filled with both potential and open questions. The advancements in understanding and utilizing indefinite causal structures have opened up exciting opportunities for quantum information processing, computation, and our fundamental understanding of quantum mechanics. However, this growing field still faces numerous technical and conceptual challenges that necessitate further investigation.

Although significant results have been achieved in the classification of quantum processes, the very notion of ``indefinite causal order'' for process matrices still has to be fully clarified. As discussed in Sec.~\ref{sectionmultiseparability}, two definitions of processes with \textit{definite} causal order exist: quantum circuits with classical control of causal order (QC-CCs) are defined constructively, as those processes that can be realized by supplementing ordinary quantum circuits with classical (possibly probabilistic and dynamic) control of the order between operations. On the other hand, causally separable processes are defined abstractly, as those for which a definite causal order can be assigned recursively for each given choice of operation. Although the two definitions coincide for bipartite and tripartite processes, the question for the arbitrary multipartite case remain open. A further direction is identifying and exploring further meaningful classes of processes with indefinite causal order. Currently, the largest such class, defined constructively, is that of quantum circuits with quantum control of causal order (QC-QC) and it remains to be understood if more general processes can be given a direct physical interpretation within classical or quantum spacetime implementations.

A related challenge is to gain a deeper understanding of ``pure'' processes. The process matrix formalism is an extension of ``mixed'' quantum theory, based on density operators and quantum channels. However, quintessential features of quantum theory are best understood in terms of pure states and unitary evolution, and   ``going to the Church of the Larger Hilbert Space'' \citep{Gottesman2000} is a powerful methodology in quantum information. In particular, unitary processes, namely, processes that preserve reversibility of quantum operations, are only well understood for up to ``two-slot'' processes (with open past and future). As discussed in Sec.~\ref{subsec:unitary_processes}, all such unitary processes are either causally ordered or, essentially, equivalent to the quantum switch, which also implies that it is not possible to superpose two-slot processes with different causal order without the assistance of a control system. It is still unclear---and it is an important open question---whether similar results extend to more general unitary processes. 

A limitation of the current formulation of the process matrix formalism is that it only models a finite set of events. An extension to a continuous manifold of events, modeling a---possibly quantum---spacetime, presents both conceptual and technical challenges, as it would require to deal with effectively infinite tensor products. More formally, generalizations of tensor products would be needed, such as the type-III von Neumann algebras of operators in algebraic quantum field theory \citep{fewster2019algebraic}, but the mathematical properties and physical interpretation of such algebras would likely need to be modified and the notion of causal order, thus far relying on finite permutations of sites, would need to be revisited. Although some exploration of quantum fields with indefinite causal structure exists \citep{Jia2018a},
this still remains a largely unexplored topic. Recent promising developments in this direction include two distinct approaches to the continuum limit of causally ordered process matrices \citep{Costa2025, Wassner2026}.

On the experimental side, an open question is whether any of the currently explored platforms is potentially scalable, both in terms of dimensionality of the target and of number of switched gates. In order for an implementation to lead to an advantage, its resource cost should scale linearly with the number of gates (rather than quadratically, which is the cost of a quantum-circuit implementation) and logarithmically in the target's dimension---which is not the case for spatial modes or time bins, with the latter holding a record of encoding $12$ qubits \citep{wei2019experimental}. Resource scaling also relates to the more foundational question of whether switch experiments can be fairly interpreted as implementing each operation ``once'': within current platforms, the control system ``travels together'' the target through the operations, making which-order information potentially available locally. For a larger number of gates, this might lead to an increasing amount of fine-tuning to ensure the same operation is performed for each control state. It remains unclear whether these limitations can be overcome within laboratory settings. A further challenge is to understand whether a broader class of processes with indefinite causal order, beyond the family of quantum switches and quantum-controlled quantum circuits, can be accessed experimentally.

Another important open direction is the design of new experimental implementations that overcome limitations present in existing demonstrations. One such limitation concerns the requirement that each operation be applied exactly once to the target system, independently of the control state. If the operation differs depending on the control, the implementation no longer faithfully realizes the intended process. Although some platforms can already enforce this requirement, doing so often involves additional mechanisms that cause the resource cost to scale quadratically with the number of operations, as opposed to the linear scaling desired for many practical advantages. A more principled approach to address such limitations would be to ensure that the control and target systems are physically separate, so that only the target system enters and exits the laboratories. This avoids scenarios where the control system, traveling alongside the target, could influence the operations or provide which-order information.

{Finally, there are important conceptual and foundational questions at the heart of the study of indefinite causal structures. Despite several promising directions, it is not yet clear whether the proposed applications can lead to concrete advantages. It should be noted that indefinite causal order should not be understood as a new physical resource---any protocol would still be based on ordinary quantum mechanics as the background physical theory. Rather, an advantage can arise if the amount of resources in a given protocol scales according to the indefinite-order description rather than the causally ordered one, leading to a more efficient use of resources, or if there are applications for which performing individual operations in an indefinite order is an advantageous constraint as opposed to performing multiple operations in a definite order.

At a deeper level, the physical interpretation of processes with indefinite causal order has yet to be fully explored, together with a more general understanding of causal relations in quantum theory and its possible extensions, including quantum gravity and generalized probabilistic theories. In this respect, efforts should be made to discriminate different levels at which events can be described and causal relations established: while it can be useful to identify causal relations between operationally-defined events \textit{in} spacetime, applications to quantum-gravity scenarios should rather concern events \textit{of} spacetime---or of a suitable non-classical version of it. More generally, additional conditions or physical requirements should be seeked in order to associate a (possibly indefinite) causal structure to either (quantum) spacetime, a useful operational scenario, or an ad-hoc choice of temporally delocalized subsystems.  

\section*{Acknowledgements} We thank A.~Abbott, M.~Ara\'ujo, P.~Arrighi, \"A.~Baumeler, H.~Dourdent, F.~Giacomini, A.~Grinbaum, A.-C.~de la Hamette, D.~Jia, H.~Kristj\'ansson, R.~Kunjwal, N.~Miklin, S.~Milz, N.~M\'oller, 
N.~Paunkovi\'c, P.~Perinotti, J.~Romero, H.~Seabrook, T.~Short, T.~Str\"omberg, V.~Vilasini and M.~Wilson for valuable feedback on this manuscript.
F.C.~acknowledges support from the 
Knut and Alice Wallenberg Foundation through the Wallenberg Initiative 
for Network and Quantum Information (WINQ). G.R.~acknowledges financial support from the Royal Commission for the Exhibition of 1851 through a Research Fellowship, from the European Commission through Advanced Grant ERC-2020-ADG101021085 (FLQuant), and from EPSRC through Standard Proposal Grant EP/X016218/1 (Mono-Squeeze). C.B.~acknowledges financial support from the Agence Nationale de la Recherche (ANR) through the project ANR-22-CE47-0012. \v{C}.B.~acknowledges financial support from the Austrian Science Fund (FWF)
[10.55776/F71] and [10.55776/COE1] and from the John Templeton Foundation through the WOST grant. M.T.Q. is supported by the Agence Nationale de la Recherche (ANR) through the JCJC programme under grant number ANR-25-CE47-6396-01-HOQO-KS.
This work benefitted from network activities through the INAQT network, supported by the Engineering and Physical Sciences Research Council (grant number EP/W026910/1). This work is supported from COST Action CA23115: Relativistic Quantum Information, funded by COST (European Cooperation in Science and Technology). For the purpose of open access, the authors have applied a CC-BY license to any author-accepted manuscript version arising from this submission.

\vspace{2mm}
The authors have no conflicts of interest to disclose.


\renewcommand{\baselinestretch}{1.2}

\nocite{apsrev42Control} 
\bibliographystyle{apsrmp4-2}
\bibliography{Review_BIB.bib}


\appendix

\section{Temporally delocalized tensor product structures}
\label{app:subsystems}

Given a total Hilbert space $\H$, a tensor decomposition into factors $\H^A$, $\H^B$ is an isomorphism $\H\cong \H^A\otimes \H^B$, which can be expressed explicitly through a unitary
\begin{equation}
   U_{\textrm{TP}} : \H \rightarrow \H^A\otimes \H^B.
\end{equation}
As for most of this review, we limit ourselves to finite-dimensional systems for simplicity, although tensor product structures are well defined for infinite-dimensional systems too \citep{Liu2004}. 

To take a simple example, relevant for switch-like experiments, consider a situation where a target system is probed at one of two possible times, depending on the value of a control degree of freedom, see Fig.~\ref{Fig:controlled_time}.

\begin{figure}
\centering
\includegraphics[width=0.85\columnwidth]{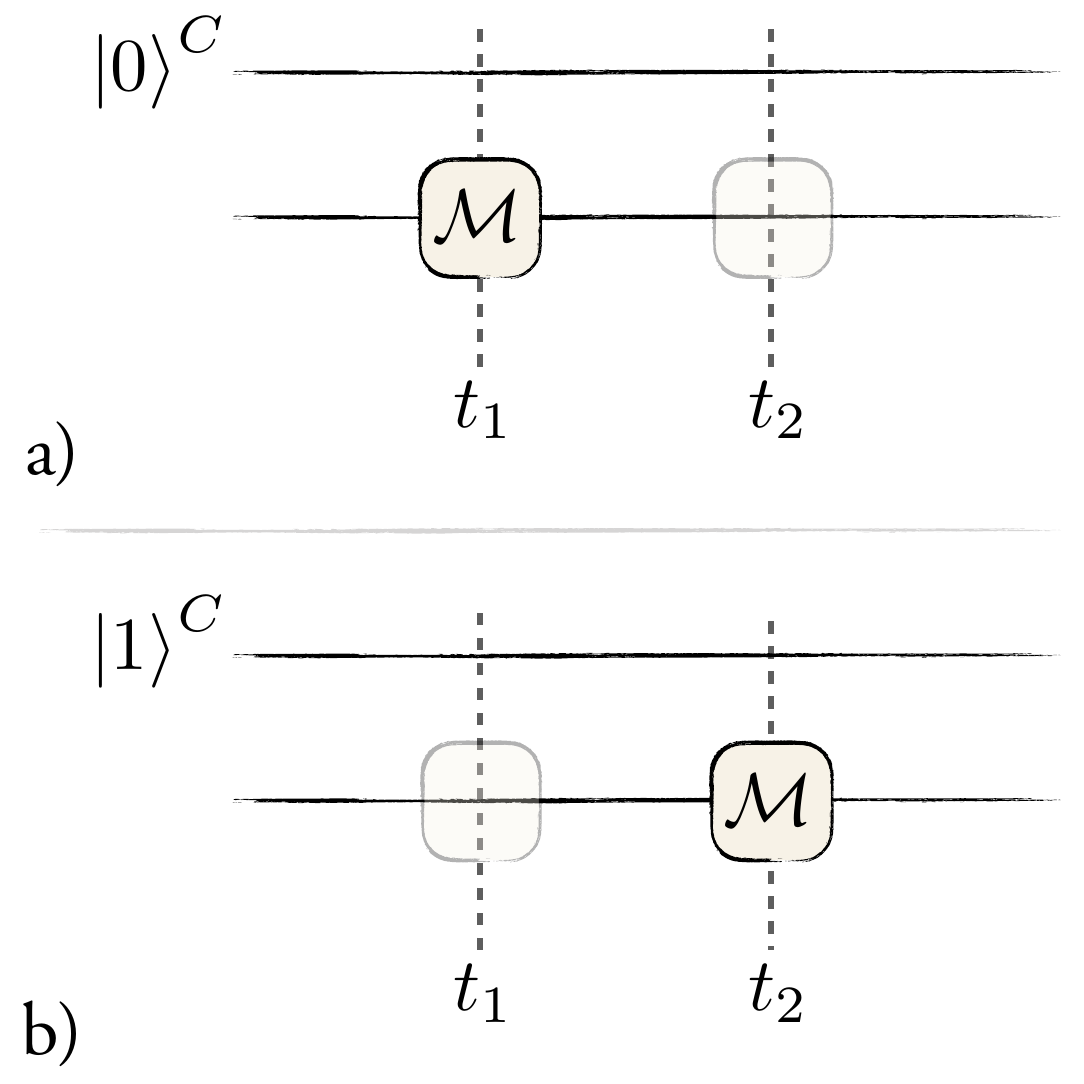}
\caption{\label{Fig:controlled_time} Operation with quantum-controlled timing. 
The time at which the operation $\M$ is applied on a target system depends on the state of the control: time $t_1$ for $\ket{0}$ and time $t_2$ for $\ket{1}$. In such a situation, it is possible to identify a subsystem decomposition such that $\M$ acts on a \textit{single} subsystem, delocalized between times $t_1$ and $t_2$.}
\label{img:TimeDeloc}
\end{figure}
In a time-local description, the operations takes place on the space $\H_{\textrm{loc}}=\H^{ C}\otimes \H^{A^1} \otimes \H^{A^2}$, where $\H^{ C}$ is a two-dimensional (input) control system and $\H^{A^{1,2}}=\H^{A^{1,2}_I}\otimes \H^{A^{1,2}_O}$ represent input and output of two isomorphic target spaces. 

We want to identify a single, temporally delocalized subsystem that is accessed by operations that act on $A^1$ when the control system is in state $\ket{0}$ and on $A^2$ when the control is $\ket{1}$.\footnote{To describe coherent control, we would need to extend the control system to an input-output pair. However, since the control is not used further, coherence is not relevant in this particular example, and a ``controlled-time'' operation is equivalent to one where the control is measured and, conditioned on the outcome, the operation is performed at the earlier or later time. This is consistent with the interpretation of ${C}$ as an input system.} This is natural, for example, for passive optical elements acting on a single-photon subspace, where the control system determines the time at which the photon passes through.  
By assumption, the delocalized subsystem on which the operation acts has to be isomorphic to the target space, and we denote it by $\H^A\cong \H^{A^{1,2}}$. Dimension counting tells us that the cofactor subsystem must be isomorphic to control times target space\footnote{In \citep{Oreshkov2019}, the cofactor was incorrectly characterized as isomorphic to the target system.}. The decomposition of the entire system into temporally delocalized ones is therefore
\begin{equation}
    \H_{\textrm{deloc}} = \H^{\tilde C}\otimes \H^A\otimes \H^B \cong \H_{\textrm{loc}},
\end{equation}
with $\H^{\tilde{C}}\cong \H^C$, $\H^B\cong \H^A\cong \H^{A^{1,2}}$. As we are about to see, even though the delocalized control system $\tilde C$ is isomorphic to the localized one, $C$, it is important to distinguish them.

We can convince ourselves, and later verify, that the unitary $U_{\textrm{TP}} : \H_{\textrm{loc}}  \rightarrow \H_{\textrm{deloc}}$ defining the isomorphism between localized and delocalized representations is
\begin{equation}
\label{UTP}
    U_{\textrm{TP}} = R\left( \proj{0}^{{C}}\otimes \id^{A^1A^2} + \proj{1}^{{C}} \otimes S^{A^1A^2}\right),
\end{equation}
where $S^{A^1A^2}$ is the swap operator between spaces $\H^{A^1}$ and $\H^{A^2}$, while $R:\H_{\textrm{loc}}\rightarrow \H_{\textrm{deloc}}$ identifies localized and delocalized Hilbert spaces through the relabeling (in chosen bases) $C\mapsto \tilde{C}$, $A^1\mapsto A$, $A^2\mapsto B$. 

To check that $U_{\textrm{TP}}$ gives the right transformation, let us consider an operation with Choi operator $M^A$. Its local representation should apply $M$ at time $t_1$, and the identity channel at time $t_2$, when the control $ C$ is in state $\ket{0}$. When  $  C$ is in state $\ket{1}$, the identity should apply at time $t_1$ and $M$ at time $t_2$. Therefore, we define the localized representation of the operation\footnote{Recall that, in our notation, the superscript only denotes the subsystem and it does not label different operators. Therefore, the expressions $M^{A^1}$ and $M^{A^2}$ denote \emph{the same} operator $M$, acting on two different (but isomorphic) subsystems $A_1$ and $A_2$, identified through their respective chosen bases.} 
\begin{multline}
\label{physicalM}
    M_{\textrm{loc}} = \proj{0}^{ C}\otimes M^{A^1}\otimes \kketbbra{\id}{\id}^{A^2}  \\
    + \proj{1}^{ C}\otimes \kketbbra{\id}{\id}^{A^1}\otimes M^{A^2},
\end{multline}
which can be easily checked to be CPTP: $\tr_{A^1_OA^2_O}M_{\textrm{loc}} = \id^{{C} A^1_IA^2_I}$.
The delocalized representation is then found by applying the isomorphism \eqref{UTP}, which, after a short calculation, gives
\begin{multline}
\label{explicitMlog}
    M_{\textrm{deloc}} \coloneqq U_{\textrm{TP}} M_{\textrm{loc}}U_{\textrm{TP}}^{\dag}
    =
    \id^{\tilde C}\otimes M^{A}\otimes \kketbbra{\id}{\id}^{B}.
\end{multline}
This confirms that an operation of the form \eqref{physicalM}, acting either at time $t_1$ or $t_2$ depending on the state of a control system, can be interpreted as an operation acting on a \textit{single} subsystem $A$. Note that, even though $M_{\textrm{loc}}$ involves the original control $ C$ in a non-trivial way, $M_{\textrm{deloc}}$ is insensitive to the delocalized control $\tilde C$.

We can also look at how the above tensor product transformation affects the process matrix. Given an arbitrary time-local process matrix $W_{\textrm{loc}}^{{C} A^1 A^2}$, compatible with the causal order ${C} \prec A^1 \prec A^2$, its temporally delocalized representation is given again through the isomorphism
\begin{equation}
    W_{\textrm{deloc}} \coloneqq U_{\textrm{TP}} W_{\textrm{loc}}U_{\textrm{TP}}^{\dag}.
\end{equation}
If we want to identify the process relevant for the subsystem $A$ alone, we have to take the reduced process over $B$ and $\tilde C$, which, as discussed in Sec.~\ref{conditionalprocesses}, generally depends on the CPTP maps performed in such systems. The scenario outlined above leads to operations of the type \eqref{explicitMlog}, where the identity channel is applied at $B$, while $\tilde C$ is traced out. Therefore, in this scenario, the reduced delocalized process matrix is
\begin{equation} \label{properreducedprocess}
    \bar{W}_{\textrm{deloc}}^{A} = \tr_{\tilde{C} B} \left[W_{\textrm{deloc}}^{\tilde{C} A B}\left(\id^{\tilde{C} A}\otimes\kketbbra{\id}{\id}^B\right)\right].
\end{equation}

As a concrete example, Let us take a process where the target system is initially in some state $\ket{\psi}$ and it evolves through the identity channel between $A^1$ and $A^2$,  without interacting with the control state, $\ket{\xi} = c_0 \ket{0} + c_1 \ket{1}$. To write this as a pure process, we add a final target subsystem $F$, with identity evolution from $A^2$ to $F$. The temporally localized process matrix is then  $W_{\textrm{loc}}^{CA^1A^2F}= \ketbra{w_{\textrm{loc}}}{w_{\textrm{loc}}}^{CA^1A^2F}$, with
\begin{equation} \label{purephysical}
    \ket{w_{\textrm{loc}}}^{CA^1A^2F}\coloneqq  \ket{\xi}^C\ket{\psi}^{A^1_{I}}\kket{\id}^{A^1_OA^2_I}\kket{\id}^{A^2_O F}.
\end{equation}
Applying the same isomorphism as before (extended trivially to the final subsystem $F$), we find the delocalized representation of the process vector:
\begin{multline}  \label{purelogical}  
   \ket{w_{\textrm{deloc}}} \coloneqq U_{\textrm{TP}} \otimes \id^F
   \ket{w_{\textrm{loc}}} = c_0\ket{0}^{\Tilde{C}}\ket{\psi}^{A_{I}}\kket{\id}^{A_O B_I} \kket{\id}^{B_O F} \\
    + c_1\ket{1}^{\Tilde{C}}\ket{\psi}^{B_{I}}\kket{\id}^{B_O A_I} \kket{\id}^{A_O F}.
\end{multline}
Taking now the reduced process as in Eq.~\eqref{properreducedprocess}, we obtain
\begin{equation}
    \bar{W}_{\textrm{deloc}}^{A} = \ketbra{\psi}{\psi}^{A_I}\otimes \kketbbra{\id}{\id}^{A_O F}.
\end{equation}
This is precisely what we should expect: in the temporal description, we are acting on the system at either time $t_1$ or time $t_2$ but, since the system evolves trivially between the two times, this is effectively the same as acting only once on a single system $A$, which is prepared in state $\ket{\psi}$ and, after the operation, is transferred to the next site $F$. This construction highlights the fact that, whenever we model a finite-time operation as a single ``slot'' in a process matrix, we are implicitly reducing an underlying multi-time (in the limit, continuous-time) description to a selection of relevant time-delocalized subsystems. In other words, time-delocalized subsystems are necessary to make sense of causally ordered processes as much as those with indefinite causal order.

Another observation is that, in the time-delocalized tensor decomposition, the full process, Eq.~\eqref{purelogical}, coincides with that of a quantum switch, with past control and target fixed to $\ket{\xi}$ and $\ket{\psi}$, respectively.
A more common description of the switch, found in the literature, typically involves four temporally-localized subsystems, two for Alice and two for Bob. Using a similar procedure to that outlined above, the class of operations available in the protocol identifies two temporally delocalized subsystems, one for Alice and one for Bob, such that the reduced process on those subsystems gives again the quantum switch. Clearly, even more ``fine-grained'' descriptions are possible, with multiple time-local sites for each party,  but as long as the overall process behaves like the switch, the effective description on the relevant delocalized subsystems will always coincide with the two-site process matrix description.
 
\end{document}